\documentclass{tufte-handout}

\title[SIQP I: Foundations]{\emph{\texttt{\textcolor{myred}{\LARGE{The Structure and
          Interpretation of Quantum Programs I:\hspace{7pt}\emph{Foundations}}}}}}

\author[David Wakeham]{\href{mailto:david@torsor.io}{David Wakeham}
 $\diamond$ \href{torsor.io}{Torsor Labs}
}

\date{}

\usepackage{graphicx} 
  \setkeys{Gin}{width=\linewidth,totalheight=\textheight,keepaspectratio}
  \graphicspath{{graphics/}} 
\usepackage{amsmath, amssymb}  
\usepackage{units}    
\usepackage{multicol} 
\usepackage{enumitem}   
\usepackage{hyperref}
\usepackage{mathtools}
\setcitestyle{numbers}
\usepackage{cjhebrew}

\DeclareMathOperator*{\argmax}{arg\,max}
\DeclareMathOperator*{\argmin}{arg\,min}

\newcommand{\rrangle}{\rangle\hspace{-3pt}\rangle}
\newcommand{\llangle}{\langle\hspace{-3pt}\langle}

\usepackage{titlesec}
\titleformat{\section}
{\normalfont\LARGE}{\thesection}
{1em}{\itshape}

\usepackage{glossaries}
\usepackage{tikz-cd}

\newcommand{\tocgroup}[1]{%
  \addtocontents{toc}{\protect\vspace{10pt}}%
  \addtocontents{toc}{\protect\noindent\protect\textsc{\protect\Large{#1}}\protect\par}%
}

\makenoidxglossaries

\newglossaryentry{cstar}{
  name={C${}^*$-algebra},
  description={A complex associative algebra with an involution $\ast$ and
    a norm satisfying the C${}^*$ identity $\|A^*A\| = \|A\|^2$. Provides
    the algebraic foundation for quantum observables}
}

\newglossaryentry{staralg}{
  name={$\ast$-algebra},
  description={A complex associative algebra with an antilinear automorphism $\ast$}
}

\newglossaryentry{GNS}{
  name={GNS},
  description={Construction due to Gelfand, Naimark and Segal which
    takes a state $\pi$ on a C${}^*$-algebra and produces a Hilbert space
    $H_\pi$ where the algebra acts as bounded operators}
}

\newglossaryentry{univ}{
  name={universal representation},
  description={An isometric isomorphism that faithfully embeds a C${}^*$-algebra into bounded operators on a Hilbert space, preserving both algebraic and metric structure}
}

\newglossaryentry{prim}{
  name={primitive spectrum},
  description={The set of pure states of a C${}^*$-algebra up to unitary equivalence, denoted $\text{Prim}(A) = \partial S(A)/U(A)$}
}

\newglossaryentry{jordan}{
  name={Jordan product},
  description={The symmetric product $\Gamma \circ \Lambda = \frac{1}{2}(\Gamma\Lambda + \Lambda\Gamma)$ for operators, which ensures self-adjointness is preserved}
}

\newglossaryentry{spectral}{
  name={spectral theorem},
  description={Fundamental result stating that bounded self-adjoint operators can be decomposed as $\Lambda = \sum_{\lambda \in \mathfrak{S}(\Lambda)} \lambda \Pi_\lambda$ where $\Pi_\lambda$ are projectors}
}

\newglossaryentry{quotient}{
  name={quotient space},
  description={The space $\mathcal{A}/\mathcal{K}$ formed by identifying elements that differ by operators in the $\mathcal{K}$, with equivalence classes $[A] = A + \mathcal{K}$}
}

\newglossaryentry{completion}{
  name={completion},
  description={The process of filling in ``holes'' in a metric space to make it complete, where every Cauchy sequence converges. Denoted $\widehat{M}$ for metric space $M$}
}

\newglossaryentry{krein-milman}{
  name={Krein-Milman theorem},
  description={States that every convex set is the convex hull of its extreme points: $\text{conv}(K) = \text{conv}(\partial K)$}
}

\newglossaryentry{awd}{
  name={awd},
  description={Acronym for Abstract Wiring Diagram, diagrammatic notation where boxes represent operators, series multiplication corresponds to composition, and parallel corresponds to addition}
}

\newglossaryentry{definite}{
  name={definite set},
  description={The set $\mathcal{D}_\pi = \mathbb{R}I + (\mathcal{K}_\pi \cap \mathcal{K}_\pi^*)$ of self-adjoint operators with zero variance under state $\pi$, representing deterministic measurements}
}

\newglossaryentry{sharp}{
  name={sharp},
  description={An operator $\Gamma$ with zero variance under a state, i.e., $\|\Gamma\|_\pi^2 = \pi(|\Gamma|^2) - |\pi(\Gamma)|^2 = 0$, allowing factorization of expectations}
}

\newglossaryentry{kernel}{
  name={kernel},
  description={The set $\mathcal{K}_\pi = \{\Theta \in A : G_\pi(\Theta, \Theta) = 0\}$ of operators with vanishing self-correlation under state $\pi$}
}

\newglossaryentry{pvm}{
  name={PVM},
  description={Acryonym for Projection-Valued Measure, a measurement described by projectors $\Pi_\lambda$
    that are orthogonal ($\Pi_\lambda \Pi_\mu = \delta_{\mu\lambda}
    \Pi_{\hat{\lambda}}$) and resolve the identity ($\sum_\lambda \Pi_\lambda = I$)}
}

\newglossaryentry{born}{
  name={Born},
  description={Rule for the probability of measuring outcome $\lambda$ in a state $\pi$, $p_\lambda =
    \pi(\Pi_\lambda)$}
}

\newglossaryentry{luders}{
  name={Lüders},
  description={Rule for the post-measurement state: if outcome $\lambda$ is observed, the state becomes $\pi'_\lambda = p_\lambda^{-1} C_{\Pi_\lambda}[\pi]$}
}

\newglossaryentry{channel}{
  name={channel},
  description={A completely positive, trace-preserving map between
    states, typically expressed as $\pi'(A) = \pi(E_k^* A E_k)$ for
    Kraus operators $E_k$}
}

\newglossaryentry{kraus}{
  name={Kraus operators},
  description={Operators $E_k$ that define a quantum channel through $\pi'(A) = \sum_k \pi(E_k^* A E_k)$, with normalization $\sum_k E_k^* E_k = I$}
}

\newglossaryentry{pure}{
  name={pure},
  description={An extreme point of the convex set of states, characterized by having a maximal definite set that cannot be further extended}
}

\newglossaryentry{mixed}{
  name={mixed},
  description={A convex combination of pure states, $\pi = p_j \pi_j$ with $p_j \geq 0$ and $\sum_j p_j = 1$}
}

\newglossaryentry{unitary-equivalence}{
  name={Unitary equivalence},
  description={States $\pi$ and $\pi'$ are unitarily equivalent if $\pi' = C_U[\pi]$ for some unitary $U$, meaning $\pi'(A) = \pi(U^* A U)$}
}

\newglossaryentry{vector-state}{
  name={Vector state},
  description={A state of the form $\pi(A) = \langle\psi|A|\psi\rangle$ for some unit vector $|\psi\rangle$ in a Hilbert space}
}

\newglossaryentry{density-matrix}{
  name={Density matrix},
  description={An operator $\varrho = \sum_a p_a |a\rangle\langle a|$ representing a mixed state, where $\pi(A) = \text{tr}[\varrho A]$}
}

\newglossaryentry{convex-hull}{
  name={Convex hull/extreme points},
  description={The convex hull $\text{conv}(K)$ is the set of all convex combinations of elements in $K$. Extreme points $\partial K$ cannot be written as non-trivial convex combinations}
}

\newglossaryentry{switch-set}{
  name={Switch set},
  description={A minimal generating set $Q$ for the definite set $D_\pi$ under real linear combinations and Jordan products: $D_\pi = \langle Q | \mathbb{R} \rangle$}
}

\newglossaryentry{rigidity}{
  name={Rigidity (of definite sets)},
  description={Property that definite sets $D_\pi$ are rigid for pure states, meaning the restriction $\pi|_{D_\pi}$ uniquely determines the full state $\pi$}
}

\newglossaryentry{pauli-exponential}{
  name={Pauli exponential},
  description={Unitary operators of the form $e^{i\delta I + i\theta\sigma(\mathbf{n})} = e^{i\delta}[I\cos\theta + i\sin\theta\,\sigma(\mathbf{n})]$ for unit vector $\mathbf{n}$}
}

\newglossaryentry{bloch-sphere}{
  name={Bloch sphere/ball},
  description={The Bloch sphere parametrizes pure states of a qubit, while the Bloch ball (its convex hull) represents all states of the Pauli algebra}
}

\newglossaryentry{operator-sum}{
  name={Operator sum decomposition},
  description={Expression of a quantum channel as $\mathcal{E}(A) = \sum_k E_k^* A E_k$ where $E_k$ are Kraus operators}
}

\newglossaryentry{observable}{
  name={observable},
  description={A self-adjoint operator representing a measurable physical quantity, with real eigenvalues corresponding to possible measurement outcomes}
}

\newglossaryentry{operator}{
  name={Operator},
  description={An element of a C${}^*$-algebra, representing a transformation or physical operation that can be applied to quantum states}
}

\newglossaryentry{self-adjoint}{
  name={Self-adjoint (operator)},
  description={An operator $A$ satisfying $A^* = A$, ensuring real eigenvalues and corresponding to observable physical quantities}
}

\newglossaryentry{unitary}{
  name={Unitary (operator)},
  description={An operator $U$ satisfying $U^*U = UU^* = I$, representing reversible quantum evolution that preserves state norms}
}

\newglossaryentry{state-quantum}{
  name={State (quantum)},
  description={A positive linear functional $\pi: A \to \mathbb{C}$ of unit norm, representing the expectation values of observables in a quantum system}
}

\newglossaryentry{functional}{
  name={Functional},
  description={A map from a vector space to its underlying field of scalars, such as a state $\pi: A \to \mathbb{C}$ on a C${}^*$-algebra}
}

\newglossaryentry{expectation-value}{
  name={Expectation value},
  description={The average value $\pi(A) = \mathbb{E}[A|\pi]$ of observable $A$ in state $\pi$, computed as a weighted sum over measurement outcomes}
}

\newglossaryentry{hilbert-space}{
  name={Hilbert space},
  description={A complete inner product space, recovered from states via the GNS construction as $H_\pi = \widehat{A/K_\pi}$}
}

\newglossaryentry{valuation}{
  name={Valuation (Boolean)},
  description={A map $v: \mathcal{B}[P] \to \mathbb{B}$ from Boolean circuits to truth values, satisfying linearity and multiplicativity}
}

\newglossaryentry{representation}{
  name={Representation (of an algebra)},
  description={A *-homomorphism $\phi: A \to B(H)$ mapping a C${}^*$-algebra to bounded operators on a Hilbert space}
}

\newglossaryentry{operator-norm}{
  name={Operator norm},
  description={The norm $\|A\|_{\text{op}} = \sup\{\|Av\| : \|v\| \leq 1\}$ measuring the maximum stretching factor of an operator}
}

\usepackage{bbm}
\usepackage{euler}

\definecolor{timelifeblue}{RGB}{51, 102, 204}
\definecolor{myred}{RGB}{193, 45, 45}
\definecolor{mylinks}{RGB}{20, 20, 20}

\usepackage{listings}
\definecolor{codegreen}{rgb}{0,0.4,0.3}
\definecolor{codegray}{rgb}{0.5,0.5,0.5}
\definecolor{codepurple}{rgb}{0.58,0,0.82}
\definecolor{backcolour}{rgb}{0.95,0.95,0.92}

\usepackage{plex-mono}

\newcommand{\measure}{\leftarrow}

\begin{document}

\maketitle
\vspace{-15pt}

\par\noindent\rule{0.95\textwidth}{0.4pt}
\vspace{10pt}

\begin{center}
  \textsc{abstract}
\end{center}

\begin{abstract}
  \noindent 
  Qubits are a great way to build a quantum computer, but a
  limited way to program one. We
  replace the usual “states and gates” formalism with a ``props and
  ops'' (propositions and operators)
  model in which
  \begin{itemize}[itemsep=-2pt]
  \item the C${}^*$-algebra of observables supplies the syntax;
  \item states, viewed as linear functionals, give the semantics; and
  \item a novel diagrammatic calculus unifies the two.
  \end{itemize}
The first part develops the basic objects of the framework, encoding
consistent patterns of operator correlation, recovering Hilbert space
via the GNS construction, and re-deriving the Bloch sphere as the set
of all consistent correlations of operators in the Pauli algebra.

We then turn to intervention, showing how measurement modifies state, proving an operator-algebraic version of the Knill-Laflamme conditions, and expressing stabilizer codes with the same diagrammatic machinery. This provides a concise, representation-agnostic account of quantum error correction. The result is a self-contained foundation in which C${}^*$-algebras, and their dual Hilbert spaces, offer a rich and universal substrate for quantum programming; forthcoming papers will build a high-level language and quantum software applications on top of this substrate.
\end{abstract}

\vfill 
\hspace{-25pt} \includegraphics[width=0.25\columnwidth]{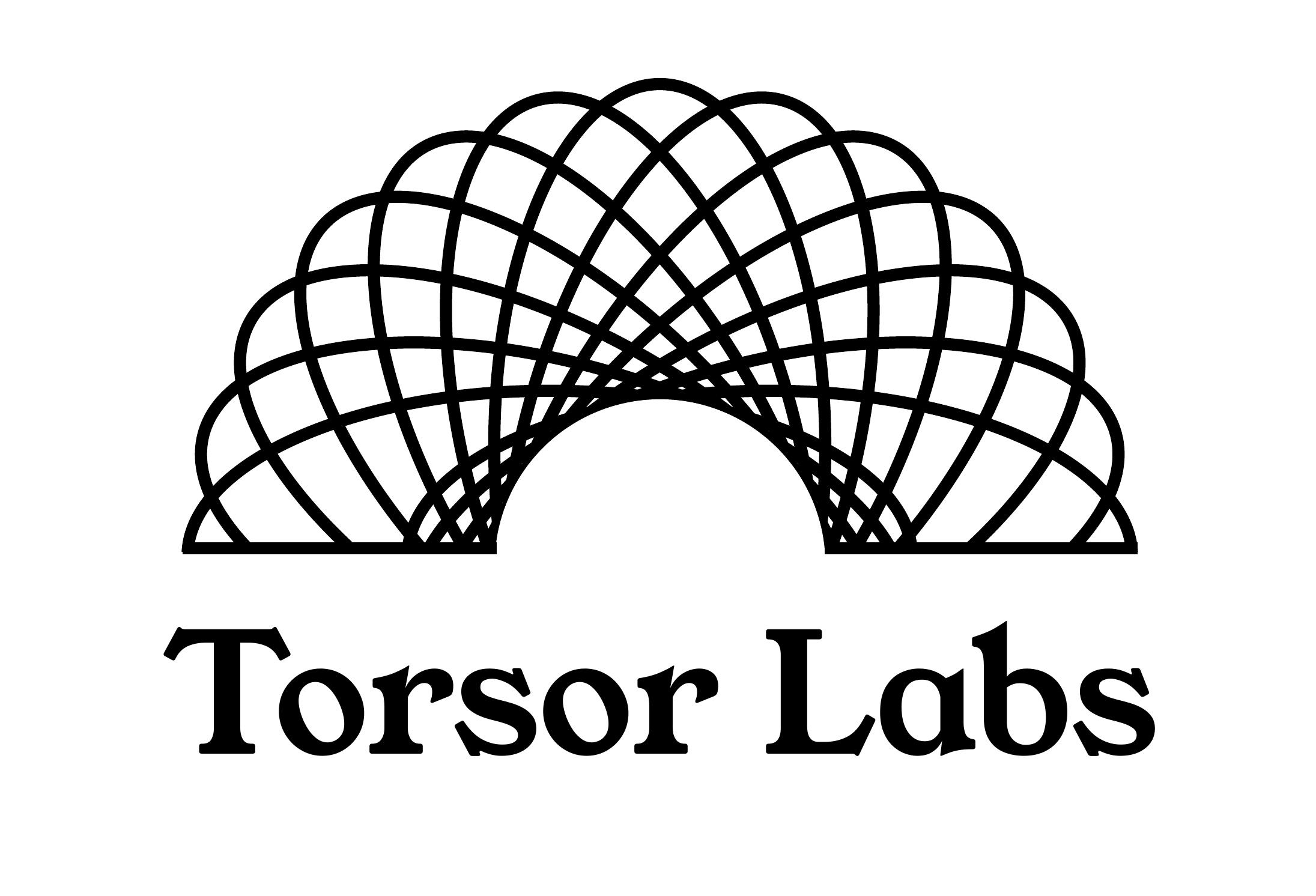}\\
\vspace{-10pt}
\noindent\hspace{0pt}\textsf{YAW}-\<b>${}_\texttt{0}$-9-25

\newpage

\section*{Reading Guide}


\textsc{What is this document?} The first installment of \emph{Structure and
Interpretation of Quantum Programming (SIQP)}, a systematic reconstruction of
quantum computing from algebraic foundations. Like its spiritual
predecessor \emph{SICP},\sidenote{\emph{Structure and Interpretation of Computer
    Programs} (1985), Harold Abelson, Gerald Jay Sussman and Julie
  Sussman.} we seek not merely to explain quantum programming, but to
understand it more deeply via the interplay of
computational pragmatics, design principles, and logical structure.

\vspace{5pt}
\noindent \textsc{Why reinvent the wheel?} The standard formulation of quantum computing—based on Hilbert space,
qubits, and circuits—plays the same role in quantum computing that truth
tables do in classical. This is an important part of the story, but by
no means all! By starting with C${}^*$-algebras and observables as our
primitive, we not only recover the standard treatment, but enable new
and flexible patterns of abstraction that stand a better chance of scaling with hardware.

\vspace{5pt}
\noindent \textsc{How do you do that?} Algebras are more flexible
than qubits. You can specify qudits, groups, harmonic oscillators, or
open systems with the same ease you work with one and zeros. You can
naturally incorporate error correction via the stabilizer formalism,
and use the same tools for simulation via Gottesman-Knill. You can
apply all this to near-term (e.g., classical shadows), medium-term
(e.g., Hadamard tests) and long-term applications (e.g. hidden subgroup problem).
\marginnote{
   \vspace{173pt}
  \begin{center}
    \hspace{20pt}\includegraphics[width=0.55\linewidth]{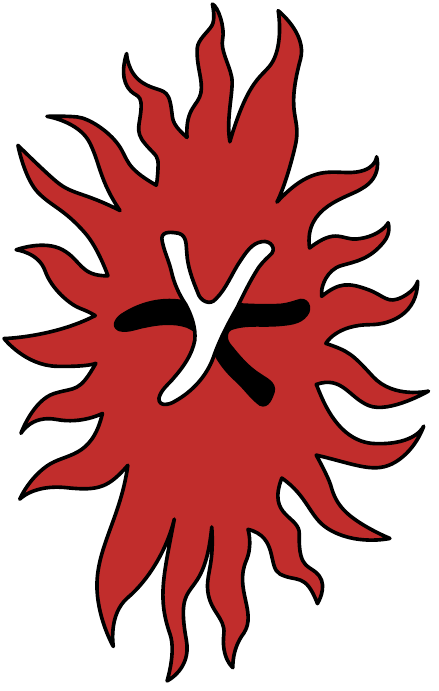}
  \end{center}  \vspace{-5pt}
  \emph{
  } \vspace{10pt}
}

\vspace{5pt}
\noindent \textsc{Who should read this?} We envision a few different segments:

\begin{itemize}[itemsep=-2pt]
\item \emph{For the traditionalist}: Start with the
  Bloch sphere (§8) to see the familiar from a new angle, then work backward to understand how it
  emerges algebraically. Who knows, you might like it!
\item \emph{For the information theorist}: Focus on the GNS
  construction (§5-7), sharpness (§10-12) and error
  correction (§13-16). Together, these tell a nice story about the
  algebraic dynamics of information.
\item   \emph{For the computer scientist}: The abstract wiring diagrams (§3) are
  the combinators of our upcoming quantum programming language. Get
  familiar with them and await Part II! 
\item   \emph{For the neophyte}: Focus on foundations (§2-7) and
  reward yourself with a deeper understanding of the Bloch sphere (§8).
  \item   \emph{For the iconoclast}: Read only marginalia and infer
    the rest.
\end{itemize}


\vspace{0pt}
\noindent \textsc{Where can I learn more?} For those eager to see the programming framework in action, visit
\href{https://torsor.io/\#community}{\texttt{torsor.io/\#community}}
to see some basic quantum algorithms implemented in pseudocode.

\vspace{5pt}
\noindent \textsc{When will the real thing be ready?} Hopefully Winter
2025! Visit \href{torsor.io}{\texttt{torsor.io}} to stay up to date.

\newpage

\tableofcontents

\thispagestyle{empty}

 


\newpage


\section{Introduction}\hypertarget{sec:summary}{}

\vspace{-6pt}

\noindent The digital age was born in a 1937 master's
  thesis by Claude Elwood Shannon,\sidenote{\emph{A Symbolic Analysis
    of Relay and Switching Circuits.}} in which he showed that the
physical operations of placing 
switches in parallel and series could
be mapped onto the Boolean operations of \texttt{AND} and
\texttt{OR}. This makes the physics of circuits
isomorphic to Boolean algebra,
with ``conducting'' corresponding to $1$ and ``non-conducting''
corresponding to $0$. 
Shannon was
interested in circuit design; 
conversely,
each time we fiddle with switches and zap them with electricity, we
are doing an experiment.
Algebraic manipulation 
lives in the
realm of \emph{proof}, called ``syntax'' by logicians.
Physical
manipulation 
belongs to the realm of \emph{truth}, also called ``semantics''. They
are two halves of a single logical coin.

Quantum computing was born in the 1981 \emph{Physics and Computation} keynote address by Richard
Feynman.\sidenote{``Simulating physics with computers'' (1982).
  Yuri Manin book, \emph{Computable and
    Uncomputable} (1980), slightly predates Feynman and contains
  similar ideas. \vspace{5pt}} There are parallels
to Shannon's work, but Feynman was motivated by the distinct
philosophy of \emph{reversible computing},
initiated by Landauer, Bennett, Fredkin and Toffoli, among
others.\sidenote{See for instance ``Irreversibility and Heat Generation
in the Computing Process'' (1961), Rolf Landauer; ``Logical Reversibility of Computation''
  (1973), Charles Bennett; ``Conservative Logic'' (1982), Edward
  Fredkin and Tomaso Toffoli. \vspace{10pt}}
They, in turn, were inspired by physics, noting that Nature was not
merely lazy but frugal,
a habit that perhaps could be adapted to computing.\marginnote{
  \vspace{-0pt}
  \begin{center}
    \hspace{-10pt}\includegraphics[width=0.9\linewidth]{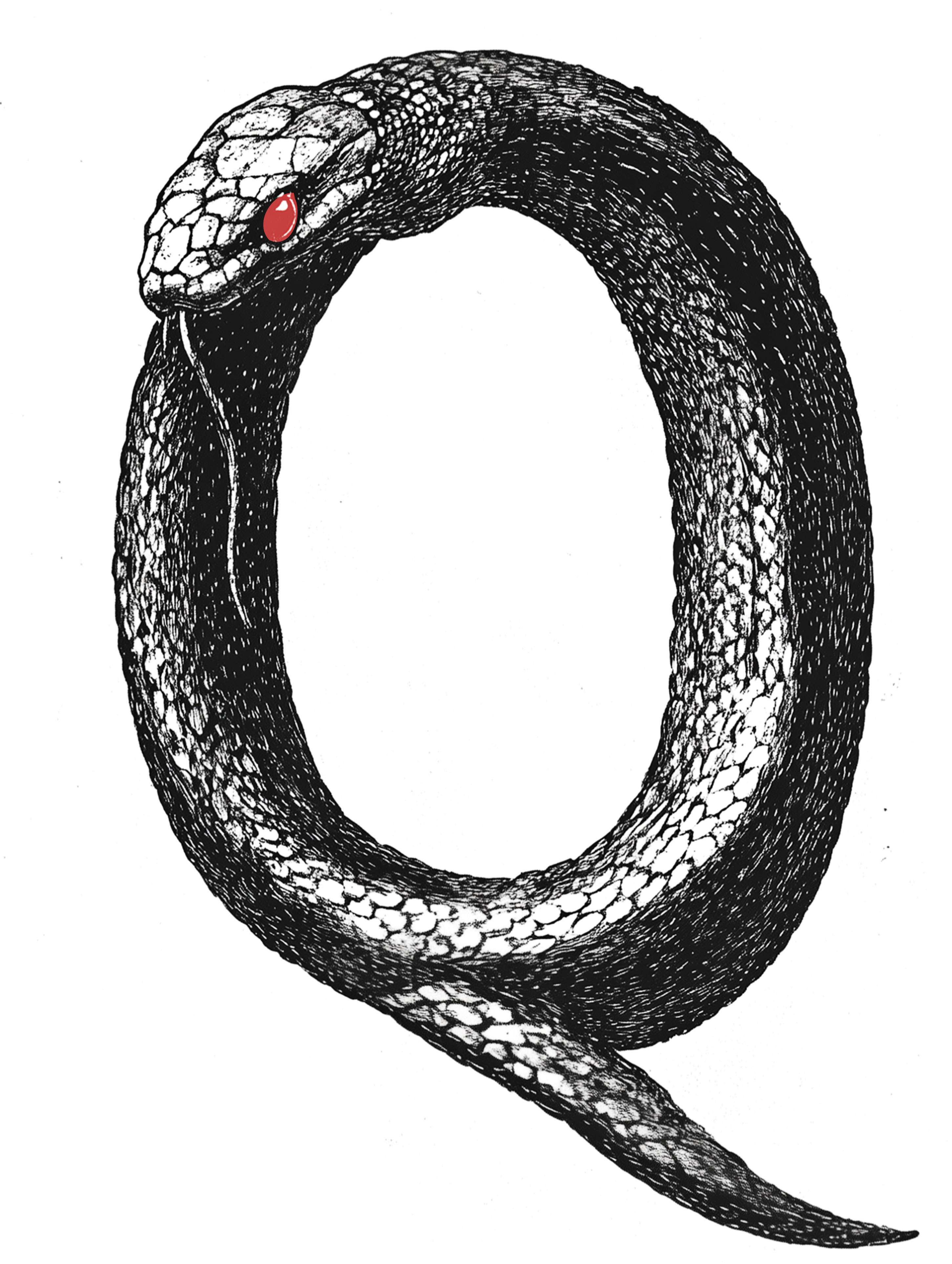}
  \end{center}  \vspace{-0pt}
  \emph{The quantum uroburos:
    great band name, limited computational formalism.
  } \vspace{10pt}
}
Feynman hung out with Fredkin long enough for some of these ideas to rub off,
and his 1981 keynote---delivered to a room full of reversibility
enthusiasts---clearly framed quantum as a generalization of
reversible computing.

In the same way classical information is built up from binary
digits (bits), Feynman proposed that quantum state should be built up
from a superposition of bits:
\begin{equation}
  |\psi\rangle = \alpha |0\rangle+\beta |1\rangle \in
  \mathbb{C}^{2}, \quad |\alpha|^2+|\beta|^2 = 1,\label{eq:qubit}
\end{equation}
better known as the \emph{quantum bit (qubit)}.
If we want to preserve
information, we should stick to \emph{unitary} operations 
until performing a final measurement.
This conventional approach 
is both inspired
and realized by quantum physics. This is partly a serendipitous fit
that required Feynman's genius and interdisciplinary synergy to realize.

At the same time, it is a snake eating its own tail.
We have been thinking for $45$ years in
terms of states, qubits, and unitary gates.
This may be a good analogy to classical reversible computing, but it
is only half of Shannon's two-part harmony; we can run the
electricity and do the experiments, but there is no algebra to fool
around with.
Hilbert space is great way to do models, or in logical terms,
\emph{semantics}, but a poor way to proofs in the same way that proof
by truth table generically involves writing an exponentially long list.

This monograph proposes that, in the context of quantum computing,
C${}^*$-algebras should play the syntactic role of Boolean algebras, and
Hilbert space the semantic dual.
This provides a structure and interpretation for quantum programs.
The paper is 
essentially linear: 

\begin{figure}[h]
  \centering
  \vspace{-5pt}
  \includegraphics[width=0.8\textwidth]{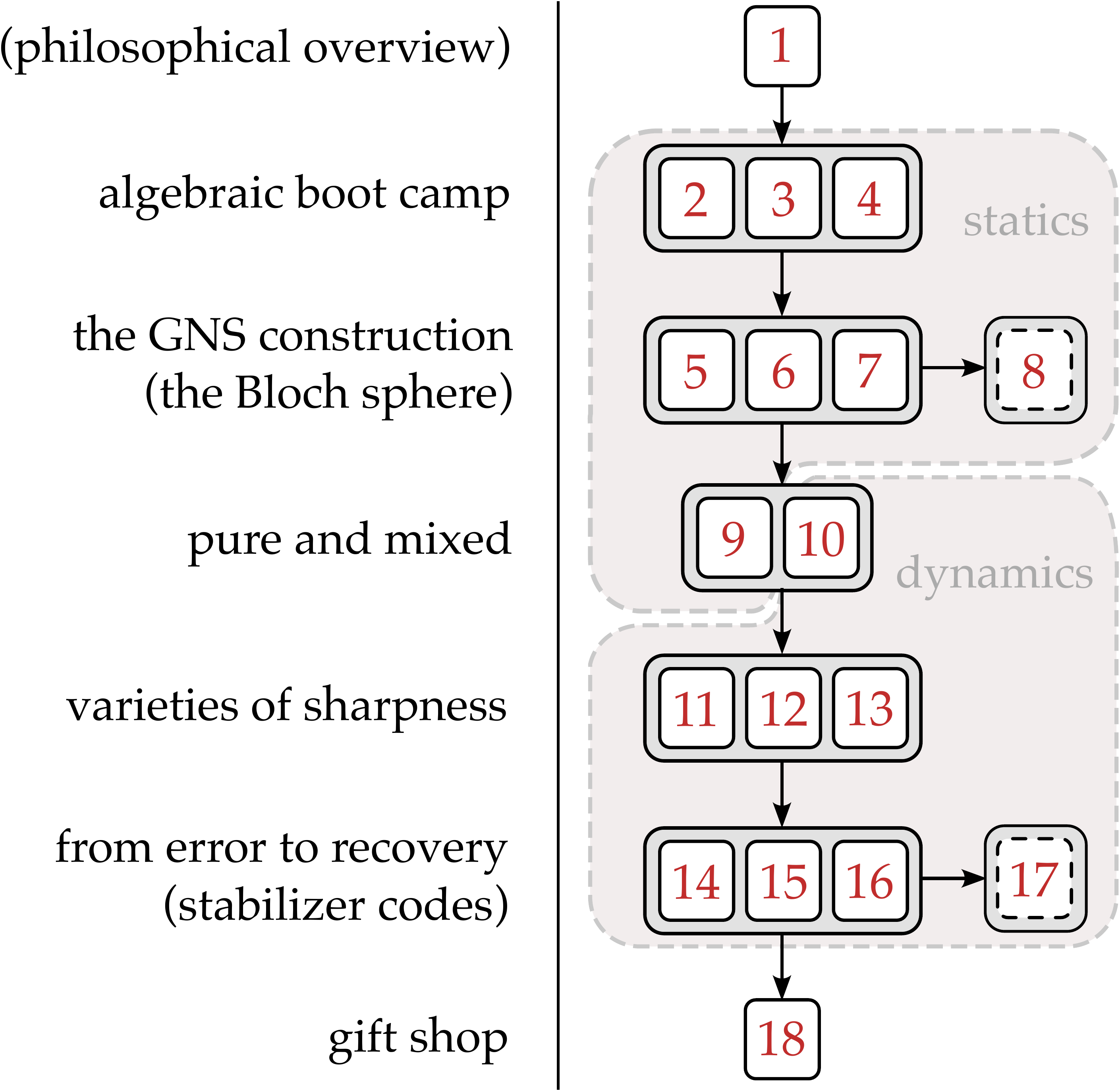}
  \caption{}
  \label{fig: flowchart}
  \vspace{-5pt}
\end{figure}

\noindent 
\begin{itemize}[itemsep=-2pt]
\item \emph{Algebraic boot camp.} Defines C${}^*$-algebras (\S
  \hyperlink{sec:3}{2}), translates algebraic expressions into
  diagrams (\S \hyperlink{sec:4}{3}), and introduces states as a way
  of encoding correlations (\S \hyperlink{sec:5}{4}).
\item \emph{The GNS construction.} Explores the correlation structure
  (\S \hyperlink{sec:6}{5}), bootstraps an associated Hilbert space (\S \hyperlink{sec:7}{6}),
  and identifies the familiar unit norm vectors (\S
  \hyperlink{sec:8}{7}). We illustrate with the Bloch sphere (\S
  \hyperlink{sec:9}{8}).
\item \emph{Pure and mixed.} Mixed states are constructed as convex
  combinations (\S \hyperlink{sec:10}{9}) and pure states as maximally
  sharp states (\S \hyperlink{sec:11}{10}).
\item \emph{Varieties of sharpness.} We introduce tensor products
  and entanglement (\S \hyperlink{sec:12}{11}), commutative
  subalgebras (\S \hyperlink{sec:13}{12}), and use this to give a
  Gleason-style ``derivation'' of Born and Lüders rules (\S \hyperlink{sec:14}{13}).
  \item \emph{From error to recovery.} Introduces quantum operations and
    channels (\S \hyperlink{sec:15}{14}), the Knill-Laflamme
    conditions for algebraic error correction (\S
    \hyperlink{sec:16}{15}), and a general construction of error
    correcting codes (\S \hyperlink{sec:17}{16}). We concretely
    demonstrate with the five-qubit code (\S \hyperlink{sec:18}{17}).
  \item \emph{Gift shop.} Finally, \S
\hyperlink{sec:19}{18} is a standalone showcase that can be consulted
for motivation, before, after, or \emph{in medias res}.
  \end{itemize}
\noindent We can loosely split the material into ``statics''
(what stuff is) and ``dynamics'' (what stuff does).

A note on prerequisites. We assume no exposure to functional
analysis, but we do take readers to have first (and maybe second)
course on quantum computing. Chapters 1, 2 and perhaps 10 of Mike and Ike\sidenote{\emph{Quantum Computation and
    Quantum Information} (2000), Michael Nielsen and Isaac Chuang.}
is more than sufficient.
We also assume some  ``mathematical maturity'' which, for lack of a
better definition, is the ability to take definitions on faith until
the evidence arrives.



\tocgroup{algebraic boot camp}

\section{1. Ways of computing}\hypertarget{sec:2}{}

In quantum computing, all roads lead back to John von Neumann.
Von Neumann started his career as a graduate student with David Hilbert in G\"{o}ttingen, where he
co-invented Hilbert space.\sidenote{``Uber die Grundlagen der
  Quantenmechanik'' (1927), David Hilbert and John von Neumann. \vspace{5pt}}
As a young mathematician, he pioneered the study of \emph{operator
  algebras},\sidenote{``On Rings of Operators I/II'' (1936/7), Murray
  and von Neumann. \vspace{5pt}}
and after ``losing faith'' in the Hilbert space formulation of
quantum mechanics,\sidenote{``Why John von Neumann did not Like the Hilbert Space Formalism of Quantum Mechanics
(and What he Liked Instead)'' (1996), Miklós Rédei.} turned to
operator algebras as an alternative foundation.
Von Neumann was particularly interested in the logic of \emph{projectors} $\Pi: \mathcal{H}\to
\mathcal{H}$, operators on a Hilbert space $\mathcal{H}$ satisfying
\begin{equation}
  \Pi^2 =\Pi, \quad \Pi^\dagger = \Pi.\label{eq:proj}
\end{equation}
Physically, these correspond to ``atomic'' yes/no measurements;
mathematically, they closely parallel propositional
variables in a Boolean algebra, which satisfy an \emph{idempotence} condition
$p^2 = p$.

Von Neumann pursued this hint rather literally, erecting a whole theory of
``quantum logic'' around it.\sidenote{``The Logic of Quantum
  Mechanics'' (1936), Garrett Birkhoff and John von Neumann. This
  gives a fragment of linear logic, see ``Linear Logic for Generalized
  Quantum Mechanics'' (1992), Vaughan Pratt. \vspace{5pt}}
Although it provides some hints, quantum
logic is too restrictive to compute with; for instance, it has no notion of
conditional, which makes programming difficult!
Von Neumann was drawn away from foundational questions by
wartime work on the Manhattan Project, where he crossed paths with
Feynman. After the war, he devoted himself energetically to
practical applications like game theory, nuclear strategy and large-scale
simulation.\sidenote{``Numerical integration of the barotropic
  vorticity equation'' (1950), Charney, Fj{\"o}rtoft and von
  Neumann.}
Things could have been different (as envisioned in the companion piece,
\href{https://arxiv.org/abs/2503.00005}{\emph{A Short History of
    Rocks}}), but the task of making quantum compute would be left to
his bongo-playing colleague from Los Alamos.

\begin{figure}[h]
  \centering
  \vspace{2pt}
  \includegraphics[width=0.47\textwidth]{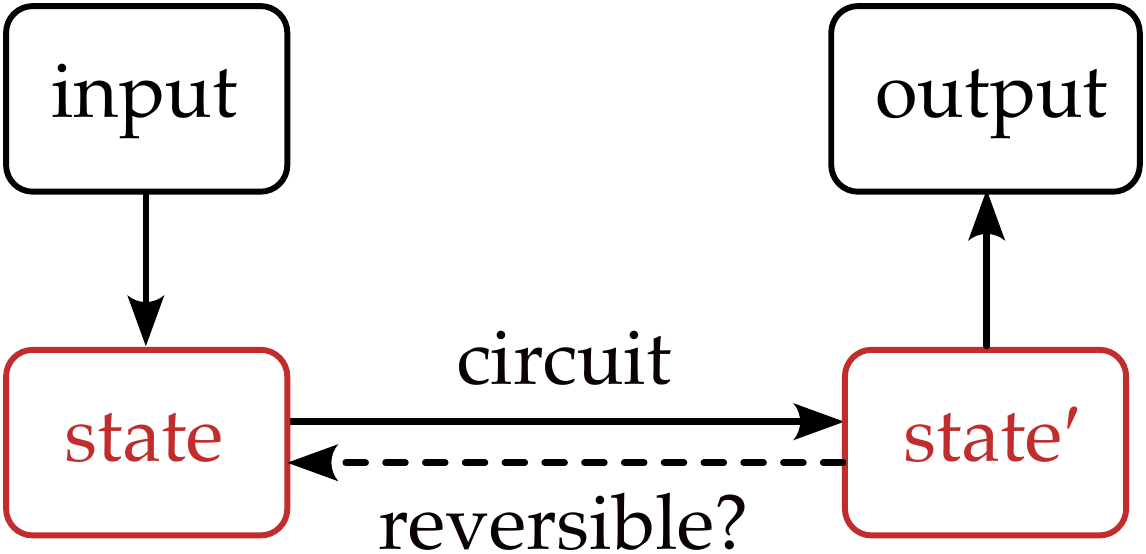}
  \caption{The basic schema of reversible computing separates circuit and state.}
  \label{fig:rev-flow}
  \vspace{-5pt}
\end{figure}

Reversible computing---either classical or quantum---is 
based on the schema of encoding input data
into a state, processing it with a circuit, then reading output from state once
more. Computation is reversible if there is a circuit which always
produces the initial from the final state. We picture this as a
flowchart in Fig. \ref{fig:rev-flow}.

\noindent By focusing on conservation or non-conservation of
information, we implicitly make the courier of that
information---state--the key object.

State is classically unproblematic.
For a single wire, the logical state is identified with whether
current is flowing ($1$) or not ($0$), a question we can answer with an ammeter.
For a collection of wires, we use a ``cross-section'' of ammeters to
obtain a list of Booleans.
From combinatorics, reversibility means we cannot allow wires to split (``fan out'') or join
(``fan in'').
Otherwise, there
isn't enough space on the small side to accomodate the
possibilities of the large.

\begin{figure}[h]
  \centering
  \vspace{-0pt}
  \includegraphics[width=0.6\textwidth]{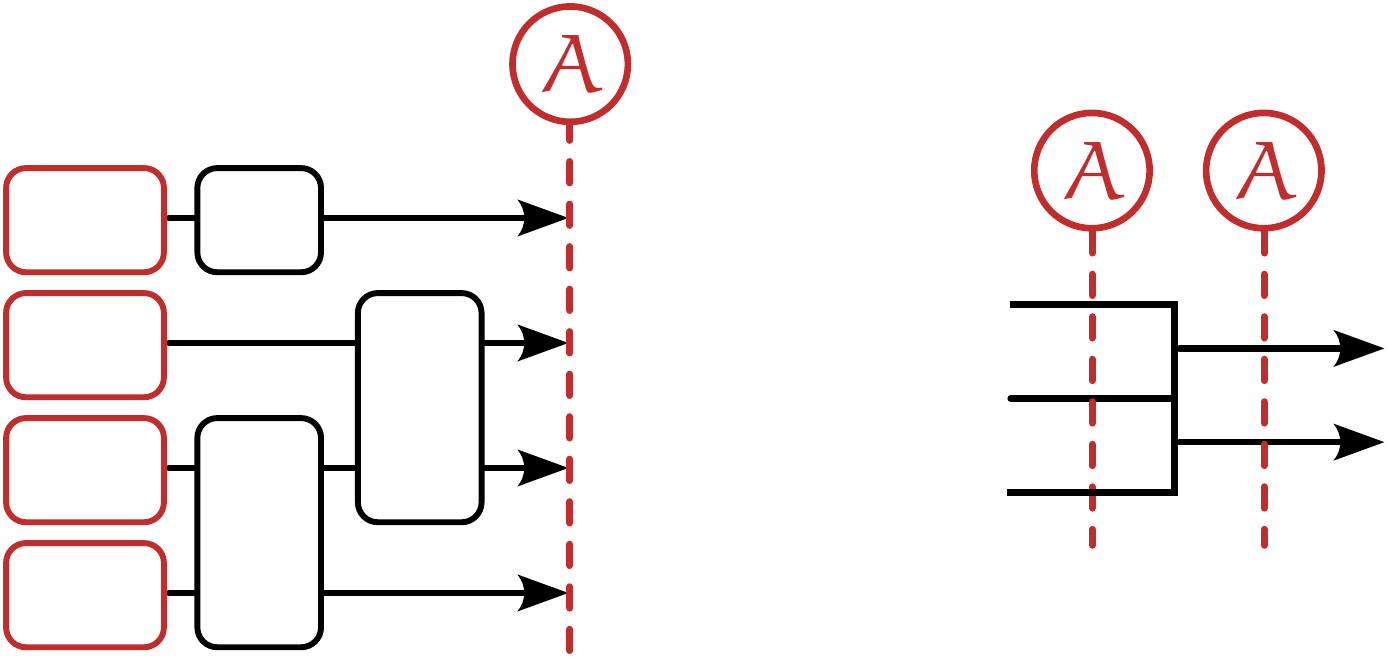}
  \caption{\textsc{Left}. Measuring state with
    ammeters. \textsc{Right}. If we branch or join, state on one side
    becomes irrecoverable from the other.}
  \label{fig:rev-cross}
  \vspace{-5pt}
\end{figure}

The quantum case is superficially similar.
All fundamental laws 
are reversible, so it seems reasonable to expect the
circuits in a quantum computer to be. We replace
wires with qubits and ammeters with measurement of the Pauli $Z$,
which yields outcomes $\lambda \pm 1$.

\begin{figure}[h]
  \centering
  \vspace{-0pt}
  \includegraphics[width=0.27\textwidth]{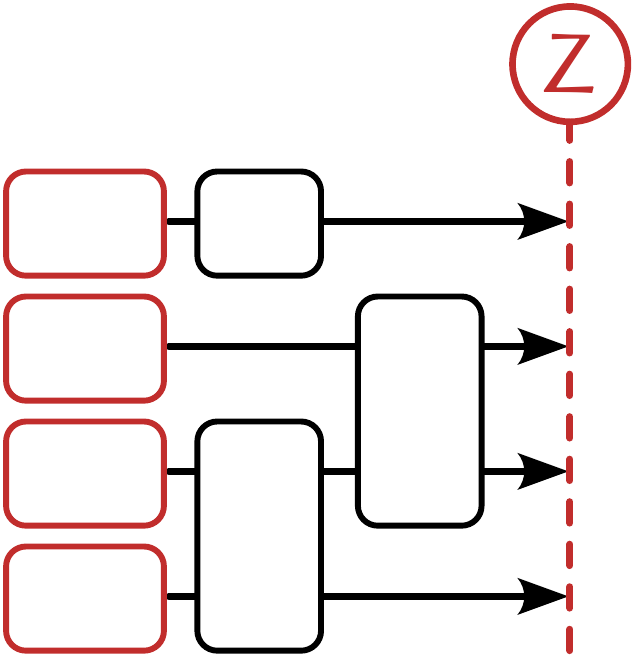}
  \caption{Measuring quantum state with the Pauli $Z$ gives binary
    outcomes for each qubit.}
  \label{fig:rev-cross}
  \vspace{-5pt}
\end{figure}

\noindent
There is a loose combinatorial intuition that fan outs and
fan ins should also be forbidden, an intuition made precise by the
\textsc{No Cloning
Theorem}.\sidenote{``The concept of transition in quantum mechanics''
  (1970), James Park; ``A
  single quantum cannot be cloned'' (1982), Wootters and
  Zurek.}
We seem to have an excellent analogy, then, between reversible computing
in classical and quantum cases; indeed, this analogy has been
guiding the field since its inception.

\begin{figure}[h]
  \centering
  \vspace{-0pt}
  \includegraphics[width=0.32\textwidth]{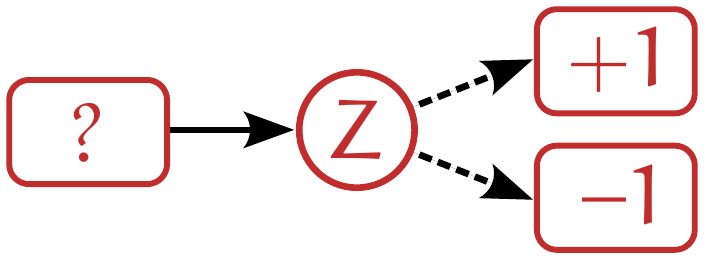}
  \caption{Measuring quantum state is a computational act.}
  \label{fig:rev-cross3}
  \vspace{-5pt}
\end{figure}

But under careful observation, the analogy breaks down.
First of all, we never see the quantum state itself, only the
measurement outcomes. A single measurement typically tells us \emph{nothing}
about the state.\marginnote{If we measure $Z = +1$, all we learn is
  that $\alpha \neq 0$ in (\ref{eq:qubit}). This is generically
  uninformative unless we do something clever, which to be fair, is
  the whole point of quantum algorithm design.
}
Next, state itself is ill-defined because of 
phase ambiguity,
i.e. the fact that $|\psi\rangle \sim e^{i\theta}|\psi\rangle$ are
equivalent. This is what led to von Neumann's crisis of faith in
Hilbert space!
Finally, measurement \emph{changes}
the state; we know what it is now, but not 
before! Measurement becomes part of the computation. See Fig. \ref{fig:rev-cross3}.

\noindent These subtleties are of course familiar to
anyone who has taken an undergraduate course in modern physics.
But familiar or not, they make the primitive abstractions of Fig. \ref{fig:rev-flow}
poorly suited to quantum computing.
This would be a quibble if no other approach existed. As it turns
out, however, we can repurpose Shannon's classical insights for the
quantum realm.

\begin{figure}[h]
  \centering
  \vspace{-2pt}
  \includegraphics[width=0.43\textwidth]{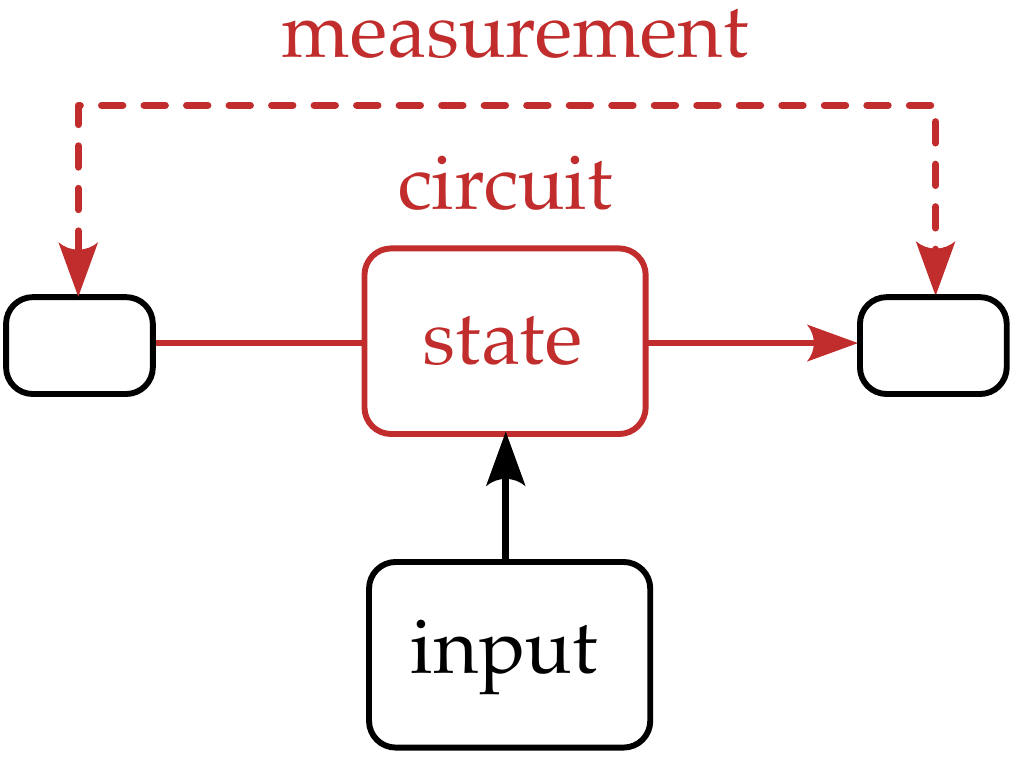}
  \caption{In a switching circuit, state and circuit are combined into
    a single object and jointly measured.}
  \label{fig:rev-flow0}
  \vspace{-3pt}
\end{figure}

\noindent Shannon used a model of computation called a \emph{switching
  circuit}, shown in Fig. \ref{fig:rev-flow0}.
This is an electrical relay which encodes state into
a configuration of \emph{switches} rather than a pattern of current.
This incorporates state into the circuit itself,
and we test if the combined structure is connected. Physically, we
can use a voltmeter\marginnote{Mathematically, the idea of using
  voltmeters rather than ammeters suggests that measurement should be
  viewed as a \emph{torsor}.} and a small test current; unlike the ammeter,
which interferes with the circuit, the voltmeter (in the limit of zero
current) has no effect.

The measurement across the whole circuit is determined by
measurement of individual switches. For instance, consider the formula $\texttt{OR}(x, y) =x + y - xy$. The switches are $x$
and $y$; if we leave $y$ open and close $x$, for instance, we get a
Boolean \emph{valuation} $v(x) = 1, v(y) = 0$. The overall value is
\[
  v[\texttt{OR}(x, y)] = v(x) + v(y) - v(x)v(y) = 1.
\]
The math faithfully reflects our intuition that the circuit is
closed.

\marginnote{
  \vspace{-50pt}
  \begin{center}
    \hspace{-10pt}\includegraphics[width=0.7\linewidth]{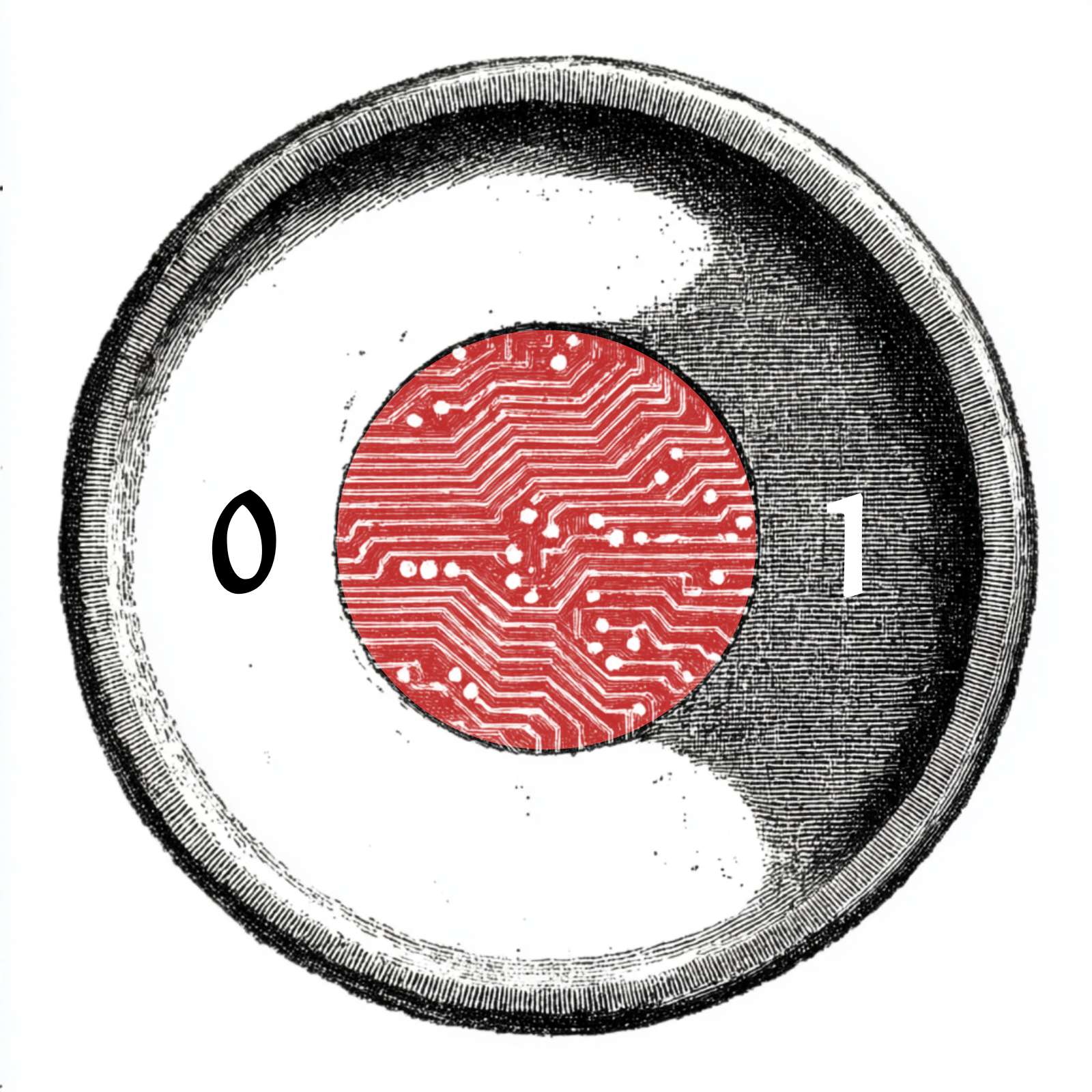}
  \end{center}  \vspace{-0pt}
  \emph{Circuits and truth tables: two sides of the same coin.
  } \vspace{-10pt}
}

Let's repackage these ideas using \emph{syntax} and \emph{semantics}.
A circuit with indeterminate switches is
represented by an expression like $\texttt{OR}(x, y)$. 
This is purely syntactic, and can be manipulated symbolically using Boolean algebra.
States, on the other hand, live in the semantic realm.
A state assigns a Boolean value in $\mathbb{B}=\{0, 1\}$ to
each propositional
variable, or equivalently, flips each switch on or off.
Measurement across the whole circuit is determined compositionally from
measurements of individual switches, as in Fig. \ref{fig:rev-truth}:

\begin{figure}[h]
  \centering
  \vspace{-2pt}
  \includegraphics[width=0.52\textwidth]{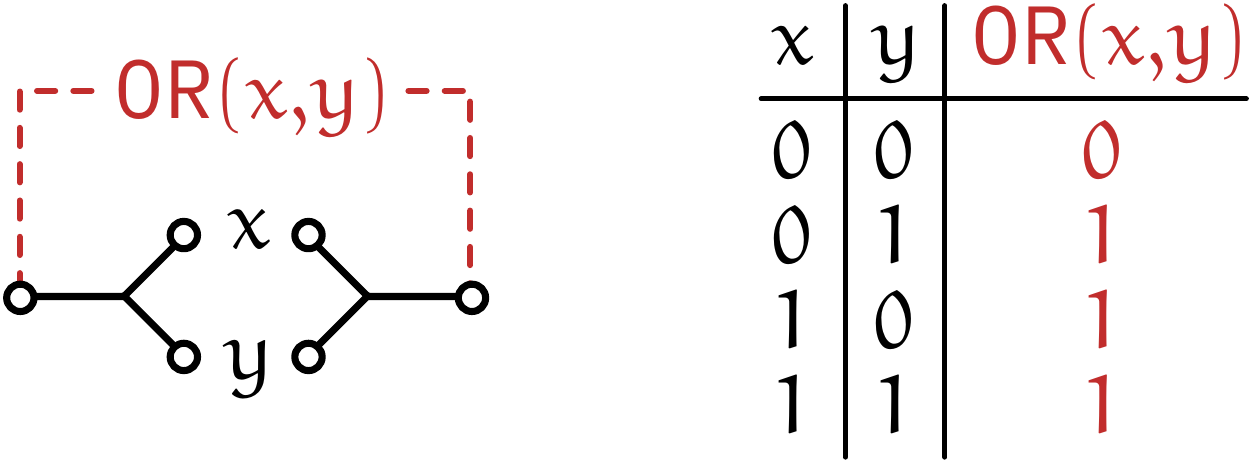}
  \caption{A truth table lists all assignment of Boolean values to
    switches (valuations), and the induced value of the circuit.}
  \label{fig:rev-truth}
  \vspace{-3pt}
\end{figure}

\noindent
The full list of states, and correponding status of the circuit,
is called a \emph{truth table}.
Our goal now is to find 
an analogous
bipartite structure
for quantum computing, with bonus points if measurement uses a voltmeter rather than an ammeter.

To keep things concrete, we'll develop a simple example in parallel
with the general case, and a
particularly revealing choice (both for its familiarity and unexpected
depth) is the qubit. 
Recall that a qubit state (\ref{eq:qubit}) lives in a two-dimensional Hilbert
space $\mathcal{H} = \mathbb{C}^2$. The Pauli $Z$ we have been
measuring with can either be viewed as living in an abstract\marginnote{``Bounded'' means
  the image of any state has bounded length. This is not a problem in
  finite dimensions, but in infinite dimensions, is equivalent to
  continuity.}
space of \emph{bounded linear operators} $\mathcal{B}(\mathcal{H})$, or concretely, the set
$\mathsf{M}_{2}(\mathbb{C})$ of $2 \times 2$ complex
matrices, with an expression 
\[
      Z =
  \begin{bmatrix}
    1 & 0 \\ 0 & -1
  \end{bmatrix}.
\]
Below, we'll explain how to think about $Z$ (and its siblings) 
without reference to matrices, qubits, or Hilbert space at all.

\section{2. A trip to Disneyland}\hypertarget{sec:3}{}

In 1947, a
century after George Boole\sidenote{\emph{The Mathematical Analysis of
    Logic}.} published his groundbreaking pamphlet on the algebra of logic,
Irving Segal 
performed an equivalent feat for physics.\sidenote{``Irreducible
  representations of operator algebras''; ``On the embedding of normed rings into the
  ring of operators in Hilbert space'' (1943), Israel Gelfand and
  Mark Naimark.
}
Before the war, Segal had worked with von Neumann and Einstein at the
IAS, where, coincidentally, he was a postdoc at the same time as
Shannon.
The war commandeered his brain for ballistics research, but in 1946,
he spent the summer at Princeton and returned
to thinking about his favourite problem: the mathematical foundations
of quantum mechanics.

Segal was particularly concerned about particle physics,\marginnote{Wave
  of hand to a physicist is sleight of hand to a mathematician.} where
theoreticians like Feynman were playing three cup
monte with infinity. 
The existing Hilbert space methods were unwieldy and ill-defined, permitting decidedly
unphysical operations; the binary projectors of quantum logic, on the
other hand, were too tightly constrained.
Segal needed something in between. He found it in
the work of two Russian mathematicians, Gelfand and Naimark,
who had studied certain rings of operators. 
Segal realized that their structure---which he baptized a
\emph{C${}^*$-algebra}---was perfect for modelling the behaviour of
quantum-mechanical observables and
measurements.\sidenote{``Irreducible representations of operator
  algebras'' (1947), Irving Segal.}
The rest of this section gently introduces the math; readers may skip
it and return later as needed.

A \emph{\gls{cstar}} $\mathcal{A}$ is a set of operators which wears many hats. First of all, it is a vector space over
$\mathbb{C}$, so closed under \emph{linear combinations}:
\[
  \alpha A + \beta B \in \mathcal{A} \,\, \text{ for } \,\, A, B \in
  \mathcal{A} \,\,\text{ and }\,\, \alpha, \beta \in\mathbb{C}.
\]
This encodes the linearity of quantum mechanics.
Unlike states, we can \emph{compose} operators, that is, apply one
after the other. We capture this with a product operation $\cdot:
\mathcal{A}\times \mathcal{A}\to \mathcal{A}$, and borrow the
following behaviours from operator composition:
\begin{itemize}[itemsep=0pt]
\item \emph{associativity}: $A\cdot(B \cdot C) = (A \cdot B)\cdot C$;
\item \emph{distributivity}: $A\cdot(B + C) = A\cdot B + A \cdot C$;
\item \emph{scalar commutativity},: $\lambda (A \cdot B) = (\lambda A)
  \cdot B = A \cdot (\lambda B)$;
\end{itemize}
for all $A, B, C\in\mathcal{A}$ and $\lambda \in \mathbb{C}$.
We can summarize everything we've listed so far by saying
$\mathcal{A}$ is an
\emph{associative algebra} over $\mathbb{C}$.
There are two further optional criteria: %
that are important:
\begin{itemize}[itemsep=0pt]
\item \emph{commutativity}: $A \cdot B = B \cdot A$;
\item \emph{unitality}: there exists $I \in \mathcal{A}$ such that
  $A\cdot I
  = I\cdot A = A$. 
\end{itemize}
In the first case, we say $\mathcal{A}$ is
\emph{commutative}, and otherwise \emph{noncommutative}; in the second
that it is \emph{unital}.
We stick to unital algebras for technical simplicity, though most of
our conclusions hold regardless.

We often use complex numbers\marginnote{Paul Halmos reputedly called
complex analysis ``the Disneyland of mathematics''. C${}^*$-algebras are the
Disneyland of functional analysis!} to model situations where a real answer
is needed; after some analytic magic, 
we return to
the real line by imposing $\overline{z} = z$.
\marginnote{
  \begin{center}
    \includegraphics[width=1.06\linewidth]{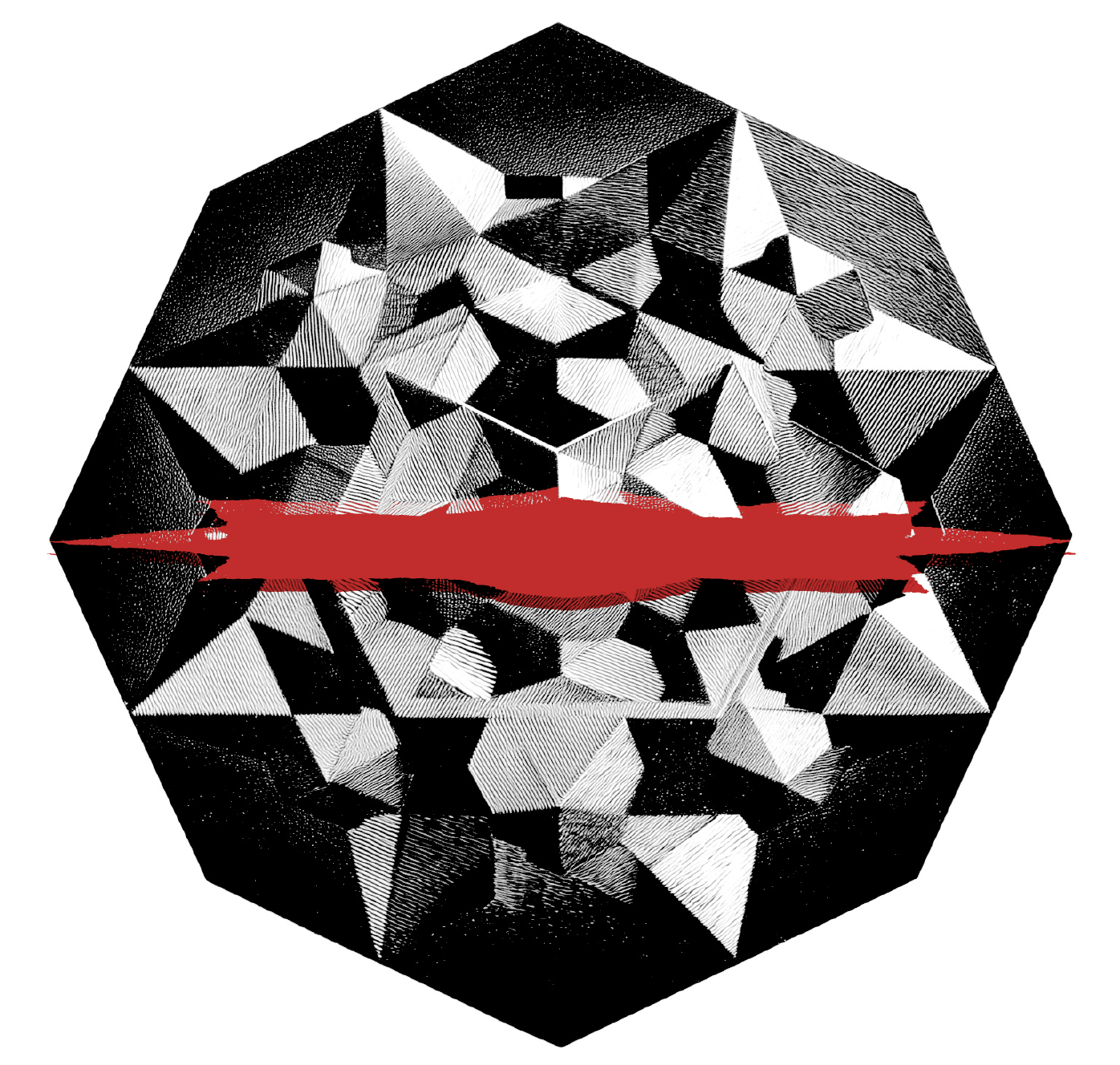}
  \end{center}  \vspace{-5pt}
  \emph{
    Disneyland $\mathcal{A}$ is a fun place to visit, but you don't live there. Reality
    is the self-adjoint cross-section $\mathcal{A}_\text{sa}$, in red.
  } \vspace{10pt}
}
Similarly, the linear structure of operators over $\mathbb{C}$ is unphysical, but
needed for ``analytic magic''; 
we project back to reality with 
$A^\dagger = A$.
Instead of a concrete Hermitian conjugate ${}^\dagger$, we use an abstract
\emph{adjoint} $\ast:\mathcal{A}\to \mathcal{A}$,
and collect self-adjoint elements into $\mathcal{A}_\text{sa}$.
The adjoint has the following properties:
\begin{itemize}[itemsep=0pt]
\item \emph{involution}: $(A^\ast)^\ast = A$;
\item \emph{antilinearity}: $(\alpha A + \beta B)^\ast =
  \overline{\alpha} A^\ast + \overline{\beta} B^\ast$;
\item \emph{antimultiplicativity}: $(A B)^\ast = B^\ast  A^\ast$;
\end{itemize}
for all $A, B \in \mathcal{A}$ and $\alpha, \beta\in \mathbb{C}$.
This makes our associative algebra into a \gls{staralg}.
So, we now have everything but the ``C'' in ``C${}^*$''!

This comes from the notion of \emph{size}.
For matrices, there are many ways to measure size, but perhaps the
most natural is the \emph{operator norm}, which is the maximum amount
by which it stretches vectors.
For matrices acting on a
Hilbert space $\mathcal{H}$, we can define
\[
  \Vert M \Vert_\text{op} = \text{inf} \,\{c \geq 0: \Vert M v\Vert \leq
  c\Vert v\Vert , \, \forall v \in \mathcal{H}\}.
\]
This means that, by definition, $\Vert Mv\Vert \leq \Vert M
\Vert_\text{op} \Vert v\Vert$.
It's not obvious how to port this over to a $\ast$-\emph{algebra}, without
vectors to act on.
But for a \emph{diagonalizable} operator, it's
simple\marginnote{In the diagonal basis, each direction gets scaled by
  at most $|\lambda_\text{max}|$.} to show that the operator norm is just the
largest (absolute) eigenvalue:
\[
\Vert M \Vert_\text{op} = |\lambda_\text{max}| = \text{sup} \big\{|\lambda|: M v = \lambda v,
\, \exists v \in \mathcal{H}- \{0\}\big\}.
\]
By the \gls{spectral}, 
\emph{normal matrices} satisfying $N^\dagger N
= N N^\dagger$ are diagonalizable, and in particular, $N = M^\dagger M = |M|^2$ is normal for any $M$.
Some caution\marginnote{For instance, $
  \begin{bsmallmatrix}
    1 & 1 \\ 0 & 1
  \end{bsmallmatrix}
$ has eigenvalue $\lambda=1$, but stretches $(1, 1)^T$ by
$c\approx 1.6$.} is needed here; if a matrix is \emph{not}
diagonalizable, the
operator norm is not always largest eigenvalue.

The next step is to define eigenvalues without eigenvectors.
This sounds tricky, but luckily, there is a tight connection between eigenvalues and \emph{invertibility} of a matrix:
\begin{equation}
  M v = \lambda v \quad \Longleftrightarrow \quad M - \lambda I \,\,
  \text{ is not invertible}.\label{eq:alg-eval}
\end{equation}
In algebraic terms, a matrix $M$ is not invertible just in case there
is not $N$ such that $MN = NM = I$.
So far, we can only port the definition of operator norm to normal operators
$A \in \mathcal{A}$.\marginnote{The ``C'' stands for
  ``closed'', since Segal showed it is \emph{topologically} closed,
  i.e. there are no missing limit
  points as measured by $\Vert \cdot \Vert$.} 
To extend this to ``abnormal'' operators, we take inspiration from the
identities $|z|^2 = |\overline{z} z|$
in complex analysis and $\Vert M\Vert_\text{op}^2 = \Vert M^\dagger
M\Vert_\text{op}$ in Hilbert space.
Since $A^*A$ is normal, we can use identify the maximum eigenvalue as
its operator norm, and then take the square root to obtain the norm of
$A$. Thus, we \emph{define} a \gls{cstar} as one in which 
\begin{equation}
  \label{eq:C${}^*$}
  \Vert A \Vert^2 = \Vert A^\ast A \Vert = \text{sup}\, \{|\lambda| :
  A^\ast A - \lambda I \,\, \text{ is not invertible}\}.
\end{equation}
This is called the \emph{C${}^*$ identity}.
Remarkably, when this norm exists, it is uniquely determined by the
algebraic structure. Pure Disney!

To actually construct \gls{cstar}s, the method of \emph{presentations}
will prove extremely useful.
The basic idea is to pick a set of unknowns $\mathcal{X}$ (akin to propositional
variables $x, y, \ldots$) which we call \emph{generators}. We subject
these to identities called
\emph{relations} or \emph{identities} $\mathcal{I}$, arranged so that each element
$R \in \mathcal{I}$ is equated to zero.
\marginnote{
  \vspace{-100pt}
  \begin{center}
    \includegraphics[width=0.9\linewidth]{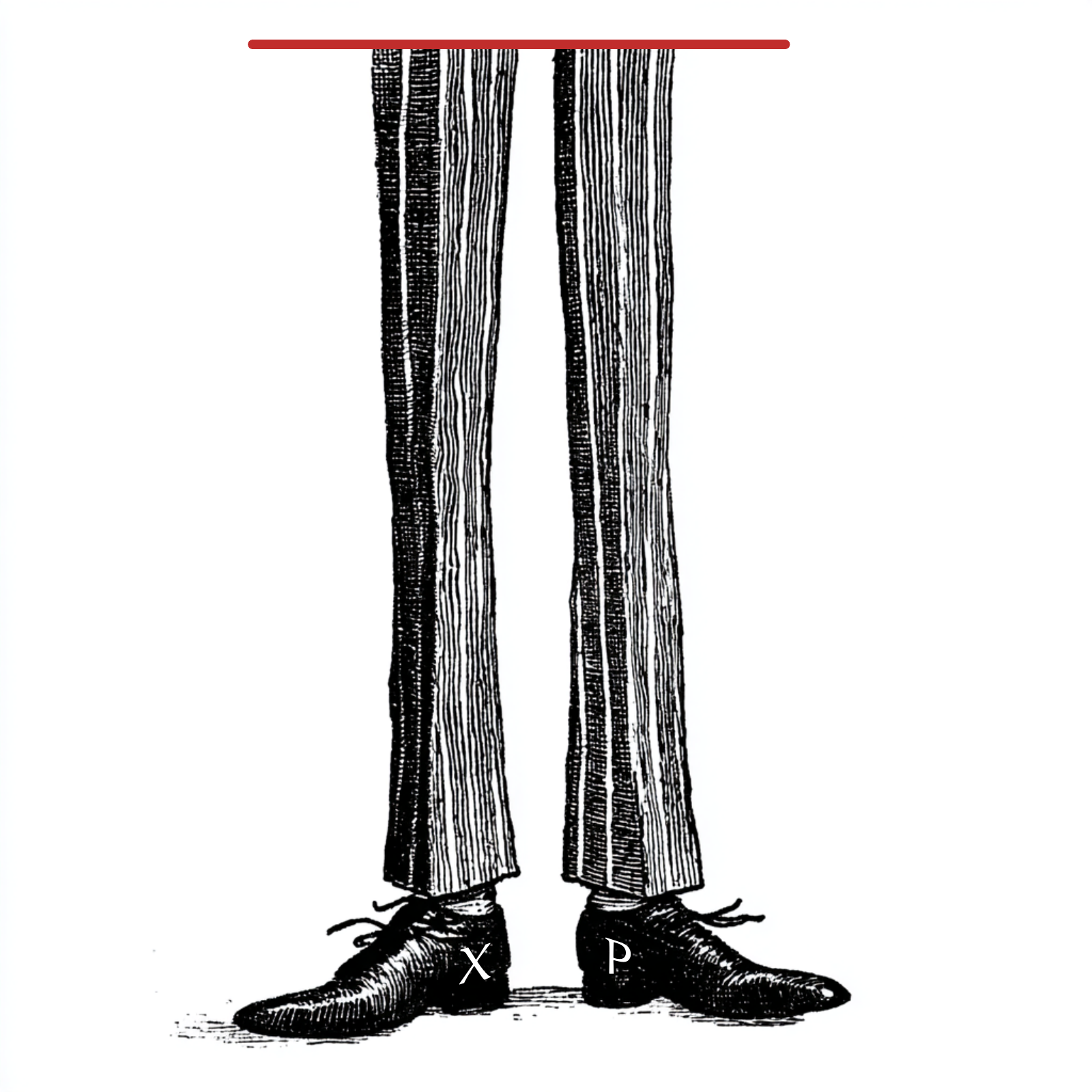}
  \end{center}  \vspace{-5pt}
  \emph{In Disneyland, you must be \emph{at most} this tall to ride. No canonical commutation relations allowed!
    \vspace{5pt}
  } \vspace{15pt}
}
For instance, the
standard canonical commutation relation (with $\hslash=1$) is
\begin{equation}
\mathcal{X} = \{X, P\}, \quad \mathcal{I} = \{XP - PX - i
I\}.\label{eq:ccr}
\end{equation}
When this algebra exists, we call it the \emph{universal C${}^*$-algebra}
$\text{C${}^*$}\langle\mathcal{X}|\mathcal{I}\rangle$.

Existence is actually nontrivial; for instance, (\ref{eq:ccr}) cannot be realized by operators with bounded
eigenvalues, and hence never exist in a C${}^*$-algebra.
We'll see Weyl's workaround for this case in \S\hyperlink{sec:19}{18} and Appendix \hyperlink{app:functional}{A}, but for
now, we note two sufficient conditions:\sidenote{See ``Quantenmechanik und Gruppentheorie'' (1927), Hermann Weyl; ``\gls{cstar} relations''
  (2010), Terry Loring. We also assume $\mathcal{X}$ and $\mathcal{I}$
are finite, and $\mathcal{I}$ is consistent. These conditions, along
with either condition on the left, ensure
that the quotient
$\mathbb{C}^\ast[\mathcal{X}]/\mathbb{C}^\ast[\mathcal{I}]$ has
well-defined norm satisfying the C${}^*$ identity (\ref{eq:C${}^*$}).}
\begin{itemize}[itemsep=0pt]
\item $\mathcal{I}$ is a set of polynomials with no constant term; or
\item $\mathcal{I}$ truncates the free $\ast$-algebra $\mathbb{C}^\ast[\mathcal{X}]$ to
  finite dimensions.
\end{itemize}
By \emph{free $\ast$-algebra} $\mathbb{C}^\ast[\mathcal{X}]$, we mean all finite $\mathbb{C}$-linear combinations of
monomials formed from $\mathcal{X}$ and adjoints.

Let's illustrate with our concrete example.
Instead of thinking of $Z$ as a matrix, we can treat it as
an unknown required to be (a) self-adjoint,
$Z = Z^\ast$, and (b) unitary $Z^2 = I$.
Unitarity has a constant term, but luckily, truncates
the polynomials $\mathbb{C}^\ast[Z]$ from arbitrary to linear degree, since
higher powers of $Z$ can be reduced. We end up with \marginnote{In
  terms of matrices, we get \[
    \begin{bmatrix}
      \alpha + \beta & \\ & \alpha - \beta
    \end{bmatrix},
\] which spans \emph{diagonal} elements of $\mathsf{M}_{2}$.}
\begin{equation}
 \mathcal{A}_Z= \text{C${}^*$}\langle Z | Z-Z^\ast, Z^2 - I\rangle = \left\{ \alpha I +
    \beta Z : \alpha, \beta \in \mathbb{C} \right\}.\label{eq:A_Z}
\end{equation}
To find the norm of $Z$, note that $Z^2 - I = (Z - I)(Z + I) = 0$
implies eigenvalues $\pm 1$, and hence a norm of $|\pm 1|=1$.

\section{3. From prop to op}\hypertarget{sec:4}{}

Shannon translated Boolean algebra into ordinary electric circuits.
We can try to translate C${}^*$-algebras into ``noncommutative'' circuits.
We start by recalling Shannon's isomorphism:
\begin{itemize}[itemsep=0pt]
\item a Boolean variable $x$ is physically encoded by a switch;
\item switches $x$ and $y$ in series corresponds
  to the product $xy$ ($\texttt{AND}$);
\item switches $x$ and $y$ in parallel corresponds to the sum $x \dot{+}
y$ ($\texttt{OR}$);\marginnote{Here, $x \dot{+} y = x + y -
  xy$ isn't quite addition but behaves enough like it for our
  purposes.}
\item scalars (in circles) multiply the adjacent proposition.
\end{itemize}
  \vspace{-8pt}
\begin{figure}[h]
  \centering
  \includegraphics[width=0.48\textwidth]{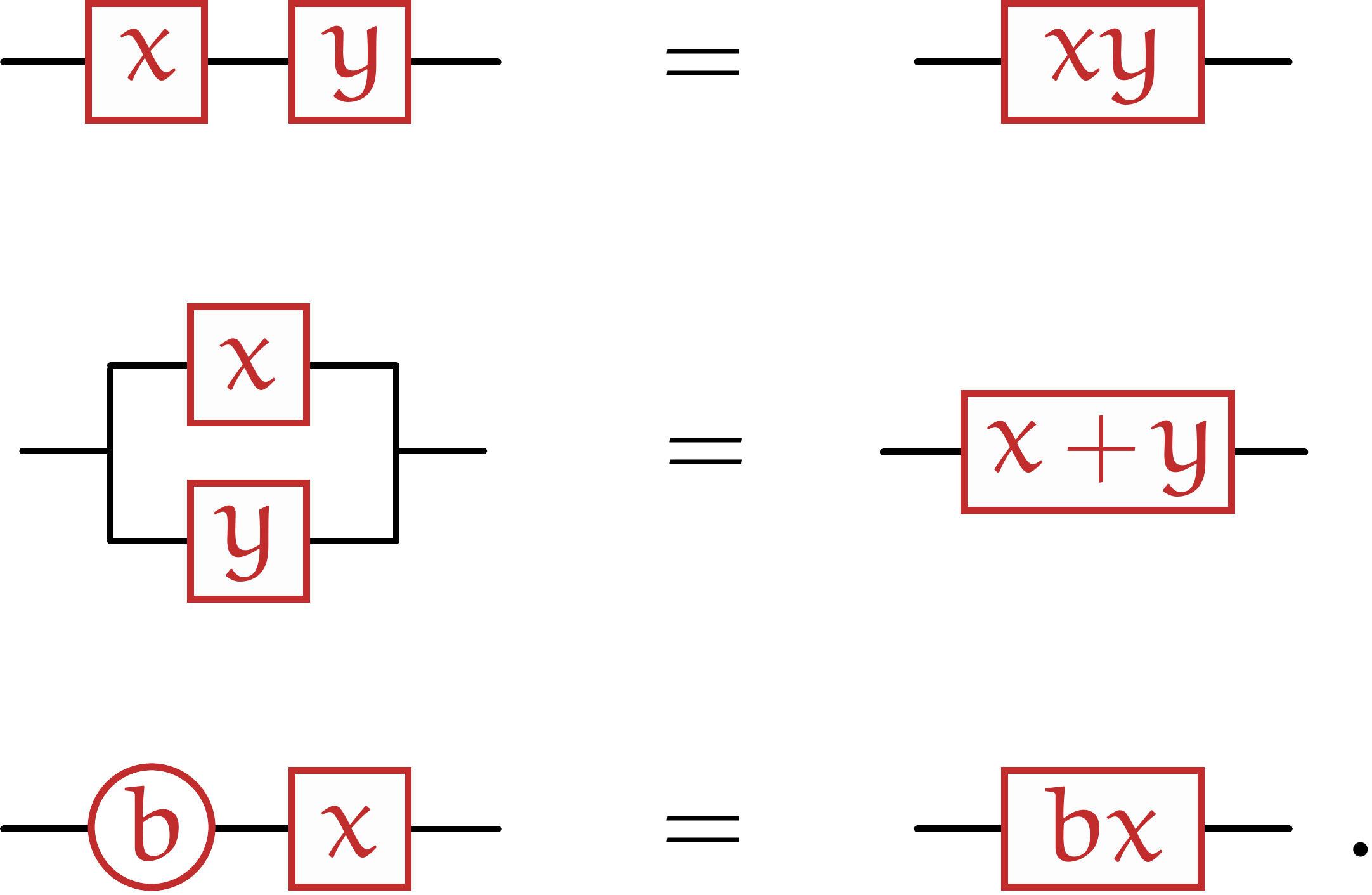}
  \caption{In a Boolean circuit, series corresponds to multiplication, parallel to
  addition, and circles to scalar multiplication.}
 \label{fig:simple1}
  \vspace{-5pt}
\end{figure}

\noindent 
We enclose formulas in squares, Booleans in circles.
  We don't usually think about scalar multiplication, but it makes the
  analogy to the noncommutative case nicer. Inserting a Boolean $b
  \in \mathbb{B}$ either does nothing ($b=1$) or snips the circuit where inserted ($b=0$).

An \emph{abstract wiring diagram (\gls{awd})}\marginnote{Pronounced
  ``ord''. We use ``wiring'' to differentiate from circuit diagrams
  (reversible quantum computing) and string diagrams (category theory).} encodes operators of a
C${}^*$-algebra in an almost identical fashion: 
\begin{itemize}[itemsep=0pt]
\item an operator variable $X$ is physically encoded by an observable;
\item expressions in series are multiplied (right to
  left);\marginnote{This means $XY$ is the same
    order as the diagram!}
\item expressions in parallel are summed;
\item scalars multiply the adjacent operator.
\end{itemize}

\begin{figure}[h]
  \centering
    \vspace{-5pt}
  \includegraphics[width=0.48\textwidth]{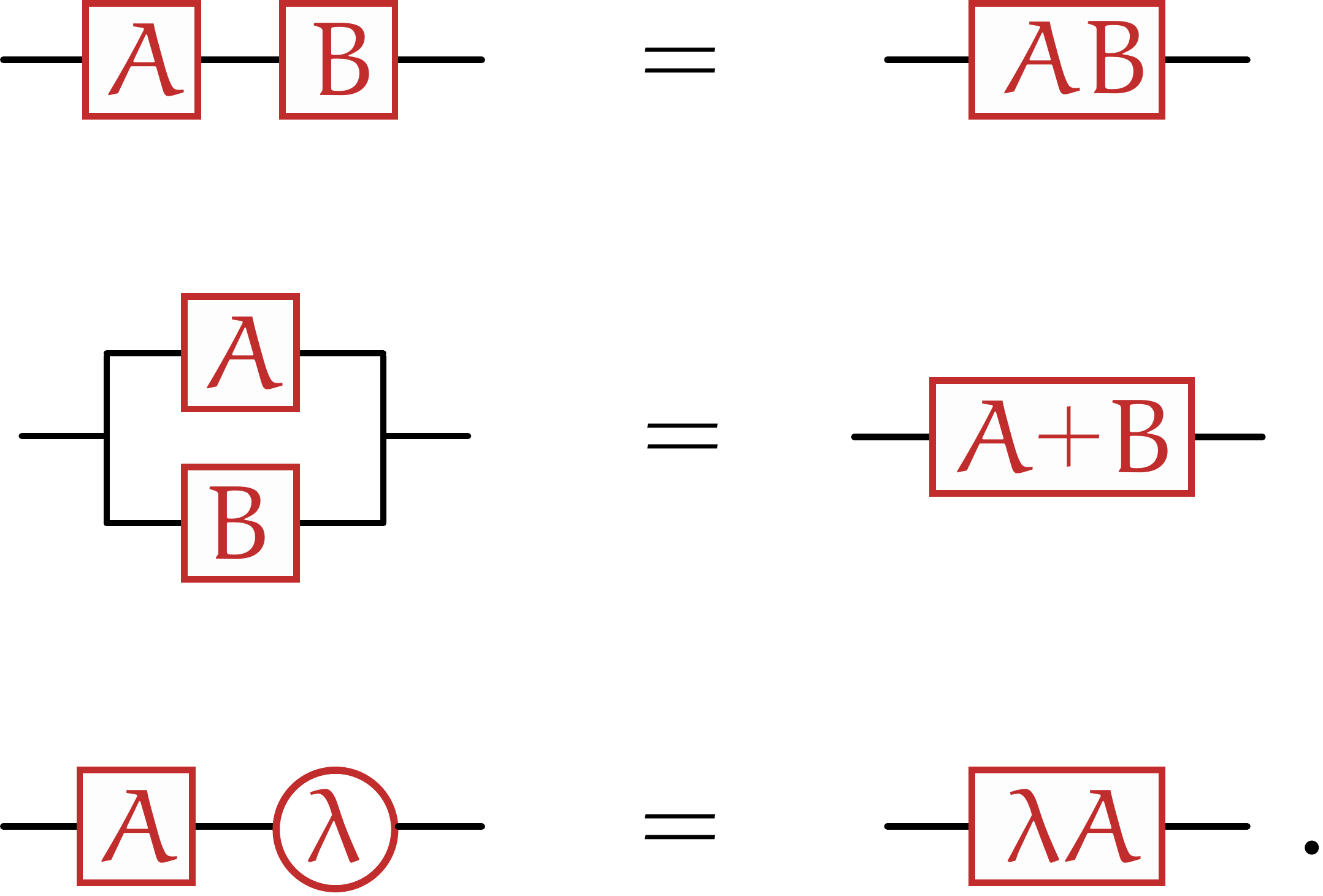}
  \label{fig:simple2}
  \vspace{-5pt}
\end{figure}

\marginnote{
  \vspace{-140pt}
  \begin{center}
    \hspace{-10pt}\includegraphics[width=1.1\linewidth]{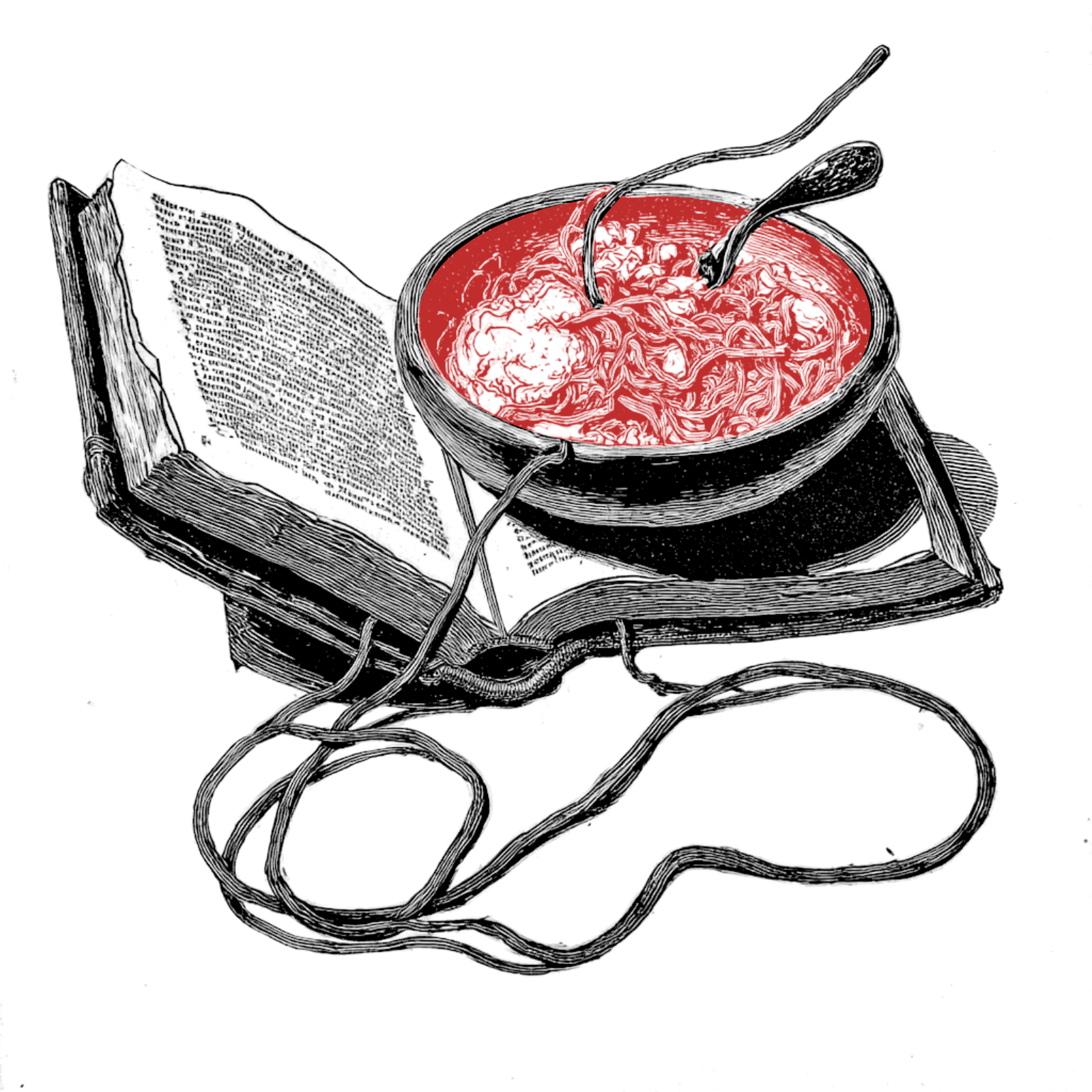}
  \end{center}  \vspace{-15pt}
  \emph{What kind of spaghetti are we cooking?
  } \vspace{-10pt}
}
\noindent 
Despite the manifest parallels, there are some differences.
Boxes don't necessarily commute, scalars are complex, and we read
right to left.
But most importantly, the word ``abstract'' signals that, unlike a
Boolean circuit, we do not have physical wires connecting
switches. The links between them are mathematical.

The simplest interpretation---that we
measure an observable at each box---isn't consistent with the
algebra.
A simple example is provided by the Stern-Gerlach experiment.\sidenote{``The
  experimental proof of directional quantization in the magnetic
  field'' (1922), Walther Gerlach and Otto Stern. We will take this behaviour as an experimental given for
now, and discuss how to reflect it in the formalism later.}
An electron approaches a magnetic field in the $\mathbf{z}$ direction
which effectively measures the orientation of its spin; the Pauli $Z$
is a idealized version of this.
But although $Z^2 = I$, measuring twice is not the same as doing
nothing!
The $Z$ deflection, though initially random, is fixed under subsequent
$Z$ measurements. 

\begin{figure}[h]
  \centering
    \vspace{-0pt}
  \includegraphics[width=0.48\textwidth]{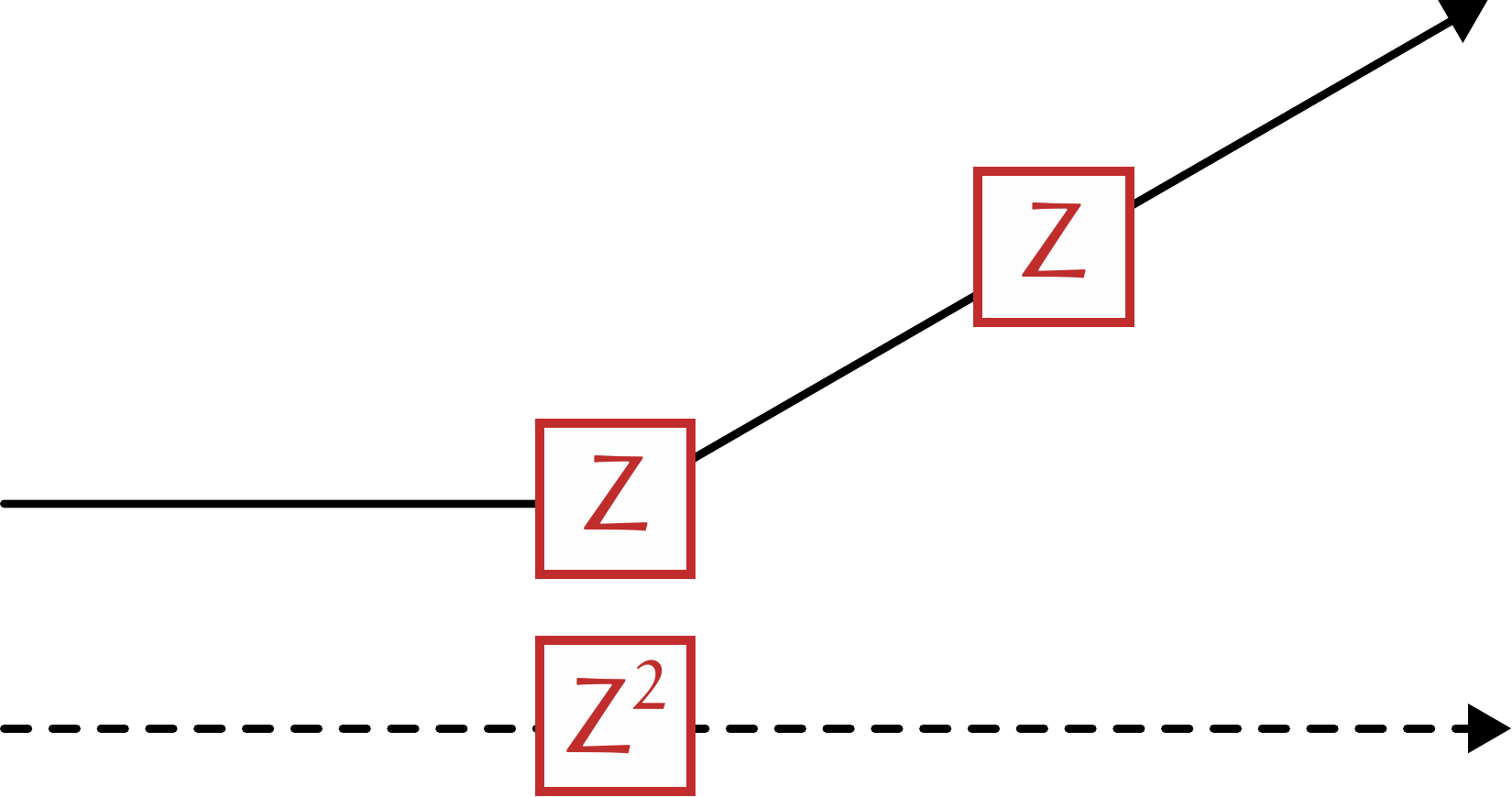}
  \caption{On the solid path, an evenly polarized spin is measured and deflects;
    the subsequent $Z$ measurement is consistent. On the dotted path,
    the spin is not measured and hence is undeflected.
  }
  \label{fig:stern}
  \vspace{-5pt}
\end{figure}

If \gls{awd}s 
are not physical layouts, what are they?
At the very least, they are \emph{symbolic models of computation}.
They let us manipulate observables algebraically.
As a simple illustration of how this can be used, we'll enlarge our
Pauli $Z$ example to include the Pauli $X$ and $Y$. As explicit matrices,
\begin{equation}
  \label{eq:pauli-gen}
  \sigma_{(1)} =
  \begin{bmatrix}
    0 & 1 \\ 1 & 0
  \end{bmatrix}, \quad
    \sigma_{(2)} =
  \begin{bmatrix}
    0 & -i \\ i & 0
  \end{bmatrix}, \quad
    \sigma_{(3)} =
  \begin{bmatrix}
    1 & 0 \\ 0 & -1
  \end{bmatrix}.
\end{equation}
Along with the identity operator $I$,\marginnote{It's straightforward to show
they are linearly independent, and since there are four, they
form a basis for $\mathsf{M}_{2}(\mathbb{C})$.} these form a basis for the full
set of single-qubit operators.
We can characterize them without matrices using the \emph{Pauli
  relations}:
\begin{equation}
  \sigma_{(i)} \sigma_{(j)} = \delta_{ij} I + i \epsilon_{ijk}\sigma_{(k)},
\label{eq:pauli-rel}
\end{equation}\marginnote{\vspace{-35pt}

\noindent We also use\emph{Einstein convention} of
summing over repeated dummy indices. In a non-linear context, it
is interpreted with smallest possible scope.}
where $\delta_{ij}$ is the Kronecker delta and $\epsilon_{ijk} = \text{Sgn}(ijk)$ the totally
antisymmetric tensor.\marginnote{Swapping two indices produces a factor
  of $-1$,
  and $\epsilon_{123} = +1$. This implies cyclic symmetry,
  $\epsilon_{ijk}=\epsilon_{jki}$.} 
For relations $R_{(ij)}$ encoding (\ref{eq:pauli-rel}), the
\emph{Pauli algebra} has presentation $\mathcal{A}_{\text{Pauli}} = \text{C}^*\langle \sigma_{(i)} |
R_{(jk)}\rangle$, where $i, j, k \in \mathfrak{I}=\{1,2,3\}$.

When objects are indexed by a set $\mathfrak{I}$, we
decorate their containers with \emph{index nodes}; usually
$\mathfrak{I}$ is clear from context.
In our Pauli example, for instance, $\mathfrak{I} = \{1,2,3\}$. A square with a single
node denotes $\sigma_{(i)}$, a circle with two nodes
$\delta_{ij}$, and a circle with three nodes
$\epsilon_{ijk}$:
\begin{figure}[h]
  \centering
  \vspace{2pt}
  \includegraphics[width=0.6\textwidth]{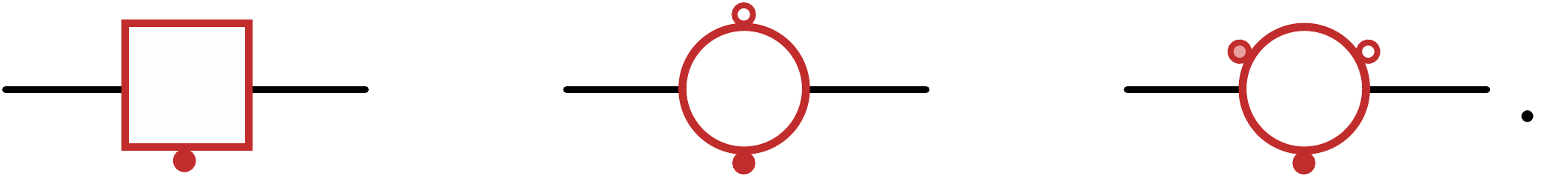}
  \caption{Special objects in the Pauli algebra, decorated by index nodes.}
  \label{fig:label}
  \vspace{-0pt}
\end{figure}

\noindent
Colour and shading act as dummy indices for index nodes, and in the
rare situation with multiple index sets at play, distinct indices get
distinct node shapes.
Containers without glyphs (as above) refer to special
families like $\sigma_{(i)}$ or $\delta_{ij}$.
To
specify other objects, or particular index values, we label explicitly.

The last operation we introduce for now is a \emph{tether}, which
pairs two indices and sums over them. In \gls{awd}s, we simply join two
nodes of the same colour and shape with a dotted line.
For instance, the Pauli relations (\ref{eq:pauli-rel}) can be
diagrammatically recast as


\begin{figure}[h]
  \centering
  \vspace{-2pt}
  \includegraphics[width=0.65\textwidth]{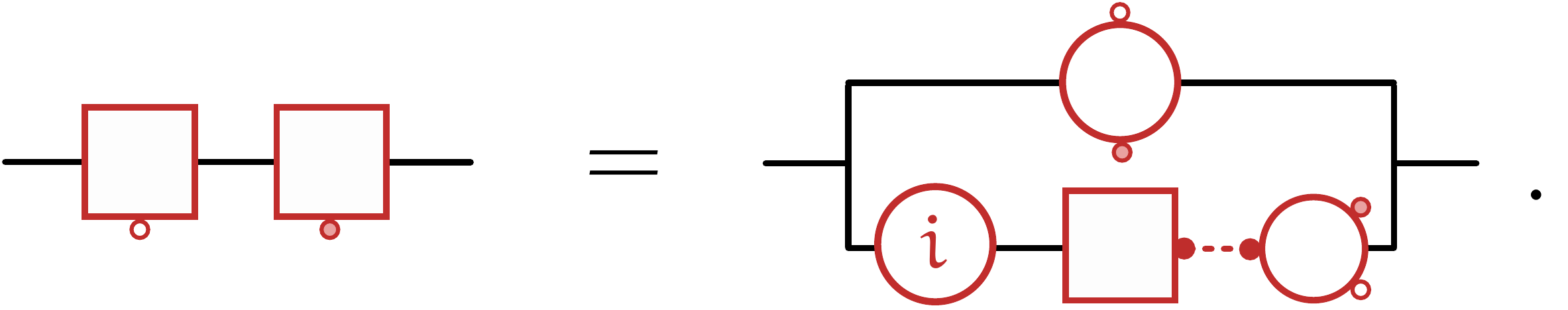}
  \label{fig:simple3}
  \vspace{-5pt}
\end{figure}

\noindent 
This trades complex of products for complexity of sums.

To illustrate, we'll show how to point a Pauli in any direction.
Let $\mathbf{n} =
(n_1, n_2, n_3)$ be a real vector of unit length, so $n_in_i = 1$:

\begin{figure}[h]
  \centering
  \vspace{-0pt}
  \includegraphics[width=0.48\textwidth]{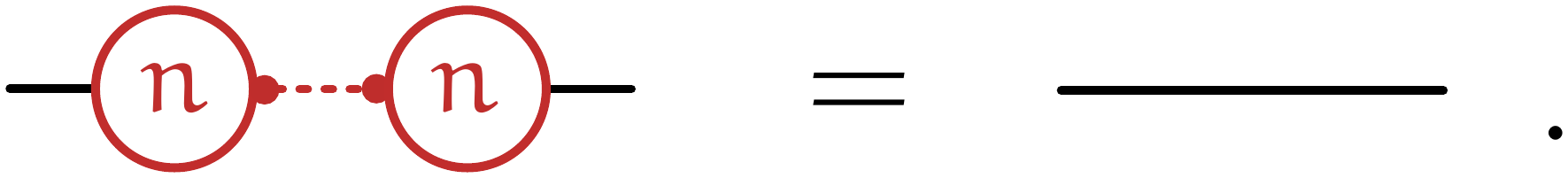}
  \label{fig:simple4}
  \vspace{-2pt}
\end{figure}

\noindent We claim the \emph{generalized Pauli operator}
$\sigma(\mathbf{n}) = n_i
\sigma_{(i)}$
is self-adjoint and
unitary. Self-adjointness is easy,\marginnote{We have a real linear
  combination of self-adjoint operators.} so it
remains to show that $\sigma(\mathbf{n})^2 = I$.
We can do this diagrammatically, indicating $\sigma(\mathbf{n})$ by a
box labelled with $\mathbf{n}$. The proof goes as follows:

\begin{figure}[h]
  \centering
  \vspace{-2pt}
  \includegraphics[width=0.75\textwidth]{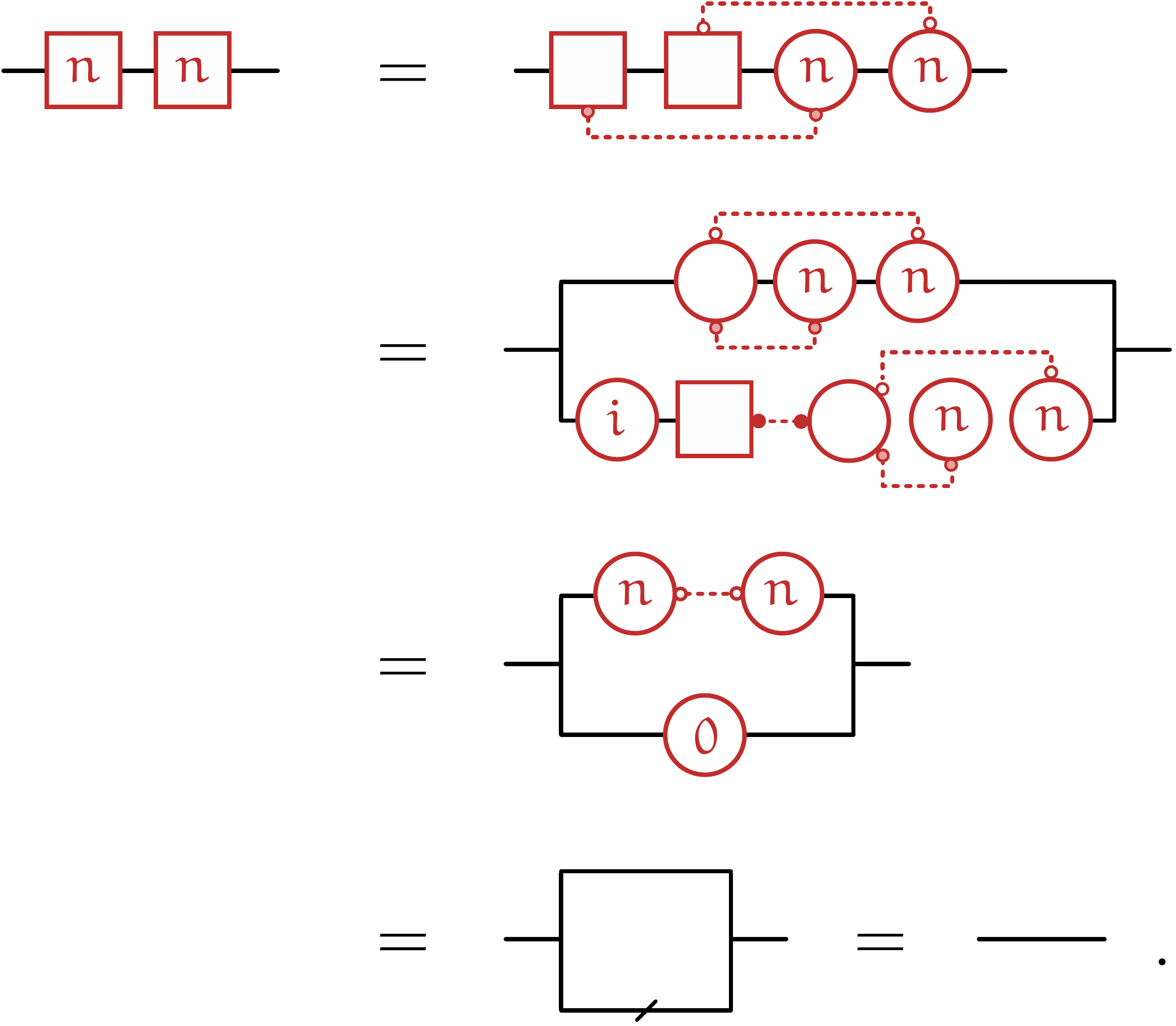}
  \caption{Proof that $\sigma(\mathbf{n})^2 = I$. A small modification
    shows that \[\sigma(\mathbf{m})
  \sigma(\mathbf{n}) = (\mathbf{m}\cdot \mathbf{n})I +
  i\sigma(\mathbf{m}\times \mathbf{n}),\] where $\mathbf{n}\cdot
\mathbf{m}=n_im_i$ is the dot and $(\mathbf{n}\times \mathbf{m})_i =
\epsilon_{ijk}n_jm_k$ the cross product.}
  \label{fig:simple5}
  \vspace{-2pt}
\end{figure}

\noindent
The role of the $\delta_{ij}$ (line $2$) is to connect incoming tether
lines,\marginnote{\vspace{-23pt}\\ \noindent Details for $\epsilon_{ijk}$: $n_i n_j \delta_{ij} = n_k n_k$, and
  $n_i n_j \epsilon_{ijk}=-n_i n_j \epsilon_{jik} = -n_i n_j
  \epsilon_{ijk}$, where we used antisymmetry, relabelled
  $i\leftrightarrow j$ and commuted $n_i, n_j$. It follows that $n_i n_j
  \epsilon_{ijk}=0$.}
while $\epsilon_{ijk}$ vanishes (line $3$) when it detects repetition.
Finally, as in the Boolean setting, inserting $0$ ``snips'' the
circuit, leaving the identity. 

This shows that \gls{awd}s can be useful for syntactic manipulation. But
what does it all \emph{mean}?\marginnote{Whoop-de-doo Basil.}
Like truth tables in the Boolean case, we will introduce a semantics of
quantum measurement, with switches we can toggle and compose.
But this apparently simple goal will take us deeper into the realms of
algebra and functional analysis than might have been expected (or hoped);
for better and worse, we feel the complications of living in a
noncommutative world.

\section{4. States as functionals}\hypertarget{sec:5}{}

In a classical circuit, a \emph{state} flips each switch into the on
or off position and sets the corresponding value of a conductance measurement. Let
$\mathcal{P} = \{x_1, x_2, \ldots, x_n\}$ denote a set of $n$
propositional switches, $\mathbb{B}[\mathcal{P}]$
the set of all circuits built using them.
We can define a state
mathematically as an
assignment $v_0: \mathcal{P} \to \mathbb{B}$, and the \emph{truth table}
as the set of all such maps, which we can view, suggestively, as a
binary vector space $\mathbb{B}^{\mathcal{P}}$.\marginnote{The
  notation $C^D$, for sets $C$ and $D$, means the set of all functions $f: D \to C$.}
States can be uniquely extended from switches to any circuit via 
\marginnote{\vspace{10pt}
  
  \begin{center}
    \includegraphics[width=0.9\linewidth]{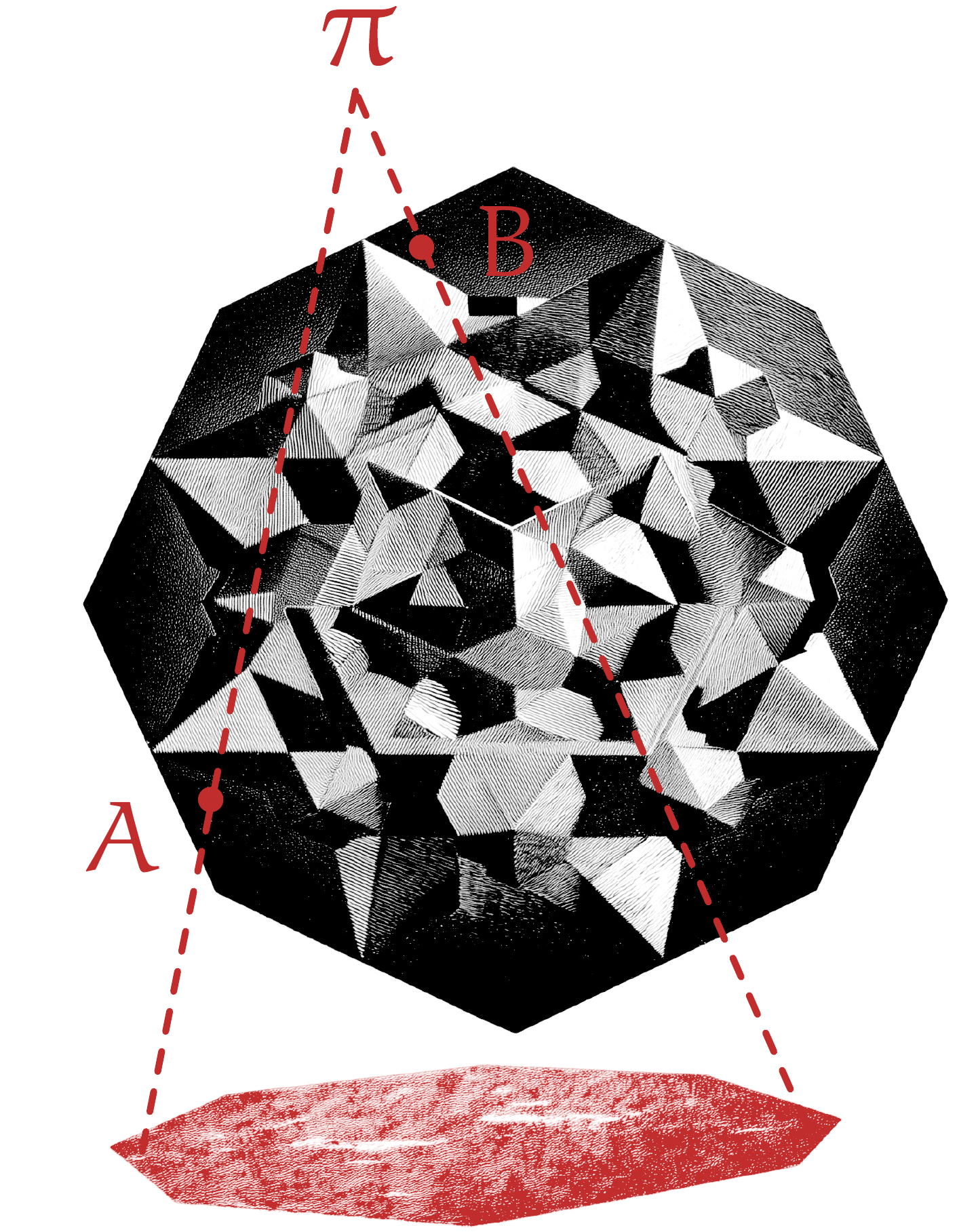}
  \end{center}  \vspace{-5pt}
  \emph{A state projects the full algebra onto a consistent set of
    expectations (red).
  }
}
\begin{equation}
  v(b \eta + b'\zeta) = b v(\eta) + b' v(\zeta), \quad v(\eta\zeta) =
  v(\eta)v(\zeta), \quad v|_{\mathcal{X}}=v_0\label{eq:bool-val}
\end{equation}
for $b, b' \in \mathbb{B}$ and $\eta, \zeta \in
\mathbb{B}[\mathcal{P}]$.
In words, $v$ must be \emph{linear} over $\mathbb{B}$,
\emph{multiplicative}, and agree on switches.
This ensures that
each state in $\mathbb{B}^{\mathcal{P}}$ uniquely determines
measurements of circuits.

Analogously, we want our noncommutative states to consistently
assign measurements to each operator $A \in \mathcal{A}$.
Our job will be to figure out what ``consistent'', ``assign'' and
``measurement'' mean!
We take as an axiom that measurement of $A$ yields a
  random element of the set of eigenvalues, or \emph{spectrum} of $A$:
\begin{equation}
  \label{eq:spectrum}
  \mathfrak{S}(A) = \{ \lambda \in \mathbb{C} : (A - \lambda I)^{-1} \text{
    does not exist}\}.
\end{equation}
For a quantum state $\pi$, we let $A|_\pi$ denote the random variable
corresponding to an operator $A$. 

Random variables are a little unwieldy to work with directly;
instead, we'll use the averages of $A_\pi$, and
think of a state as an \emph{expectation
  functional} $\pi: \mathcal{A}\to\mathbb{C}$ giving\marginnote{A
  functional is any map from a
  vector space to its underlying scalars, e.g. a Boolean valuation
$v: \mathbb{B}[\mathcal{P}] \to \mathbb{B}$.}
\begin{equation}
  \pi(A) = \mathbb{E}\big[A|_\pi\big]\label{eq:state-exp}.
\end{equation}
We picture this as an dotted circle around a circuit:

\begin{figure}[h]
  \centering
  \vspace{-5pt}
  \includegraphics[width=0.4\textwidth]{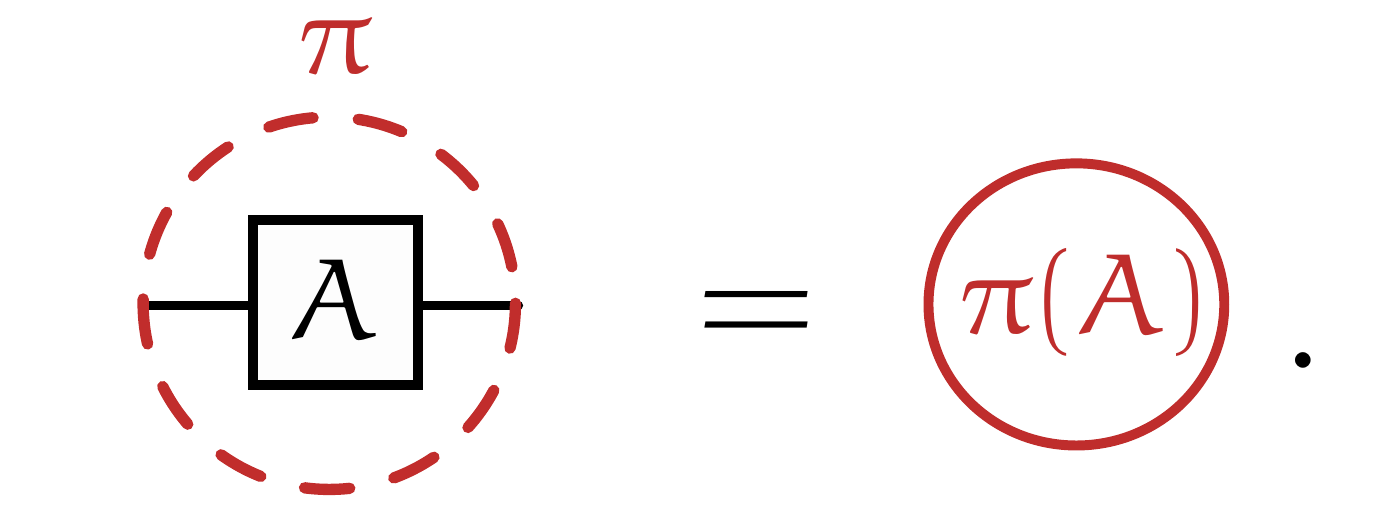}
  \label{fig:state1}
  \vspace{-5pt}
\end{figure}

\noindent The circle reminds us that the end result is a scalar, while
the dots tell us it takes an operator as a argument.\marginnote{This
  will be consistent with our later notation for a \emph{\gls{channel}},
  which takes an operator to an operator. See Fig. \ref{fig:conj2}.}

This isn't a precise definition, since we haven't explained how to determine
the random variable $A|_\pi$; instead, it is an ``intuition pump'' we
can use to motivate properties we want $\pi$ to have.
First of all, expectation is linear, so
\begin{equation}
  \pi (\alpha A + \beta B) = \alpha \pi (A) + \beta \pi
  (B).\label{eq:state-linearity}
\end{equation}
We represent by extending our \gls{awd} notation:

\begin{figure}[h]
  \centering
  \vspace{-5pt}
  \includegraphics[width=0.57\textwidth]{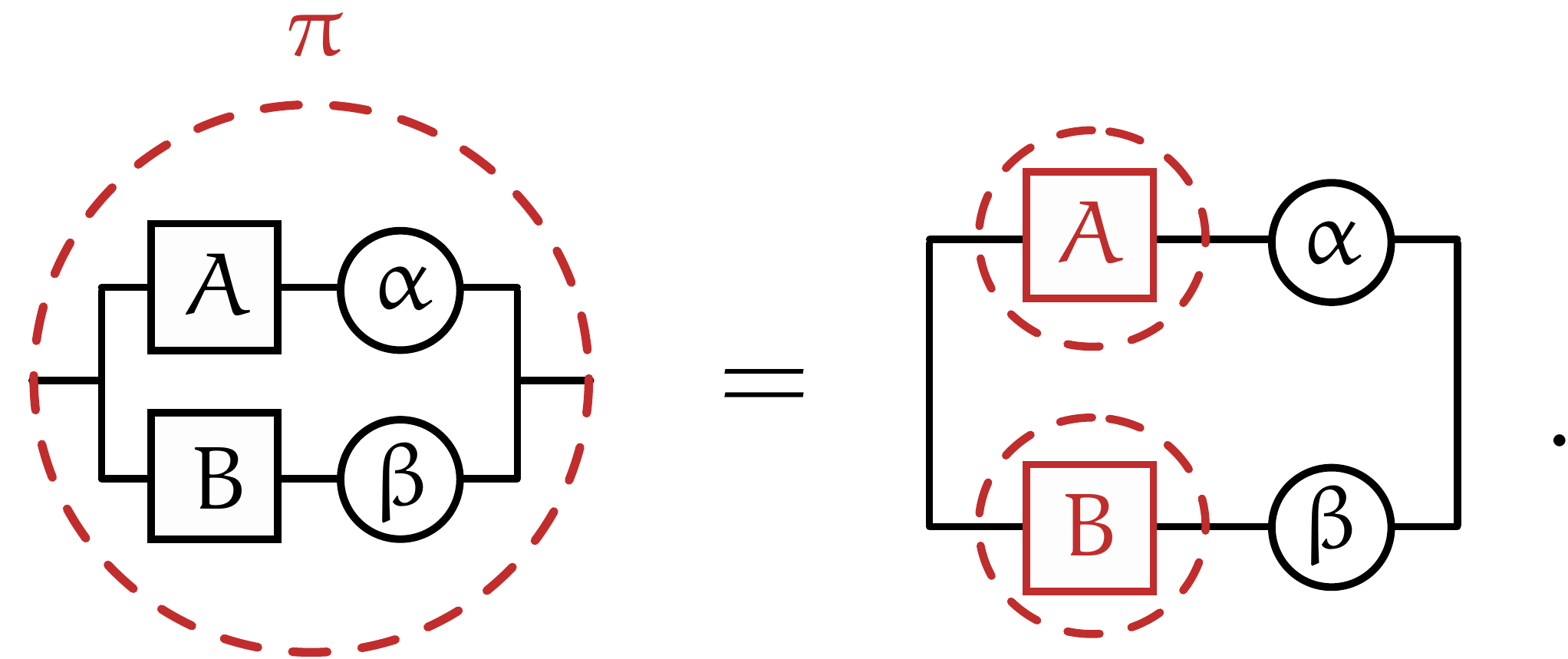}
  \label{fig:state2}
  \vspace{-8pt}
\end{figure}

\noindent The sum is still indicated by the parallel structure, but we remove the
trailing wires to indicate this is no longer an operator.

Since the identity $I$ 
  has a unique eigenvalue $1$, we measure it with
  certainty.
Thus, the fact that probabilities are normalized requires\marginnote{Equivalently, we say $\pi$ has \emph{unit norm} since the
  operator norm on linear functionals is given by
  \begin{equation}
    \Vert \pi\Vert_* = \text{sup}_{\Vert A\Vert \leq 1}
    \pi(A).\label{eq:pi_norm}
  \end{equation}
  This obeys $|\pi(A)| \leq \Vert \pi\Vert_* \Vert A\Vert$, and hence
  for a unital algebra, $\Vert \pi\Vert_* = \pi(I)$.
}
\begin{equation}
  \pi (I) = 1.\label{eq:state-norm}
\end{equation}
Finally, the average over a set of positive values should be
positive. A positive operator is one with nonnegative
eigenvalues, or equivalently, of the form $B = A^\ast A=|A|^2$, often
written $B \geq 0$.
Since the associated measurements are nonnegative, so is the expectation:
\begin{equation}
  \label{eq:state-pos}
  \pi(|A|^2) \geq 0.
\end{equation}
Again, we depict normalization and positivity as follows:
\begin{figure}[h]
  \centering
  \vspace{-3pt}
  \includegraphics[width=0.4\textwidth]{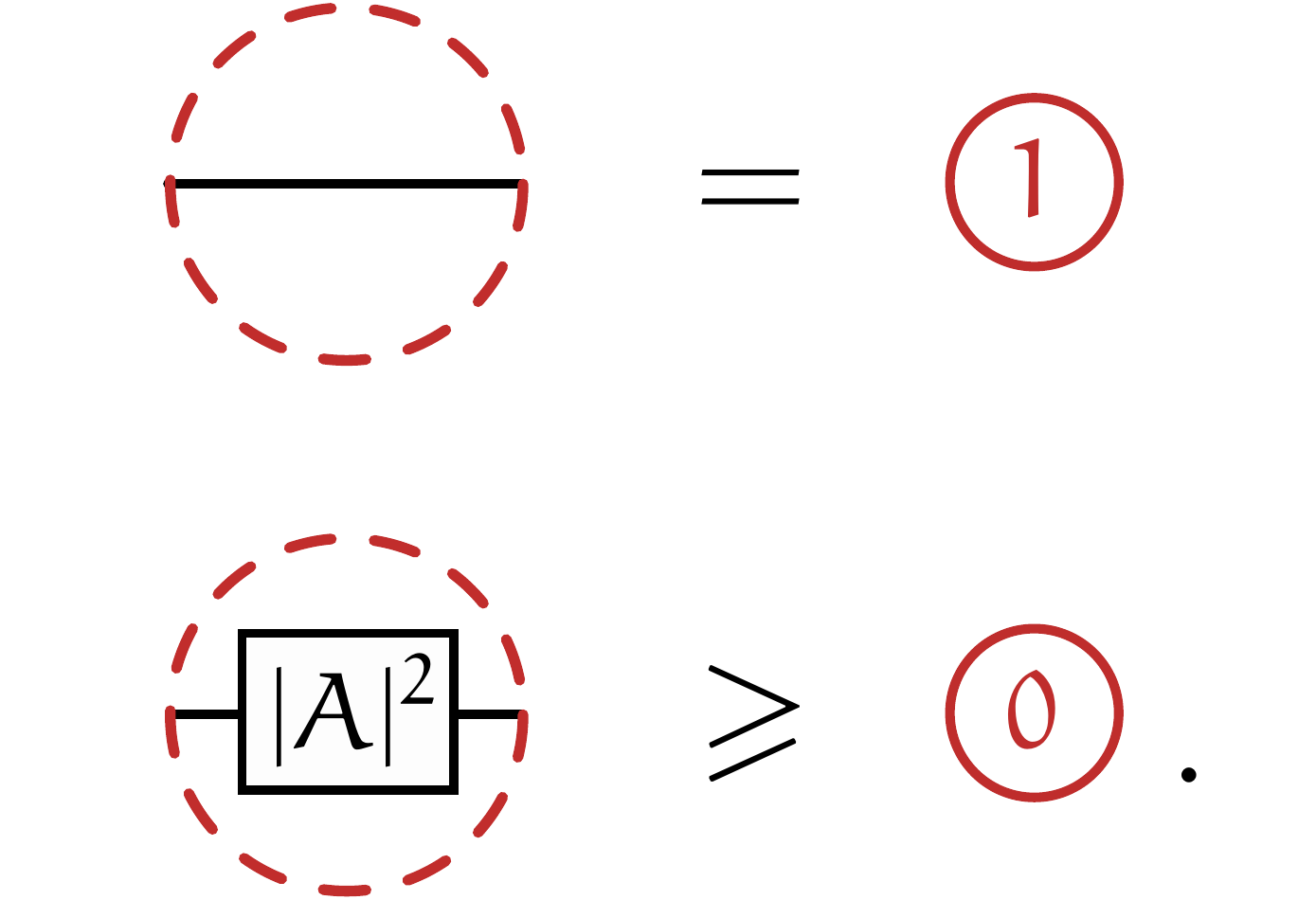}
  \label{fig:state2}
  \vspace{-5pt}
\end{figure}

\noindent Nonpositive operators can have  
arbitrary complex expectations.

\begin{figure}[h]
  \centering
  \vspace{2pt}
  \includegraphics[width=0.9\textwidth]{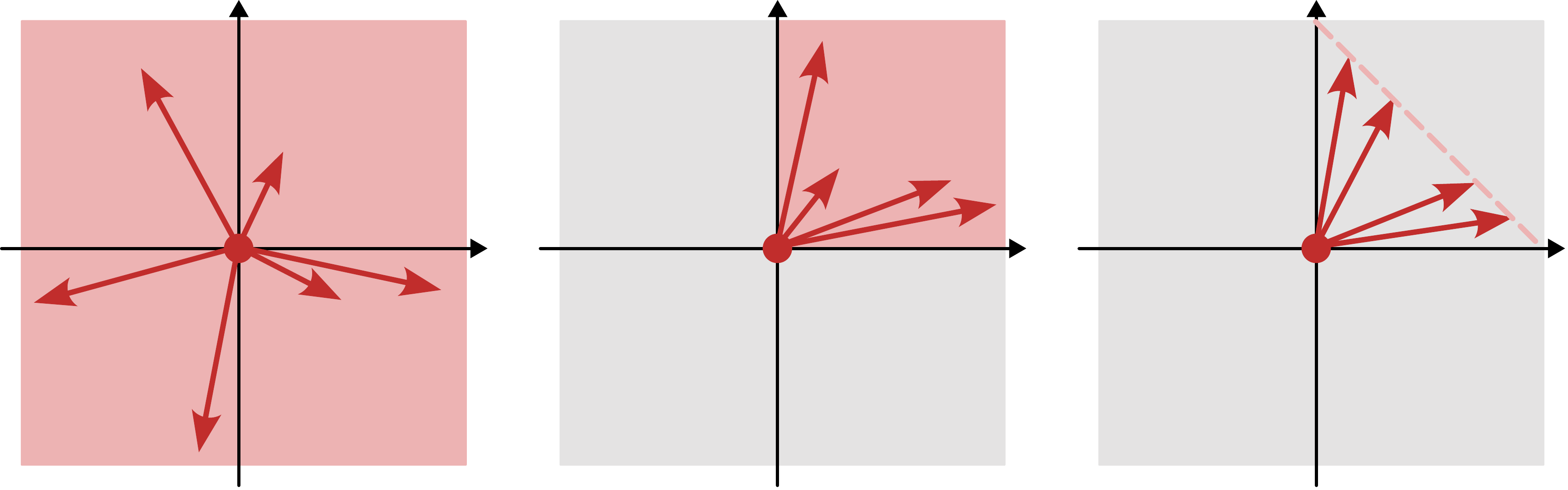}
  \caption{A cartoon of real linear functionals. \textsc{Left}. Linear functionals as vectors
    in the plane. \textsc{Middle}. Positive functionals, where each
    component $\geq 0$. \textsc{Right}. Unit norm positive
    functionals.
  }
  \label{fig:plf}
  \vspace{-5pt}
\end{figure}

In axiomatic approaches\sidenote{See for instance \emph{Foundations of
    the theory of probability} (1933), Andrei Kolmogorov. These
  follow from the probability axioms and the measure-theoretic definition of expectation.} to expectation,
the properties (\ref{eq:state-linearity})--(\ref{eq:state-pos}) are
not only reasonable but complete; we can use them to define
which functions behave like expected values of random variables.
This suggests that we can define a state on a C${}^*$-algebra $\mathcal{A}$ as any map $\pi: \mathcal{A} \to \mathbb{C}$ satisfying
(\ref{eq:state-linearity})--(\ref{eq:state-pos}), or in words, a
\emph{positive linear functional of
  unit norm}. See Fig. \ref{fig:plf} for a cartoon.

\noindent We call the full collection of states
$S(\mathcal{A})$. Note that, unlike Boolean valuations, where we
demand $v(xy) = v(x)v(y)$, we say
\emph{nothing} about products in the noncommutative case. What gives?

The basic problem is that it is not clear how to define the value of $\pi$ on some
set of ``basic switches'' and extend it globally.
This happens for a variety of reasons.
First, because measurement of a product does not
equal a product of measurements, the random variable structure is not
multiplicative:
\begin{equation}
  AB|_\pi \neq A|_\pi \cdot B|_\pi.\label{eq:rv-pi}
\end{equation}
Even if we had equality in (\ref{eq:rv-pi}), the
expectations wouldn't factorize unless $A|_\pi$ and $B|_\pi$ were
uncorrelated as random variables.\marginnote{To illustrate, consider an
invertible $A$ with eigenvalues $\lambda_i > 0$, and inverse $A^{-1}$
with eigenvalues $\lambda_i^{-1} > 0$. The product $\mathfrak{S}(AA^{-1}) = \{1\}$.}
Finally, we might hope for some clever way to compose or relate the
two sides of (\ref{eq:rv-pi}), short of equality. Unfortunately, there
is \emph{no simple relationship} between the spectra, i.e. the
outcomes $\mathfrak{S}(AB)$ and $\mathfrak{S}(A)$, $\mathfrak{S}(B)$.

Evidently, we're thinking about this the wrong way.
Progress will require us to zoom out and think about the global shape
of any functional satisfying
(\ref{eq:state-linearity})--(\ref{eq:state-pos}).
By carefully studying the multiplicative patterns in such a
functional, we can reverse engineer a switch.


\addtocontents{toc}{\protect\vspace{-20pt}\protect\contentsline{part}{\textsc{\Large{the gns construction}}}{}{}}

\section{5. Patterns of correlation}\hypertarget{sec:6}{}

We'll start by defining a simple object which encodes those
multiplicative patterns.
The \emph{correlation} of $A$ and $B$ in state $\pi$ is the expectation
of the random variable associated with their product:
\begin{equation}
  \label{eq:corr}
  G_\pi(B, A) = 
  \pi (B^\ast A).
\end{equation}
We use the convention\marginnote{If necessary, we add a
  dot or other marker in the corner to disambiguate.} that reflecting an object
in the $x$-axis corresponds to taking the adjoint. Thus:

\begin{figure}[h]
  \centering
  \vspace{-5pt}
  \includegraphics[width=0.42\textwidth]{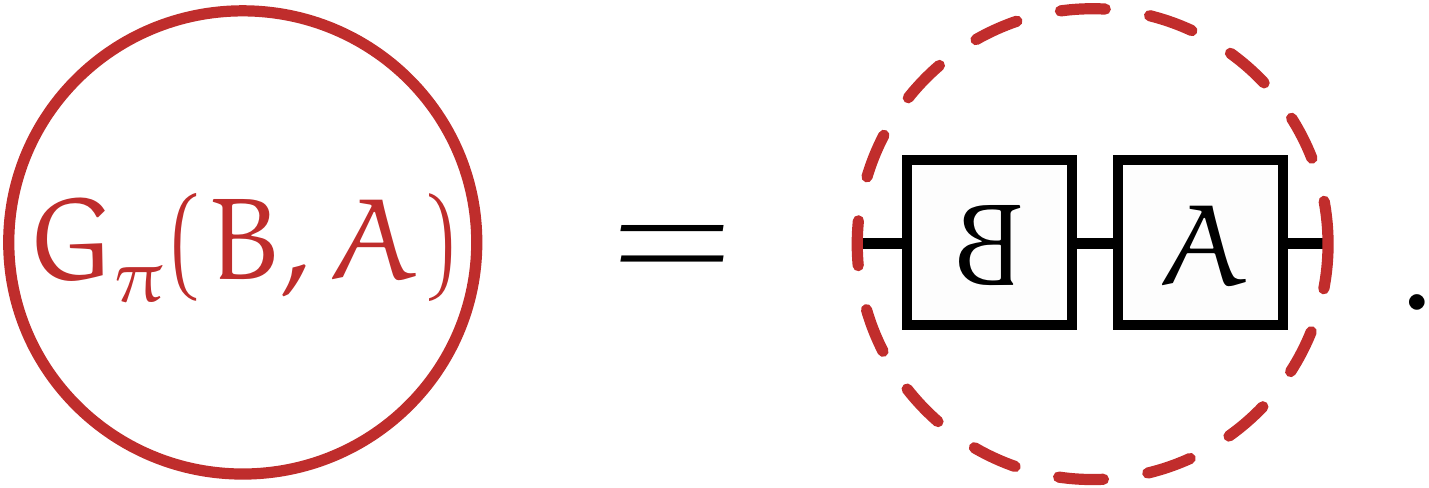}
  \label{fig:state4}
  \vspace{-8pt}
\end{figure}

\noindent Similarly, a reversed scalar $\lambda$ is the complex
conjugate $\lambda^*$.

We use $B^\ast$ in definition (\ref{eq:corr}) rather than $B$ to ensure that an observable is
positively correlated with itself, i.e. 
(\ref{eq:state-pos}) tells us that
\begin{equation}
  G_\pi(A, A) = \pi (|A|^2) \geq 0.\label{eq:sesq1}
\end{equation}
The correlation is linear in the second argument,\marginnote{Note that
  this is consistent with the physics literature, but the opposite of
  the math literature. Notation is the ultimate zero sum game!}
\begin{align}
  G_\pi(B, \alpha_1 A_1 + \alpha_1 A_1) & = \pi[B^\ast (\alpha_1 A_1 + \alpha_2 A_2) ]
                                 \notag \\ & =
  \alpha_1 \pi(B^\ast A_1) + \alpha_2 \pi(B^\ast A_2) \notag \\ & = \alpha_1
                                                           G_\pi(B,
                                                           A_1) + \alpha_2
                                                           G_\pi(B, A_2), \label{eq:sesq2}
\end{align}
and similarly \emph{antilinear} in the first.
These three properties---nonnegativity, linearity, and antilinearity---make correlation a
``positive-semidefinite sesquilinear form''.
The name is a mess, but the properties guarantee something neat: the
\emph{Cauchy-Schwarz
  inequality}.\sidenote{A nice blog post:
  \href{https://terrytao.wordpress.com/2007/09/05/amplification-arbitrage-and-the-tensor-power-trick/}{``Amplification,
    arbitrage, and the tensor power trick''} (2007), Terry Tao. Even
  with nonnegativity, the proof goes through. We can also give a
  visual ``proof'':
  \vspace{3pt}
  \begin{center}
    \includegraphics[width=0.26\textwidth]{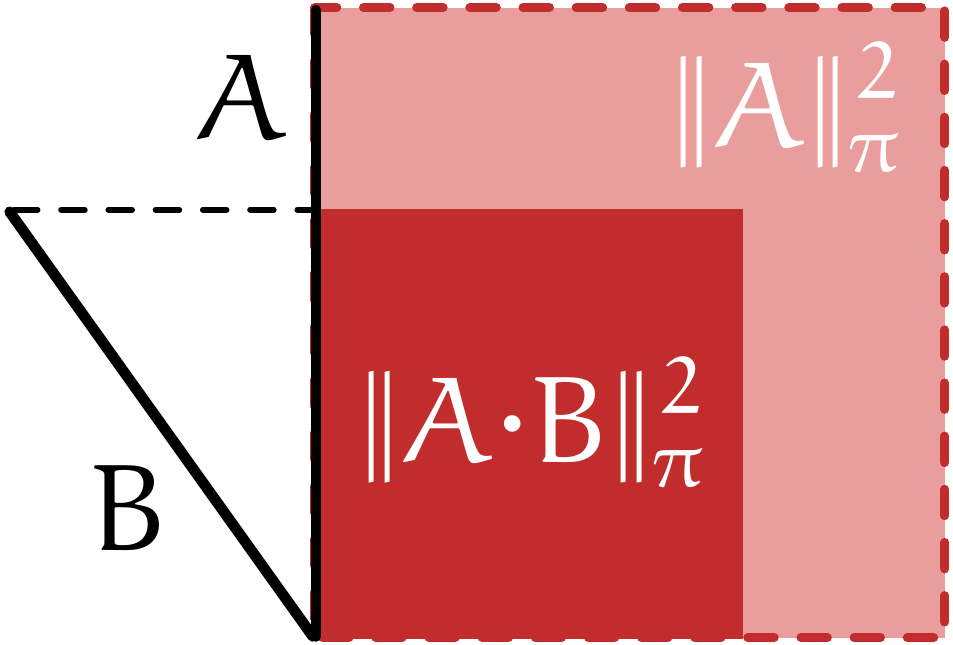}
  \end{center}
where (without loss of generality) $B$ is unit norm. The intuition is that projecting is always
norm-decreasing.}
Using metric-style notation, $G_\pi(A, A) = \Vert A\Vert_\pi^2$, we have
\begin{equation}
  \label{eq:sesq3}
  |G_\pi(B, A)|^2 \leq \Vert A\Vert_\pi^2 \Vert B\Vert_\pi^2. 
\end{equation}

\noindent The most important application of (\ref{eq:sesq3}) is to operators $\Theta$ with
\emph{vanishing} self-correlation, $G_\pi(\Theta, \Theta) =0$.
We call these operators \emph{null}, and the collection of such
operators the \emph{null space} or \emph{\gls{kernel}} of $\pi$, $\mathcal{K}_\pi$. 

It follows that the correlation with any operator $A \in \mathcal{A}$
vanishes:
\begin{equation}
  \Theta\in\mathcal{K}_\pi \quad \Longleftrightarrow \quad |G_\pi(\Theta, A)|^2 \leq \Vert A\Vert_\pi^2 \Vert \Theta\Vert_\pi^2
  = 
  0.\label{eq:sharp1}
\end{equation}
Equivalently, if $\Delta \Gamma = \Gamma - \pi(\Gamma)I$ is null, then linearity of $G_\pi$
implies that expectations factorize:
\begin{equation}
  \label{eq:sharp2}
  \pi(\Gamma A) = \pi(\Gamma) \pi(A) = \pi(A \Gamma).
\end{equation}
We call such a $\Gamma$ \emph{\gls{sharp}} or \emph{definite}. In \gls{awd}s, we
indicate this with a dotted channel passing through $\Gamma$, to
suggest we can factorize the state contour through the operator. For instance, (\ref{eq:sharp2})
corresponds to slicing, swapping, and sewing the operators back together:

\begin{figure}[h]
  \centering
  \vspace{-3pt} 
  \includegraphics[width=0.82\textwidth]{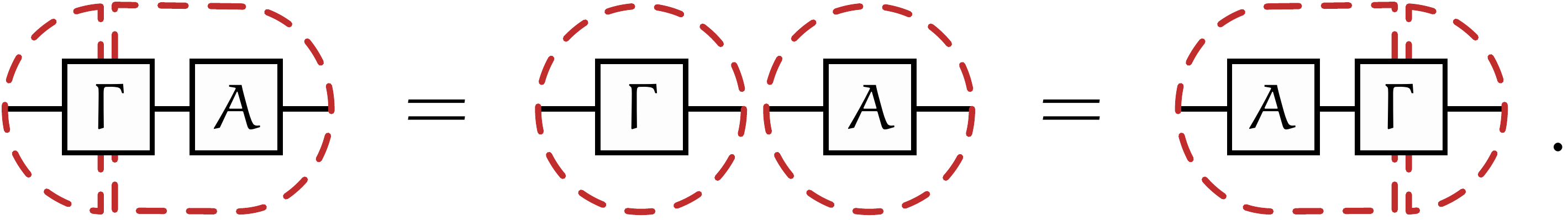}
  \caption{Factorizing an expectation through a \gls{sharp} operator.}
  \label{fig:null}
  \vspace{-5pt}
\end{figure}

\noindent We have to be careful, though:\marginnote{For instance, for an
  eigenstate of $Z$ and the Hadamard $H=\tfrac{1}{\sqrt{2}}(X+Z)$, we have
  $\pi(HZH)=\pi(X)=0$, but $\pi(H)\pi(Z) \pi(H)=\tfrac{1}{2}$. For the
  correct intepretation, see (\ref{eq:sharp4}).
} we cannot factorize any
product along $\Gamma$, since $\pi(A\Gamma\hspace{1pt}B) \neq
\pi(A)\pi(\Gamma)\pi(B)$ in general. We'll discuss the general
interpretation below.

\begin{figure}[h]
  \centering
  \vspace{-3pt}
  \includegraphics[width=0.33\textwidth]{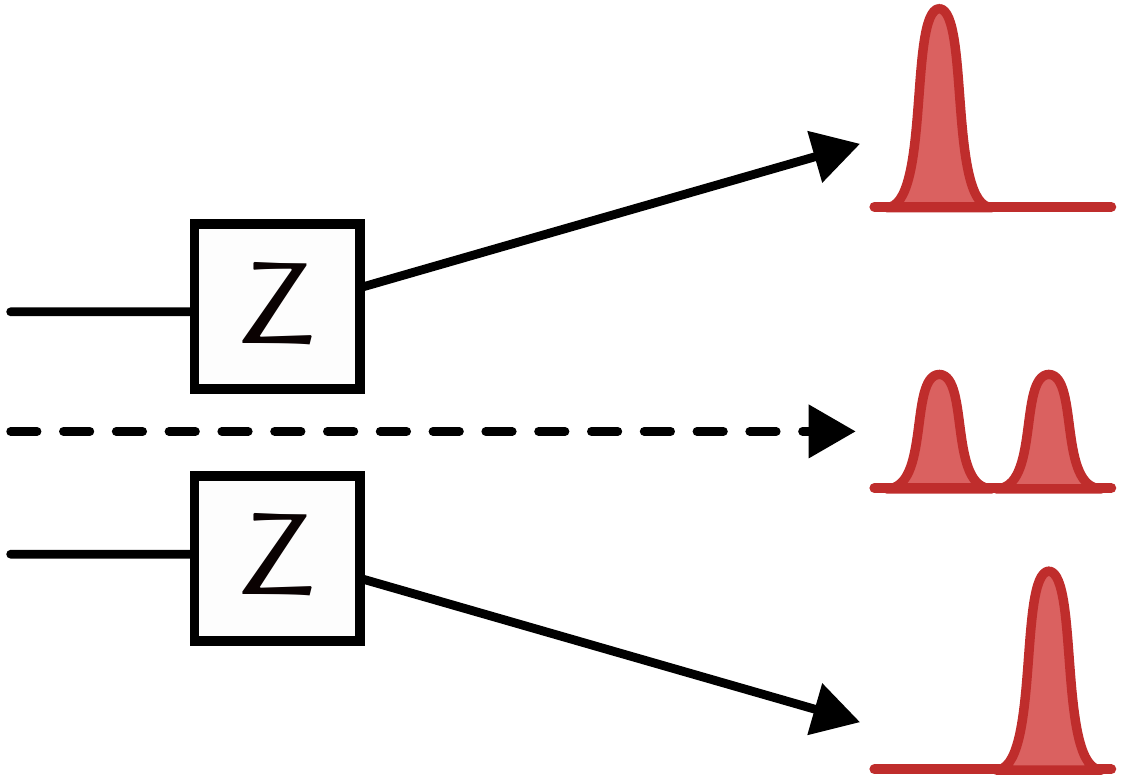}
  \caption{Cartoons of zero variance measurements when $Z \measure \pm 1$ has
  a deterministic value.}
  \label{fig:simple7}
  \vspace{-3pt}
\end{figure}

\Gls{sharp} operators arise from deterministic measurements.
If $\Gamma$ takes value $\pi(\Gamma)$ with certainty, then $\Delta\Gamma$ is null:
\begin{equation}
  \Vert \Delta \Gamma\Vert_\pi^2 = \pi(|\Gamma|^2) - |\pi(\Gamma)|^2 =
  \text{var}_\pi(\Gamma) = 0,\label{eq:sharp3}
\end{equation}
since $\Gamma$ has zero variance.
The \emph{\gls{definite}} is the set of \gls{sharp}
operators which are also self-adjoint:
\begin{equation}
  \mathcal{D}_\pi = 
  \big(\mathbb{C}I + \mathcal{K}_\pi\big) \cap \mathcal{A}_\text{sa}
  = \mathbb{R}I + \left(\mathcal{K}_\pi
  \cap \mathcal{K}_\pi^*\right),
\end{equation}
where the first equality is a definition, and the second a
theorem.\sidenote{``Extensions of pure states'' (1959), Kadison and
  Singer. This paper is famous for spawning the \emph{Kadison-Singer
    problem}, only resolved in 2017.}
Let's see how this plays out in the Stern-Gerlach experiment.
Suppose we measure $Z \measure +1$ with probability
$p$, where $\Lambda \measure \lambda$ means that measuring
$\Lambda$ yields outcome $\lambda$.\marginnote{\vspace{-2pt}\\

  \noindent Measurement has
  many similarities to variable assignment, so we borrow the
  assignment operator notation $\measure$ used in pseudocode.}
The average is
\[
  \pi(Z) = (+1)p + (-1)(1 - p) = 2p - 1.
\]
A little algebra then
shows 
\begin{equation}
  \pi(\Delta Z^2) = 
  \pi(I) - (2p - 1)^2 = 1 - (2p - 1)^2 = 4p(1-p).\label{eq:pi_pZ}
\end{equation}
We find that $Z$ is definite just in case $p=0$ or
$p=1$,
i.e. we measure $Z \measure \pm 1$ with certainty. This makes sense! We
picture this in Fig. \ref{fig:simple7}.

Returning to the main thread, how can we use these tools to understand
the behaviour of measurements?
The first step will be to unpack the structure of the \gls{kernel}.
First, note that since $G_\pi$ satisfies Cauchy-Schwarz (\ref{eq:sesq3}), it also
obeys the \emph{triangle inequality}:\marginnote{Expand
  $\Vert A+B\Vert_\pi^2$, upper
  bound $\text{Re} [G_\pi(A, B)]$ with Cauchy-Schwarz, and factorize
  into $(\Vert A
  \Vert_\pi^2+\Vert B\Vert_\pi^2)^2$. There is an even
  simpler visual ``proof'':
  \vspace{-2pt}
  \begin{center}
    \includegraphics[width=0.22\textwidth]{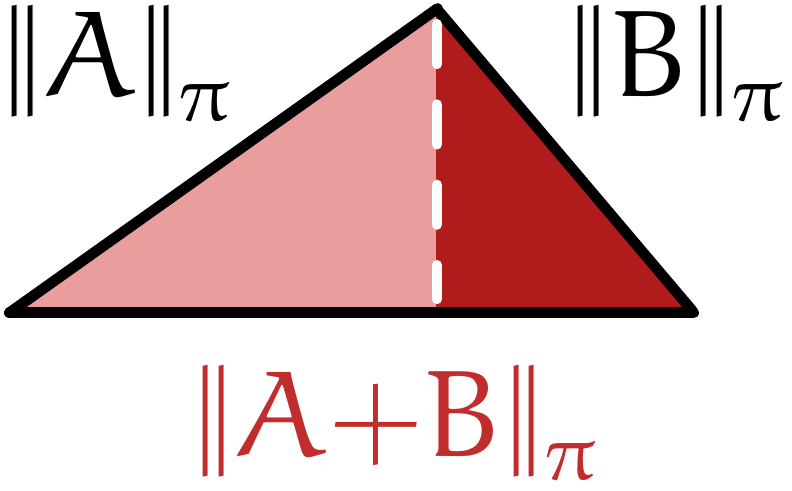}
  \end{center}
  The sum projects $A$ and $B$ in a common direction, hence decreases both norms.
}
\begin{equation}
  \label{eq:triangle}
  \Vert A + B \Vert_\pi^2 \leq \Vert A\Vert_\pi^2+\Vert B\Vert_\pi^2.
\end{equation}
It follows immediately that $\mathcal{K}_\pi$ is closed under sums,
and hence any linear combination. This makes it a
vector subspace of $\mathcal{A}$.
Similarly, $\mathcal{K}_\pi$ is closed under products. In fact, it is
closed under left multiplication by \emph{arbitrary} operators. Taking
$\Theta\in \mathcal{K}_\pi, A \in \mathcal{A}$,
\begin{align}
  \Vert A\Theta \Vert_\pi^2 & = G_\pi(A\Theta, A\Theta)
                                 = \pi[(A\Theta )^\ast A\Theta] = 0
    \label{eq:sharp-ideal}
\end{align}
using (\ref{eq:sharp1}).
Thus, $\mathcal{K}_\pi$ is a vector subspace (closed under linear
combinations), a \emph{subalgebra} (closed under products), and a
\emph{left ideal} (closed under left-multiplication by any element of $\mathcal{A}$).

To see why this might be useful,
suppose we can write a projector $\Pi_\mathcal{K}$ onto the
\gls{kernel} of $\pi$, with an orthogonal projector
$\Pi_\mathcal{K}^\perp$.\marginnote{Consult (\ref{eq:proj}) if you've
forgotten what a projector is.}
The null projector has the property that, for any $A \in\mathcal{A}$,
\[
  \pi (\Pi_\mathcal{K} A ) = \pi(A \Pi_\mathcal{K}) = 0.
\]
Notice that we can split $A$ into four pieces using these projectors:
\[
  A = \Pi_\mathcal{K} A \Pi_\mathcal{K} + \Pi_\mathcal{K} A
  \Pi_\mathcal{K}^\perp + \Pi_\mathcal{K}^\perp A \Pi_\mathcal{K} +
  \Pi_\mathcal{K}^\perp A \Pi_\mathcal{K}^\perp.
\]
Since anything with $\Pi_\mathcal{K}^\perp$ on either side vanishes
inside $\pi$, we can write
\begin{equation}
  \label{eq:restrict-proj}
  \pi(A) = \pi(\Pi_\mathcal{K}^\perp A \Pi_\mathcal{K}^\perp).
\end{equation}
More prosaically, this restricts $A$ to the part that acts
nontrivially on the state $\pi$.
Incidentally, this leads to the generalization of (\ref{eq:sharp2}):
if $\Delta \Gamma$ is null, then $\Delta \Gamma =
\Pi_{\mathcal{K}}(\Delta \Gamma) \Pi_{\mathcal{K}}$, and hence (with a
little algebra)
\begin{equation}
  \label{eq:sharp4}
  \Gamma = \Pi_{\mathcal{K}} \Gamma \,\Pi_{\mathcal{K}} +
  \Pi_{\mathcal{K}}^\perp \pi(\Gamma) \Pi_{\mathcal{K}}^\perp.
\end{equation}
This is what the gap in (\ref{fig:null}) instructs us to sum over!

On the outside
of a product, the product of $\Pi_{\mathcal{K}}$ and
$\Pi_{\mathcal{K}}^\perp$ vanishes, and we can replace $\Gamma$ with
its expectation. But for $\pi(A\Gamma\,
B)$, both components of (\ref{eq:sharp4}) survive.
Shuffling around these projectors seems to involve a lot of
bookkeeping!
But we are in luck: we can revive Hilbert space to do our bookkeeping
for us.

\section{6. Hilbert space redux}\hypertarget{sec:7}{}

Let's assemble what we know so far. A state $\pi \in S(\mathcal{A})$ is a
positive, linear functional of unit norm, $\pi: \mathcal{A} \to
\mathbb{C}$. For any state, the correlation $G_\pi(B, A)$
is \emph{almost} an inner product, but has vanishing norm on the \gls{kernel} $\mathcal{K}_\pi$.
What it lacks is \emph{positive-definiteness}:
\[
  \Vert v \Vert = 0 \quad \Longleftrightarrow \quad v = 0.
\]
To make it positive-definite, we need to somehow get rid of $\mathcal{K}_\pi$, or identify it with the zero operator.

Luckily, there is a mathematical procedure for identifying things with zero
called a \emph{quotient}.
First, we lump together everything that differs by a null
operator, called a \emph{null equivalence class}:\marginnote{An \emph{equivalence
  relation} $R \subseteq S \times S$ is a relation which is (a)
  \emph{reflexive}, $(s, s) \in R$ for all $s \in S$; (b) \emph{symmetric}, $(s, t) \in
  R$ implies $(t, s) \in R$; and (c) \emph{transitive}, $(s, t) \in R$
and $(t, u)\in R$ implies $(s, u) \in R$. An \emph{equivalence class}
under $R$ is a set of objects related to each other. We write $s
\equiv_R t$ for $(s, t) \in R$.}
\begin{equation}
  [A]_\pi = \{B \in \mathcal{A} : A - B
\in \mathcal{K}_\pi\} = A + \mathcal{K}_\pi. \label{eq:class2}
\end{equation}
The \emph{\gls{quotient}} $\mathcal{A}/\mathcal{K}_\pi = \{[A]_\pi\}_{A
\in \mathcal{A}}$ is the set of all equivalence classes. 
This is a vector space, and a linear combination of classes 
\begin{align}
  \label{eq:class-lin}
\alpha [A]_\pi + \beta [B]_\pi & = [\alpha A + \beta B]_\pi,
\end{align}
is well-defined because $\mathcal{K}_\pi$ is itself closed under
linear combinations.
On this vector space, $G_\pi$ ``lifts'' to a nondegenerate inner product:
\begin{equation}
  \langle [B]_\pi, [A]_\pi\rangle =
  G_\pi(B, A) = \pi(B^\ast A),\label{eq:inner}
\end{equation}
with $[0]_\pi = \mathcal{K}_\pi$ playing the role of zero.
The first equality follows because we can replace $B$ with any element
of its null equivalence class and get the same correlation, due to (\ref{eq:sharp1}).


A vector space with an inner product is almost, but not quite, a
Hilbert space, since it may have ``holes'' with
respect to the metric $\Vert A\Vert_\pi$.
We just used one math hack---quotients---to get rid of the pesky null
operators.
We can use another hack---\emph{\gls{completion}}---to fill in the holes.\sidenote{See, e.g. \emph{Principles of Mathematical Analysis} (1953),
  Walter Rudin. The basic idea is to form equivalence classes of
  Cauchy sequences in $\mathcal{M}$, the same way we can build $\mathbb{R}$ from
  sequences of approximations in $\mathbb{Q}$. This yields a space
  $\overline{\mathcal{M}}$ which is complete since (roughly speaking) a Cauchy sequence converges to itself!}
The result is a Hilbert space
\begin{equation}
  \label{eq:8}
  \mathcal{H}_\pi = \overline{\mathcal{A}/\mathcal{K}_\pi}^{\Vert
    \cdot \Vert_\pi},
\end{equation}
where $\overline{\mathcal{M}}^{\Vert\cdot\Vert}$ denotes the \gls{completion} of a
metric space $\mathcal{M}$ with respect to a norm $\Vert\cdot\Vert$.
This \emph{completion} is unique.
Before you get too worried about all the hacks at play, you can take
solace in the fact that \emph{finite-dimensional} examples never have any
holes to fill!

Since $\mathcal{K}_\pi$ is a left ideal,
$A\mathcal{K}_\pi \subseteq \mathcal{K}_\pi$ for any $A \in \mathcal{A}$,
and hence we can left-multiply an equivalence class:
\begin{align}
  A \cdot [B]_\pi & = A(B + \mathcal{K}_\pi) \notag \\
  & = AB + A\mathcal{K}_\pi \notag \\
  & = AB + \mathcal{K}_\pi = [AB]_\pi. \label{eq:rep-A}
\end{align}
Since $A$ is linear\marginnote{Boundedness follows from the fact that
  $|\pi(A)| \leq \Vert A\Vert$; see margins near (\ref{eq:state-norm}).
} and maps $\mathcal{H}_{\pi}$ to
itself, it acts as a linear operator we denote $A_\pi$ (not to be
confused with the random variable $A|_\pi$) on
$\mathcal{H}_{\pi}$. It's not hard to see that it's bounded, so each
$A_\pi \in \mathcal{B}(\mathcal{H}_\pi)$.\marginnote{In contrast to random
variables (\ref{eq:rv-pi}), this map \emph{is} multiplicative,
\[
  (AB)_\pi = A_\pi B_\pi,
\] which follows immediately from (\ref{eq:rep-A}):
\begin{align*}
  (AB)_\pi [C]_\pi & = [ABC]_\pi \\
  & = A_\pi [BC]_\pi = A_\pi B_\pi[C]_\pi,
\end{align*}
for an arbitrary vector $[C]_\pi \in\mathcal{H}_\pi$.
}

We can restore some of our Hilbert space intuition by using bra-ket
notation for (not necessarily normalized) vectors $[A]_\pi$. The kets live in the second slot of
the inner product, and the bras in the first:
\begin{equation}
  |A\rangle_\pi = G_\pi(\cdot, A) = \langle [\cdot]_\pi,
  [A]_\pi\rangle, \quad {}_\pi\langle B| = G_\pi(B, \cdot) = \langle
  [B]_\pi, [\cdot]_\pi\rangle.\label{eq:braket}
\end{equation}
This makes kets linear and bras antilinear.
We can capture these visually by splitting a state into two dotted semicircles:

\begin{figure}[h]
  \centering
  \vspace{-5pt}
  \includegraphics[width=0.32\textwidth]{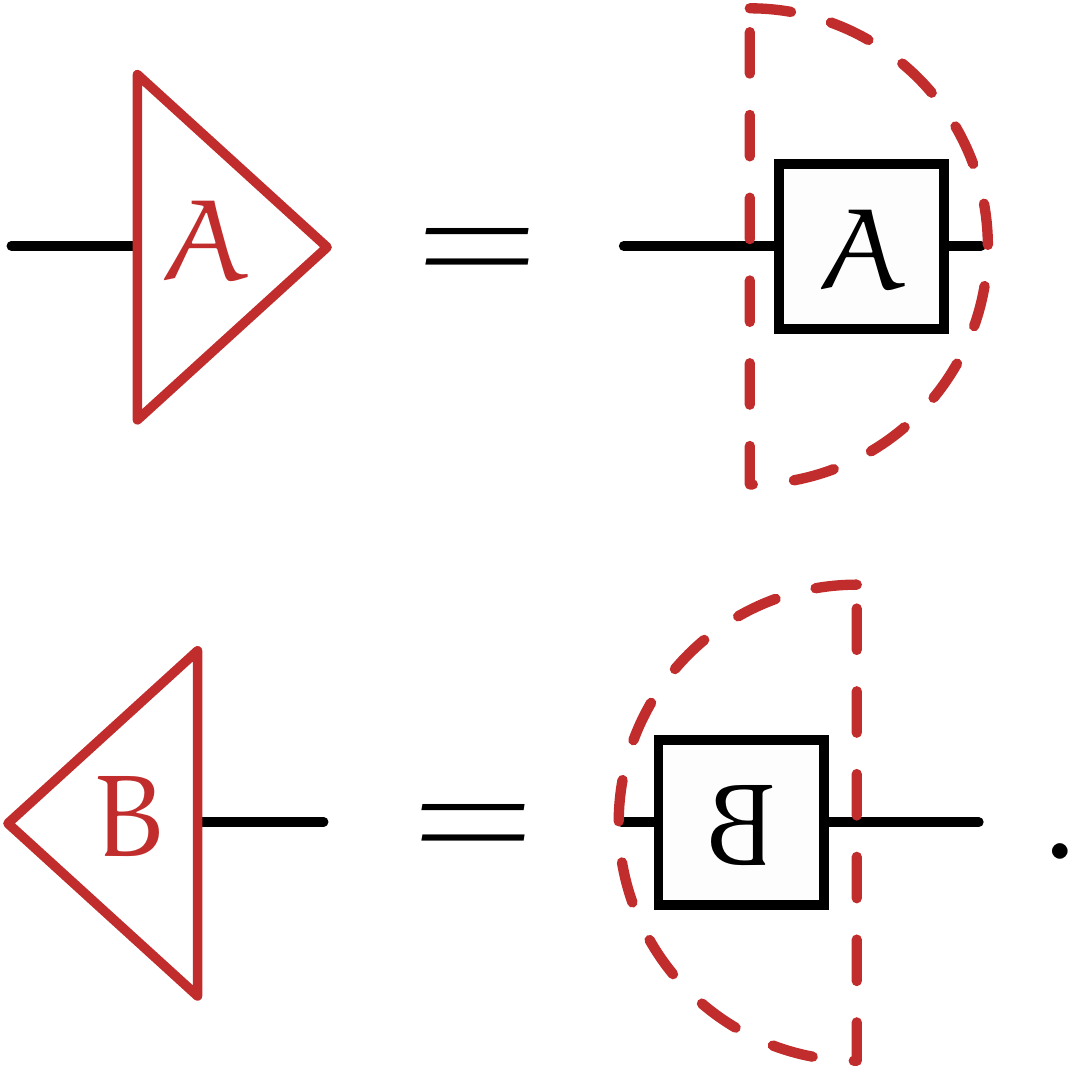}
  \label{fig:state8}
  \vspace{-5pt}
\end{figure}


\noindent 
A general matrix element is then written as follows:

\begin{figure}[h]
  \centering
  \vspace{-3pt}
  \includegraphics[width=0.75\textwidth]{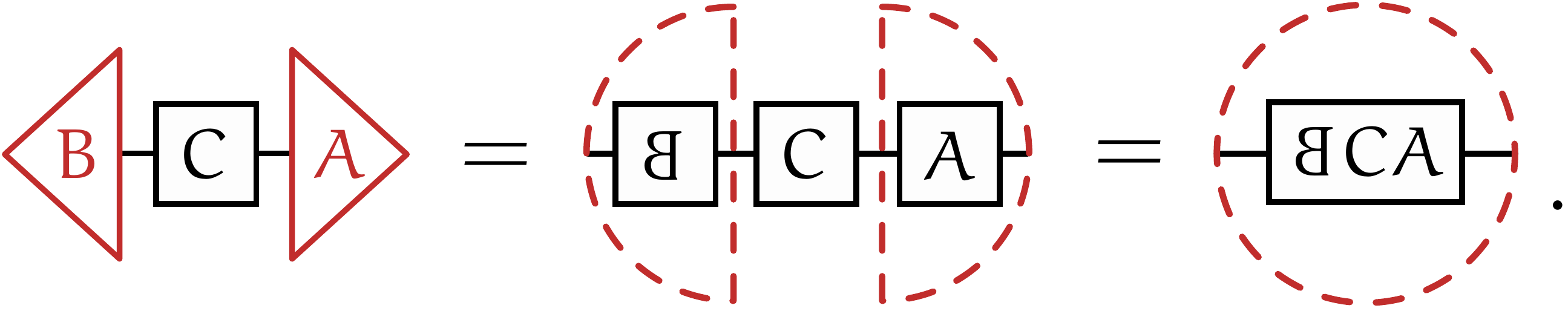}
  \label{fig:state9}
  \vspace{-5pt}
\end{figure}

\noindent This is consistent with our notation for \gls{sharp} operators in
Fig. \ref{fig:null}.\marginnote{You can shift $\Gamma$ to either side
  of the channel to form a bra or ket, with \[\langle
  [\Gamma^*]_\pi,[I]_\pi\rangle = \langle [I]_\pi,
  [\Gamma]_\pi\rangle=\pi(\Gamma)\]telling us the results are consistent.}

The procedure we've just outlined is the
\textsc{Gelfand-Naimark-Segal (\gls{GNS}) Construction}. It takes a state $\pi$ on
an abstract space of operators $\mathcal{A}$, and cooks up a concrete
Hilbert space $\mathcal{H}_\pi$ on
which those operators act.
Since each operator is bounded, we can say more precisely that it maps
$\mathcal{A}$ to a subalgebra of $\mathcal{B}(\mathcal{H}_\pi)$ in a
structure-preserving\marginnote{It preserves linear combinations,
  products, and adjoints:
  \begin{align*}
    \Phi_\pi(\alpha A + \beta B) & = \alpha \Phi_\pi(A) + \beta
                                   \Phi_\pi(B) \\
    \Phi_\pi(A B) & = \Phi_\pi(A)\Phi_B(B) \\
    \Phi_\pi(A^*) & = \Phi_\pi(A)^*.
  \end{align*}
} way called a $*$-\emph{homomorphism}.
Thus, instead of a positive, normed linear functional, a state can
also be viewed as a $*$-homomorphism $\Phi_\pi: \mathcal{A} \to
\mathcal{B}(\mathcal{H}_\pi)$ given by
\begin{equation}
  \label{eq:gns5}
  \Phi_\pi(A)[B]_\pi = A_\pi [B]_\pi = [AB]_\pi.
\end{equation}
The map $\Phi_\pi$ is called a \emph{representation} of $\mathcal{A}$.
We won't emphasize this perspective here, and simply write
$\Phi_\pi(A)=A_\pi$. 

Let's see how this works for the Pauli algebra $\mathcal{A}_\text{Pauli}$.
Since we can anticommute and square any repeated operators to
unity by (\ref{eq:pauli-rel}), 
each
element takes the form
\begin{equation}
  \label{eq:pauli-form}
  A = \alpha_0 I + \alpha_1 X + \alpha_2 Y + \alpha_3 Z, \quad
  \alpha_0, \alpha_1, \alpha_2, \alpha_3 \in \mathbb{C}.
\end{equation}
Thus, as a vector space $\text{dim}(\mathcal{A}_\text{Pauli}) =
4$.
Measuring $Z \measure +1$ in the Stern-Gerlach experiment leads
to a post-measurement state $\pi_{(0)}$.\marginnote{We choose $Z
  \measure (-1)^b$ to correspond to a functional $\pi_{(b)}$, and
  later, a ket $|b\rangle$.}
The \gls{kernel} $\mathcal{K}_{(0)}$ contains, at a minimum, $\Delta Z = Z -
I$ and all its scalar multiples.
But since it is a left ideal, it also contains the results of
left-multiplying by $X$ and $Y$:\marginnote{The $Z$ product 
$Z\Delta Z = -\Delta Z$ is boring.}
\[
  X \Delta Z = XZ - X = -iY - X = i (iX - Y) = i Y\Delta Z.
\]
Both products are proportional to $iX + Y$, so the \gls{kernel} is 
\begin{equation}
  \label{eq:pauli-sharp}
  \mathcal{K}_{(0)} = \left\{ \tfrac{1}{2}\delta (I - Z) + \tfrac{1}{2}\gamma (X + iY) :
    \gamma,\delta \in \mathbb{C}\right\}.
\end{equation}
If we express $(Z - I)/2$ and $(X+iY)/2$ as matrices\marginnote{We add
the factor of $2$ so operators are unit norm. For instance, $Z - I$
has maximum absolute eigenvalue $|-2| = 2$.}
in the usual basis, we find they correspond to the pink entries: 

\begin{figure}[h]
  \centering
  \vspace{-2pt}
  \includegraphics[width=0.27\textwidth]{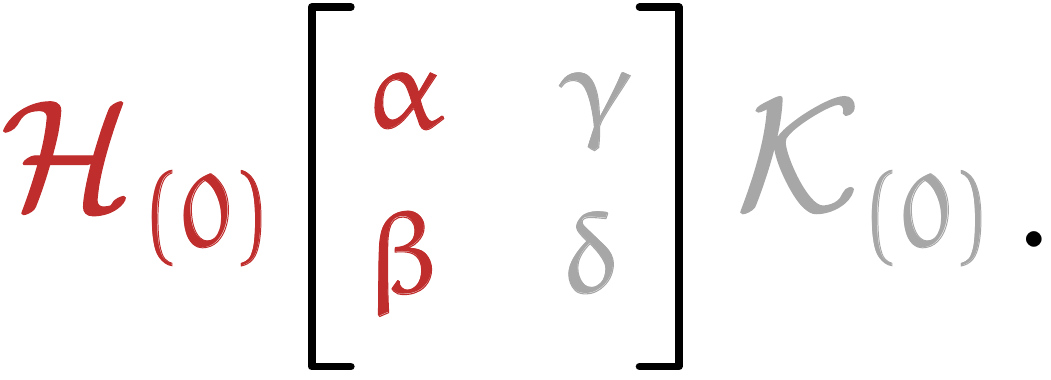}
  \caption{The \gls{GNS} construction for $Z \measure +1$ in the Pauli algebra.}
  \label{fig:gns1}
  \vspace{-7pt}
\end{figure}

\noindent
    %
This leads us to expect that the \gls{quotient} $\mathcal{H}_{(0)} =
\mathcal{A}_\text{Pauli}/\mathcal{K}_{(0)}$
lives in the first column, and therefore looks like a qubit!

For completeness, let's check.
The generators of (\ref{eq:pauli-sharp})
let us equate $Z$ with
$I$, and $Y$ with $-iX$.
This modifies the coefficients in (\ref{eq:pauli-form}):
\[
  \alpha_0 \mapsto \alpha_0 + \alpha_3, \quad \alpha_1 \mapsto \alpha_1 -
  i\alpha_2.
\]
Defining $\alpha = (\alpha_0 + \alpha_3)/2$ and $\beta = (\alpha_1 -
i\alpha_2)/2$, we obtain
\begin{equation}
  \label{eq:pauli-hilbert}
  \mathcal{H}_{(0)} = \left\{\tfrac{1}{2}\alpha (I + Z) + \tfrac{1}{2}\beta (X - iY) : \alpha,
  \beta \in \mathbb{C}\right\}.
\end{equation}
In terms of matrices, this is indeed the first column of the matrix!
To show this without using the matrix representation is straightforward.
From (\ref{eq:pauli-hilbert}), the computational
basis states will be\marginnote{These have unit norm,
  e.g.\[
    \Vert X\Vert_{(0)}^2 = G_{(0)}(X^*, X) = \pi_{(0)}(I) = 1.
  \]}
\begin{equation}
|0\rangle = \frac{1}{2}[Z+I]_{(0)} = [I]_{(0)}, \quad |1\rangle = 
\frac{1}{2}[X - iY]_{(0)} = [X]_{(0)}.\label{eq:6}
\end{equation}
The Pauli algebra $\mathcal{A}_\text{Pauli}$ 
acts\marginnote{Here is
  the work:
  \begin{align*}
    X|b\rangle & = X[X^b]_{(0)} = [X^{\neg b}]_{(0)} = |\neg b\rangle \\
Z |b\rangle & = Z[X^b]_{(0)}
                           = [(-X)^b]_{(0)} = (-1)^b|b\rangle
  \end{align*}
  where $\neg b = 1 -b$, we use $ZX = iY$ and the null relation $X + iY \equiv_+ 0$. 
} by left-multiplication, so after a little work, one obtains
$X|b\rangle = |\neg b\rangle$ and $Z|b\rangle = (-1)^b|b\rangle$.
This not only looks but behaves like a qubit!

\section{7. Changing basis}\hypertarget{sec:8}{}

We've shown how to construct a state $\pi$ and the associated Hilbert
space for which $\Gamma$ is definite; we call $\pi$ an \emph{eigenfunctional}
of $\Gamma$.\marginnote{We use \emph{eigenvector} for
  elements of $\mathcal{H}_\pi$, and \emph{eigenstate} when we want to
  hedge.}
But how much depends on the precise eigenvalue it takes?
Let's repeat our Hilbert space construction for $Z \measure -1$, which
swaps columns:

\begin{figure}[h]
  \centering
  \vspace{-3pt}
  \includegraphics[width=0.26\textwidth]{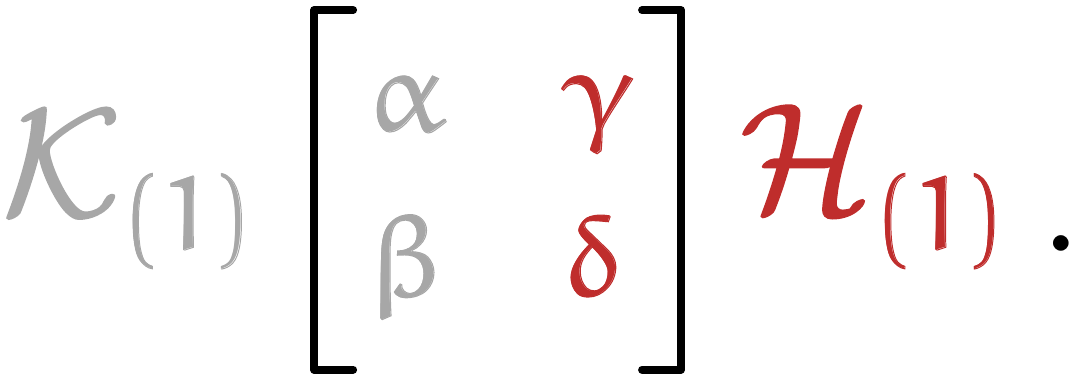}
    \caption{The \gls{GNS} construction for $Z = -1$ in the Pauli algebra.}
  \label{fig:gns2}
  \vspace{-6pt}
\end{figure}

\begin{figure}[h]
  \centering
  \vspace{-0pt}
  \includegraphics[width=0.87\textwidth]{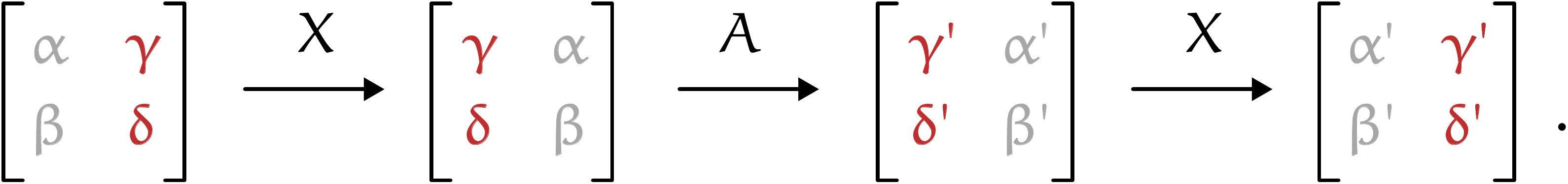}
  \caption{We can ``simulate'' $\pi_{(1)}$ with $\pi_{(0)}$ using a change of basis.}
  \label{fig:gns3}
  \vspace{-2pt}
\end{figure}

\noindent This state, which we call $\pi_{(1)}$, embeds the qubit
differently but seems to act the same way.
To make ``looks the same'' more precise, note that we translate the
$\pi_{(1)}$ embedding into the $\pi_{(0)}$ embedding by moving second column to first with a Pauli
$X$, applying a target operator, then swapping back with $X$ once
more, as in Fig. \ref{fig:gns3}.

\noindent For target operators $Y$ and $Z$, this yields $XZX = -Z$ and $XYX =
-Y$, exchanging (\ref{eq:pauli-sharp})
and (\ref{eq:pauli-hilbert}) as expected.

\marginnote{
  \vspace{-0pt}
  \begin{center}
    \hspace{-10pt}\includegraphics[width=0.77\linewidth]{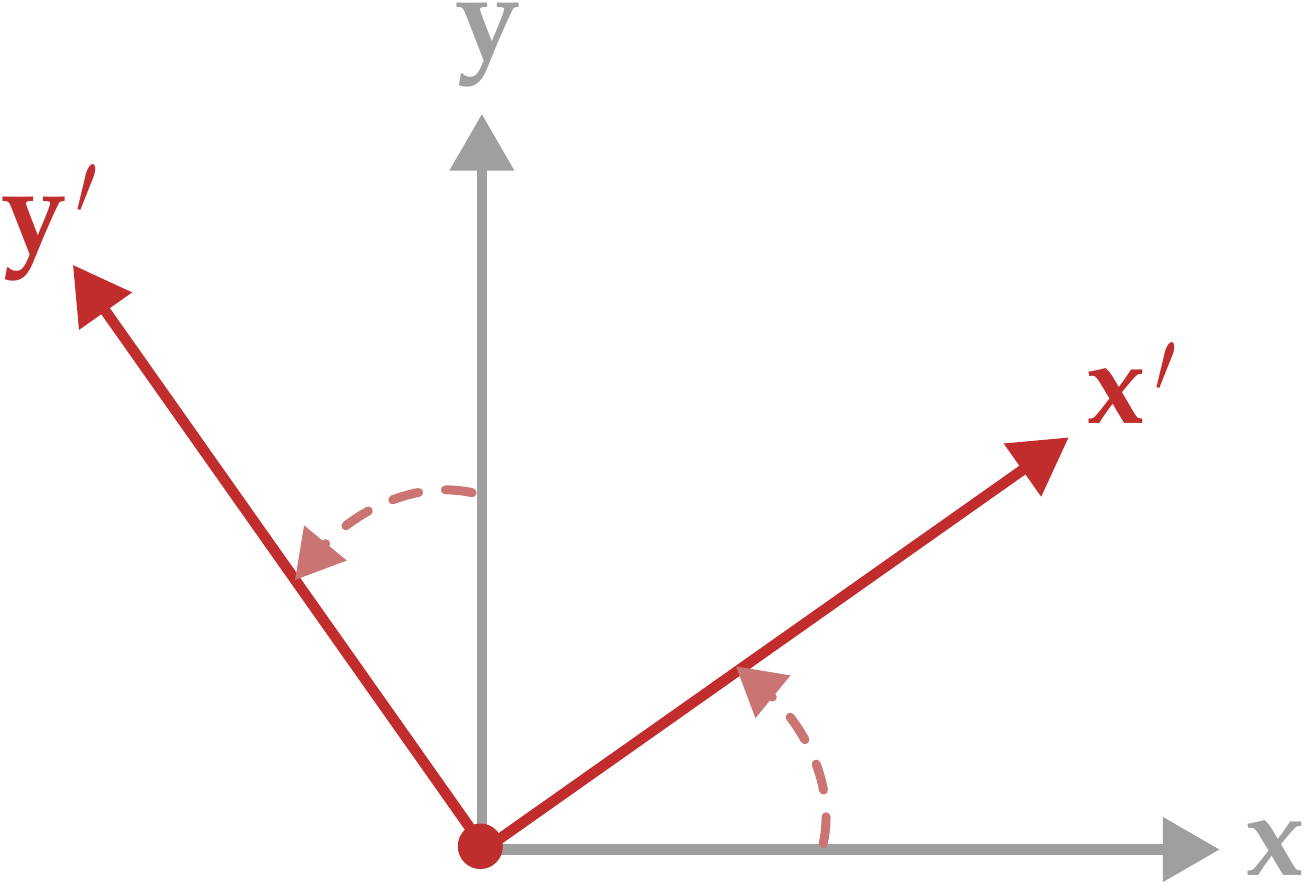}
  \end{center}  \vspace{-0pt}
  \emph{Changing basis from $(\mathbf{x}, \mathbf{y})$ to
    $(\mathbf{x}', \mathbf{y}')$ with a rotation matrix.
  } \vspace{10pt}
}
This is the phenomenon of basis change, familiar from linear algebra. 
It's often convenient to map vectors to
a new coordinate system where an operation is simpler, e.g. an
eigenbasis. If $U$ is the \emph{change of basis matrix}, then $A' = U^{-1} A U$
applies the simple operation in the old basis.
Instead of shifting operators, we can shift states:
\[
  \pi'(A) = \pi(A') = \pi(U^{-1} A U).
\]
We can think of $\pi'$ as the state $\pi$ in the new basis. But we have
to be careful; while every invertible $U$ gives rise to well-defined
operators $A' = U^{-1}A U$, $\pi'$ is not always a well-defined state.
Although linearity and the norm condition are
guaranteed,\marginnote{The norm condition follows because $\pi'(I) =
  \pi(U^{-1}U) = \pi(I) = 1.$}
positivity can fail. Evaluating $\pi'$ on a positive operator gives 
\begin{equation}
  \pi'(A^*A) = \pi[(A^*A)'] = \pi(U^{-1}A^*A U),\label{eq:uni1}
\end{equation}
and $U^{-1}A^*A U$ need not be positive.
We can fix this by requiring $U$ to be \emph{unitary}, so the adjoint
and inverse coincide: $U^* = U^{-1}$. In this case, the argument $U^{-1}A^*A U =
|AU|^2$ of (\ref{eq:uni1})\marginnote{We could consider $A' = U^*
  A U$ from the outset, which would ensure positivity, but then the
  norm condition might fail.} is positive, and hence the new state
$\pi'$ is positive by the positivity of $\pi$.

Let $\mathcal{U}(\mathcal{A})$ denote the set of unitary operators in $\mathcal{A}$,
and define \emph{conjugation} by any $U \in \mathcal{U}(\mathcal{A})$
for operators and hence states as\marginnote{We call $\mathcal{C}^U$ the
  ``pullback'' of $\mathcal{C}_U$, since it ``pulls $\mathcal{C}_U$ back'' onto states.}
\begin{align}
  \label{eq:conj1}
  \mathcal{C}_U[A] = U^* A U, \quad \mathcal{C}^U[\pi](A) =
  \pi(\mathcal{C}_U[A]).
\end{align}
\noindent These maps are ``multiplicative'' over the change of basis:
\begin{equation}
  \label{eq:conj2}
  \mathcal{C}_{UV} = \mathcal{C}_V \circ \mathcal{C}_U, \quad \mathcal{C}^{UV} = \mathcal{C}^U \circ \mathcal{C}^V,
\end{equation}
with an identity transformation $U = I$.
Returning to our original question, we say $\pi$ and $\pi'$ are \emph{unitarily
  equivalent} if they are conjugate, $\pi'=
\mathcal{C}^U[\pi]$.
\marginnote{
  \vspace{-100pt}
  \begin{center}
    \hspace{-10pt}\includegraphics[width=0.8\linewidth]{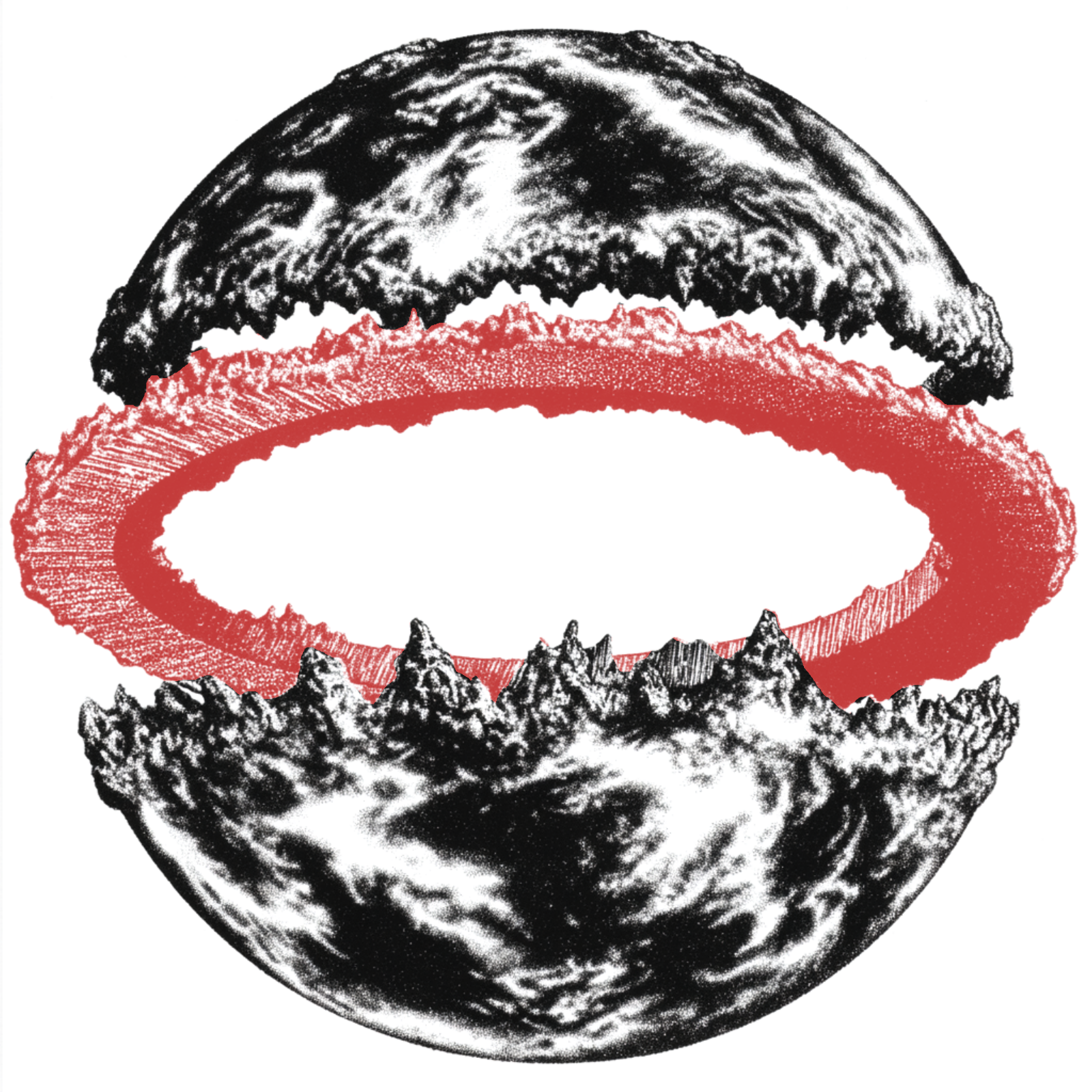}
  \end{center}  \vspace{-5pt}
  \emph{Much confusion about phase symmetry is avoided by
    thinking about the unitary equivalence from which it is born. 
  } \vspace{-10pt}
}
As the  name suggests, being unitarily equivalent is an equivalence
relation between states, as is easily checked:
\begin{itemize}[itemsep=0pt]
\item for reflexivity, setting $U=I$ gives $\mathcal{C}^I[\pi] = \pi$;
\item for symmetry, if $\mathcal{C}^U[\pi] = \pi'$, then
  $\mathcal{C}^{U^*}[\pi'] = \pi$;
\item for transitivity, if $\mathcal{C}^U[\pi] =
\pi'$ and $\mathcal{C}^{U'}[\pi'] = \pi''$, then $\mathcal{C}^{UU'}[\pi]=\pi''$.
\end{itemize} 
This lumps states into \emph{unitary equivalence classes} which have
the same operator correlations, up to a change of basis.

Diagrammatically, we'll attach $U$ outside a dotted box to conjugate
an operator. Within a state, we can detach it from the box and
attach to the inside of the dotted circle to indicate state conjugation:
\begin{figure}[h]
  \centering
  \vspace{-3pt}
  \includegraphics[width=0.49\textwidth]{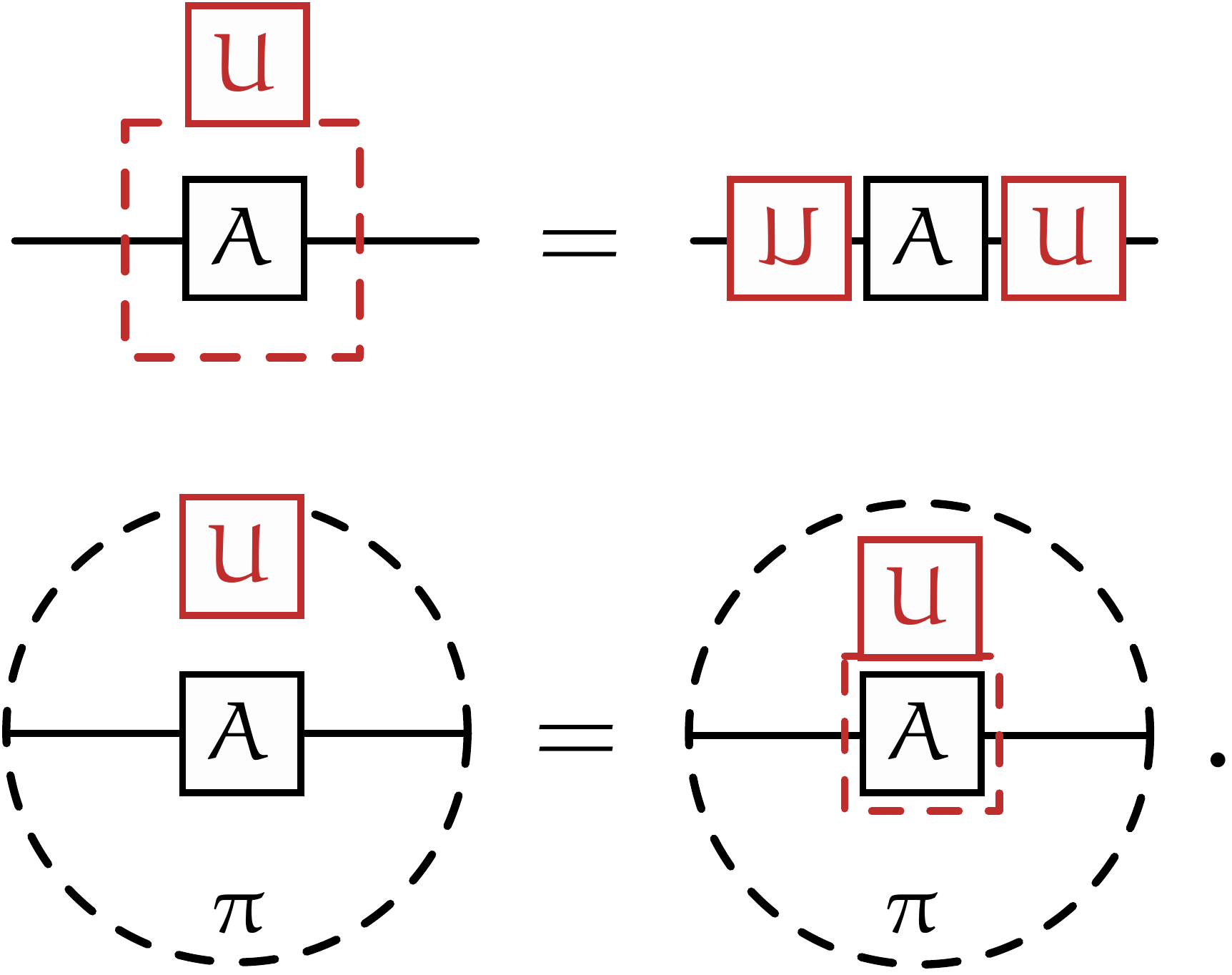}
  \label{fig:state9}
  \vspace{-5pt}
\end{figure}

\begin{figure}[h]
  \centering
  \vspace{-2pt}
  \includegraphics[width=0.51\textwidth]{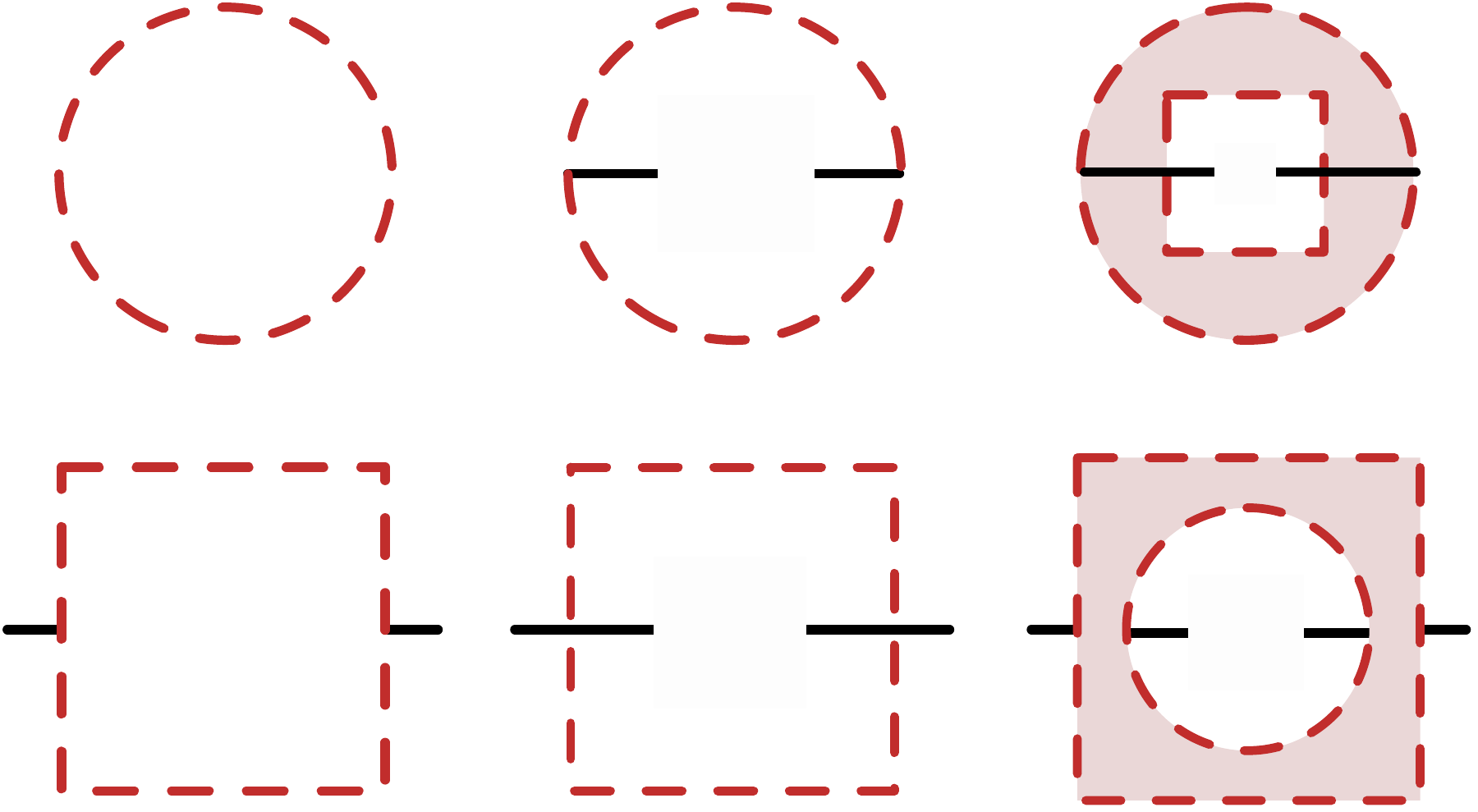}
  \caption{\textsc{Left.} Scalar and operator-valued functions of
    $\mathbb{C}$, e.g. $\lambda \mapsto 2\lambda$ and $\lambda\mapsto
    \lambda I$. \textsc{Middle.}
    Functions of $\mathcal{A}$, e.g. a state $\pi$ or a channel like
    $\mathcal{C}_U$. \textsc{Right.} Scalar function of a channel,
    e.g. channel capacity, and operator-valued function of a state,
    e.g. $\pi \mapsto A_\pi$.}
  \label{fig:conj2}
  \vspace{-4pt}
\end{figure}

\noindent 
This may seem a bit ad hoc, but is an instance of a more general
convention that dotted contours are functions, with contour shape
indicating output type. We give various examples
in Fig. \ref{fig:conj2}.

\noindent Input type is a bit trickier, but we use leads for
operators, blank space for scalars, and otherwise nest contours.

\marginnote{
  \vspace{-0pt}
  \begin{center}
    \hspace{-10pt}\includegraphics[width=0.68\linewidth]{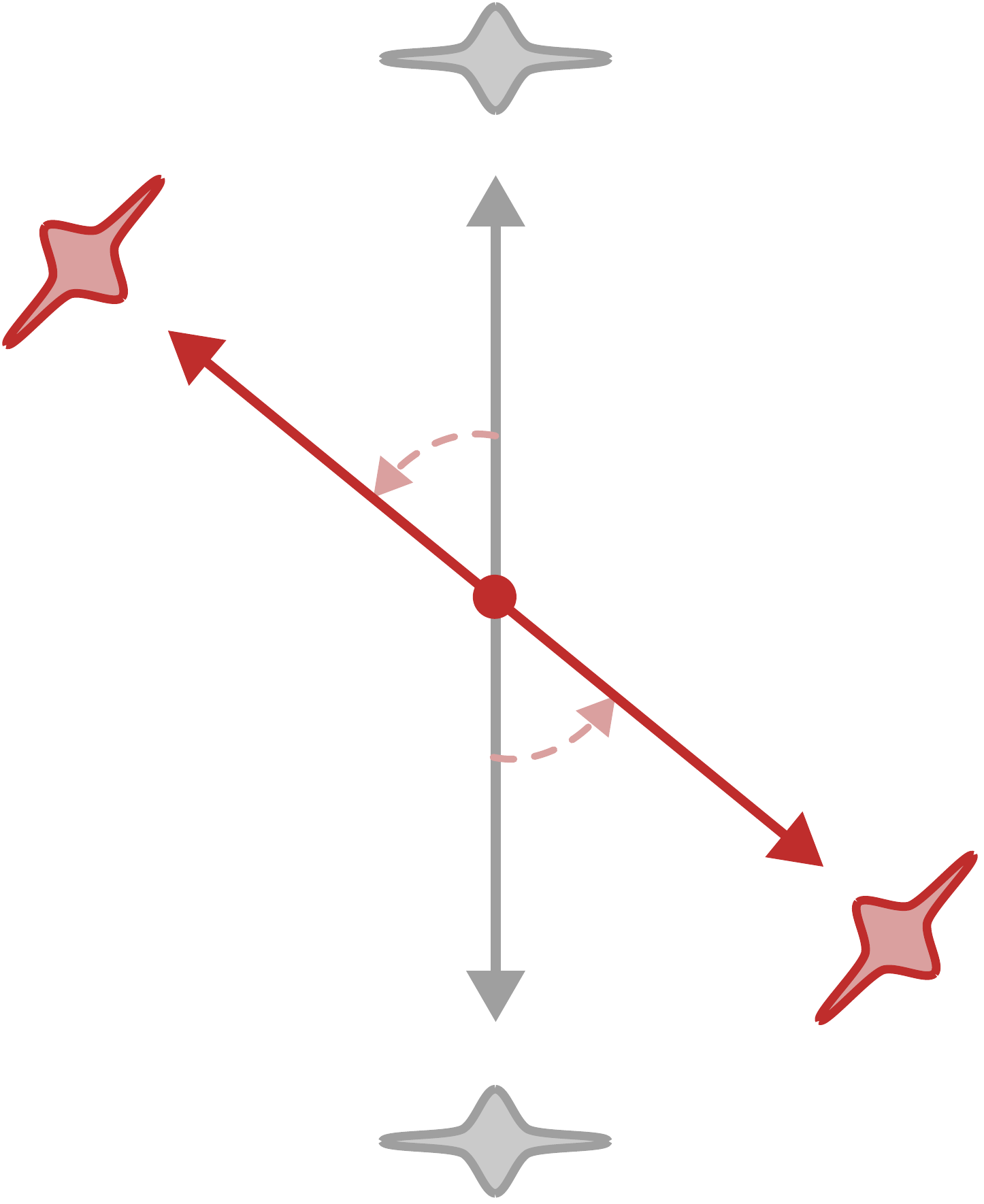}
  \end{center}  \vspace{-0pt}
  \emph{Rotating the Stern-Gerlach apparatus changes the physical
    outcomes.
  } \vspace{10pt}
}After this pictorial detour, let's return to the question of
equivalence.
It's clear that distinct but unitarily related states are different;
they lead to different experimental outcomes.
But it's easy to plonk them into the \emph{same} Hilbert space!
For instance, consider the \gls{GNS} Hilbert space $\mathcal{H}_\pi$
associated with $\pi$, and another state $\pi' = \mathcal{C}^U[\pi]$
in the same unitary equivalence class.
We can associate these states with respective unit vectors $\pi \mapsto [I]_\pi, \pi' \mapsto [U]_\pi$,
since, using the bra-ket notation (\ref{eq:braket}),
\[
  \pi(A) = {}_\pi\langle I|A|I\rangle_\pi, \quad \pi'(A) =
  {}_\pi\langle I|U^*AU|I\rangle_\pi = {}_\pi\langle U| A |U\rangle_\pi.
\]
Thus, we recover the standard notion of expectations with respect to a
unit norm vector inside a Hilbert space!
The unitary orbit
\begin{equation}
\mathcal{V}_\pi = \mathcal{U}(\mathcal{A}) \cdot [I]_\pi \subseteq
\mathcal{H}_\pi\label{eq:folium}
\end{equation}
\marginnote{Clearly ${}_\pi\langle U|U\rangle_\pi = \pi(U^*U)
= \pi(I) = 1$, and though we won't show it, every unit vector has this
form.}is the set of \emph{vector states} in $\mathcal{H}_\pi$,
since these are precisely the vectors of unit norm.
In the Hilbert space formulation, these are what we usually think of
as states!

\section{8. The Bloch sphere}\hypertarget{sec:9}{}

\marginnote{
  \begin{center}
    \includegraphics[width=0.9\linewidth]{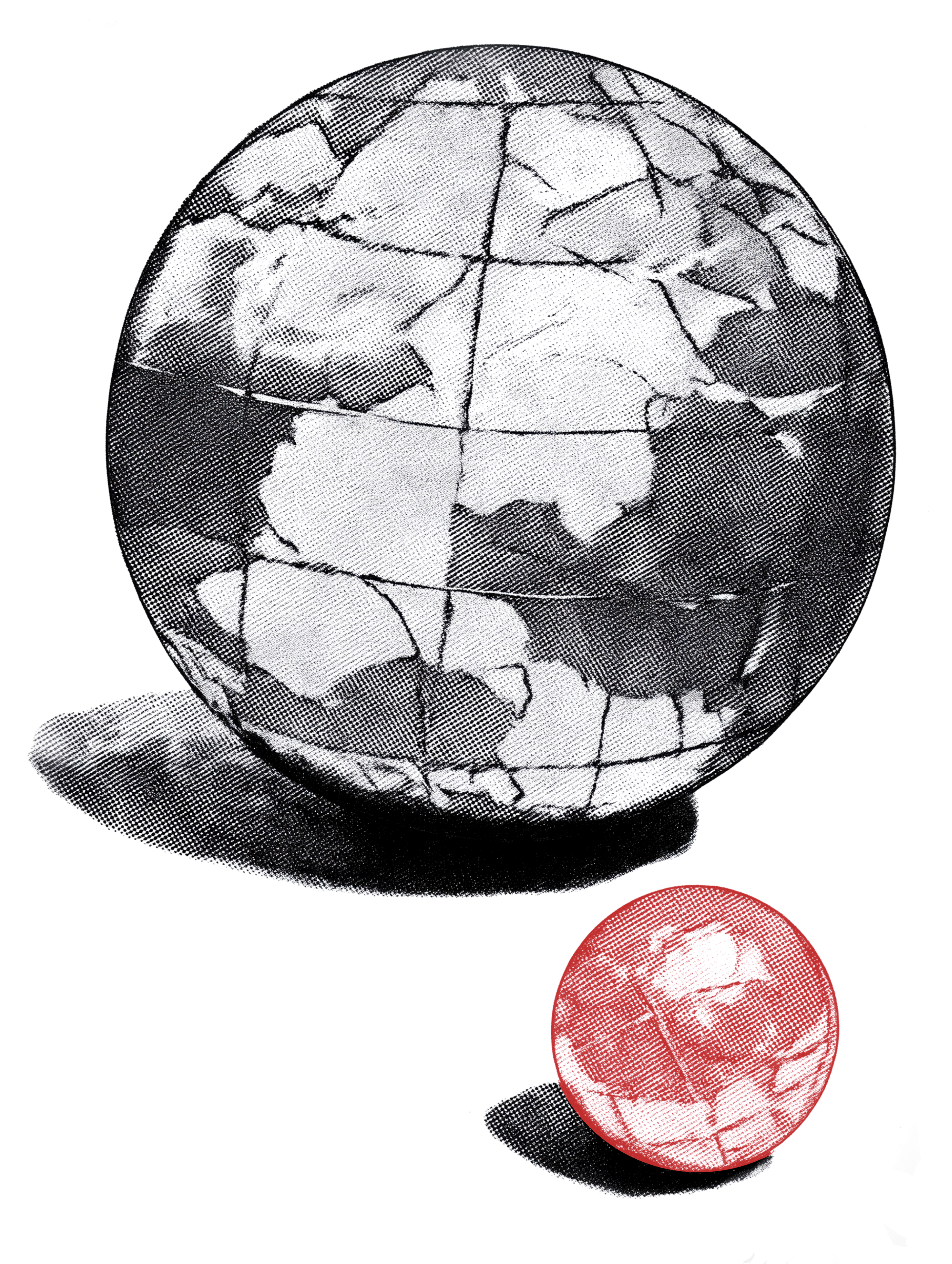}
  \end{center}  \vspace{-5pt}
  \emph{
    Unitary operators beget vector states.
  } \vspace{10pt}
}
As a particularly elegant illustration,
we will show that the unitary equivalence class of
$\pi_{(0)}$ looks like a 
sphere.
To do this, we will:
\begin{itemize}[itemsep=0pt]
\item describe all unitaries $\mathcal{U}(\mathcal{A}_\text{Pauli})$;
\item show that conjugation maps $Z$ to operators of
  the form
  $\sigma(\mathbf{n})$;
\item conclude $\pi_{(0)}$ is conjugate to the
  $\sigma(\mathbf{n})\measure +1$ eigenfunctional $\pi_{(\mathbf{n})}$;
\item finally, interpret these statements geometrically.
\end{itemize}
This will give us a particularly clear derivation of the Bloch sphere
and the mysterious half-angles associated with it. 

We start by parameterizing the unitaries.
Let's extend our previous notation and write
$\sigma(\mathbf{v}) = v_i\sigma_{(i)}$ for arbitrary
$\mathbf{v}\in\mathbb{C}^3$. Then any $A \in
\mathcal{A}_\text{Pauli}$ can be written $A = \alpha I +
\sigma(\mathbf{v})$, with square
\begin{align*}
  |A|^2 & = \big(\alpha I + \sigma(\mathbf{v})\big)^* \big(\alpha I +
          \sigma(\mathbf{v})\big)\\
  & = |\alpha|^2 I + \sigma(\overline{\alpha}\mathbf{v} + \alpha
    \mathbf{v}^*)  + |\sigma(\mathbf{v})|^2.
\end{align*}
The proof in Fig. \ref{fig:simple5} goes through as before, except
that $I$ is replaced by $|\mathbf{v}|^2 = v_iv_i^*$.
For $A$ to be unitary, this means the middle term must vanish and the
remaining coefficients sum to unity:
\[
  \text{Re}[\overline{\alpha}\mathbf{v}] = 0, \quad |\alpha|^2 + |\mathbf{v}|^2 = 1.
\]
We can solve these constraints with $\alpha = e^{i\delta}\cos\theta$ and $\mathbf{v} =
i e^{i\delta}\sin\theta \, \mathbf{n}$ for phases $\delta, \theta\in[0,2\pi)$ and
unit vector $\mathbf{n}\in\mathbb{R}^3$.
We call the resulting expression a \emph{Pauli exponential} and write
\begin{equation}
  \label{eq:pauli-exp1}
  e^{i\delta I + i\theta \sigma(\mathbf{n})} = e^{i\delta}\big(I\cos\theta
  + i \sin\theta \, \sigma(\mathbf{n})\big),
\end{equation}
since it agrees with the results of exponentiating $i\delta I
+ i\theta \sigma(\mathbf{n})$:
\begin{align}
  e^{i\delta I + i\theta \sigma(\mathbf{n})} & = e^{i\delta}\sum_{k=0}^\infty\frac{[i\theta
                                   \sigma(\mathbf{n})]^k}{k!} \notag
  \\
  & = e^{i\delta}\sum_{k=0}^\infty\frac{(-1)^k\theta^{2k}}{(2k)!} + i
    \sum_{k=0}^\infty\frac{(-1)^k\theta^{2k+1}}{(2k+1)!}\sigma(\mathbf{n})\notag \\
& = e^{i\delta} \big(I\cos\theta + i \sin\theta \,
                                                                               \sigma(\mathbf{n})\big).   \notag 
\end{align}
This follows from\marginnote{\vspace{-80pt}

  \noindent We split the 
  series into even and odd terms, simplify using
  $\sigma(\mathbf{n})^{2k} = I$ and $\sigma(\mathbf{n})^{2k+1} =
\sigma(\mathbf{n})$, and invoke the Taylor series for sine and
cosine. Note that we pull $e^{i\delta I}$ out the front since $I$ commutes
with everything else.} straightforward power
series manipulation.
More generally, if $f(x)$ has a power series, $f(A)$ denotes the result of
formally replacing $x$ with $A$:
\begin{equation}
  \label{eq:pow-series}
f(x) = \sum_{k=0}^\infty c_k x^k \quad \Longrightarrow \quad f(A) = \sum_{k=0}^\infty c_k A^k.
\end{equation}
We don't worry about convergence or other analytic niceties
here.\marginnote{\vspace{-30pt}

\newpage
  
  \noindent These niceties lead to the \emph{continuous
  functional calculus}. We give details and applications in Appendix \hyperlink{app:functional}{A}.}

Now that we have a general unitary of the form $U = e^{-i\theta\sigma(\mathbf{n})}$, \marginnote{We can ignore the $e^{i\delta}$ in
  (\ref{eq:pauli-exp1}) since it is cancelled by the adjoint phase
  $e^{-i\delta}$.} we can use it to build the equivalence
class of $Z$.
For simplicity, we'll assume $\mathbf{n} = (n_1, n_2, 0)$ is
orthogonal to the $\mathbf{z}$-axis.
Algebraically, this implies that $Z$ and
$\sigma(\mathbf{n})$ \emph{anticommute}:
\[
  Z \sigma(\mathbf{n}) = Z (n_1 X + n_2 Y) = -(n_1 X + n_2 Y)Z = -\sigma(\mathbf{n})Z.
\]
Conjugating by $U = e^{-i\theta \sigma(\mathbf{n})}$ then gives
\begin{align}
  \mathcal{C}_U[Z] & = \big(I\cos\theta + i
                                      \sin\theta\,
                                      \sigma(\mathbf{n})\big)Z
                                      \big(I\cos\theta - i \sin\theta\,
                                                   \sigma(\mathbf{n})\big) \notag\\
  & = \big(\cos^2\theta - \sin^2\theta\big) Z + 2i\sin\theta\cos\theta
    \sigma(\mathbf{n}) Z\notag\\
                                                 & = \cos(2\theta)Z +
                                                   i \sin(2\theta)
                                                   \sigma(\mathbf{n})Z
                                                   \notag\\
  & = \sin(2\theta)\left[ -n_2 X + n_1 Y+\cot(2\theta) Z\right] \label{eq:cis2}= \sigma(\mathbf{n}'),
\end{align}
where $\mathbf{n}' =
    \sin(2\theta)(-n_2, n_1, \cot(2\theta))$ is a unit
    vector:\marginnote{Remember that $\mathbf{n}=(n_1, n_2, 0)$ is a
      unit vector so $n_1^2 + n_2^2 = 1$.}
\begin{align*}
  |\mathbf{n}'|^2& = \sin^2(2\theta)\big(n_1^2 + n_2^2 + \cot^2(2\theta)\big) =1.
\end{align*}
Geometrically, conjugation rotates $\mathbf{z}$ by an angle $2\theta$ towards
the vector $(-n_2, n_1, 0)$ orthogonal to both $\mathbf{z}$
and $\mathbf{n}$ (Fig. \ref{fig:cis1}, left):

\begin{figure}[h]
  \centering
  \vspace{-2pt}
  \includegraphics[width=0.75\textwidth]{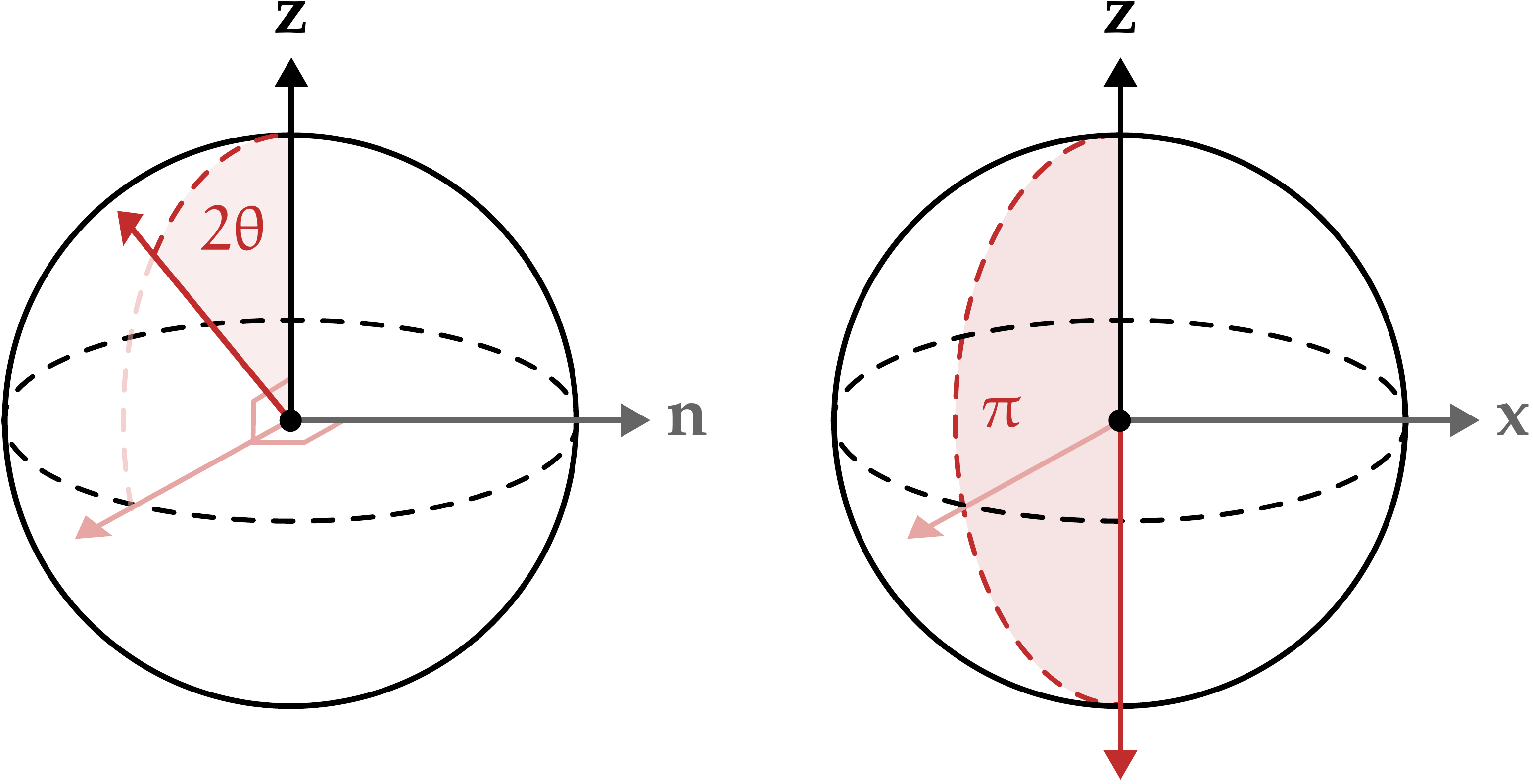}
  \caption{
    \textsc{Left}. Conjugating $Z$ by $e^{-i\theta\sigma(\mathbf{n})}$
    (for orthogonal $\mathbf{n}$)
    rotates the unit vector $\mathbf{z}$ ccw by $2\theta$ in the plane
    perpendicular to $\mathbf{n}$. \textsc{Right}. A special case $XZX
    = -Z$.
  }
  \label{fig:cis1}
  \vspace{-5pt}
\end{figure}

\noindent As a special case, setting $U = e^{-i(\pi/2)X} = X$ recovers
the column-swap change of basis $\mathcal{C}_U[Z] = -Z$ from the previous section.

\begin{figure}[h]
  \centering
  \vspace{-5pt}
  \includegraphics[width=0.28\textwidth]{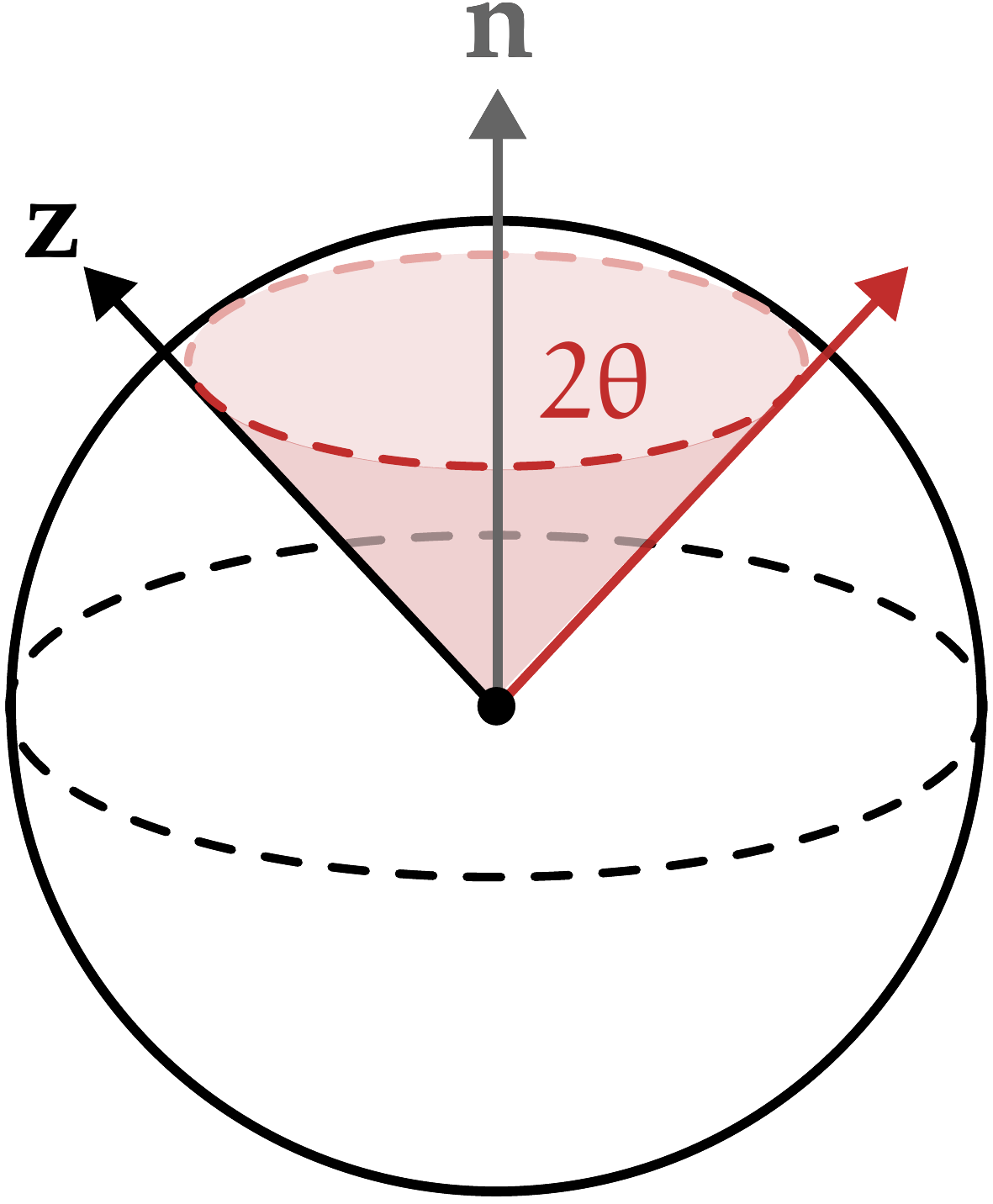}
  \caption{
    Conjugation by $e^{-i\theta\sigma(\mathbf{n})}$ rotates
    $\mathbf{z}$ counter-clockwise by $2\theta$ in the plane perpendicular to
    $\mathbf{n}$.
  }
  \label{fig:cis6}
  \vspace{-10pt}
\end{figure}

We see on geometric grounds that we can map $\mathbf{z}$ to any unit
vector $\mathbf{m}$, and hence $Z$ to any generalized Pauli
$\sigma(\mathbf{n})$: choose $\mathbf{n}$ orthogonal to our
target vector $\mathbf{m}$ and $\theta$ half the angle between
$\mathbf{z}$ and $\mathbf{m}$. Thus, every generalized Pauli operator
is conjugate.
We've only considered $\mathbf{n} \perp \mathbf{z}$, but the
geometry of the general case is similar: conjugation by
$U = e^{-i\theta\sigma(\mathbf{n})}$ rotates $\mathbf{z}$ an angle
$\theta$ counter-clockwise around the $\mathbf{n}$-axis, tracing out a cone at fixed
azimuthal angle to $\mathbf{n}$, shown in Fig. \ref{fig:cis6}.
It doesn't add much, but if you're curious, here is the general
result\marginnote{To derive this, you can extend the proof in
  Fig. \ref{fig:simple5} to obtain the neat identity \[\sigma(\mathbf{m})
  \sigma(\mathbf{n}) = (\mathbf{m}\cdot \mathbf{n})I +
  i\sigma(\mathbf{m}\times \mathbf{n}).\] Applying this a few times
gives the result on the left.}
in algebraic form:
\begin{align*}
  \mathcal{C}_U[\sigma(\mathbf{m})] = \cos(2\theta)\sigma(\mathbf{m}) &+
  i\sin(2\theta)\sigma(\mathbf{m}\times \mathbf{n})\\ & + 2\sin^2\theta (\mathbf{m}\cdot \mathbf{n})\,\sigma(\mathbf{n}),
\end{align*}
where $\mathbf{m}\cdot\mathbf{n} = m_in_i$ and $(\mathbf{m}\times
\mathbf{n})_i=\epsilon_{ijk} m_j n_k$ are our old friends from vector analysis, the dot and
cross product. We recover (\ref{eq:cis2}) by setting
$\mathbf{m}=\mathbf{z}$ and $\mathbf{m}\cdot \mathbf{n}=0$.
\newpage

Equivalence of operators translates into equivalence of states.
To prove this, we actually need to give a more careful definition of
the \gls{kernel} $\mathcal{K}_\pi$, which fully characterizes the state $\pi$.
Recall that $\mathcal{K}_\pi$ consists of operators uncorrelated with
anything else in $\pi$, as per (\ref{eq:sharp1}).
There is some minimal set $\mathcal{I}_\pi$ of
identifications that
produces all such operators, usually corresponding to definite
measurement outcomes, e.g. $Z \equiv_{(0)} I$
for $\pi_{(0)}$. We can regard $\mathcal{K}_\pi$ as the
left ideal generated by $\mathcal{I}_\pi$, 
\begin{equation}
  \mathcal{K}_\pi = \mathcal{A}\mathcal{I}_\pi \subseteq \mathcal{A},
  \label{eq:left-ideal}
\end{equation}
consisting of all terms of the form $AR \in \mathcal{A}
\mathcal{I}_\pi$, simplified using the ``ambient'' algebra
$\mathcal{A}$.
Formally, we can view this as the $\ast$-algebra generated by the
(redundant) set $\mathcal{A} \mathcal{I}_\pi$, subject to any complete
set of relations for $\mathcal{A}$.

The result of this somewhat abstract characterization is a simple
sufficient condition for the equivalence of states:
\begin{equation}
  \mathcal{C}_U[\mathcal{I}_\pi] = \mathcal{I}_{\pi'} \,\,\, \Longrightarrow \,\,\,
  \mathcal{C}_U[\mathcal{K}_\pi] = \mathcal{K}_{\pi'} \,\,\,
  \Longleftrightarrow \,\,\, \mathcal{C}^U[\pi] = \pi'.\label{eq:R_pi}
\end{equation}
It's important to note that this condition isn't \emph{necessary}: the sets $\mathcal{I}_\pi$ are not
unique,\marginnote{For instance, $Z \equiv_{(0)} I$ is equivalent to
  $\alpha Z\equiv_{(0)} \alpha I$.} so $\mathcal{C}_U[\mathcal{I}_\pi]
\neq \mathcal{I}_{\pi'}$ does not mean $\pi$ and $\pi'$ are
inequivalent.
But once we have a set $\mathcal{I}_\pi$ we can ``sweep out'' the
vector states by seeing what left ideals are generated by
$\mathcal{C}_U[\mathcal{I}_\pi]$.
So, returning to Pauli operators, the $Z \measure +1$ eigenfunctional has a
generating set $\mathcal{I}_{(0)} = \{Z - I\}$.
The hard work we did
conjugating operators above now pays dividends, since
\[
  \mathcal{C}_U[\mathcal{I}_{(0)}] = \{\mathcal{C}_U[Z] - I\} =
  \{\sigma(\mathbf{m}) - I\}
\]
for some unit vector $\mathbf{m}$ determined by $U$.
It follows from (\ref{eq:R_pi}) that $\mathcal{C}^U[\pi_{(0)}] =
\pi_{(\mathbf{m})}$, the $\sigma(\mathbf{m}) \measure +1$ eigenfunctional, just as
we expect.

To cast this in a more familiar form, let's work out the vector states
(\ref{eq:folium}) of $\mathcal{H}_{(0)}$, i.e. the unitary orbit of $[I]_{(0)} = |0\rangle$.
Recall
that we can rotate $\mathbf{z}$ to any target vector using an
orthogonal vector
\[
  \mathbf{n} = (n_1, n_2, 0) = (\cos\phi, \sin\phi, 0)
\]
for some polar angle $\phi \in [0, 2\pi)$.
Applying a general Pauli exponential and simplifying with $[X]_+ =
i[Y]_+ = |1\rangle$, we find
\begin{align}
  e^{i\delta}e^{-i\theta\sigma(\mathbf{n})}|0\rangle & =e^{i\delta}\big[\cos \theta |0\rangle + i
                                           \sin\theta (\cos\phi \,X +
                                                       \sin\phi \,Y)
                                           |0\rangle\big] \notag \\
  & = 
                                           e^{i\delta}\big[\cos \theta |0\rangle + i
                                           \sin\theta (\cos\phi + i\sin\phi)
                                           |1\rangle\big] \notag \\
    & = e^{i\delta}\left[\cos\theta |0\rangle + i
                                           \sin\theta e^{i\phi}
                                           |1\rangle\right]\notag = |\psi\rangle.
\end{align}
Now for some bookkeeping.
We want to make sure we don't double count vector states, and a
crucial observation is that $\theta$ is \emph{doubled}
on the sphere of unit vectors (see Fig. \ref{fig:cis1}). We therefore
define a new angle $\vartheta = 2\theta$ restricted to $[0,
2\pi]$. This leads to the familiar expression
\begin{equation}
  \label{eq:bloch}
  |\psi\rangle = e^{i\delta}\left[\cos\left(\tfrac{1}{2}\vartheta\right) |0\rangle + i
                                           \sin\left(\tfrac{1}{2}\vartheta\right) e^{i\phi}
                                           |1\rangle\right],
\end{equation}
which parametrizes the \textsc{Bloch Sphere},\sidenote{``Nuclear
  induction'' (1946), Felix Bloch.} shown in Fig. \ref{fig:cis2}:

\begin{figure}[h]
  \centering
  \vspace{-5pt}
  \includegraphics[width=0.35\textwidth]{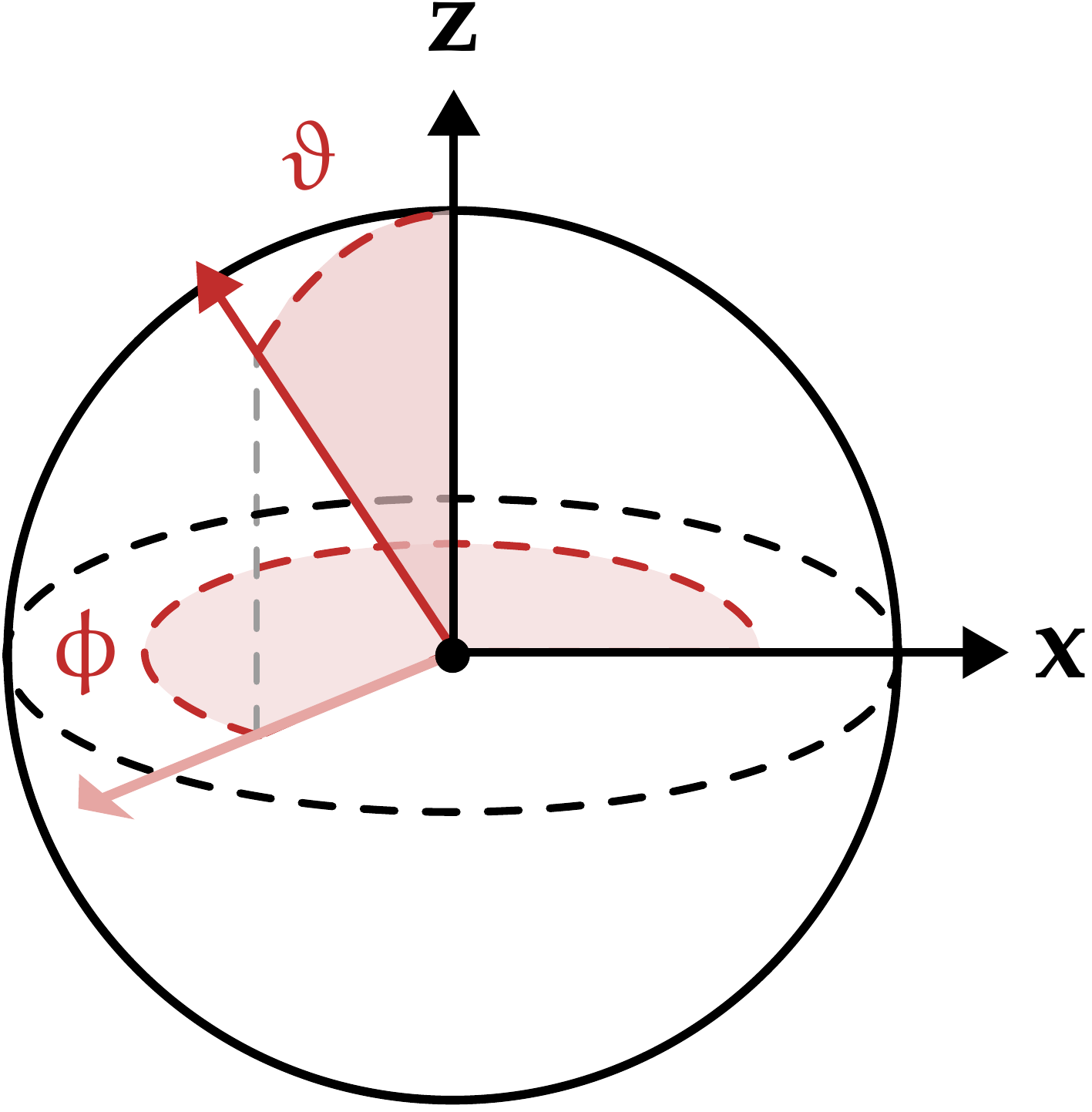}
  \caption{The Bloch sphere of state vectors $\mathcal{V}_{(0)}
    \subseteq \mathcal{H}_{(0)}$, with global
    phase suppressed.}
  \label{fig:cis2}
  \vspace{-5pt}
\end{figure}
\noindent As usual, we suppress global phase,\marginnote{If you like,
  you can imagine a circle attached to each point of the sphere. If
  you are particularly adept at higher-dimensional visualization, you can braid those to
  form a three-sphere $\mathbb{S}^3$! This is called the \emph{Hopf fibration}.}
but it still has mathematical
utility. Noting that $\cos[(2\pi - \vartheta)/2]
= -\cos(\vartheta/2)$, and similarly for $\sin$,
we see that substituting $\vartheta \mapsto 2\pi - \vartheta$ adds an overall
minus sign. We can absorb this into the
phase $\delta \mapsto \delta + \pi$, 
and thus reduce the range of $\vartheta \in [0,\pi]$. It now
properly resembles an azimuthal angle!

Hopefully, the origin of the half-angles in (\ref{eq:bloch}) is now
clear: when we conjugate an operator
$\sigma(\mathbf{m})$ by a Pauli exponential
$e^{-i\theta\sigma(\mathbf{n})}$, the vector $\mathbf{m}$ is rotated
by $2\theta$ since we have a rotation on either side. Ultimately, this
means the vector states $\mathcal{V}_{(\mathbf{m})}$ of $\sigma(\mathbf{m})$ covers the unit sphere
\emph{twice}. The second cover is associated with a factor of $-1$,
but when we have a global phase we can absorb it.


\addtocontents{toc}{\protect\vspace{-20pt}\protect\contentsline{part}{\textsc{\Large{pure and mixed}}}{}{}}

\section{9. Mixed states}\hypertarget{sec:10}{}

In building a truth table for a noncommutative circuit, we want to make sure we find all the
states.\marginnote{The argument goes as follows: if $\Lambda = \Lambda^*$ is self-adjoint, then it is a
real linear combination:
\begin{align}
  \Lambda = \alpha
  I+\sigma(\mathbf{v}) \label{eq:bloch-adjoint}
\end{align}
for some $(\alpha, \mathbf{v}) \in\mathbb{R}^4$. Writing $\mathbf{v}=|v|\hat{\mathbf{v}}$,
  then a \gls{sharp} measurement $\Lambda \measure \lambda$ is equivalent to $\sigma(\hat{\mathbf{v}})
  \equiv_\pi |v|^{-1}(\lambda-\alpha)I$, so we are done.} We've just explained, for instance, how to populate a whole
sphere of states in the case of the Pauli algebra. You can show that
all states of $\mathcal{A}_\text{Pauli}$ given by definite
measurements of (self-adjoint) observables are captured on the Bloch
sphere.
But while this lists all states with \emph{definite}
measurements, we have completely neglected the \emph{indefinite}
states, where measurements are fuzzy or information is partial. These
are no less important!

Luckily, we can build indefinite states by mixing definite ones.
\marginnote{
  \vspace{-0pt}
  \begin{center}
    \hspace{-10pt}\includegraphics[width=0.95\linewidth]{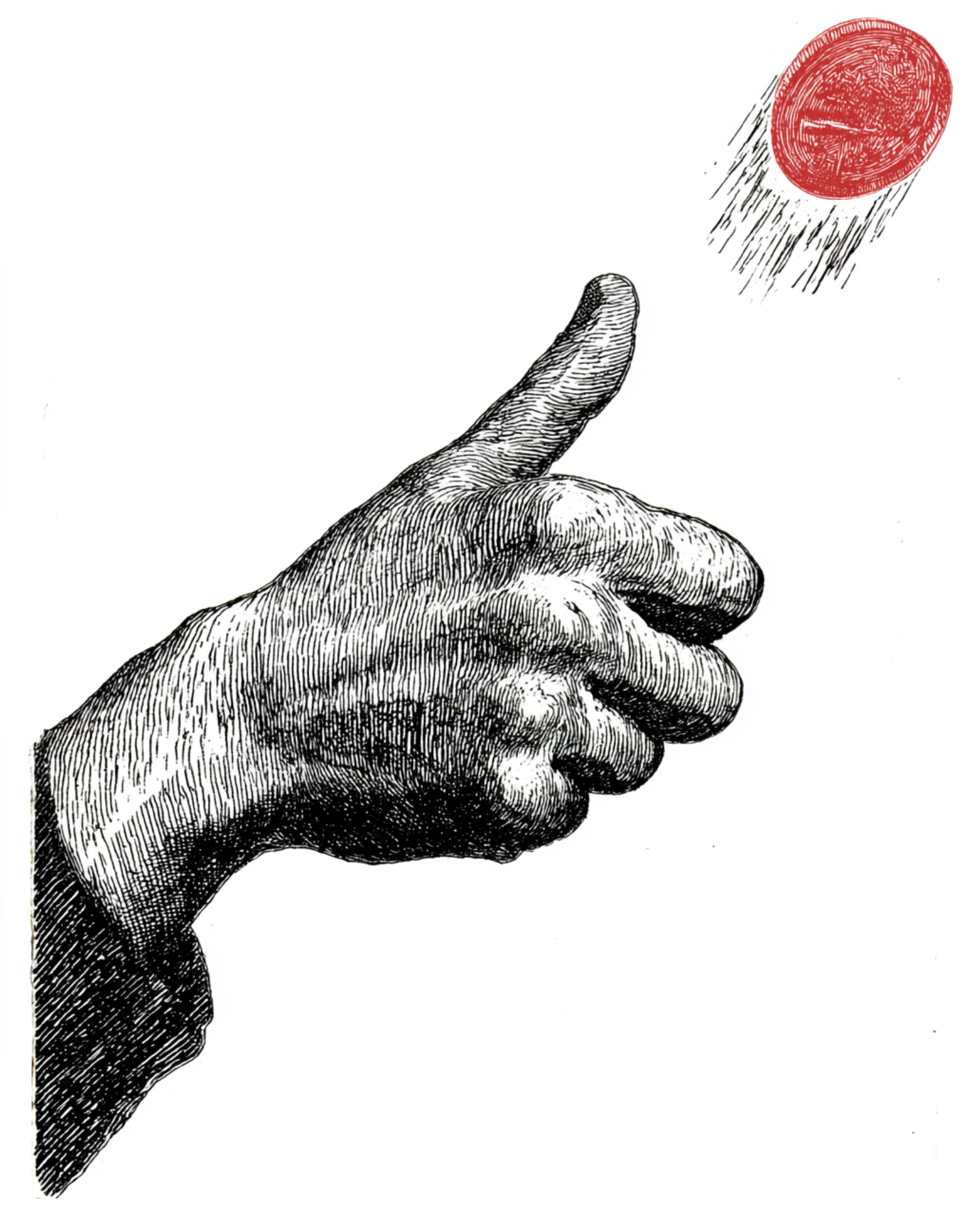}
  \end{center}  \vspace{-5pt}
  \emph{To make a mixture, just toss a (potentially biased) coin and pick a state.
  } \vspace{-5pt}
}In the language of expectation functionals, any ``probabilistic''
combination of two states forms a state, i.e.
\begin{equation}
  \pi = p_1 \pi + p_2\pi' 
  \label{eq:mix1}
\end{equation}
is a state provided the coefficients are positive (to ensure
$\pi$ is a positive functional) and sum to one (to ensure $\pi$ is
normalized).
More generally, a positive, normalized linear combination is called a \emph{convex combination}, and the resulting
functional a \emph{\gls{mixed} state}. 
We can view this as randomly choosing a state $\pi_{(j)}$ with probability $p_j$, or equivalently, associating any operator $A$ to a random
variable $A|_{\pi_{(j)}}$.

Consider any set of elements $K$ in a vector space. We call the set of all convex
combinations of elements in $K$ the \emph{convex hull}:
\begin{equation}
  \text{conv}(K) = \left\{\sum_j p_i k_j : k_j \in K, p_j \geq 0,
    \sum_j p_j = 1\right\}.\label{eq:conv1}
\end{equation}
An \emph{extreme point} is one that cannot be obtained by mixing, or
equivalently, such that removing it from $K$ gives a different
hull.
We will abuse notation and write
\begin{equation}
 \partial K = \left\{k \in K : \text{conv}(K - \{\pi\}) \neq \text{conv}(K)\right\}.\label{eq:conv2}
\end{equation}
In the algebraic setting, and $K = S(\mathcal{A})$ is the set of
states, the extreme points $\partial S(\mathcal{A})$ are called \emph{\gls{pure}
states}.

\begin{figure}[h]
  \centering
  \vspace{-4pt}
  \includegraphics[width=0.58\textwidth]{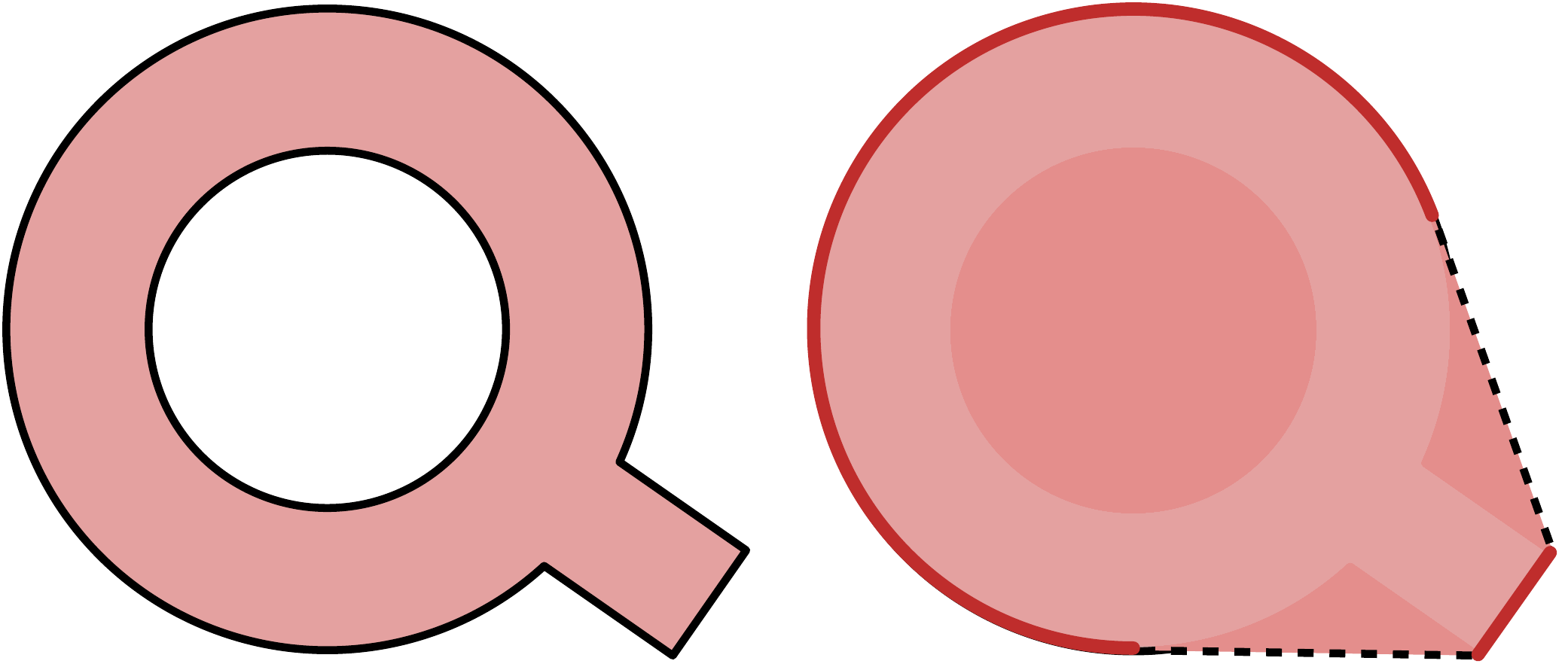}
  \caption{A set and its convex hull, with extreme points in red.}
  \label{fig:krein}
  \vspace{-4pt}
\end{figure}

Fig. \ref{fig:krein} gives a simple example.
Note that the convex hull is in fact the hull of
the extreme points. This fact is true in general, a result
called the \textsc{\gls{krein-milman}}:\sidenote{``On extreme points of regular
  convex sets'' (1940), Mark Krein and David Milman.} 
\begin{equation}
  \label{eq:conv3}
  \text{conv}(K) = \text{conv}(\partial K).
\end{equation}
The intuition is that you can ``unmix'' $k \in \text{conv}(K)$
to obtain ``more extreme'' points that combined to give it, unmix those, and
so on, in such a way that the process terminates with a set of extreme
points. In finite dimensions this argument works nicely, but\marginnote{The
  Axiom of Choice 
  postulates that you can pick an element from a nonempty
  set. Surprisingly, this is one of the most controversial statements
  in mathematics!}
in
infinite dimensions, you need the Axiom of Choice.

The \gls{krein-milman} (\ref{eq:conv3}) tells us that we can obtain
everything by combining \gls{pure} states, so in a sense we are done.
With unitary orbits, we were also ``done'' once we had single
representative; it turned out to be helpful to explicitly construct
the $\mathcal{V}_\pi$, and in particular we recovered the standard
Hilbert space description of states.
Similarly, it is useful here to explicitly construct and work with
\gls{mixed} states, and recover the conventional formalism.
To guide us, we'll work with a concrete example on $\mathcal{A}_\text{Pauli}$:
\begin{equation}
  \label{eq:pi_p}
  \pi_{(p)} = p\pi_{(0)} + (1-p)\pi_{(1)} = p_{b}\pi_{(b)}, 
\end{equation}
which flips a coin and uses $\pi_{(0)}$ with probability $p_0=p$ and
$\pi_{(1)}$ with $p_1= 1-p$.
These states form an axis inside the Bloch sphere:

\begin{figure}[h]
  \centering
  \vspace{-0pt}
  \includegraphics[width=0.28\textwidth]{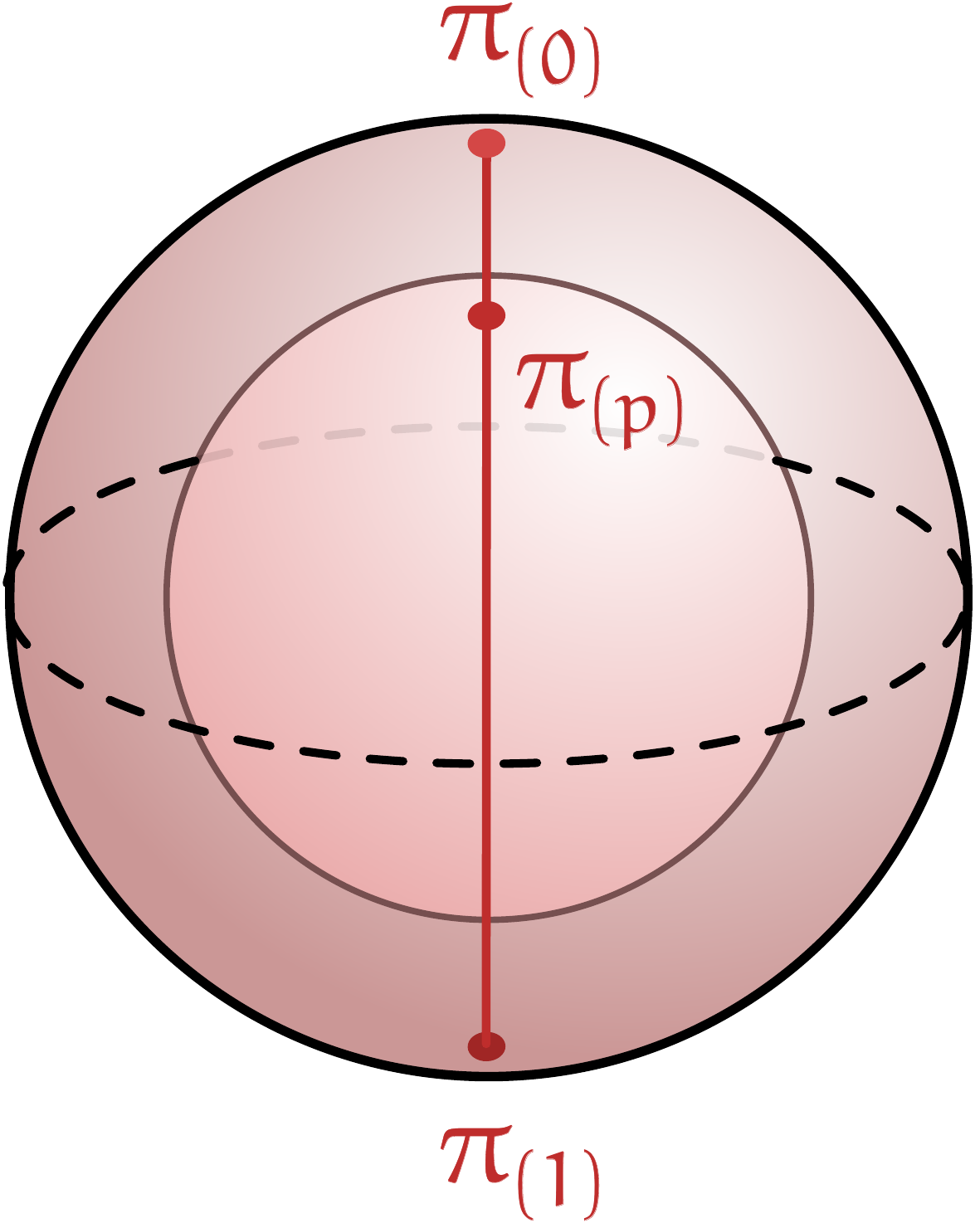}
  \caption{The \emph{Bloch ball} is the convex hull of the Bloch sphere, and
    encodes all states of $\mathcal{A}_\text{Pauli}$. The state $\pi_{(0)}$ is at the north
    pole, $\pi_{(1)}$ at the south, and $\pi_{(1)}$ lies on the axis
    between. The orbit of $\pi_{(p)}$ is a sphere of mixed states with
    equivalent mixedness.}
  \label{fig:cis3}
  \vspace{-2pt}
\end{figure}

\noindent Since every convex line lies
inside, we know from the \gls{krein-milman} that the \emph{Bloch ball} captures all
states of $\mathcal{A}_\text{Pauli}$.

As we showed in (\ref{eq:pi_pZ}), in this state $Z$ has variance
$4p(1-p)$, so it is not \gls{sharp} when $0 < p < 1$.
In fact, you can show\marginnote{The \emph{Law of
    Total Variance} implies that, for \gls{mixed} states $\pi_p=p_i\pi_{(i)}$,
  we have
\[
  \text{var}_p(\Gamma) = p_i\text{var}_i(\Gamma) + \text{var}_{\pi_{(i)} \sim p_i}[\pi_{(i)}(\Gamma)].
\]
The variance of operators in a mixture is the mean
variance plus the variance of means.
For coin flips, this gives (\ref{eq:op-var-mix}).
} that for any operator $A \in
\mathcal{A}_\text{Pauli}$, the variance is
\begin{equation}
  \pi_{(p)}(|\Delta A|^2) = p_b \pi_{(b)}(|\Delta A|^2) +
  p_0p_1|\Delta\pi(A)|^2,\label{eq:op-var-mix}
\end{equation}
where $\Delta \pi = \pi_{(0)} - \pi_{(1)}$.
The only way for an operator to be definite, then, is to be definite
for both $\pi_{(0)}$ and $\pi_{(1)}$, with the same average. But the only such
operator is the identity!\marginnote{Since $Z$ is \gls{sharp} but the
  means differ, while $X$ and $Y$ both have vanishing mean but nonzero
variance.
}
Thus, the \gls{kernel} is trivial, $\mathcal{K}_{(p)} =
\{0\}$,
and the \gls{GNS} Hilbert space is the algebra itself:
\[
  \mathcal{H}_{(p)} = \mathcal{A}_\text{Pauli}/\mathcal{K}_{(p)} \cong \mathcal{A}_\text{Pauli}.
\]
Effectively, this lumps together the columns from
Figs. \ref{fig:gns1} and \ref{fig:gns2} to form a matrix. This is what
we might have guessed would result from combining $\pi_{(0)}$ and $\pi_{(1)}$!

Lumping together two vector spaces $V$ and $W$ (over the same field) is called the \emph{direct
  sum}. We just stack vectors from each space on top of each other,
and define operations component-wise:
\begin{align}
  V \oplus W & = \{(v, w) : v \in V, w \in W\}   \label{eq:dirsum1}\\
  \alpha_i (v_i, w_{\hat\imath}) & = (\alpha_i v_i, \alpha_{\hat\imath} w_{\hat\imath}).   \label{eq:dirsum2}
\end{align}
In (\ref{eq:dirsum2}), we introduce the convention that hatted
indices like are \emph{not} summed over, so that $(v_i, w_{\hat\imath})$ represents a single vector in the direct sum, and
only once we place $\alpha_i$ out front do we sum over $i$.
Returning to our story, the Hilbert space associated with $\pi_p$ is then
\begin{equation}
  \mathcal{H}_{(p)} = \mathcal{H}_{(0)} \oplus \mathcal{H}_{(1)}.\label{eq:p_dirsum}
\end{equation}
Since $\mathcal{H}_{(p)}$ can be broken down into smaller pieces, we say
it is \emph{reducible}; this holds for mixed states in general, since
we can break them down into the component pure Hilbert spaces.
The pure state Hilbert spaces $\mathcal{H}_{(b)}$
are \emph{irreducible}, i.e. cannot be so reduced.\marginnote{To
  reduce them, we'd need a three-dimensional kernel, which we show is
  impossible in the next section.
  \vspace{5pt}}
Again, this statement generalizes, so a state $\pi \in S(\mathcal{A})$
is pure just in case $\mathcal{H}_\pi$ is irreducible.\sidenote{If you
  don't believe me, you may believe Theorem 10.2.3 of \emph{Fundamentals of the Theory of
    Operator Algebras II} (1997), Richard Kadison and John Ringrose.}
We give a more useful purity criterion below.

Although $\pi_{(1)}$ is associated with a distinct \gls{GNS} Hilbert space $\mathcal{H}_{(1)}$, 
recall from Fig. \ref{fig:gns3} that it can be embedded in
$\mathcal{H}_{(1)}$ using a change-of-basis matrix $X$, or equivalently,
as a vector state $X|0\rangle = |1\rangle$.
This lets us evaluate $\pi_{(p)}$ entirely within $\mathcal{H}_{(0)}$:
\begin{equation}
  \pi_{(p)}(A) = p_0 \pi_{(0)}(A) + p_1 \pi_{(1)}(XAX) = p_b\langle
  b|A|\hat{b}\rangle, \label{eq:exp_p}
\end{equation}
where $b \in \{0, 1\}$ and we apply our summation convention to basis
vector labels.
It's convenient to rewrite this using two new operations. First, given
two vector states $|\phi\rangle=U|0\rangle, |\psi\rangle= U'|0\rangle$, we define an operator on
$\mathcal{H}_{(0)}$ called the \emph{outer product} which scales
$|\psi\rangle$ by the overlap with $|\phi\rangle$:
\begin{equation}
  \label{eq:outer-pauli}
  |\psi\rangle\langle \phi| (|\zeta\rangle) = \langle \phi|\zeta\rangle | \psi\rangle.
\end{equation}
This is linear in $|\zeta\rangle$ because the inner product is
linear. A closely related operation is the \emph{trace}, defined by
\begin{equation}
  \label{eq:trace1}
  \mbox{tr}\big[|\psi\rangle \langle \phi| A \big] = \langle \phi | A
  |\psi\rangle = \langle 0 | U^* A U' |0\rangle = \pi(U^*A U'), 
\end{equation}
where $|0\rangle$ is the vector state associated with $\pi$.

\begin{figure}[h]
  \centering
  \vspace{-0pt}
  \includegraphics[width=0.65\textwidth]{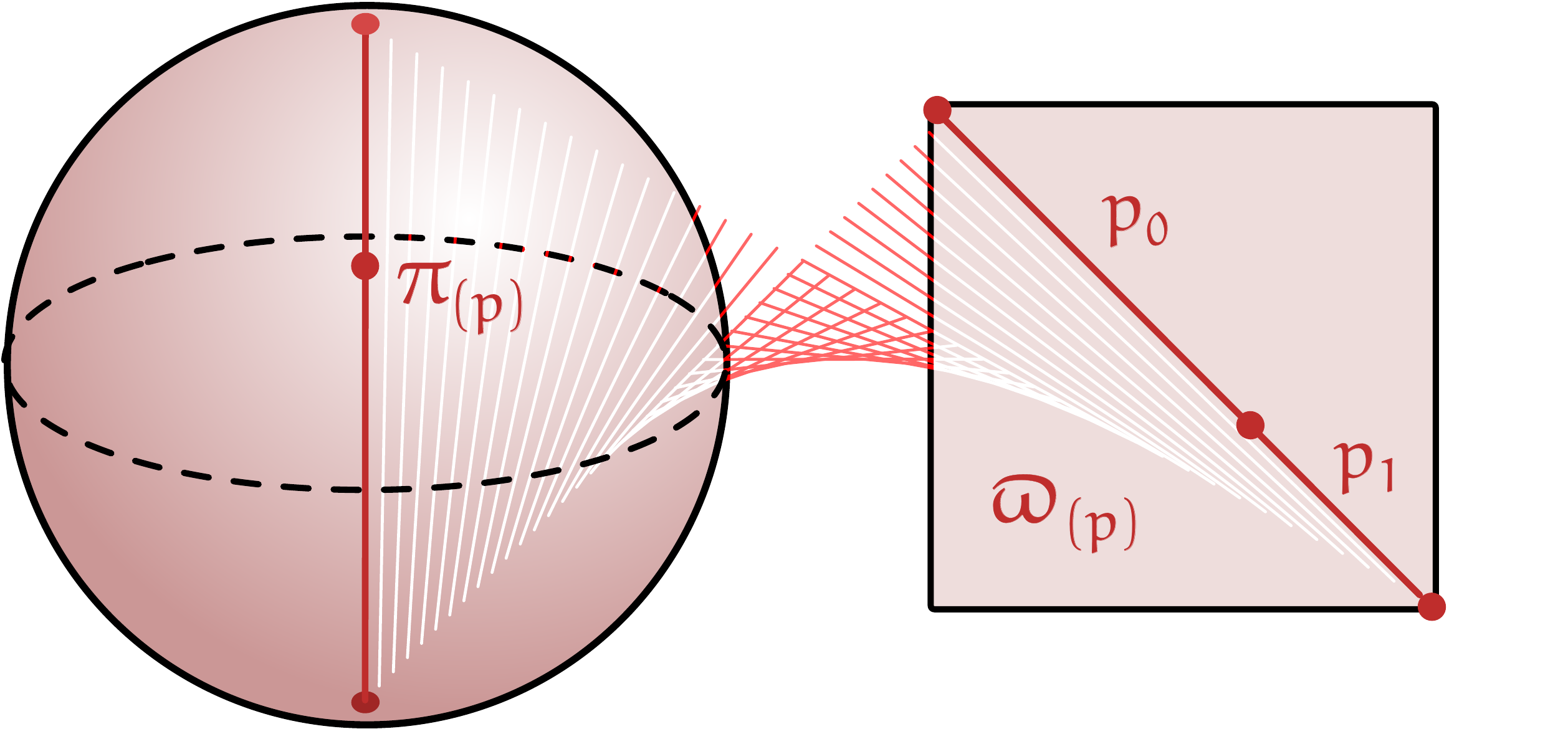}
  \caption{A fanciful depiction of how a \gls{mixed} state $\pi_{(p)}$ in
    the Bloch ball transforms into a density matrix $\varpi_{(p)}$: invert the
    Bloch axis and map onto the principal diagonal. In higher
    dimensions, we unravel a simplex; left as an exercise to readers
    who successfully wove the Hopf fibration.
    }
  \label{fig:cis5}
  \vspace{-2pt}
\end{figure}

Using outer products and the trace, we can write (\ref{eq:exp_p}) as
\begin{align}
  \pi_{(p)}(A) = \mbox{tr}\left[\big(p_b |b\rangle\langle
  \hat{b}|\big)A\right] = \mbox{tr}[\varpi_{(p)} A],
\end{align}
where $\varpi_{(p)}$ is an operator
called the \emph{density matrix}.\marginnote{If
  $\{|i\rangle\}_{i\in\mathfrak{I}}$ is an orthonormal basis of $\mathcal{H}$, then
$\{|i\rangle\langle i|\}_{i,i'\in\mathfrak{I}}$ is an orthonormal basis of
$\mathcal{B}(\mathcal{H})$. Orthonormality is easy, and there are
enough to span the space.}
Fig. \ref{fig:cis5} visualizes how this matrix is formed. Note that, in general, the set of bounded operators
$\mathcal{B}(\mathcal{H})$ is spanned by outer products. This implies
that the trace is defined for arbitrary matrices and
is \emph{cyclic}, in the sense that
\begin{equation}
  \mbox{tr}[AB] = A_{ij}B_{ji} = B_{ji} A_{ij} =
  \mbox{tr}[BA]\label{eq:cyclic}
\end{equation}
for $A = A_{ij}|i\rangle\langle j|$ and $B = B_{ij} |i\rangle \langle j|$.

\newpage 
These results generalize nicely.
Suppose we mix $\pi_{(p)} = p_{i} \pi_{(i)}$, where the states $\pi_{(i)} =
\mathcal{C}^{U_{(i)}}[\pi]$ are unitarily equivalent to some fiducial
representative $\pi$, analogous to $\pi_{(0)}$.\marginnote{``Fiducial'' means
we choose it; ``canonical'' means God chooses it.}
We can write
\begin{equation}
  \pi_{(p)}(A) = p_i\mathcal{C}^{U_{(i)}}[\pi](A) = p_i \pi\big(U_{(i)}^* A
  U_{(\hat\imath)}\big) = \mbox{tr}\big[\varpi_{(p)} A\big],\label{eq:mixed-U}
\end{equation}
for a density matrix
\begin{equation}
  \label{eq:mixed-gen}
  \varpi_{(p)} = p_i  |U_{(i)}\rangle\langle U_{(\hat\imath)}|.
\end{equation}
This is the conventional formalism of \gls{mixed} states.

We can repeat the argument in the \gls{awd} formalism.
To mix unitarily equivalent states, we tether $p_i$ to $U_i$
and conjugate a ficudial state.
It's then straightforward to obtain $\pi_{(p)}(A) = p_i \langle U_{(i)}|A|U_{(\hat\imath)}\rangle$:

\begin{figure}[h]
  \centering
  \vspace{-6pt}
  \includegraphics[width=0.85\textwidth]{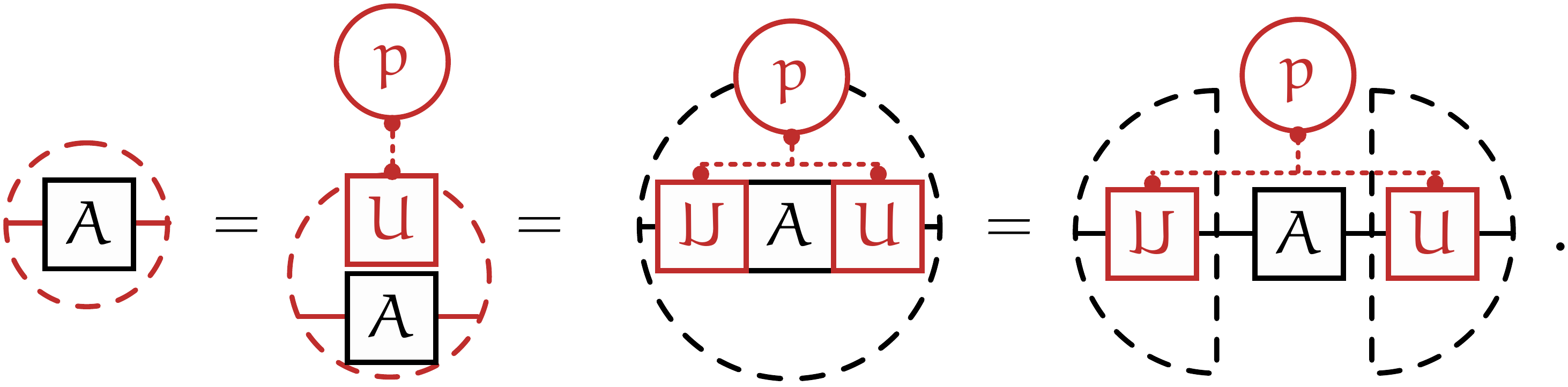}
  \caption{For $\pi_i = \mathcal{C}^{U_{(i)}}[\pi_0]$, the state $\pi_{(p)} =p_i\pi_{(i)}$
    takes a simple form. Note that the tether line can ``split'' when different objects
carry the same index.
  }
  \label{fig:mixed2}
  \vspace{-5pt}
\end{figure}

\noindent Finally, we drag
$|U_{(i)}\rangle$ around to form an outer product:

\begin{figure}[h]
  \centering
  \vspace{-6pt}
  \includegraphics[width=0.57\textwidth]{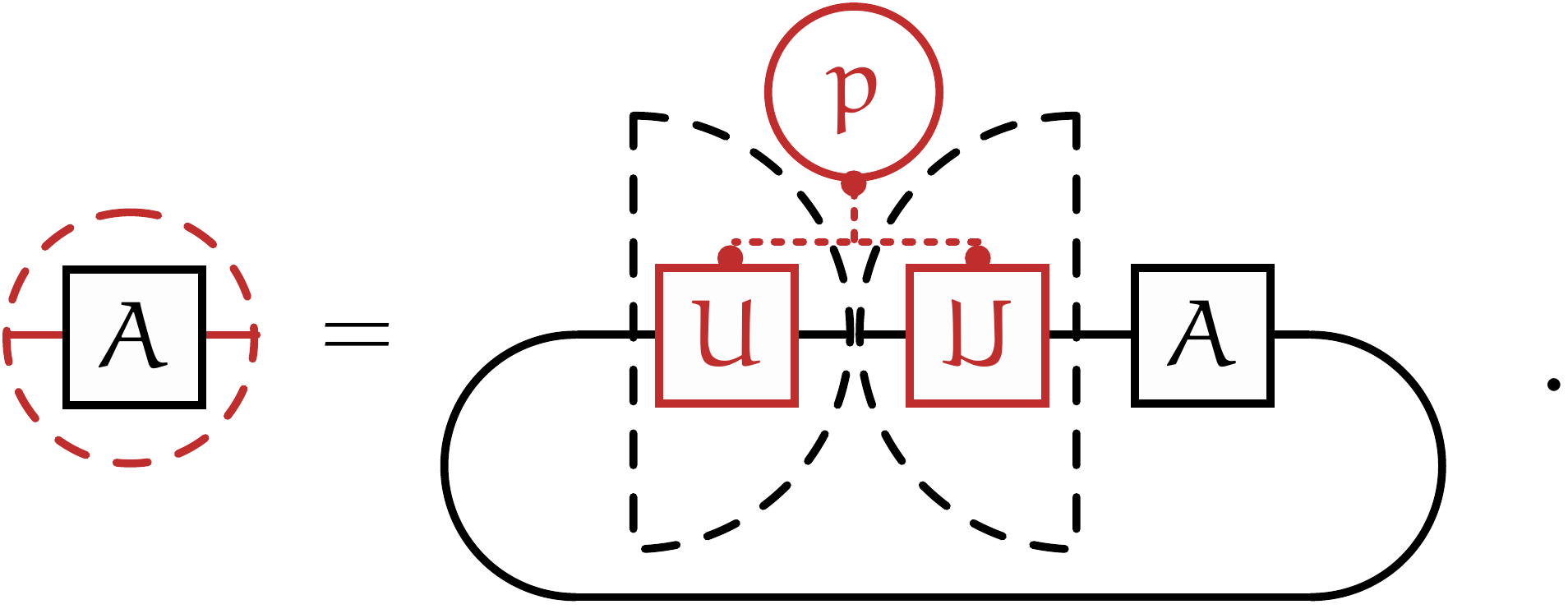}
  \caption{
    Deforming Fig. \ref{fig:mixed2} to obtain
    (\ref{eq:mixed-gen}), with the \gls{awd} of a
    trace. 
  }
  \label{fig:mixed3}
  \vspace{-5pt}
\end{figure}

\noindent This is (\ref{eq:mixed-U}), which embeds
(\ref{eq:mixed-gen}) as the operator next to $A$,
and represents the trace as a solid loop with\marginnote{
  For now, think of this as a convention for the trace; we will
  interpret the loop in terms of entanglement below.
}
operators strung along it anticlockwise; this manifests
(\ref{eq:cyclic}) visually.

\section{10. Pure states}\hypertarget{sec:11}{}

According to the \gls{krein-milman} (\ref{eq:conv3}), everything is made of
\gls{pure} states.
We stated above that these had irreducible GNS Hilbert spaces, but
this turns out to be a cumbersome way of testing purity.\marginnote{Particularly given
  our efforts to escape the inconvenience of Hilbert space!}
We will introduce a method which uses observables directly, led, as
usual, by the Pauli algebra.
In this case, the \gls{pure} states live on the Bloch sphere,
and correspond to \gls{sharp} measurements of a self-adjoint
operator $\sigma(\mathbf{n}) \measure \pm 1$.
This measurement fixes all expectations;\marginnote{An ad hoc way to see this:
  \begin{align*}\pi_{(0)}(X)=\pi_{(0)}(XZ^2)=-\pi_{(0)}(ZXZ),\end{align*} using Pauli
  relations. Since $Z$ is sharp, it factors out to give
  $-\pi_{(0)}(X)$, hence
  $\pi_{(0)}(X)=0$. We give a more general argument below.} e.g., $Z \measure +1$ implies $\pi_{(0)}(X)
=\pi_{(0)}(Y)=0$, so
all expectations can be computed.
Once you fix everything, nothing more can be made sharp,
so $\mathcal{D}_{(\mathbf{n})}$ is \emph{maximal}.
Conversely, in a \gls{mixed} state $\pi_{(p)}$, the \gls{definite} $\mathcal{D}_{(p)}
= \mathbb{R} I$ \emph{can} be extended. Any measurement will do!
                                             
\begin{figure}[h]
  \centering
  \vspace{-4pt}
  \includegraphics[width=0.6\textwidth]{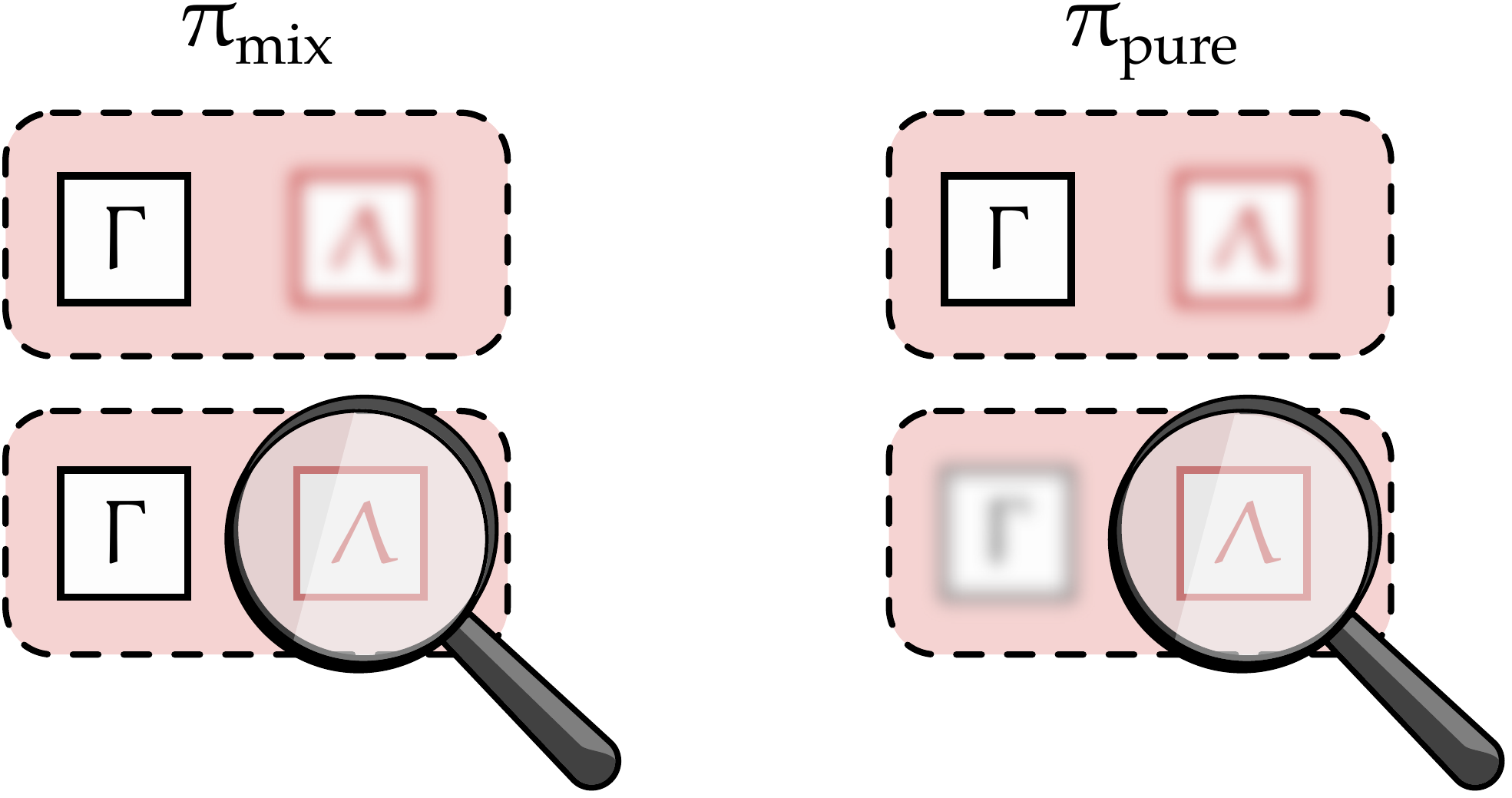}
  \caption{\textsc{Left.} In a \gls{mixed} state, we can increase the set of
    \gls{sharp} observables.
    \textsc{Right.} In \gls{pure} states, defining one observable blurs an
    observable elsewhere.
    }
  \label{fig:stormer1}
  \vspace{-6pt}
\end{figure}

 This equivalence holds in general.
To state it precisely, recall that the \gls{definite} $\mathcal{D}_\pi$
consists of all self-adjoint operators with zero variance under
$\pi$, i.e. $\pi(\Gamma^2)=\pi(\Gamma)^2$.
\textsc{Størmer's Theorem}\sidenote{``A characterization of pure states of
  \gls{cstar}s'' (1967), Erling Størmer.} shows that $\pi$ is \gls{pure} just in case $\mathcal{D}_\pi$ is
\emph{maximal}, that is, there is no
other state $\pi'$ such that $\mathcal{D}_{\pi'} \supsetneq
\mathcal{D}_{\pi}$.
Operationally, if we try to sharpen a blurry observable, another will
become indefinite!
This is an
uncertainty principle, but for a whole collection of operators
$\mathcal{D}_\pi$; see Fig. \ref{fig:stormer1}.

In contrast to the \gls{kernel} $\mathcal{K}_\pi$, which is closed under
products and complex linear combinations, the \gls{definite}
$\mathcal{D}_\pi$ is closed under neither.\marginnote{Self-adjointness is
easily spoiled! For $\Gamma, \Lambda \in
\mathcal{D}_\pi$, $(i\Gamma)^*=-i\Gamma
  \neq i\Gamma$ unless $\Gamma=0$,
and $(\Gamma\Lambda)^*=\Lambda\Gamma \neq \Gamma\Lambda$ unless
$\Gamma$ and $\Lambda$ commute. We'll see in a moment that the Jordan
product gets around this by symmetrizing.}
However, it is closed under \emph{real} linear combinations, since $\Gamma, \Lambda \in
\mathcal{D}_\pi$ and $a, b\in\mathbb{R}$ implies
\[
  (a \Gamma + b \Lambda)^* = \overline{a}\Gamma^* +
  \overline{b}\Lambda^*  = a \Gamma + b \Lambda,
\]
and linear combinations of sharp observables are sharp by the triangle
inequality (\ref{eq:triangle}).
And although $\mathcal{D}_\pi$ is not closed under products, it is closed under a peculiar operation called the
\emph{\gls{jordan}}:\marginnote{``Peculiar'' because it is
  \emph{non-associative}, i.e. $A\circ(B\circ C)\neq (A \circ B)\circ
  C$ in general.}
\begin{equation}
  \label{eq:5}
  \Gamma \circ \Lambda = \frac{1}{2}(\Gamma \Lambda + \Lambda\Gamma) =
  \frac{1}{2}\{\Gamma, \Lambda\},
\end{equation}
where $\{\cdot, \cdot\}$ is the \emph{anticommutator}.
This is commutative by construction, and ensures that
$(\Gamma\circ\Lambda)^*=\Gamma\circ\Lambda$ for self-adjoint
$\Gamma,\Lambda$.
Moreoever, the \gls{jordan} is sharp by equation (\ref{eq:sharp1}), with
\begin{equation}
  \pi(\Gamma \circ \Lambda) = \frac{1}{2} \pi(\Gamma\Lambda) +
  \frac{1}{2} \pi(\Lambda\Gamma) =
  \pi(\Gamma)\pi(\Lambda).\label{eq:jordan-mult}
\end{equation}
Crucially, (\ref{eq:jordan-mult}) tells us that if we restrict the
state $\pi$ to its \gls{definite} $\mathcal{D}_\pi$, the resulting
real-valued functional
$\chi_\pi = \pi|_{\mathcal{D}_\pi}$ is also
\emph{multiplicative} with respect to the \gls{jordan}.

Remarkably, the functionals $\chi_\pi$ fully determine the state when
it is \gls{pure}, as we prove in Appendix \hyperlink{app:stormer}{B}.
We say that $\mathcal{D}_\pi$ is \emph{rigid} for \gls{pure} states.
\marginnote{
  \vspace{-140pt}
  \begin{center}
    \hspace{-10pt}\includegraphics[width=0.99\linewidth]{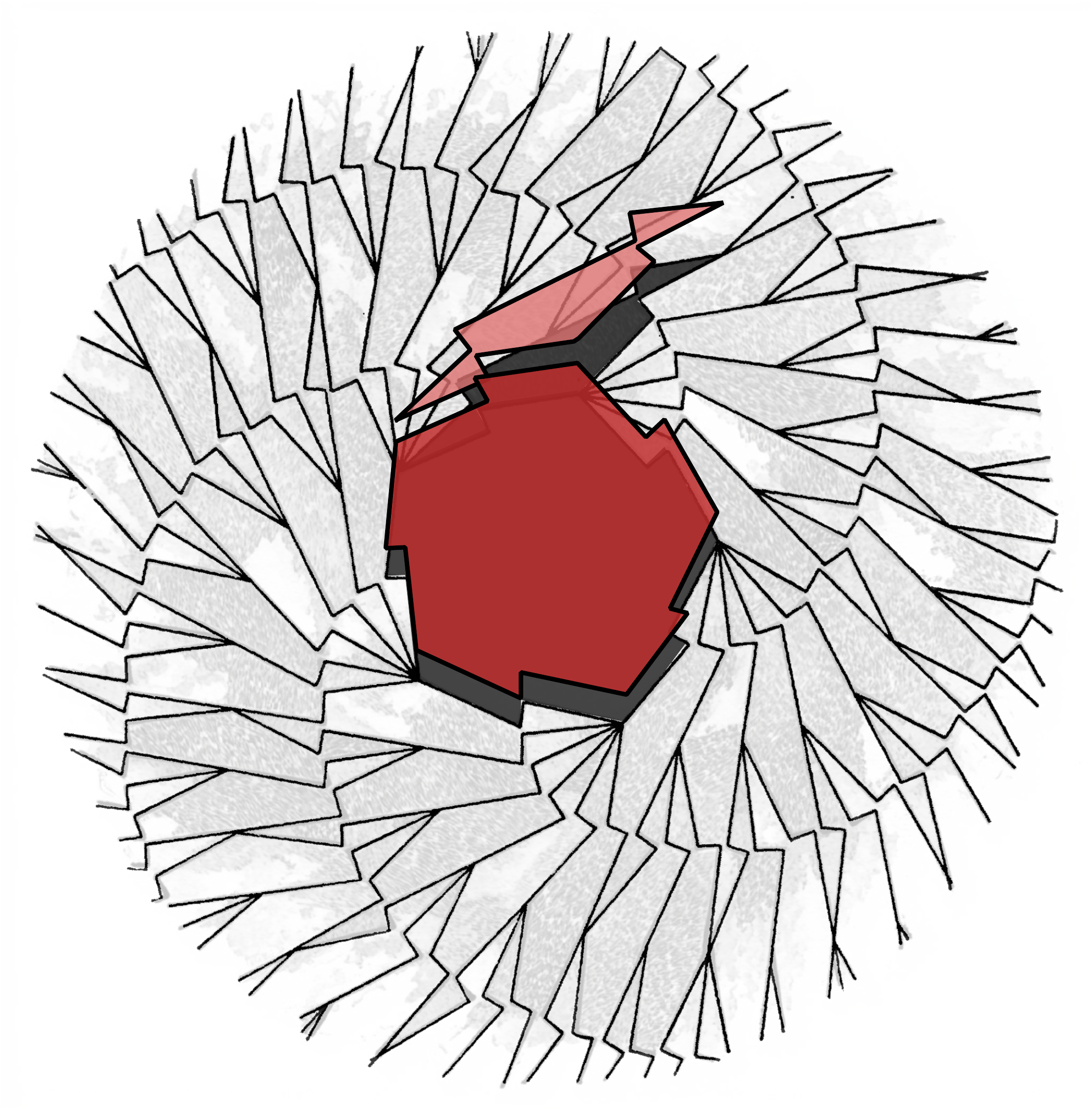}
  \end{center}  \vspace{-0pt}
  \vspace{-10pt}
}
This terminology is borrowed from complex analysis and hyperbolic
geometry, but simple illustrations are provided by planar tilings\sidenote{``Patch-determined tilings'' (1977), Grünbaum and
  Shephard. Above, we modify their Fig. 10: a tiling is
  determined by the placement of a single sawtooth pentagon and
  rhomb.} and even recurrence relations. In the Fibonacci sequence,
$F_0 = 0$ and $F_1=1$ determine everything else!
Consider a family of functions
$\mathcal{F} = \{f: S\to T\}$ with restrictions
$\mathcal{F}|_{\mathcal{I}} = \{f|_{R}: R\to
T\}$.
We say $R\subseteq S$ is \emph{rigid for}
$\mathcal{F}$ if, for all $f' \in \mathcal{F}|_{R}$, there
is a unique $f \in \mathcal{F}$ such that $f' = f|_{R}$.
Thus, \gls{definite}s $\mathcal{D}$ are rigid for \gls{pure} states,
$\mathcal{F} = \partial S(\mathcal{A})$, using our notation
for the convex extreme points.

Rigidity is central to Størmer's theorem. 
Intriguingly, rigidity also characterizes \emph{truth tables}.
Recall from (\ref{eq:bool-val}) that lines of the table are Boolean vectors $v_0
\in \mathbb{B}^{\mathcal{P}}$,
and they uniquely extend to valuations
$v: \mathbb{B}[\mathcal{P}]\to \mathbb{B}$. Put differently,
$\mathcal{P}$ is rigid for the family of Boolean valuations.
The set $\mathcal{P}$ of propositional variables, or more physically,
basic switches, is also \emph{minimal}, since removing a
switch $x$ will leave the truth value of many circuits (such as $x$
itself!) undefined.
This would be analogous to a \gls{mixed} state.
Unlike $\mathcal{X}$, a \gls{definite} $\mathcal{D}$ is a whole vector space
over $\mathbb{R}$ and hence far from minimal;
$\chi$ is determined by the basis of $\mathcal{D}$. Since
we can also multiply elements of the basis, this is not usually
minimal either!
We will define a \emph{switch set} $\mathcal{Q}$ for $\mathcal{D}$ as
one which minimally generates it in the algebraic sense:
\begin{equation}
  \label{eq:switches}
  \mathcal{Q} \in \min \argmax_{\mathcal{G}\subseteq\mathcal{D}}
  \text{J}^\circ\langle \mathcal{G}\rangle,  
\end{equation}
where\marginnote{Note
``min'' indicates the collection of inclusion-minimal elements.
There is always more than one (we can rescale generators), and these
need not have the same size!} the argmax means that $\mathcal{G}$ generates $\mathcal{D}$, the
J${}^\circ$ indicates that generation is with respect to real linear
combinations and the Jordan product, and finally, the min means that
we pick a \emph{minimal} generating set, such that removing any element no longer
generates $\mathcal{D}$.

\begin{figure}[h]
  \centering
  \vspace{-3pt} 
  \includegraphics[width=0.65\textwidth]{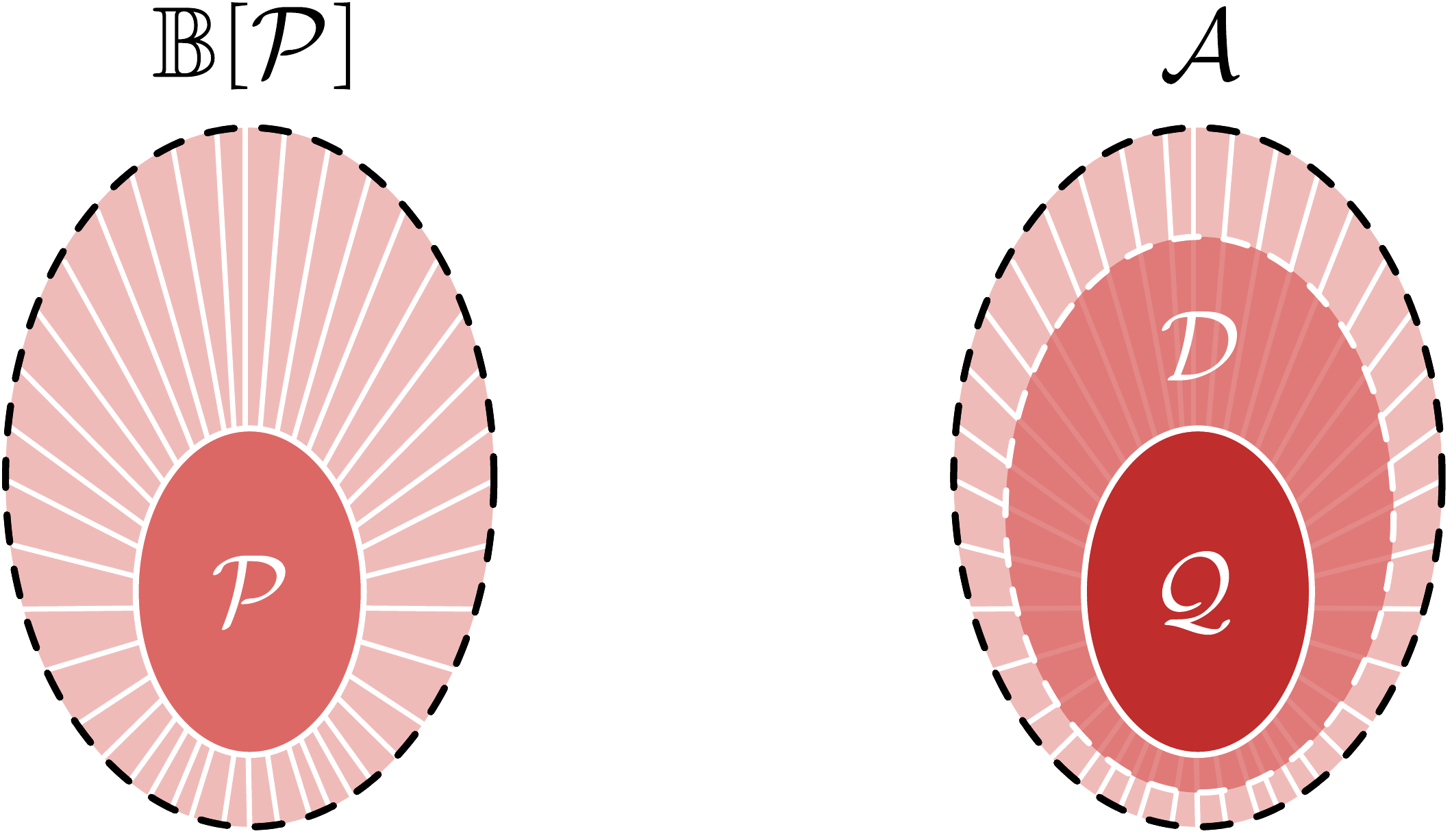}
  \caption{\textsc{Left.} Values on the Boolean
    switches $\mathcal{P}$ determine the value of any
    circuit. \textsc{Right.} Quantum switches
    $\mathcal{Q}$ determine $\mathcal{D}$ determine the full state
    for \gls{pure} $\pi$.}
  \label{fig:rigid}
  \vspace{-6pt}
\end{figure}

Let's flesh this out with an example.
Suppose we rotate our Stern-Gerlach apparatus in the $\mathbf{n}$-direction and measure $\sigma(\mathbf{n}) \measure
+1$. Pauli rotating the $Z\measure +1$ result (\ref{eq:pauli-sharp}), the \gls{kernel} takes the form
\[
  \mathcal{K}_{(\mathbf{n})} =
  \mbox{span}_{\mathbb{C}}\{\sigma(\mathbf{n}) - I, \Theta_{(\mathbf{n})}\},
  \quad \Theta_{(\mathbf{n})} = \sigma(\mathbf{n}_\perp)+ i\sigma(\mathbf{n}\times \mathbf{n}_\perp),
\]
where $\mathbf{n}_\perp$ is any unit vector orthogonal to $\mathbf{n}$.
Since $\Theta_{(\mathbf{n})}$ is not self-adjoint, the \gls{definite} is $\mathcal{D}_{(\mathbf{n})}=\mbox{span}_{\mathbb{R}}\{I,\sigma(\mathbf{n})\}$.
To check rigidity, note that since $\Theta_{(\mathbf{n})}$ has vanishing
mean,\marginnote{To be clear, real and imaginary components
  are \emph{separately} determined due to positivity of states.} 
\[
\text{Re}\left\{\pi_{(\mathbf{n})}[\Theta_{(\mathbf{n})}]\right\}=
\frac{1}{2}\pi_{(\mathbf{n})}[\Theta_{(\mathbf{n})} + \Theta_{(\mathbf{n})}^*] =
\pi_{(\mathbf{n})}[\sigma(\mathbf{n}_\perp)] = 0,
\]
and similarly $\pi_{\pm\mathbf{n}}[\sigma(\mathbf{n}\times\mathbf{n}_\perp)]=0$.
Since $I$, $\sigma(\mathbf{n})$, $\sigma(\mathbf{n}_\perp)$ and
$\sigma(\mathbf{n}\times \mathbf{n}_\perp)$ span the whole
Pauli algebra, the values are indeed determined! The last thing to do
is find a switch set.
Since $\sigma(\mathbf{n})\circ \sigma(\mathbf{n})= I$ is the other basis element of
$\mathcal{D}_{(\mathbf{n})}$, we can just take $\mathcal{Q}_{(\mathbf{n})} =
\{\sigma(\mathbf{n})\}$.
Unsurprisingly, we learn that the value $\sigma(\mathbf{n}) \measure \pm1$
determines everything else.

\addtocontents{toc}{\protect\vspace{-20pt}\protect\contentsline{part}{\textsc{\Large{varieties
      of sharpness}}}{}{}}

\section{11. Tensor products}\hypertarget{sec:12}{}

In a Boolean circuit, we can configure a switch by hand, and each
switch is independent.
In a quantum circuit, we configure a switch by measurement.
Measurement is more unruly than manual configuration, since outcomes are random, and later measurements
can mess up the result of earlier measurements.
We can't configure $\mathcal{Q}$ at will!

It will be instructive to consider a situation where
measurements are, like classical switches, completely independent.
Suppose our system consists of separate parts called
\emph{factors} or \emph{locations}, indexed by $\ell \in \mathfrak{L}$, each of which has 
associated ``local'' observables $A^{(\ell)} \in
\mathcal{A}^{(\ell)}$.\marginnote{We use the
bracketed superscript $(\ell)$ to indicate locations, as distinguished
from unbracketed superscripts (powers) or bracketed subscripts
(labelled operator family). We omit these location
superscripts in explicit tensor products.}
Local measurement is \emph{guaranteed} to be uncorrelated in the sense that
\begin{equation}
\pi_\text{CDP}\left(\prod_{\ell\in\mathfrak{L}} \Lambda^{(\ell)}\right) =
\prod_{\ell\in\mathfrak{L}}\pi_\text{CDP}\left(\Lambda^{(\ell)}\right), \quad \Lambda^{(\ell)} \in
\mathcal{A}^{(\ell)},\label{eq:CDP1}
\end{equation}
provided the systems are far enough away
from each other and have always been so. This is called the \textsc{Cluster
Decomposition Principle}.\sidenote{See ``Cluster
  Decomposition Properties of the $S$ Matrix'' (1963), E.H. Wichmann
  and J.H. Crichton, or Weinberg's magisterial
  \emph{The Quantum Theory of Fields, Volume 1: Foundations}.}
Since a product of scalars commutes, the order in which we measure on each factor is irrelevant.
Hence, if we measure a set of observables $\Lambda^{(\ell)} \in
\mathcal{A}^{(\ell)}$, they will be \emph{simultaneously sharp}, since
we can suppose that each was the last to be measured. We will make
this argument more precise in the \hyperlink{sec:13}{next section}.

\begin{figure}[h]
  \centering
  \vspace{-3pt} 
  \includegraphics[width=0.45\textwidth]{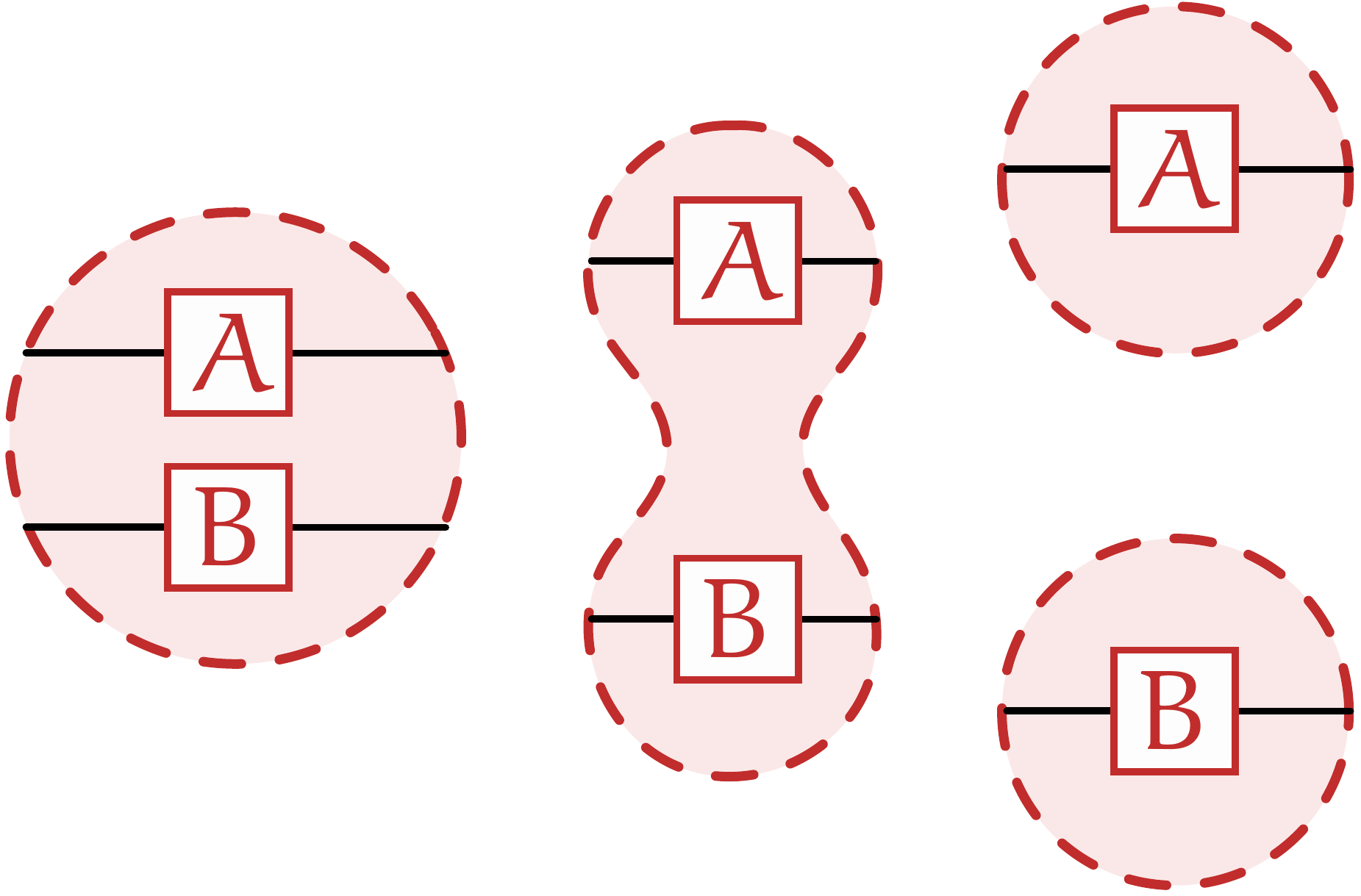}
  \caption{As we separate systems (and their histories), their expectations factorize
    according to the cluster decomposition principle. This implies
    measurements on the different systems are jointly sharp.
  }
  \label{fig:local1}
  \vspace{-6pt}
\end{figure}

\begin{figure}[h]
  \centering
  \vspace{-3pt} 
  \includegraphics[width=0.49\textwidth]{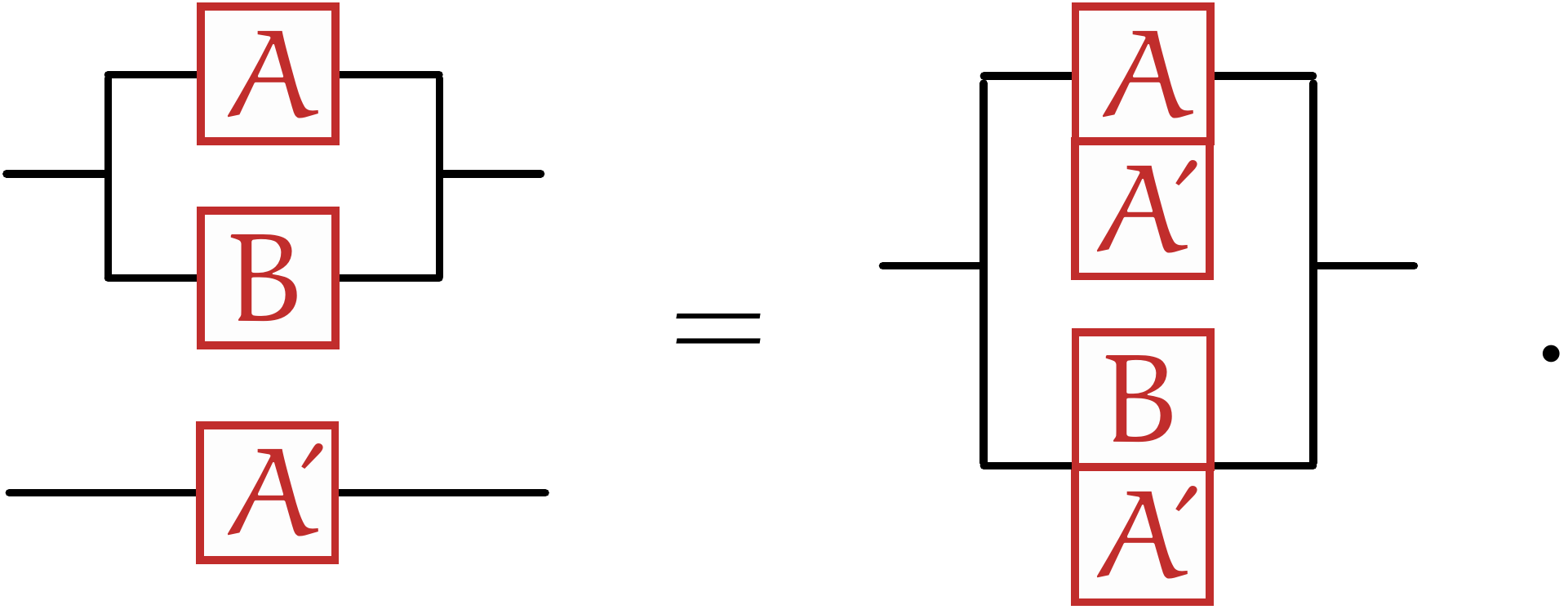}
  \caption{In awds, the tensor product is vertical
    concatenation, read downwards. Linearity ``copies'' from one wire into the
    additive structure of another.
  }
  \label{fig:local2}
  \vspace{-6pt}
\end{figure}

We can describe the algebra acting on this collection of systems in two stages.
For two C${}^*$-algebras $\mathcal{A}^{(1)}$ and $\mathcal{A}^{(2)}$, we first define the
\emph{algebraic tensor product} $\mathcal{A}^{(1)} \odot
\mathcal{A}^{(2)}$ as 
the set of pairs $\mathcal{A}^{(1)}\times \mathcal{A}^{(2)}$ subject to a
\emph{bilinearity relation} (see Fig. \ref{fig:local2}):
\begin{align}
  \label{eq:tensor1}
\mathcal{A}^{(1)}
\odot \mathcal{A}^{(2)} & = \mbox{span}_{\mathbb{C}}(\mathcal{A}^{(1)}\times \mathcal{A}^{(2)})/\mathcal{I}_\text{BL},
\end{align}
where $\mathcal{I}_\text{BL}$ encodes linearity in each
component:
\begin{align}
  (\alpha A_1 + \beta B_1, A_2) & \equiv_{\text{BL}} \alpha (A_1, A_2) + \beta (B_1, A_2) \notag\\
    (A_1, \alpha A_1 + \beta B_2) & \equiv_{\text{BL}} \alpha (A_1,A_2) + \beta (A_1,B_2).  \label{eq:tensor2}
\end{align}
We picture additivity in
Fig. \ref{fig:local2}. We write $A_1 \odot A_2$ for the equivalence class $[(A_1,A_2)
]_{\mathcal{I}_\text{BL}}$.
The product is defined component-wise:
\begin{align}
  \label{eq:tensor3}
  (A_1 \odot A_2) (B_1 \odot B_2) = A_1B_1 \odot A_2B_2.
\end{align}
With these definitions, $\mathcal{A}^{(1)} \odot \mathcal{A}^{(2)}$ is a
\gls{staralg}.

\newpage

To promote this to a genuine C${}^*$-algebra, we need a norm satisfying
(\ref{eq:C${}^*$}).
Here is a place where Hilbert space is very handy! We can use the GNS construction to identify abstract operators $A \in
\mathcal{A}$ with bounded operators $\Phi_\pi(A) = A_\pi \in
\mathcal{B}(\mathcal{H}_\pi)$, and the tensor product with the familiar Hilbert
space tensor product $\otimes$, so
\begin{equation}
\Vert A_1 \odot A_2\Vert_{(\pi_1, \pi_2)} = \Vert \Phi_{\pi_1}(A_1)\otimes
\Phi_{\pi_2}(A_2)\Vert_\text{op},\label{eq:tensor4}
\end{equation}
where $\Vert\cdot \Vert_\text{op}$ is the operator norm.
An arbitrary state can lose information about operators, and thereby decrease the operator norm; however, it can
never \emph{increase} it.\marginnote{Since it involves a supremum over
vectors, losing vectors only decreases it.}
We are thus led to define the \emph{spatial norm} as the supremum of
(\ref{eq:tensor4}) over choice of states:
\begin{align}
  \label{eq:10}
  \Vert A_{1(i)} \odot A_{2(i)} \Vert_{\text{min}} = \sup_{(\pi_1,\pi_2)} \left\{ \Vert
  \Phi_{\pi_1}(A_{1(i)})\otimes \Phi_{\pi_2}(A_{2(i)}) \Vert_\text{op}\right\},
\end{align}
where, as usual, we sum over the repeated index $i$, and unbracketed
subscripts denote copies.
Finally, we define the \emph{spatial tensor product}
\begin{align}
  \label{eq:3}
  \mathcal{A}^{(1)} \otimes \mathcal{A}^{(2)} = \overline{\mathcal{A}^{(1)} \odot \mathcal{A}^{(2)}}^{\Vert\cdot\Vert_\text{min}},
\end{align}
i.e. the completion of the algebraic tensor product with respect to
the spatial norm.
It can be shown\sidenote{\emph{C${}^*$-Algebras and
    Finite-Dimensional Approximations} (2008), N. Brown and N. Ozawa.} not only that this is a C${}^*$-norm, but the smallest
possible norm, hence the subscripted ``min''.

The state space for $\mathcal{A}^{(1)}\otimes \mathcal{A}^{(2)}$ consists of
positive, normalized functionals, with linearity on the tensor
product\marginnote{The identity is $I^{(1)} \otimes I^{(2)}$, and since $(\pi_1
  \otimes \pi_2)(I^{(1)}I^{(2)}) = \pi_1(I^{(1)})\pi_2(I^{(2)})$,
  if we
rescale $\pi_1$ to ensure $\pi_1(I^{(1)}) = 1$, then $\pi_2(I^{(2)}) = 1$. A similar
argument works for positivity.}
equivalent to linearity in each slot. We can always rescale a
positive, normalized product so that the factors are positive and
normalized. Hence:
\begin{equation}
  S(\mathcal{A}^{(1)}\otimes \mathcal{A}^{(2)}) \cong S(\mathcal{A}^{(1)})\otimes
  S(\mathcal{A}^{(2)}).\label{eq:22}
\end{equation}
Iterating the tensor product to multiple
factors $\mathcal{A} =
\bigotimes_{\ell\in\mathfrak{L}}\mathcal{A}^{(\ell)}$ gives an
associated state space $S(\mathcal{A}) = \bigotimes_{\ell\in\mathfrak{L}}S(\mathcal{A}^{(\ell)})$.
In the language of tensor products, cluster decomposition (\ref{eq:CDP1}) says that systems with historically well-separated factors
have states of the form
\begin{equation}
  \label{eq:17}
  \pi = \bigotimes_{\ell\in\mathfrak{L}} \pi^{(\ell)},
\end{equation}
called \emph{separable} or \emph{product states}. A state which is not
separable (with respect to the labelling $\ell \in \mathfrak{L}$) is
called \emph{entangled}.\marginnote{
  \vspace{-75pt}
  \begin{center}
    \hspace{-25pt}\includegraphics[width=1.2\linewidth]{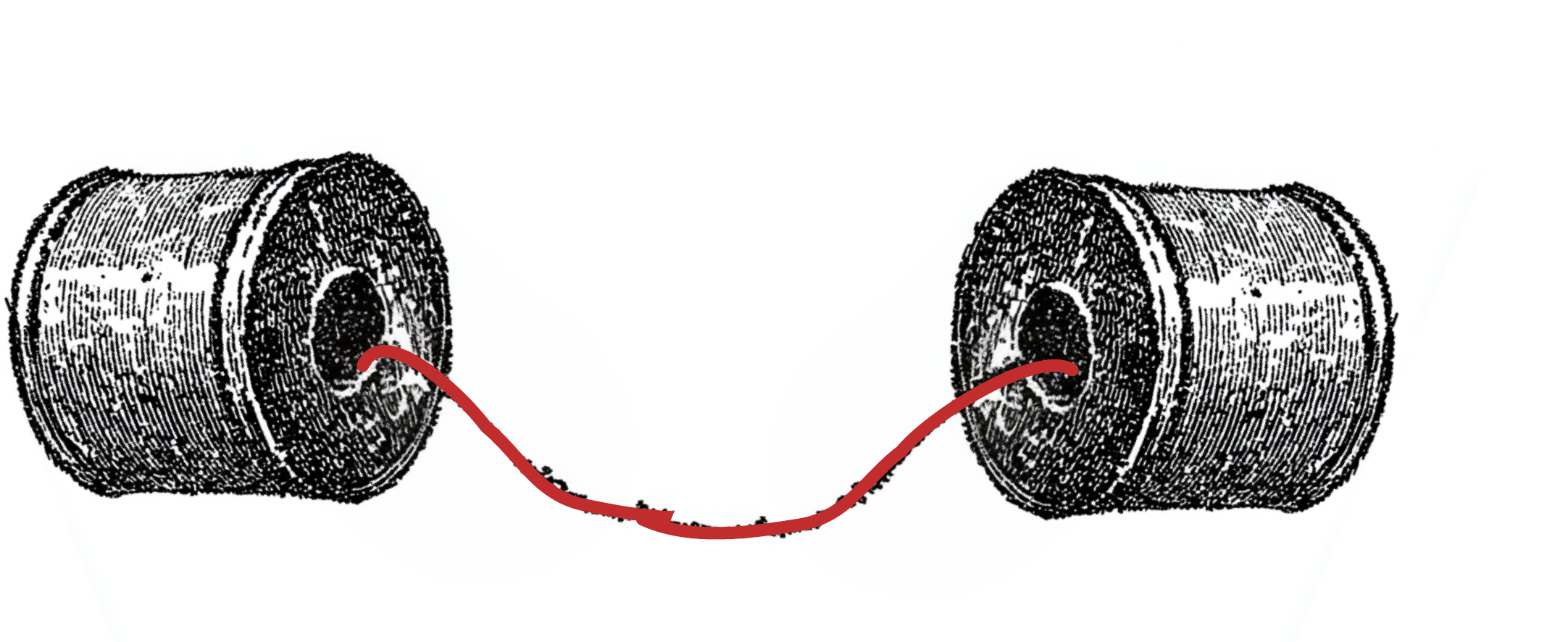}
  \end{center}  \vspace{-5pt}
  \emph{An entangled pair, aka Bell telephones.
  } \vspace{-5pt}
}

To see how it works, let's take two copies of the Pauli algebra.
If we choose a state $\pi^{(1)}$ on the first copy, the algebra becomes isomorphic
to $\mathsf{M}_2(\mathbb{C})$.
For a basis matrix $E_{(ab)} = |a\rangle\langle b|$ in the first copy,
we think of $E_{(ab)} \otimes A$ as placing $A$ in entry $(a, b)$, and
hence
\[
  \begin{bmatrix}
    \alpha & \gamma \\
    \beta & \delta
  \end{bmatrix} \otimes A \cong   \begin{bmatrix}
    \alpha A & \gamma A \\
    \beta A & \delta A
  \end{bmatrix}.
\]
This is our usual nested matrix notation.
This generalizes so that the tensor product of
$\mathcal{A}$ and complex-valued square matrices gives
$\mathcal{A}$-valued squares matrices:
\begin{equation}
  \label{eq:19}
  \mathsf{M}_n(\mathbb{C}) \otimes \mathcal{A} \cong \mathsf{M}_n(\mathcal{A}).
\end{equation}
Back to our example, choosing $\pi^{(1)}$ on the second copy makes
it, too, equivalent to $\mathsf{M}_2(\mathbb{C})$, so we get a
matrix-valued matrix equivalent to $\mathsf{M}_4(\mathbb{C})$.\marginnote{You can literally put an $X$ in your
  $X$, so you can swap when you swap.
}
This identification does not depend on having an entangled state, but
simply makes the isomorphism more complicated.

Just as we concatenate boxes, 
we concatenate (or even fuse)
wires, as in Fig. \ref{fig:local2}. 
For instance, consider the \emph{CNOT operator}. We can express this in
terms of the \emph{projectors}\marginnote{These projectors already showed up as
  generators of the kernel (\ref{eq:pauli-sharp}) and Hilbert space
  (\ref{eq:pauli-hilbert}) for $\pi_{(0)}$.}
$E_{(00)} = |0\rangle\langle 0|$ and $E_{(11)} =
|1\rangle \langle 1|$ associated with $Z$.
We define
\begin{equation}
  \label{eq:13}
  \text{CNOT} = E_{(b\hat{b})} \otimes X^{b} = |0\rangle \langle
  0|\otimes I + |1\rangle \langle 1| \otimes X \cong
  \begin{bmatrix}
    I & \\
    & X 
  \end{bmatrix}
,
\end{equation}
where $\hat{b}$ means that this occurrence of $b$ is exempt from the
Einstein convention.
CNOT applies $X^b$ to the second qubit
if the first is $|b\rangle = X^b |0\rangle$.
Since $E_{(b\hat{b})} = \mathcal{C}_{X^b}\big[|0\rangle\langle 0|\big]$, we can compactly
write

\begin{figure}[h]
  \centering
  \vspace{-3pt} 
  \includegraphics[width=0.53\textwidth]{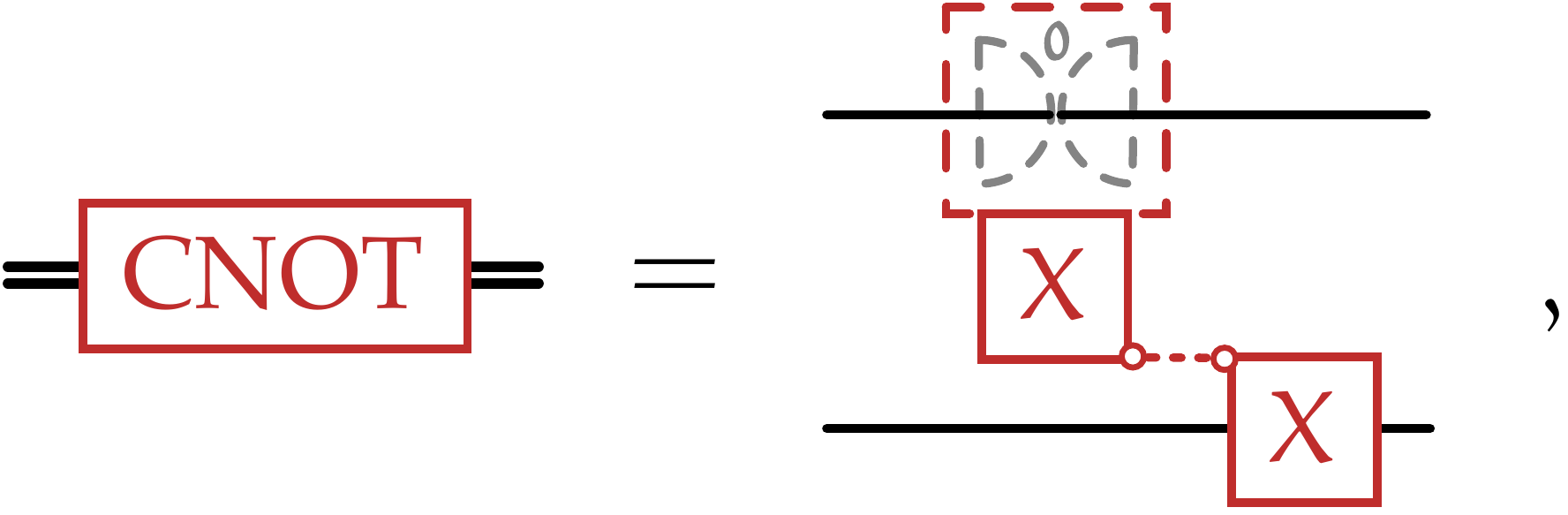}
  \caption{CNOT in awd notation. Index nodes on corners
    indicate a power $X^b$. Above, $|0\rangle
    \langle 0|$ is conjugated by $X^b$, tethered 
    $X^b$ below. Note that we give a more explicit way to bunch and unbunch wires
    in \S 15.
  }
  \label{fig:tensor2}
  \vspace{-6pt}
\end{figure}

\noindent where, by convention, index nodes on box
corners stand for powers, and the
$0$ indicates fiducial state $\pi_{(0)}$.

\begin{figure}[h]
  \centering
  \vspace{-3pt} 
  \includegraphics[width=0.67\textwidth]{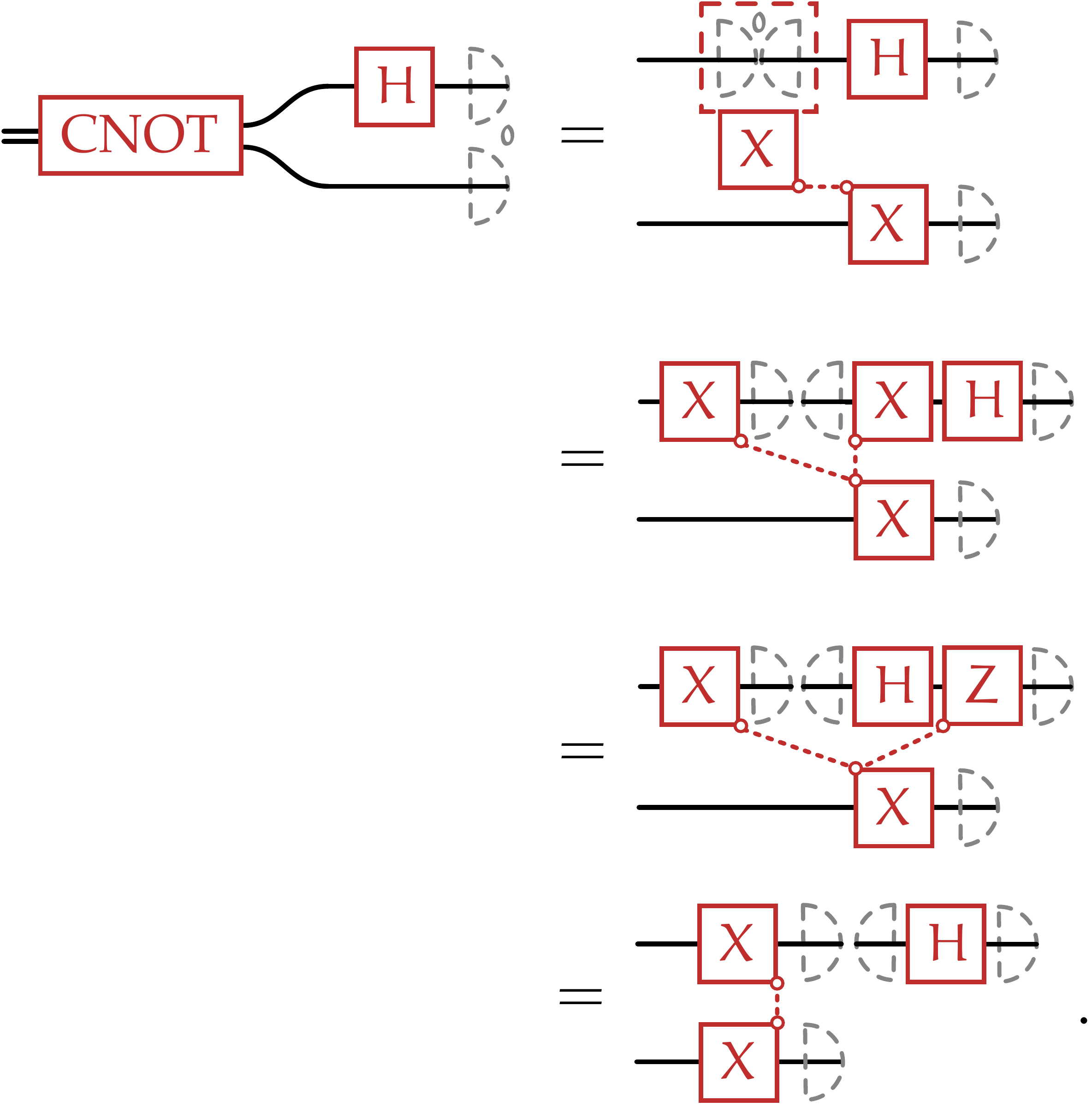}
  \caption{Creation of the Bell state. \textsc{Line 1.} We insert
    the definition of CNOT from Fig. \ref{fig:tensor2}. \textsc{Line
      2}. We expand the red conjugator, keeping track of tether
    lines. \textsc{Line 3.} We use the identity $XH =
    HZ$. \textsc{Line 4.} We absorb $Z^b |0\rangle = |0\rangle$,
    leaving $\langle 0|H|0\rangle =
    \tfrac{1}{\sqrt{2}}$.
  }
  \label{fig:tensor3}
  \vspace{-6pt}
\end{figure}

A crucial application of the CNOT is building entangled states. The
\emph{Hadamard operator} $H = \tfrac{1}{\sqrt{2}}(X + Z)$ is a generalized Pauli
operator which obeys $XH = HZ$.\marginnote{Since $X(X + Z) = I+ XZ
  = (Z+X)Z$.}
Starting both qubits in state $|0\rangle$, applying the Hadamard to
the first, then CNOT, and simplifying, gives Fig. \ref{fig:tensor3}.
There is a fair bit going on in the diagram (see caption for
a blow-by-blow summary), but the final result is a vector
\begin{equation}
  \label{eq:bell}
  |\Psi_\text{Bell}\rangle = 
  \frac{1}{\sqrt{2}}|bb\rangle
\end{equation}
we call the \emph{Bell (vector) state},\sidenote{``On the Einstein
  Podolsky Rosen paradox'' (1964), John Stewart Bell; see also ``Can
  Quantum-Mechanical Description of Physical Reality be Considered
  Complete?'' (1935), Einstein, Podolsky and Rosen.} or
the equivalent functional
\begin{equation}
  \pi_\text{Bell}(A)=\mbox{tr}\big[\varpi_\text{Bell}A\big], \quad \varpi_\text{Bell} = 
  \frac{1}{2}|b\rangle\langle a|\otimes |b\rangle\langle a|.\label{eq:varpi-bell}
\end{equation}
To confirm this state is entangled, there is a nifty shortcut.
Recall that the trace (\ref{eq:trace1}) was
defined for outer products $|\psi\rangle\langle \phi|$ and extended
to arbitrary operators by linearity, since the outer products span
$\mathcal{B}(\mathcal{H})$.
Similarly, we can define the \emph{partial trace} for $A_1 \otimes
A_2 \in \mathcal{A}^{(1)} \otimes \mathcal{A}^{(2)}$ by
\begin{equation}
  \label{eq:part-trace}
  \mbox{tr}_{(1)} \big[A_1 \otimes
A_2\big] = \mbox{tr}[A_1] A_2, \quad \mbox{tr}_{(2)} \big[A_1 \otimes
A_2\big] = \mbox{tr}[A_2] A_1,
\end{equation}
where the subscript indicates the system we trace over.
We then extend to arbitrary operators in
$\mathcal{A}^{(1)}\otimes\mathcal{A}^{(2)}$ by linearity.

The \emph{reduced density matrix} is the result of performing a
partial trace (\ref{eq:part-trace}) on a density matrix $\varpi$ and
seeing what's left over.
We write
  \begin{equation}
  \varpi^{(1)} = \mbox{tr}_{(2)}[\varpi], \quad \varpi^{(2)} =
  \mbox{tr}_{(1)}[\varpi],\label{eq:part-trace2}
\end{equation}
so $\varpi^{(\ell)}$ traces over the locations $\ell' \neq \ell$.
It follows from (\ref{eq:part-trace}) and (\ref{eq:part-trace2}) that,
if $\varpi = \rho_1 \otimes \rho_2$ is separable,
then 
\[
  \varpi^{(1)} \otimes \varpi^{(2)} = \mbox{tr}[\rho_1]\mbox{tr}[\rho_2]
  \rho_1 \otimes \rho_2.
\]
From (\ref{eq:mixed-U}), it follows that $\mbox{tr}[\rho_1] =
\pi_{\rho_1}(I^{(1)}) = 1$ for any density matrix $\rho_1$. Thus, a separable
density $\varpi$ is the product of its reduced densities:
\begin{equation}
  \label{eq:21}
  \varpi^{(1)} \otimes \varpi^{(2)} = \rho_1 \otimes \rho_2 = \varpi.
\end{equation}
Applying this prescription to (\ref{eq:varpi-bell}), we find that the
reduced density
\begin{equation}
  \varpi_\text{Bell}^{(\ell)} = \frac{1}{2}
  \mbox{tr}\big[|a\rangle\langle b|\big]|a\rangle\langle b| =
\frac{1}{2} \delta_{ab}|a\rangle\langle b| = \frac{1}{2}I^{(\ell)}.
  \label{eq:max-mix}
\end{equation}
This is proportional to the identity, and reveals nothing about the
state. It is called the \emph{maximally mixed state}, since
measurements of it are maximally random, and corresponds to the
centre of the Bloch ball.
The tensor product of $I^{(1)}$ and  $I^{(2)}$ is just the identity $I
\in \mathcal{A}^{\otimes 2}_\text{Pauli}$,
\emph{not} the Bell density, so $|\Psi_\text{Bell}\rangle$ is
entangled as claimed. 

\section{12. Commuting factors}\hypertarget{sec:13}{}

Tensor products are a good way to describe systems 
spread across different locations, when state is separable as in cluster
decomposition (\ref{eq:CDP1}), or entangled like Bell (\ref{eq:bell}).
We explained in the former case how to make jointly sharp observables,
simply by measuring on each factor.
In fact, this remains true even when a state is entangled! But the
expectations need not factorize.\marginnote{
  \vspace{-80pt}
  \begin{center}
    \hspace{-10pt}\includegraphics[width=0.65
    \linewidth]{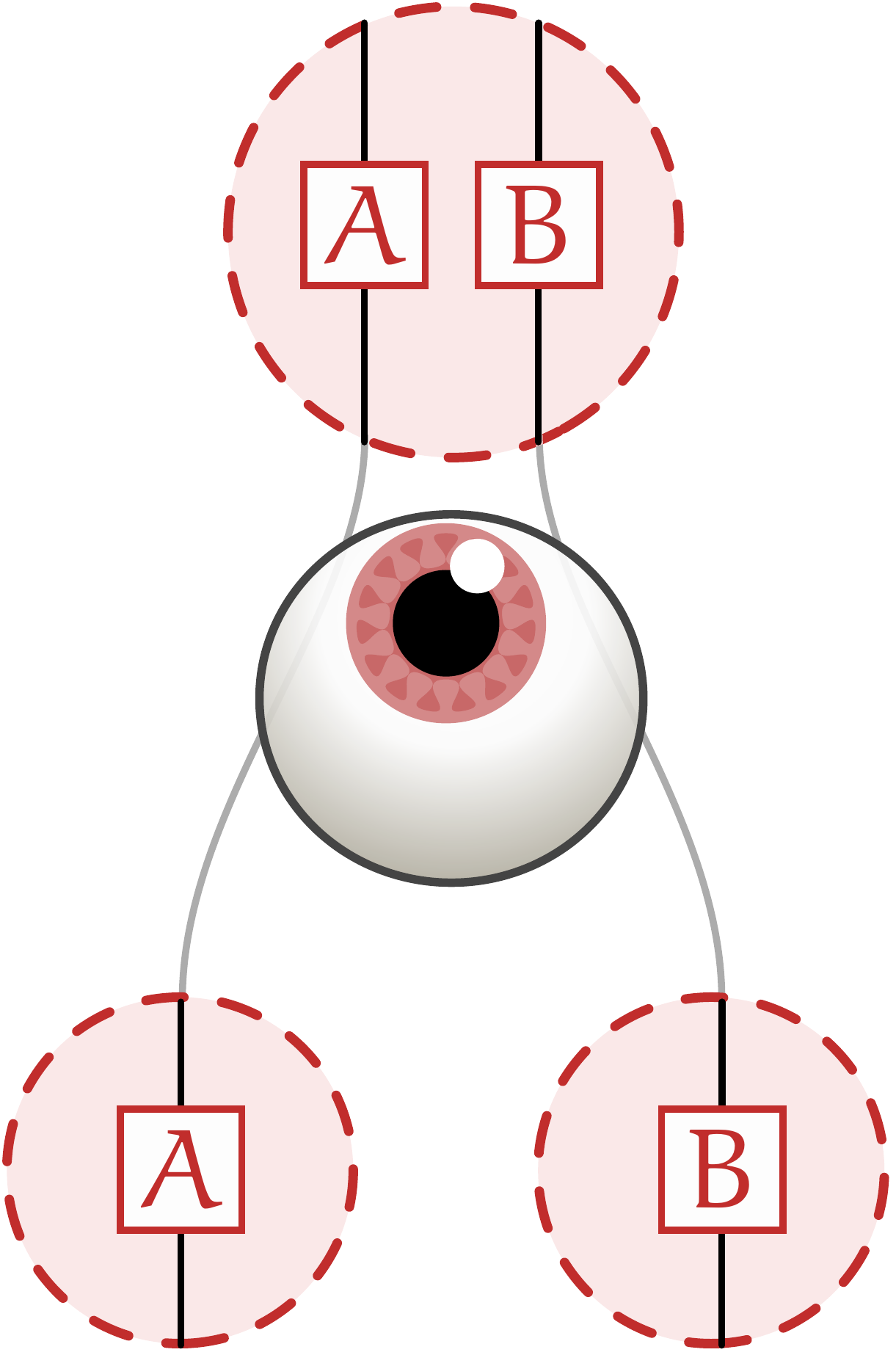}
  \end{center}  \vspace{-0pt}
  \emph{Observer-induced factorization.
  } \vspace{10pt}
}
To illustrate this subtlety, we can use the Bell state
(\ref{eq:varpi-bell}), and measure $Z \otimes Z$ with average
\begin{align*}
  \pi_\text{Bell}(Z \otimes Z) & = \frac{1}{2} \langle a| Z
                                 |b\rangle\langle a|Z|b\rangle
    = \frac{1}{2}\delta_{ab}\delta_{ab} = 1. 
\end{align*}
On the other hand, the individual measurements $Z^{(1)} = Z \otimes I$
and\marginnote{When we superscript an operator like $Z$ which is
  associated with a single factor, we will in general tensor it with
  identities in the remaining locations. This is in contrast to an
  expression like $\varpi^{(\ell)}$, which lives on factor $\ell$.}
$Z^{(2)} = I \otimes Z$ have vanishing expectation:
\[
  \pi_\text{Bell}(Z^{(\ell)}) = \frac{1}{2} \langle a| Z
  |b\rangle\langle a|b\rangle = \frac{1}{2} (-1)^b\langle
  a|b\rangle\delta_{ab} = 0.
\]
Thus, $\pi_\text{Bell}(Z^{(1)}Z^{(2)}) \neq
\pi_\text{Bell}(Z^{(1)})\pi_\text{Bell}(Z^{(2)})$, and entanglement
manifests as non-factorization.

To see why measuring on each factor nevertheless leads to jointly
sharp $Z^{(\ell)}$, we need to explain how to measure a subsystem.
We give full details in the \hyperlink{sec:14}{next section}, but
for now, we get by with a rule for \emph{post-selected partial
  measurement (PPM)}, consisting of two steps. First, for a
global state $\pi$, measuring location $\ell \in
\mathfrak{L}$ factorizes the density into the local reduced density
and its complement:\sidenote{Why? Loosely speaking, because measurement
  \emph{maximally entangles} the measuring apparatus with system $\ell$, and
  entanglement is \emph{monogamous}. See for instance ``Distributed
  Entanglement'' (2000), Coffman, Kundu and Wootters.}
\begin{equation}
  \label{eq:ppm1}
  \varpi_\pi \mapsto 
  \mbox{tr}_{(\overline{\ell})} \varpi_\pi \otimes \mbox{tr}_{(\ell)}
  \varpi_\pi = \varpi_\pi^{(\ell)}\otimes
  \varpi_\pi^{(\overline{\ell})},
\end{equation}
generalizing the partial trace notation of (\ref{eq:part-trace2}).
For the second stage, measuring $\Lambda \measure
\lambda$ on $\ell$ further modifies (\ref{eq:ppm1}) to
\begin{equation}
  \label{eq:ppm2}
  \varpi_\pi^{(\ell)}\otimes
  \varpi_\pi^{(\overline{\ell})} \mapsto p_\lambda^{-1}\Pi_\lambda \otimes
  \varpi_\pi^{(\overline{\ell})}
\end{equation}
where $p_\lambda = \mbox{tr}[\varpi^{(\ell)}_\pi \Pi_\lambda]$, and $\Pi_\lambda$ projects onto the $\lambda$
eigenspace, i.e.
\begin{equation}
  \label{eq:ppm3}
 \Pi_\lambda \Lambda = \lambda \Pi_\lambda = \Lambda\Pi_\lambda.
\end{equation}
Equation (\ref{eq:ppm1}) is the ``partial'' and (\ref{eq:ppm2}) the
``post-selected'' of PPM.
From (\ref{eq:ppm1})--(\ref{eq:ppm3}), we see that measurements of
system $\ell$ disentangle it from $\overline{\ell}$, and no
measurement of $\overline{\ell}$ can re-entangle it.
Moreoever, the measured local observables $\Lambda^{(\ell)}$ will be
sharp, since\marginnote{Using idempotence of $\Pi_\lambda$, cyclicity
  of the trace, and the definition of $p_\lambda$.}
\[
  \pi^{(\ell)}_\text{PPM}(\Lambda^k) = p_\lambda^{-1}\mbox{tr}\big[\Pi_\lambda \Lambda^k
 \big] = p_\lambda^{-1}\mbox{tr}\big[\Pi_\lambda \Lambda^k
 \Pi_\lambda\big] = \lambda^k,
\]
and hence $\pi^{(\ell)}_\text{PPM}(\Delta \Lambda^2) = \pi^{(\ell)}_\text{PPM}(\Lambda^2)-\pi^{(\ell)}_\text{PPM}(\Lambda)^2 = 0$.
This shows that, given a tensor factorization
$\bigotimes_{\ell \in\mathfrak{L}} \mathcal{A}^{(\ell)}$, measurement of
a set of local observables $\Lambda^{(\ell)}$ makes them jointly
sharp. 

\newpage

\marginnote{
  \vspace{-20pt}
  \begin{center}
    \hspace{15pt}\includegraphics[width=0.95\linewidth]{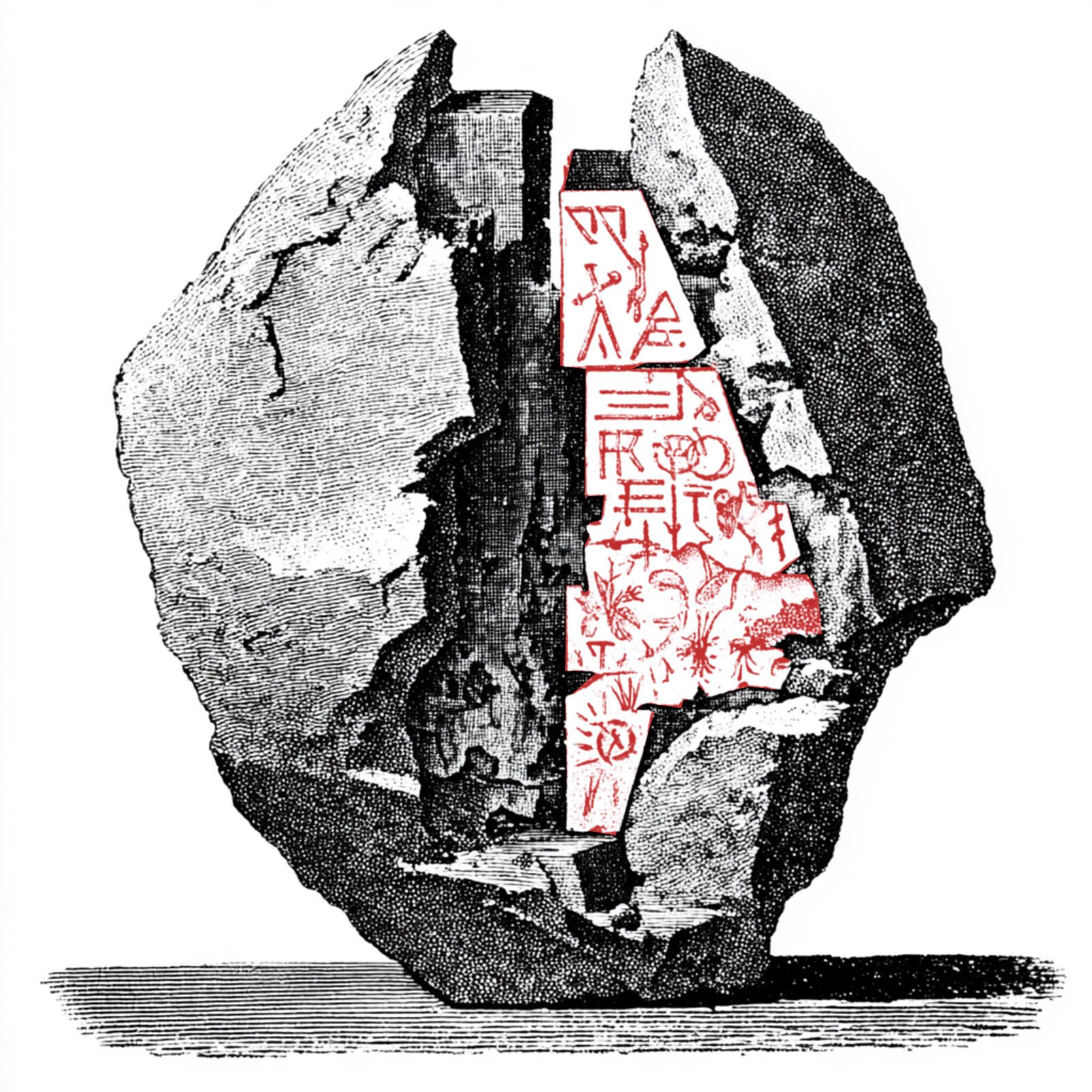}
  \end{center}  \vspace{-5pt}
  \emph{To carve an algebra into pieces, we first need to understand its structure.
  } \vspace{-5pt}
}So, like cluster decomposition (\ref{eq:CDP1}), we seems to have a good way to engineer
switches. But also like cluster decomposition, there is a catch: we need our
system to be carved into tensor factors \emph{already}.
Typically, Nature hands us a big messy algebra $\mathcal{A}$ and \emph{we}
do the carving, and since all we have is algebra, we must exploit
algebraic clues.
Our work with tensor products gives us one such clue: observables
associated with different local algebras \emph{commute}, e.g.
\begin{equation}
  \label{eq:commute-alg}
  Z^{(1)}Z^{(2)} = 
  Z \otimes Z = 
  Z^{(2)}Z^{(1)}.
\end{equation}
We say the local algebras $\mathcal{A}_\text{Pauli}^{(1)}$ and $\mathcal{A}_\text{Pauli}^{(2)}$
themselves commute, since each observable does. This is an entirely
algebraic statement; hopefully it prescribes sensible pieces
to carve the algebra into.

Let's see what joint measurement
of commuting observables
yields.
Suppose that $\Lambda$ and $\Gamma$ are self-adjoint, with vanishing
\emph{commutator} $[\Lambda,\Gamma] = \Lambda\Gamma- \Gamma\,
\Lambda = 0$. 
In the finite-dimensional case, there is a standard
argument\sidenote{See \emph{Linear Algebra Done Right} (1995), Sheldon
  Axler, for instance. Instead of defining a unitary $U$,
  use project onto the eigenbasis with $\Pi_i$
  directly.} that
they can be simultaneously diagonalized:
\begin{equation}
 \Lambda = \sum_{i\in\mathfrak{I}} \lambda_i \Pi_i, \quad \Gamma  =
 \sum_{i\in\mathfrak{I}} \gamma_i \Pi_i,\label{eq:32}
\end{equation}
for some set of orthogonal and hence commuting projectors $\Pi_i,
i\in\mathfrak{I}$.\marginnote{Since $\Pi_i\Pi_{i'} = \delta_{ii'}\Pi_{\hat{i}}
  = \Pi_{i'}\Pi_i$.}
Thus, any sum of projectors $\Pi = \sum_{i'\in\mathfrak{I}'}
\Pi_{i'}$, where $\mathfrak{I}'\subseteq \mathfrak{I}$ is a subset of
the full index set, commutes with both $\Lambda$ and
$\Gamma$. We prove $\Lambda$:
\begin{align*}
  \Pi \Lambda & = \sum_{i'} \Pi_{i'} \sum_i \lambda_i \Pi_i 
              = \sum_{i,i'} \lambda_i \Pi_i \Pi_{i'} 
                     = \Lambda \Pi,
\end{align*}
with a similar result for $\Gamma$.
In the general case, the same conclusion follows from the
continuous functional calculus (Appendix
\hyperlink{app:functional}{A}).

There is a simple generalization of PPM
(\ref{eq:ppm1})--(\ref{eq:ppm3}) that captures the structure of
measurements in this case.
If we measure $\Lambda \measure \lambda$, the associated projector
is just the sum of those projectors with the matching
eigenvalue. Using the indicator function $\mathbb{I}$, we can
write\marginnote{The indicator function $\mathbb{I}[P] = 1$ if the
  condition $P$ is true, and $0$ otherwise.}
\begin{equation}
  \label{eq:proj_lambda}
  \Pi_\lambda = \sum_{i \in \mathfrak{I}} \Pi_i \mathbb{I}[\lambda = \lambda_i],
\end{equation}
and note that $\Lambda$ satisfies (\ref{eq:ppm3}).
Then the post-measurement state is
\begin{equation}
  \label{eq:ppm4}
  \pi \overset{\lambda}{\longmapsto} \pi' = p_\lambda^{-1}\mathcal{C}^{\lambda}[\pi],
  \quad p_\lambda = \pi(\Pi_\lambda),
\end{equation}
where $\mathcal{C}^{\lambda} = \mathcal{C}^{\Pi_\lambda}$.
We explain in more detail below, but let us check this satisfies our
requirements that $\Lambda$ is sharp with average $\lambda$:
\begin{align*}
  \pi'(\Lambda^k) & = p_\lambda^{-1}\pi (\Pi_\lambda \Lambda^k\Pi_\lambda) =
  p_\lambda^{-1}\pi (\Pi_\lambda^2 \lambda^k) = \lambda^k
                    p_\lambda^{-1}\pi (\Pi_\lambda) = \lambda^k,
\end{align*}
where dragging $\Pi_\lambda$ through $\Lambda$ $k$ times creates
$\lambda^k$ via (\ref{eq:ppm3}). From $k=1$ we get mean $\lambda$, and
from $k=2$ we get sharpness.

\newpage

Returning to our observables $\Lambda$ and
$\Gamma$, suppose we measure $(\Lambda, \Gamma) \measure (\lambda,
\gamma)$.\sidenote{We use a tuple rather than a set because order
  matters. At a deeper level, this is suggested by the
  \emph{Curry–Howard correspondence}, but that is beyond the scope of
  this marginalium. See \emph{Combinatory Logic} (1958), Haskell
  B. Curry; ``The Formulae-as-Types Notion of
  Construction'' (1980), William A. Howard.}
The projectors $\Pi_\lambda, \Pi_\gamma$, given by (\ref{eq:proj_lambda}), are built from the
same set of commuting projectors and therefore commute. Let's see what
happens when we measure in either order:
\begin{align}
  \pi & \,\,\,\overset{\lambda}{\longmapsto}\,\,\, \pi^\lambda=
        p_\lambda^{-1}\mathcal{C}^{\lambda}[\pi]
    \,\,\,    \overset{\gamma}{\longmapsto}\,\,\, \pi^{\gamma \lambda}=
        p_{\lambda\gamma}^{-1}p_\lambda^{-1}\mathcal{C}^{\gamma \lambda}[\pi] \notag \\
  \pi & \,\,\,\overset{\gamma}{\longmapsto}\,\,\, \pi^{\gamma}=
        p_\gamma^{-1}\mathcal{C}^{\gamma}[\pi]
    \,\,\,    \overset{\lambda}{\longmapsto}\,\,\, \pi^{\lambda\gamma}=
        p_{\gamma
        \lambda}^{-1}p_\gamma^{-1}\mathcal{C}^{\lambda\gamma}[\pi], \label{eq:pi_lambda_gamma}
\end{align}
using (\ref{eq:conj2}). The superscripts stand for $\Pi_\gamma
\Pi_{\lambda}$ and $\Pi_\lambda \Pi_{\gamma}$, which are equal as
commented above. This means that the states are the same up to a
constant, and if normalized, these
constants are equal.\marginnote{Alternatively, you can
verify that\[
  p_{\lambda\gamma}p_{\lambda}=p_{\gamma \lambda}p_{\gamma} = \pi (\Pi_\gamma\Pi_\lambda),
\] so the constants are the same.} 

We learn that when observables
commute, 
their measurements commute.
In contrast,
non-commuting observables
have values of $\lambda$ and $\gamma$ where $\Pi_\lambda$ and
$\Pi_\gamma$ do not commute, so the states in (\ref{eq:pi_lambda_gamma}) are not equal.
This introduces path dependence into the process,

\begin{figure}[h]
  \centering
  \vspace{- 2pt}
  \includegraphics[width=0.48
  \textwidth]{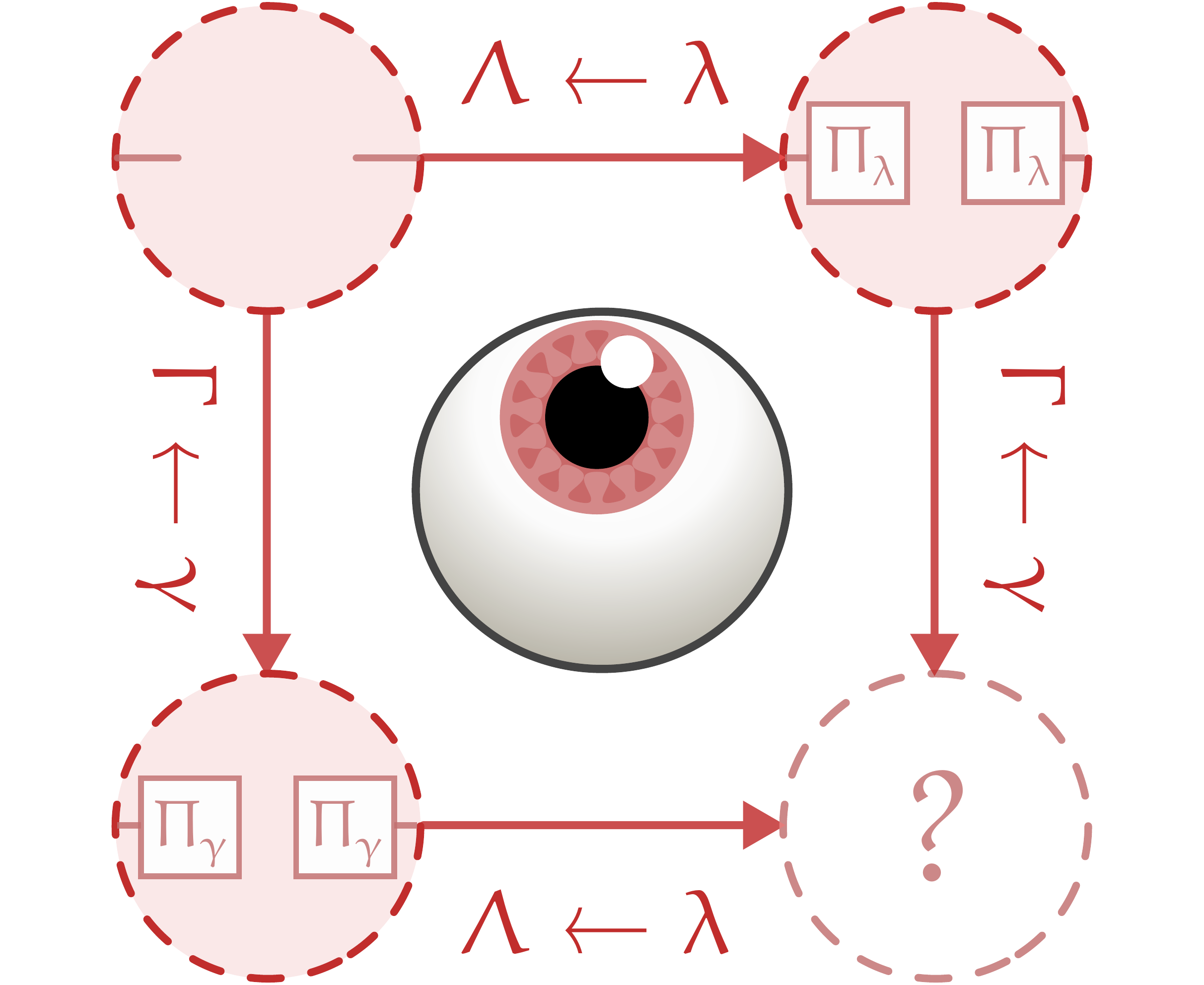}
  \caption{The measurement square associated with two observables. Whether measurements commute (the square closes) depends on whether the
    observables commute.}
  \label{fig:local5}
  \vspace{-3pt}
\end{figure}

\noindent so the ``measurement square'' of Fig. \ref{fig:local5}
does not close. Take the Pauli algebra, for instance. The projectors for $Z \measure -1$, $X \measure +1$ are
\[
  \Pi_{Z,-1} = \tfrac{1}{2} (I + Z), \quad \Pi_{X,-1} = \tfrac{1}{2} (I
  - X).
\]
Since $X$ and $Z$ have a nonzero commutator $[Z, X]=2iY$,
\[
  \Pi_{Z,-1} \Pi_{X,-1} - \Pi_{X,-1}\Pi_{Z,-1} = -\tfrac{1}{4}[Z, X] = -\tfrac{1}{2} i Y.
\]
Hence, $Z \measure -1$ does not commmute with $X \measure +1$.

Sometimes we get lucky, and measure two non-commuting observables
in such a way that both are sharp.
To bound \emph{how} sharp they can simultaneously be, we can use the \emph{Robertson–Schrödinger
  relation}
\begin{equation}
  \label{eq:rs4}
|\pi([\Lambda, \Gamma])|^2 +
4\left|\pi(\Lambda \circ \Gamma) - \pi(\Lambda)\pi(\Gamma)\right|^2 \leq 4\Vert
\Delta \Lambda\Vert_\pi^2 \Vert\Delta \Gamma\Vert_\pi^2
\end{equation}
proved in Appendix \hyperlink{app:stormer}{B}.
The commutator term tells us that, for joint sharpness, $\Lambda$ and $\Gamma$ must \emph{commute in expectation}:
  \begin{equation}
|\pi([\Lambda, \Gamma])|^2 = 0 \quad \Longrightarrow \quad
\pi(\Lambda\Gamma) = \pi(\Gamma\Lambda).\label{eq:26}
\end{equation}
This is equivalent to requiring $\Pi_\lambda$ and $\Pi_\gamma$ to
commute.
The anticommutator term, on the other hand, is guaranteed to vanish
by (\ref{eq:jordan-mult}).

\newpage

Measurement is fundamentally
random; to maximize control over sharpness, it's therefore best to work with
commuting variables,
and focus on \emph{abelian switch sets} 
$\mathcal{Q}$.
Along with the definite set $\mathcal{D}$, we can consider the
\emph{abelian C${}^*$-subalgebra} $\mathcal{M}$ built from elements of
$\mathcal{Q}$:\marginnote{$\mathcal{M}$ is also abelian because the property of
  commuting with everything is closed under linear combinations and products.}
\begin{equation}
  \mathcal{D} = \text{J}^\circ\langle \mathcal{Q}\rangle, \quad
  \mathcal{M} = \text{C${}^*$}\langle \mathcal{Q}\rangle. \label{eq:28}
\end{equation}
A state $\pi(\mathcal{Q})$ associated with measuring all switches in $\mathcal{Q}$ is
pure just in case $\mathcal{D}_{\pi(\mathcal{Q})}$ is maximal, by Størmer's
theorem; in turn, this is equivalent to $\mathcal{M}_{\pi(\mathcal{Q})}$ being a \emph{maximal abelian
subalgebra (MASA)}, since $\mathcal{M}_{\pi(\mathcal{Q})}$ is
just the ``complexification'' of
$\mathcal{D}_{\pi(\mathcal{Q})}$.\sidenote{
Complexification just means we replace
$\mathbb{R}$ with $\mathbb{C}$; the Jordan product is the same as the
regular product when operators commute, $\Lambda \circ \Gamma = \Lambda\Gamma$.}

``Maximal'' means nothing else can be added to $\mathcal{M}$ while
remaining abelian, and hence the only operators to commute with everything in
$\mathcal{M}$ are the elements of $\mathcal{M}$ itself.
We capture this neatly with the \emph{commutant} $\mathcal{M}'$, the
set of all operators commute with all of $\mathcal{M}$:
\begin{equation}
  \label{eq:29}
  \mathcal{M}' = \{A \in \mathcal{A} : [M, A] = 0 \text{ for all } M
  \in \mathcal{M}\}.
\end{equation}
Then $\mathcal{M}$ is abelian just in case $\mathcal{M} \subseteq
\mathcal{M}'$, and a MASA when $\mathcal{M}=\mathcal{M}'$.
For instance, as we saw in \S \hyperlink{sec:11}{11}, the switch
set $\mathcal{Q} = \{\sigma(\mathbf{n})\}$ corresponds to measuring
$\sigma(\mathbf{n})$, with definite set and hence
MASA\marginnote{Clearly, $\mathcal{M}_{\mathbf{n}}$ is abelian, but it
is also maximal since no other $\sigma(\mathbf{m})$ commutes with $\sigma(\mathbf{n})$.}
\[
  \mathcal{D}_{\mathbf{n}} = \text{span}_{\mathbb{R}}\{I,
  \sigma(\mathbf{n})\}, \quad \mathcal{M}_{\mathbf{n}} = \text{span}_{\mathbb{C}}\{I,
  \sigma(\mathbf{n})\},
\]
since $\mathcal{M}_{\mathbf{n}}$ an abelian subalgebra that
complexifies $\mathcal{D}_{\mathbf{n}}$.

\begin{figure}[h]
  \centering
  \vspace{-3pt} 
  \includegraphics[width=0.65\textwidth]{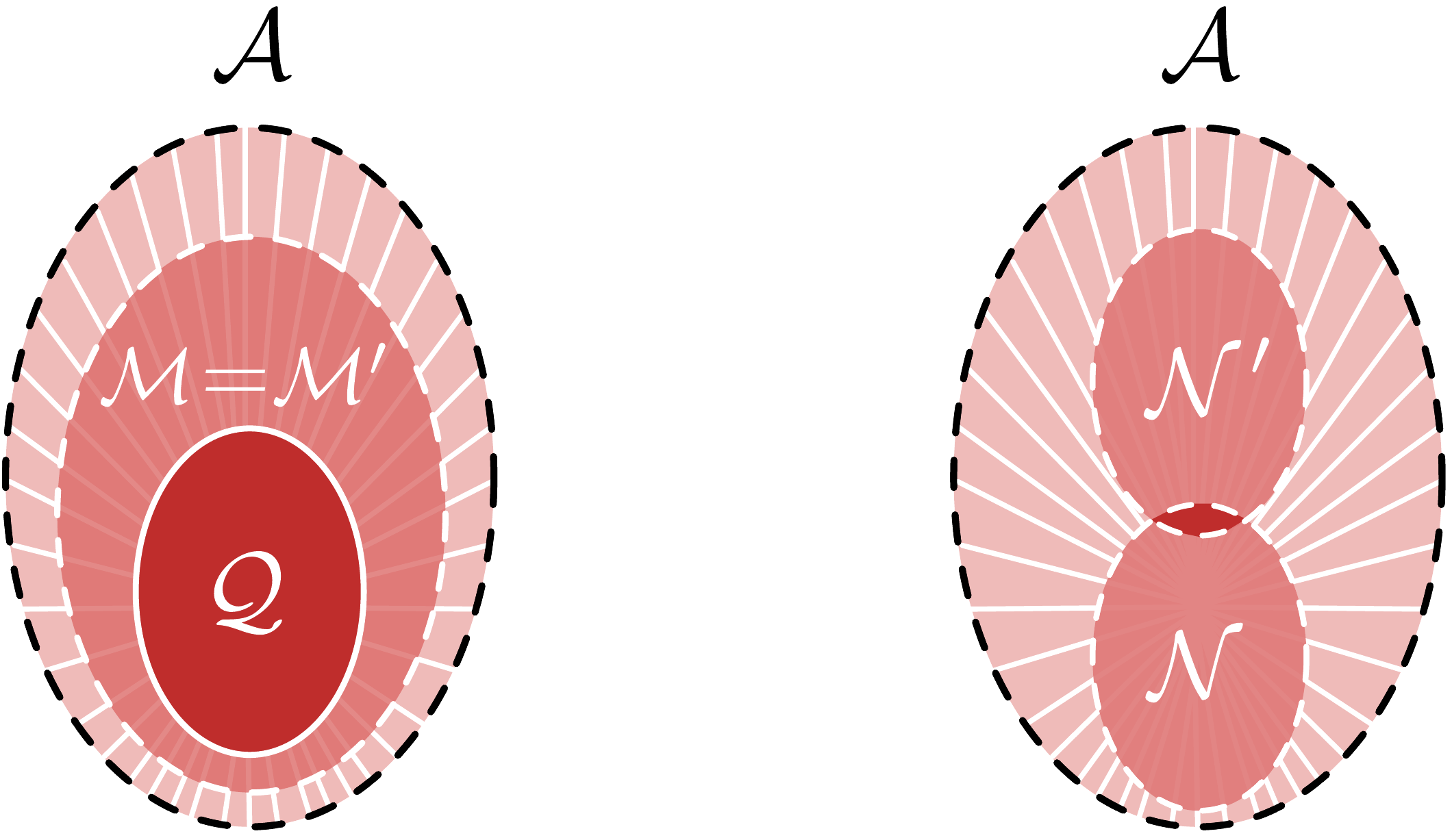}
  \caption{\textsc{Left.} The MASA $\mathcal{M}$
    associated to an abelian switch set $\mathcal{Q}$ is its own
    commutant. \textsc{Right.} In finite dimensions, an
    arbitrary subalgebra $\mathcal{N}$ is a tensor factor if it spans
    the full algebra with its commutant; in this case, the overlap is trivial.}
  \label{fig:rigid2}
  \vspace{-6pt}
\end{figure}

Curiously, tensor products like (\ref{eq:commute-alg}) are on the
other end of the spectrum.
Consider arbitrary subalgebras $\mathcal{N}_1, \mathcal{N}_2 \subseteq \mathcal{A}$,
and define their \emph{join} as the subalgebra they generate:
\begin{equation}
  \label{eq:24}
  \mathcal{N}_1 \vee \mathcal{N}_2 = \text{C${}^*$} \langle
  \mathcal{N}_1\cup \mathcal{N}_1\rangle.
\end{equation}
The \textsc{Bicommutant Theorem} of von Neumann\sidenote{``Zur Algebra der Funktionaloperatoren und Theorie
  der normalen Operatoren'' (1929). This result only holds for
   finite-dimensional systems.
} implies that, if
$\mathcal{N} \vee \mathcal{N}'$ is the full algebra, they tensor
factorize $\mathcal{A}$:
\begin{equation}
  \label{eq:bicommutant}
  \mathcal{N} \vee \mathcal{N}' = \mathcal{A} \quad \Longrightarrow
  \quad \mathcal{A} \cong \mathcal{N} \otimes \mathcal{N}'.
\end{equation}
It additionally follows that $\mathcal{N} \cap \mathcal{N}'
=\mathbb{C} I$ is trivial.
As a sanity check, in $\mathcal{A}_\text{Pauli}^{\otimes 2} = \mathcal{A}_\text{Pauli}
\otimes \mathcal{A}_\text{Pauli}$, the first factor
$\mathcal{A}_\text{Pauli}^{(1)}$ has commutant
$\mathcal{A}_\text{Pauli}^{(2)}$, they join to give the full space, and
overlap trivially.
Since it gives reliable switches (via MASAs) and useful algebraic chunks (tensor factors), commutation is clearly the right
structural lens.

\section{13. How to measure}\hypertarget{sec:14}{}


We have argued that quantum mechanics is about observables and
measurement; states are a bookkeeping device for
self-consistently keeping track of things.
We've been coy about the measurements
themselves, however.
We presented post-selected partial measurement in equations
(\ref{eq:ppm1})--(\ref{eq:ppm3}), and a general post-selected
measurement in (\ref{eq:ppm4}). But what induces the value we
post-select onto? The answer involves \emph{quantum randomness} we
have yet to explain.

To rectify this, let's return---as we so often do---to von Neumann,
who argued that the true atoms of quantum mechanics are the ``yes-no'' measurements, or projectors
$\Pi$ satisfying (\ref{eq:proj}). They have a simple binary spectrum, obtained
from the defining equation:
\begin{equation}
  \Pi^2 - \Pi = \Pi (\Pi - I) = 0 \quad \Longrightarrow \quad
  \mathfrak{S}(\Pi) = \{0, 1\}.\label{eq:spec-proj}
\end{equation}
Physically, a projector represents a \emph{filter}, allowing some things
through unchanged (``yes'', $\Pi \measure 1$) and blocking others
(``no'', $\Pi \measure 0$).
A classic example is a polarizing filter, which lets through
only the light aligned with the filter.
If we observe the light on the other side, we know that $\Pi \measure
1$; otherwise $\Pi \measure 0$.

\begin{figure}[h]
  \centering
  \vspace{-5pt} 
  \includegraphics[width=0.57\textwidth]{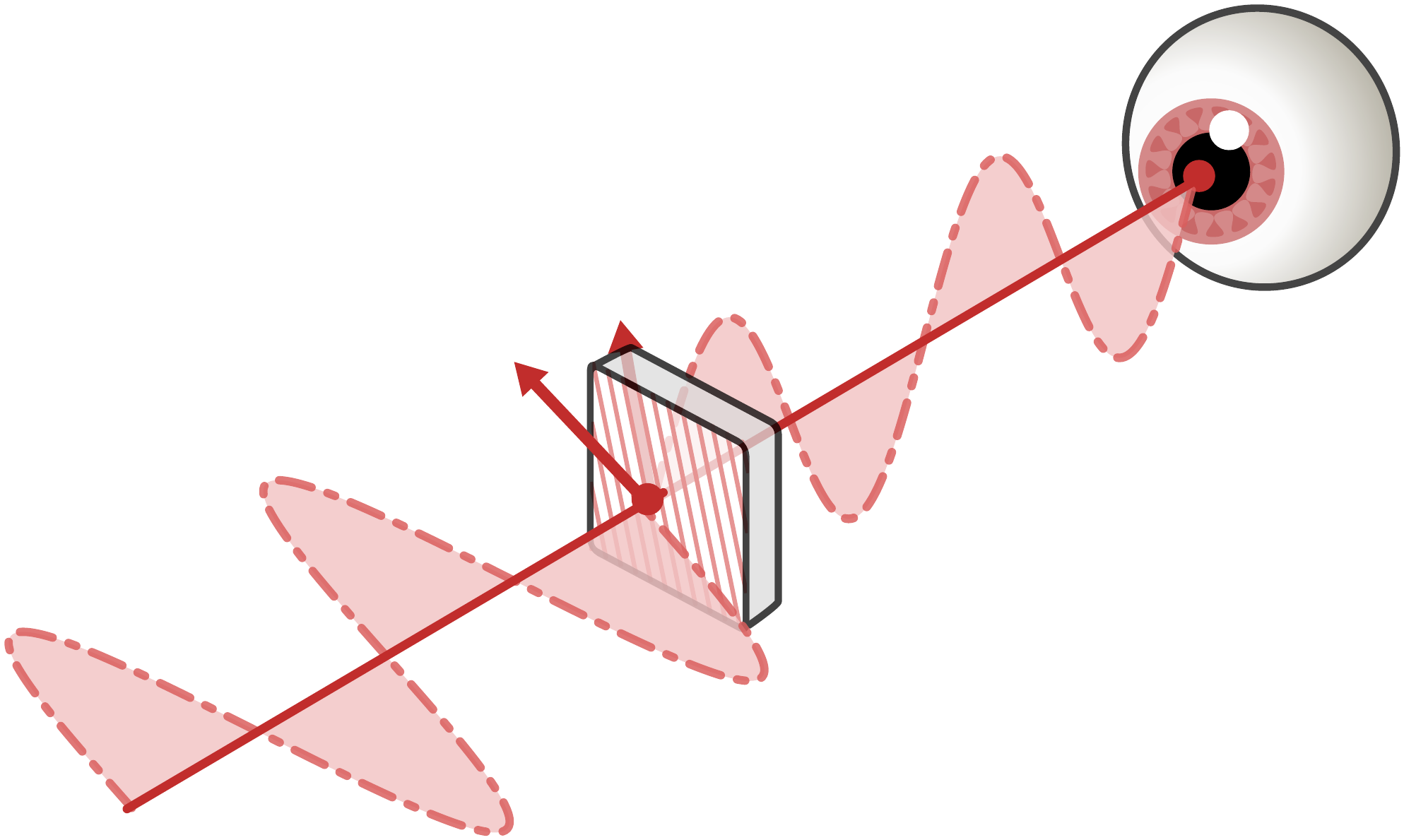}
  \caption{
    A polarizing filter where light is partially aligned and therefore
    passes through ($\Pi \to 1$) to the observer.
  }
  \label{fig:filter1}
  \vspace{-8pt}
\end{figure}

If we see the light has been polarized, we
can post-select, i.e. use our
knowledge of the measurement outcome to conditionally update the state.
Recall from (\ref{eq:restrict-proj}) that, for any state $\pi$, we can
identify
\[
  \pi(A) = \pi (\Pi^\perp_\mathcal{K} A\Pi_{\mathcal{K}}^\perp),
\]
where $\Pi^\perp_\mathcal{K}$ projects onto the orthogonal complement
of the kernel $\mathcal{K}_\pi$.
Once the light passes through the filter, the kernel includes states
of light that did \emph{not} pass through, so we update
\[
  \pi(A) \,\,\,\overset{}{\longmapsto} \,\,\, \pi'(A) =
  p_1^{-1}\pi (\Pi^\perp_\mathcal{K} \Pi A \Pi\Pi_{\mathcal{K}}^\perp),
\]
since $\Pi_{\mathcal{K}}^\perp$ projects away from the kernel of
$\pi$, and $\Pi$ projects us away from the kernel $\Pi^\perp =I -
\Pi$ of the latest measurement.
We're\marginnote{``Not doing anything funny'' is an important physical
  criterion we return to below.} assuming that the filtering doesn't do anything funny
to
$\Pi_{\mathcal{K}}^\perp$; we also need a normalization constant $p_1 =
\pi(\Pi)$ so that $\pi'(I) = 1$.

Filtering is not quite measurement in the usual sense: when light
doesn't make it through the polarizer, it gets absorbed.
This is 
bad for the lab budget, so instead of throwing it
away, we 
can make a copy\marginnote{``Making a copy'' really means sampling
  multiple times from a mixed state, which is a distribution like a
  coin flip where outcomes are pure states.} and pass it through the second filter if it doesn't
make it through the first that it is \emph{guaranteed} to get through.
If the first filter is $\Pi$, the second, orthogonal filter guaranteed
to work is $\Pi^\perp$.
The measurement outcome tells us which filter we applied.

\begin{figure}[h]
  \centering
  \vspace{-0pt} 
  \includegraphics[width=0.64\textwidth]{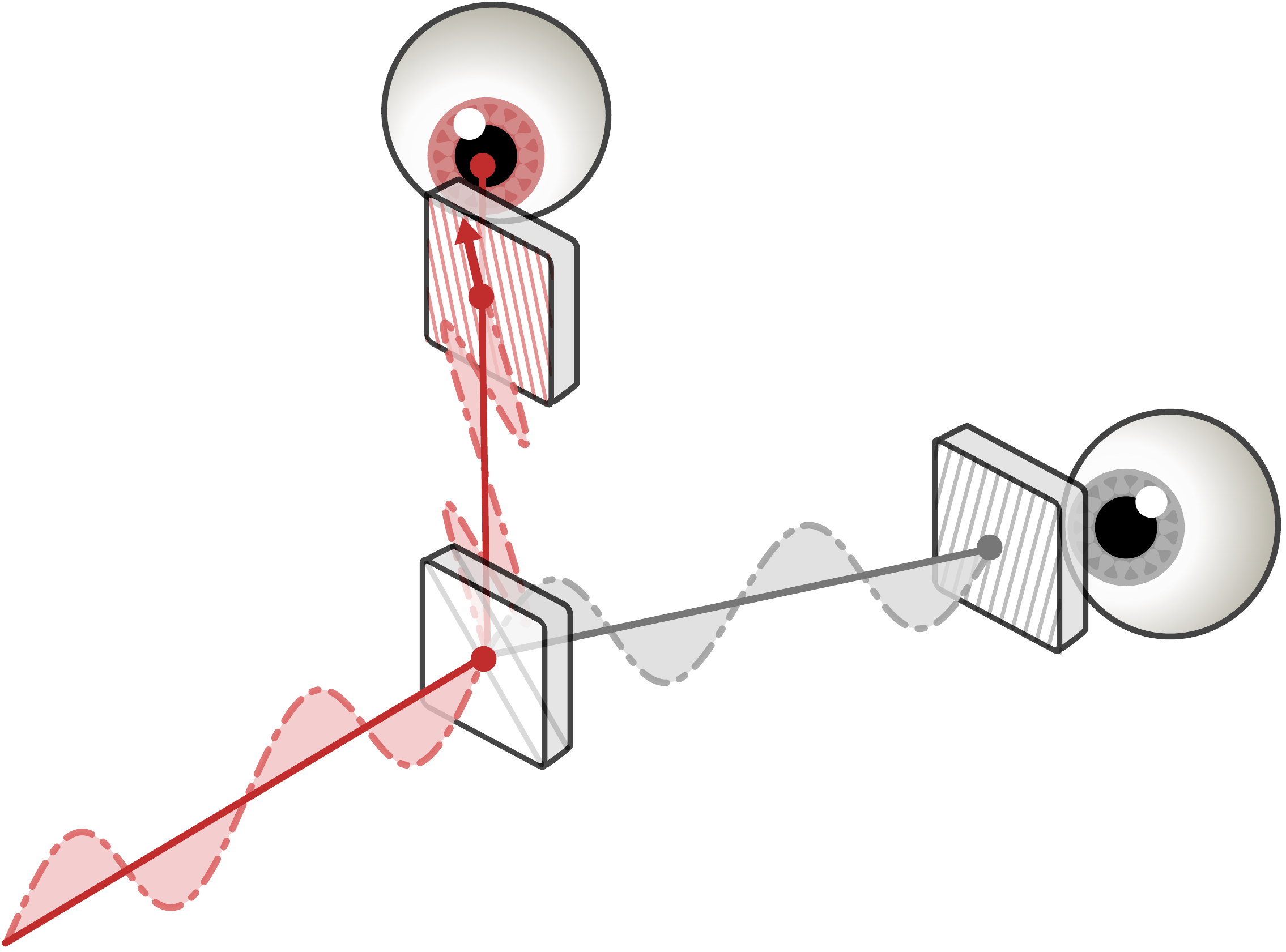}
  \caption{
    Measuring polarization with orthogonal filters: if it doesn't get
    through the first, we apply the second.
  }
  \label{fig:filter2}
  \vspace{-3pt}
\end{figure}


\noindent Suppose we make a measurement, but \emph{don't} observe the
outcome, a little like Schr\"{o}dinger's cat.\sidenote{``Die gegenwärtige
  Situation in der Quantenmechanik'' (1935), Erwin Schr\"{o}dinger.}
The state conditioned on filter $\Pi_{(b)}$, where $\Pi_{(0)} = \Pi^\perp$ and
$\Pi_{(1)} = \Pi$, is 
\begin{equation}
  \label{eq:baby-luders}
  \pi_{(b)}(A) = p_b^{-1}\pi(\Pi_{(\hat{b})} A \Pi_{(\hat{b})}).
\end{equation}
Thus, the mixed state corresponding to our unseen measurement is
\begin{equation}
  \pi' = q_0 \pi_{(0)} + q_1\pi_{(1)},\label{eq:mixed-unseen}
\end{equation}
where $q_b = \mathbb{P}[\Pi \measure b]$ is the probability it passes through $\Pi_{(b)}$.
Since any complementary probabilities leads to a valid mixed state, it seems like we need to
know more about the dynamics to find $\pi'$.

The extra piece of information, ironically, is that \emph{we need nothing else}.
We assume that Nature is indifferent between the outcomes, 
in the sense that $q_b$ should only be built out of the ingredient $\pi$
and $\Pi_b$ at hand.
The unique solution is the normalization constant
  \begin{equation}
  q_b = \pi(\Pi_{(b)}) = p_b.\label{eq:baby-born}
\end{equation}
\marginnote{
  \vspace{-200pt}
  \begin{center}
    \hspace{15pt}\includegraphics[width=0.95\linewidth]{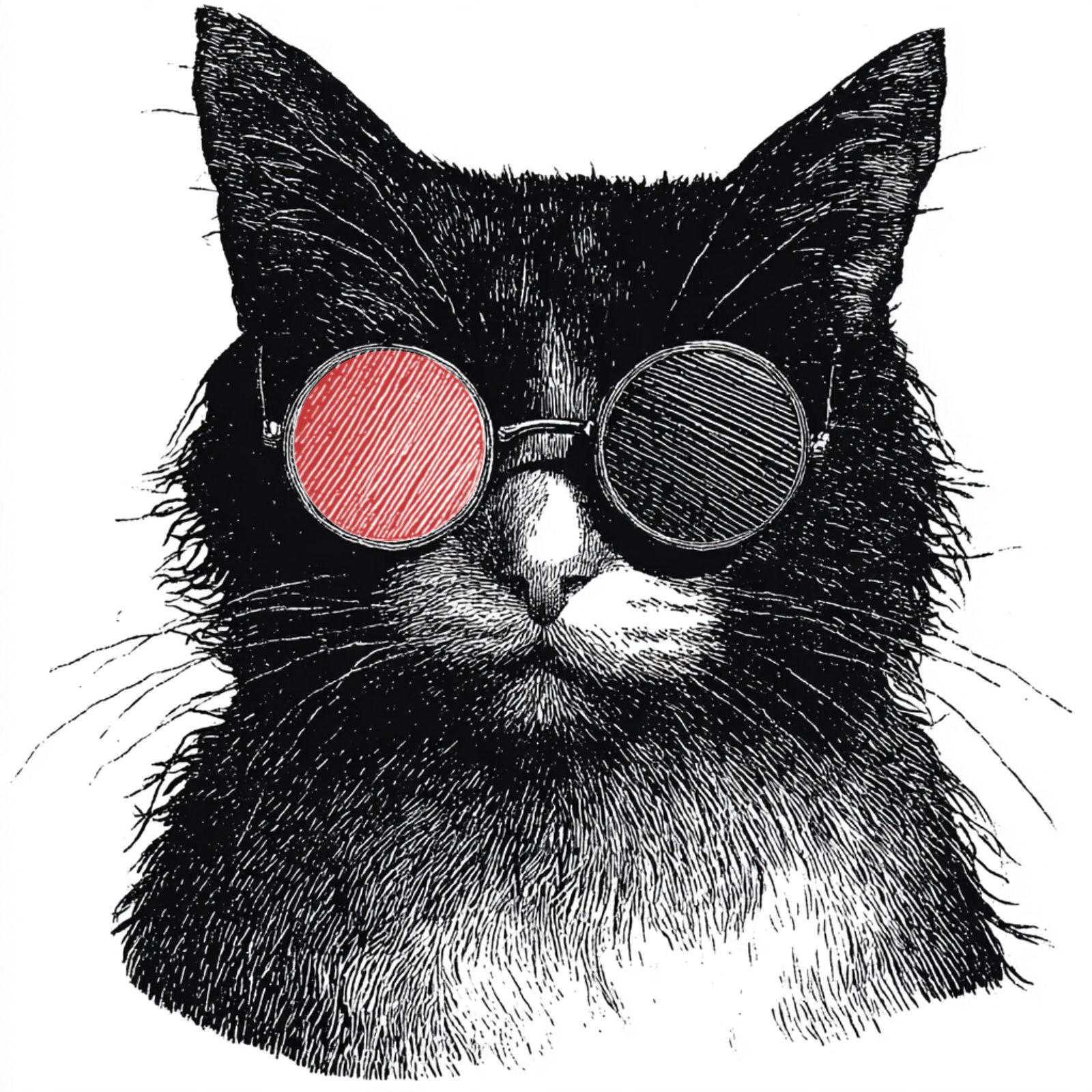}
  \end{center}  \vspace{-5pt}
  \emph{Only Schr\"{o}dinger's cat knows which filter lit up.
  } \vspace{-5pt}
}Why does this follow? By assumption, the only state we have is $\pi$,
and the only operators are the $\Pi_{(b)}$. The $\pi(\Pi_{(b)}) = p_b$ are
complementary, and when $p_b = 0$, setting $q_b = p_b$ cancels the
divergence in $p_b^{-1}$, so we are done.
It follows that the measured state is
\begin{equation}
  \pi'(A) = 
  \pi(\Pi_{(0)} A \Pi_{(0)}) + \pi(\Pi_{(1)} A \Pi_{(1)}) = \sum_{b=0,1}\mathcal{C}^b[\pi](A),\label{eq:baby-measure}
\end{equation}
where $\mathcal{C}^b$ conjugates by $\pi_{(b)}$.
The full update rule is very simple!

\newpage

At this point, we should reveal our cards.
You may have already recognized (\ref{eq:baby-born}) as the celebrated \textsc{\gls{born} Rule}\sidenote{``On the quantum mechanics of
  collisions'' (1926), Max Born; ``Concerning the
  state-change due to the measurement process'' (1950), Gerhart
  Lüders; ``Measures on the closed subspaces of
  a Hilbert space'' (1957), Andrew Gleason.},
while the post-selected update (\ref{eq:baby-luders}) is \textsc{L\"{u}ders Rule}.
The indifference between the eigenvalues of $\Pi$ that led to Born's
rule a baby version of what is called
\emph{noncontextuality}---no additional context need apply---and makes our
argument a dramatically simplified version of
\textsc{Gleason's Theorem}, 
which derives Born's rule without post-selection.

We can use the \gls{awd} conjugation notation (Fig. \ref{fig:conj2}) to
capture the effect of measurement. Nested circles take
states to states, and by convention, circles are normalized. The \gls{born}
rule is represented by the first circle on the RHS, the \gls{luders} rule by
the second:

\begin{figure}[h]
  \centering
  \vspace{-0pt} 
  \includegraphics[width=0.5\textwidth]{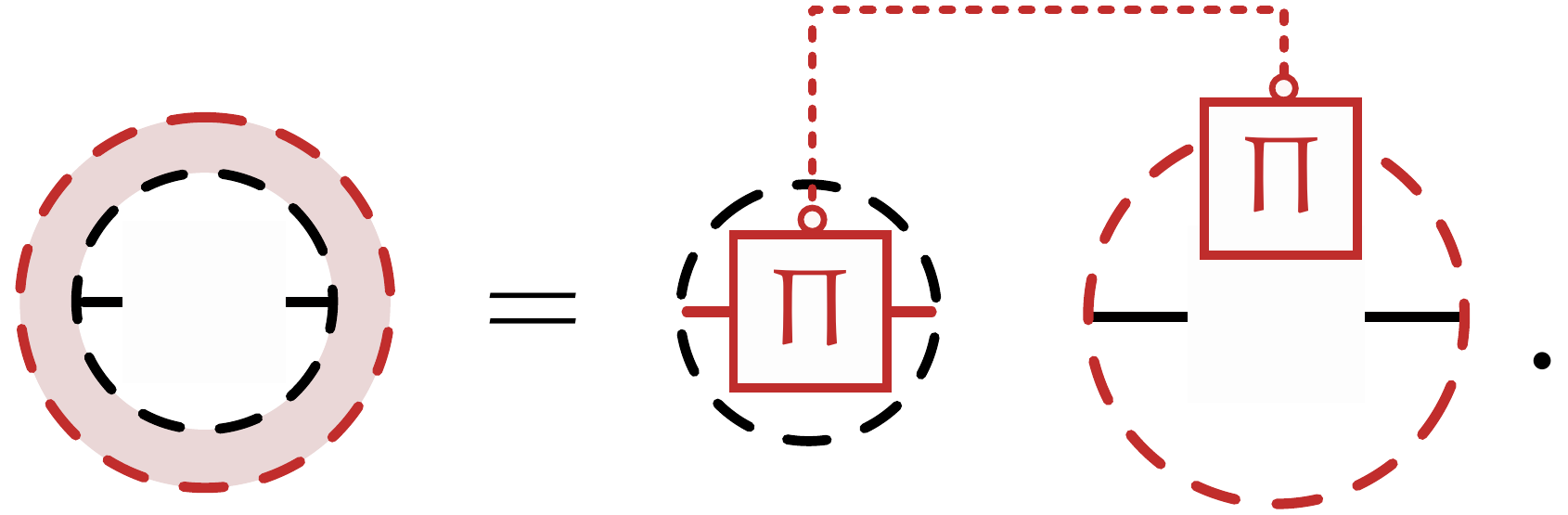}
  \caption{\gls{born} and \gls{luders} in \gls{awd} form. Note that
    the black leads on the LHS are ``transferred'' to the red state.}
  \label{fig:meas1}
  \vspace{-3pt}
\end{figure}

\noindent It's easy to generalize all this to
\emph{projection-valued measures (\gls{pvm}s)}. For a complete,
orthogonal set of filters labelled by $\lambda
\in \mathfrak{F}$, we have
\begin{equation}
  \label{eq:PVM}
  \sum_{\lambda\in\mathfrak{F}} \Pi_{(\lambda)} = I, \quad \Pi_{(\mu)}
  \Pi_{(\lambda)} = \delta_{\mu\lambda}\Pi_{(\hat{\lambda})}.
\end{equation}
We'll write $\mathcal{C}^{\mathfrak{F}}$ for the sum of filters acting
on states by conjugations, and $\mathcal{C}_{\mathfrak{F}}$ for
operators. 
Assuming indifference between eigenvalues, as above, gives
  \begin{equation}
  \pi'=\mathcal{C}^{\mathfrak{F}}[\pi], 
  \quad
  \pi_{(\lambda)} =
  p_\lambda^{-1}\mathcal{C}^\lambda[\pi], \quad p_\lambda = \pi(\Pi_{(\lambda)}).\label{eq:opri-meas}
\end{equation}
This is exactly the \gls{born} and \gls{luders}
rules suggested by Fig. \ref{fig:meas1}.

This would all be academic if these \gls{pvm}s were hard to find, but Nature
makes them for us!
Every self-adjoint operator secretly encodes a set $\mathfrak{F}$ by virtue of the \textsc{Spectral Theorem}.\sidenote{\emph{Mathematical Foundations of Quantum
    Mechanics} (1932), John von Neumann. The
  Hilbert space $\mathcal{H}$ must be \emph{separable}, meaning it has
  a dense countable subset.
} This guarantees that in a well-behaved Hilbert space $\mathcal{H}$, a bounded, self-adjoint
operator $\Lambda \in \mathcal{B}(\mathcal{H})_\text{sa}$ can 
be decomposed into an ``eigenweighted'' PVM:
\begin{equation}
  \label{eq:spectral}
  \Lambda = \sum_{\lambda \in \mathfrak{S}(\Lambda)} \lambda \Pi_{(\lambda)},
  \quad \Pi_{(\mu)}
  \Pi_{(\lambda)} = \delta_{\mu\lambda}\Pi_{(\hat{\lambda})}.
\end{equation}
The index set is the spectrum
$\mathfrak{S}(\Lambda) \subseteq \mathbb{R}$. 
To port this back to an abstract \gls{cstar} $\mathcal{A}$, we 
need to ensure our 
Hilbert space faithfully embeds $\mathcal{A}$.
Gelfand and Naimark showed\sidenote{``On the embedding of normed rings into the
  ring of operators in Hilbert space'' (1943), Israel Gelfand and
  Mark Naimark. Recall that $\partial S(\mathcal{A})$ is the set of
  extreme points, i.e. \gls{pure} states, and $\mathcal{U}(\mathcal{A})$
  the unitaries which act by conjugation on $S(\mathcal{A})$.} that the direct sum of \gls{pure} states, identified
up to unitary equivalence, is enough:
\begin{equation}
  \label{eq:14}
  \tilde{\pi} = \bigoplus_{\pi \in \text{Prim}(\mathcal{A})}\pi, 
  \quad \text{Prim}(\mathcal{A})=\partial S(\mathcal{A})/\mathcal{U}(\mathcal{A}),
\end{equation}
where $\text{Prim}(\mathcal{A})$ is called the \emph{\gls{prim}} of $\mathcal{A}$.
For the Pauli algebra, $\partial S(\mathcal{A}_\text{Pauli})$ is the
Bloch sphere, and it has a single equivalence class, so any \gls{pure}
state will do.

We can spell out in more detail what it means for this construction to ``work''.
The state $\tilde{\pi}$ induces a map $\tilde{\phi}: \mathcal{A} \to
\tilde{\phi}(\mathcal{A}) \subseteq \mathcal{B}(\tilde{\mathcal{H}})$,
where $\tilde{\mathcal{H}}=\mathcal{H}_{\tilde{\pi}}$. This map
$\tilde{\phi}$ has two key properties:
\vspace{-2pt}
\begin{itemize}[itemsep=-2pt]
\item it is an \emph{isomorphism}, i.e. linear, multiplicative, and invertible;
\item it is an \emph{isometry}, so $\Vert A\Vert = \Vert
  \tilde{\phi}(A)\Vert_\text{op}$ for the operator norm $\Vert\cdot\Vert_\text{op}$.
\end{itemize}
\vspace{-2pt}
Thus, $\tilde{\phi}$ is an \emph{isometric isomorphism} which
faithfully embeds both the algebraic and metric structure as a subalgebra of $\mathcal{B}(\tilde{\mathcal{H}})$.
For this reason, it is called the \emph{\gls{univ}}.
To apply the \gls{spectral} to $\Lambda \in \mathcal{A}_\text{sa}$,
we just map to Hilbert space and back:
\begin{equation}
  \begin{tikzcd}
    \Lambda \arrow[mapsto]{r}{\text{spec}} \arrow[swap,
    mapsto]{d}{\tilde{\phi}} & \sum_\lambda \lambda \Pi_{(\lambda)} \arrow[mapsfrom]{d}{\tilde{\phi}^{-1}} \\
    \Lambda \arrow[swap, mapsto]{r}{\widetilde{\text{spec}}} & \sum_{\tilde{\lambda}} \tilde{\lambda} \tilde{\Pi}_{(\tilde{\lambda})}
  \end{tikzcd}.\label{eq:qec2}
\end{equation}

\noindent 
The main obstacle 
to this argument is that, if $\text{Prim}(\mathcal{A})$ is
  uncountable, the universal Hilbert space $\tilde{\mathcal{H}}$ is
  too big for the \gls{spectral} to hold.
  Luckily, we are interested in less exotic spaces.
  We are not so lucky with measurement, where even the simplest
  experiments are more exotic than the standard formalism.


\addtocontents{toc}{\protect\vspace{-20pt}\protect\contentsline{part}{\textsc{\Large{from
      error to recovery}}}{}{}}
  
\section{14. How to err}\hypertarget{sec:15}{}

\begin{figure}[h]
  \centering
  \vspace{-5pt} 
  \includegraphics[width=0.6\textwidth]{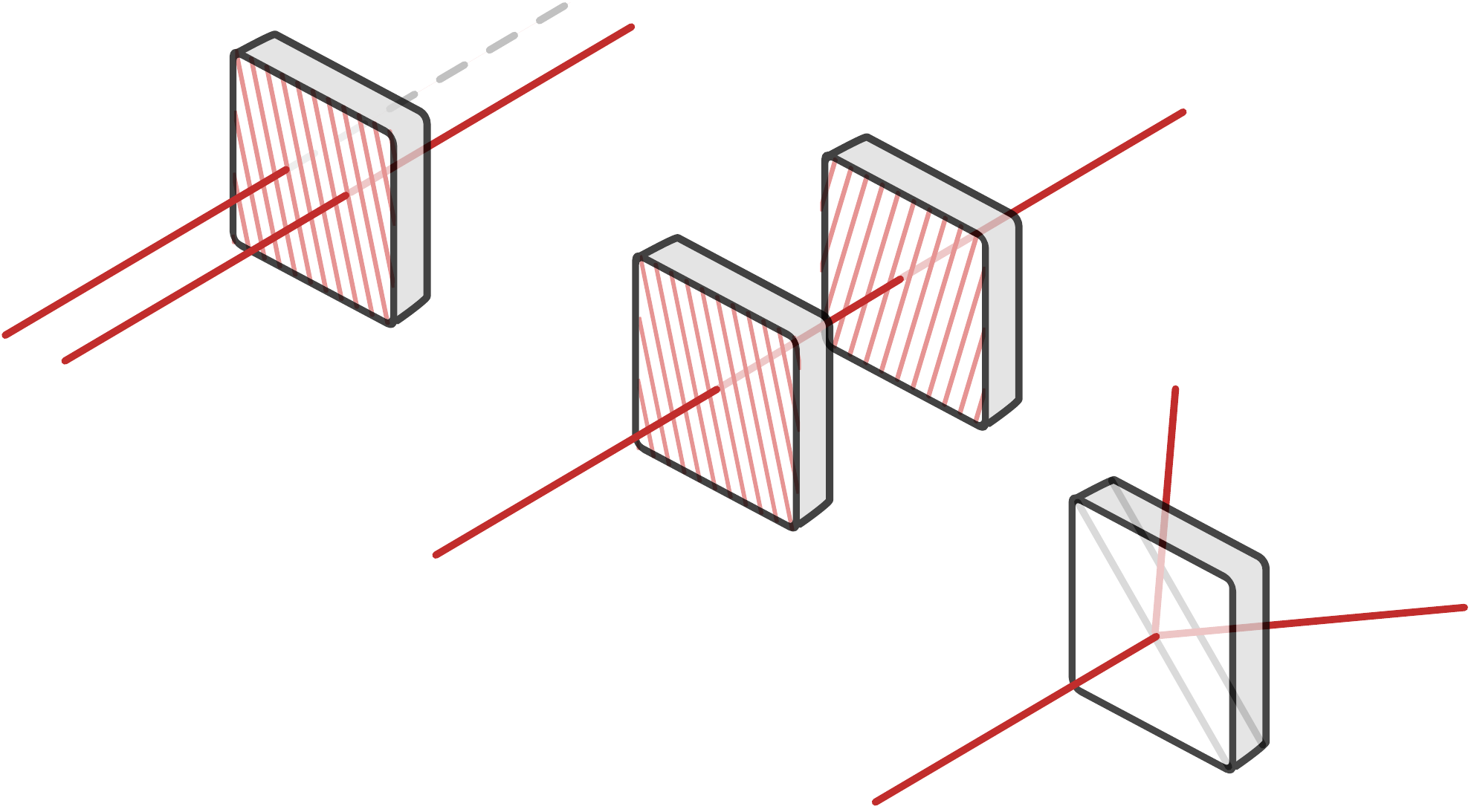}
  \caption{Problems with PVMs. \textsc{Left}. Some photons are
    absorbed. \textsc{Middle}. Some filters are
    compatible. \textsc{Right}. Some things are not filters.}
  \label{fig:filter4}
  \vspace{-5pt}
\end{figure}

\noindent 
The spectral theorem associates to each observable $\Lambda$ a
complete,
orthogonal set of projectors, or PVM (\ref{eq:PVM}).
But 
even for something as simple as a polarizer or a beamsplitter,
we need a richer
formalism! 
We call these more general measurements \emph{quantum
  operations}. They can 
be incomplete (e.g. a
lone polarizer), non-orthogonal (polarizers which are not orthogonal),
and non-projective (the beamsplitter itself). We picture some these behaviours in Fig. \ref{fig:filter4}.

\noindent Happily, the formalism for PVMs generalizes without much effort.
Instead of projectors $\Pi_{(\lambda)}$ labelled by $\lambda \in
\mathfrak{S}(\Lambda)$, we will suppose a collection of operators
$B_{(k)}$ labelled by $k\in\mathfrak{K}$. Our goal will be to
figure out what properties the $B_{(k)}$ should have.

First off, we define $\mathfrak{K}$ as a \emph{classical sample space},
from which we randomly select a single element $k$ each time we
perform the\marginnote{There is no loss of generality in sampling a
  single $k$. If you want to sample a pair, for instance, you
  replace $\mathfrak{K}$ by the Cartesian product $\mathfrak{K}\times\mathfrak{K}$.}
operation; this is what $\mathfrak{K} \measure k$ means.
If $k$ is observed, then the only way to update a state $\pi$ using
$B_{(k)}$ is to conjugate (to preserve positivity) and normalize (to
preserve unitality, i.e. mapping $I \mapsto 1$):
\begin{equation}
  \label{eq:op-luders}
  \pi_{(k)} = p_k^{-1}\mathcal{C}^{\hat{k}}[\pi], \quad p_k =
  \mathcal{C}^{k}[\pi](I) = \pi(B_{(k)}^*B_{(\hat{k})}).
\end{equation}
This is a generalization of Lüders rule to arbitrary 
$B_{(k)}$, called \emph{Kraus operators}.\sidenote{See \emph{Effects and Operations: Fundamental Notions of
    Quantum Theory} (1983), Karl Kraus.}
The Born rule is subtler. We start with a general
mixture of unobserved operations of the form (\ref{eq:op-luders}):
\[
  \chi_\pi = q_k  \pi_{(k)} = q_k p_{\hat{k}}^{-1} \mathcal{C}^k[\pi].
\]
Since each term is individually normalized, the whole is normalized provided the $q_k$ sum to
unity. But normalizing the state is not necessarily what the Born rule
suggests!\marginnote{This is why we have $\chi_\pi$ rather than $\pi'$; we do not assume
$\chi_\pi$ is a state.}

\begin{figure}[h]
  \centering
  \vspace{-0pt} 
  \includegraphics[width=0.5\textwidth]{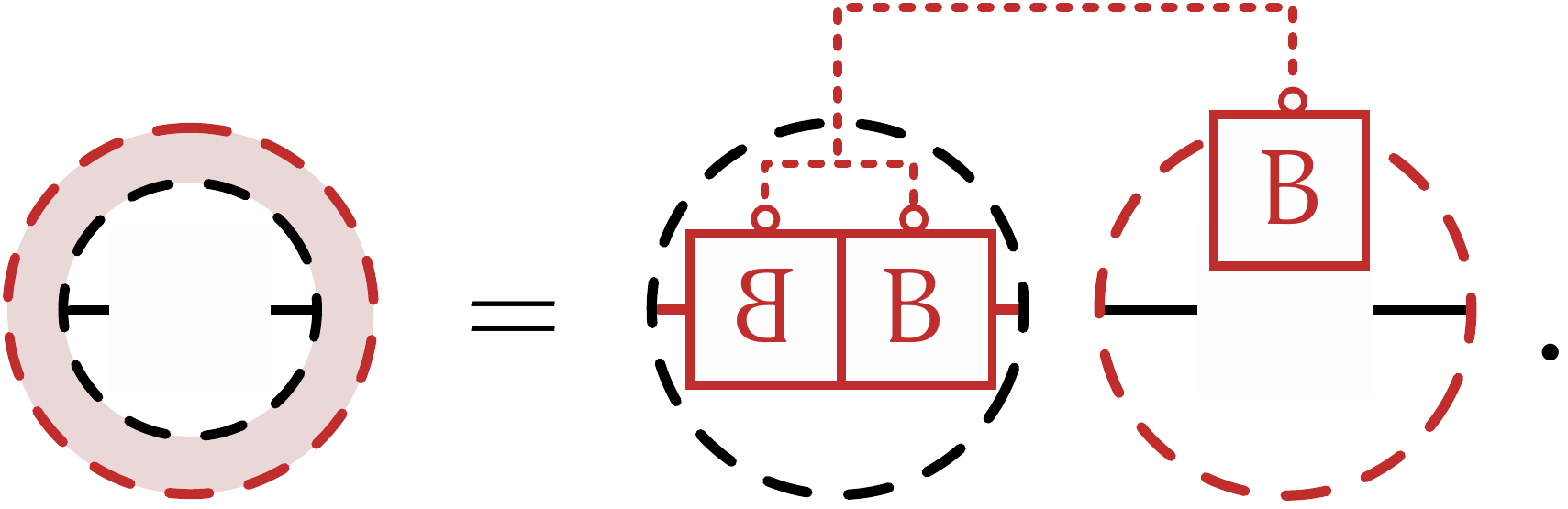}
  \caption{Born and \gls{luders} rules for a quantum operation
    $\mathcal{C}^{\mathfrak{K}}$.}
  \label{fig:meas2}
  \vspace{-3pt}
\end{figure}

To see why, consider the case of the lone polarizer. Removing the
orthogonal filter from (\ref{eq:mixed-unseen})
gives a single term, $q_1 \pi_{(1)}$,
and normalization would suggest we set $q_1 = 1$. But this means light
passes through the filter with certainty, which
is nonsense; in fact, we know it gets through with
probability $q_1 = \pi_{(1)}(\Pi)$. Extending this, we can make the
argument from indifference that
\begin{equation}
  \label{eq:op-born}
  \pi \longmapsto \chi_\pi = q_k \pi_{(k)} =
  \mathcal{C}^{\mathfrak{K}}[\pi], \quad q_k = p_k = \pi(B^*_{(k)} B_{(k)}).
\end{equation}
The sum of conjugators is called the \emph{Kraus form}.
We depict the generalized Born and \gls{luders} rules in Fig. \ref{fig:meas2}.

It's clear that $\chi_\pi$ is a positive linear functional (it is a
positive linear combination of positive functionals), but not
necessarily normalized. We quantify this with the \emph{trace}:\marginnote{This equals the trace of the corresponding density
  matrix. It
  helps to keep the notation separate even if we muddy the namespace.}
\begin{equation}
  \label{eq:op-trace}
  \tau_{\mathfrak{K}}(\pi) = \chi_\pi(1) = \sum_{k\in\mathcal{K}} p_k = \pi (B_{(k)}^* B_{(k)}),
\end{equation}
We can interpret $\tau_{\mathfrak{K}}(\pi)$ as the probability we
don't destroy our system, like a photon getting absorbed by a
polarizer, for input state
$\pi$.
A sensible operation therefore obeys $\tau_{\mathfrak{K}}(\pi) \leq 1$, or
\[
  \tau_{\mathfrak{K}}(\pi) = \pi(B_{(k)}^* B_{(k)}) \leq 1 \quad
  \Longrightarrow \quad \pi(I - B_{(k)}^* B_{(k)}) \geq 0.
\]
Only positive operators have positive
expectation in all states; there is a negative eigenvalue, hence a
negative eigenvector. Thus,
\begin{equation}
  \label{eq:op-sum}
  I - B^*_{(k)}B_{(k)} \geq 0.
\end{equation}
Together, (\ref{eq:op-luders}), (\ref{eq:op-born}) and
(\ref{eq:op-sum}) define quantum operations.
\newpage

If $\tau_\mathfrak{K}(\pi)=1$ for
all $\pi$ (i.e. our experiments don't destroy 
lab supplies) 
the operation is called a \emph{quantum channel}. Channels linearly map states to
states, but not every such map is a channel. 
An example is provided by the familiar matrix transpose, $A \mapsto
A^\top$.
To see how it acts on states, we pull it back in the usual way:
\[
  \pi^\top\!(A) = \pi (A^\top).
\]
The functional $\pi^\top$ is linear by construction, and normalized
since $\pi^\top(I) = \pi(I^\top) = 1$.
To show that it's positive, we write $A = R^* R$ for some $R$ and
transpose. The catch is that, to invoke the usual properties of ${}^\top$,
we must work in the universal representation
$\tilde{\mathcal{H}}$ where everything is a matrix. In this case,\marginnote{We omit the $\sim$ hat on
  operators to avoid notational clutter. It's a convenient sin to
  pretend they are matrices.}
\[
  A^\top = (R^* R)^\top = R^\top \overline{R} = \overline{R}^*\overline{R},
\]
which implies that $A^\top$ is positive.
This shows the transpose is indeed a linear map from states to
states.

Showing that the transpose is not a quantum operation
   (\ref{eq:op-born}) requires more effort.
First, we pick an orthonormal basis
$\{|i\rangle\}_{i\in\mathfrak{I}}$ of the universal Hilbert space
$\tilde{\mathcal{H}}$, and
form a corresponding basis of outer products $E_{(ij)} =
|i\rangle\langle j|$ for
$\mathcal{B}(\tilde{\mathcal{H}})$.
On these outer products, transposition
swaps indices, $E_{(ij)}^\top = E_{(ji)}$, so  
\[
  \Delta = E_{(ij)} - E_{(ji)} \quad \longmapsto \quad
  E_{(ji)} - E_{(ij)} = -\Delta.
\]
\marginnote{
  \vspace{-80pt}
  \begin{center}
\includegraphics[width=0.65\linewidth]{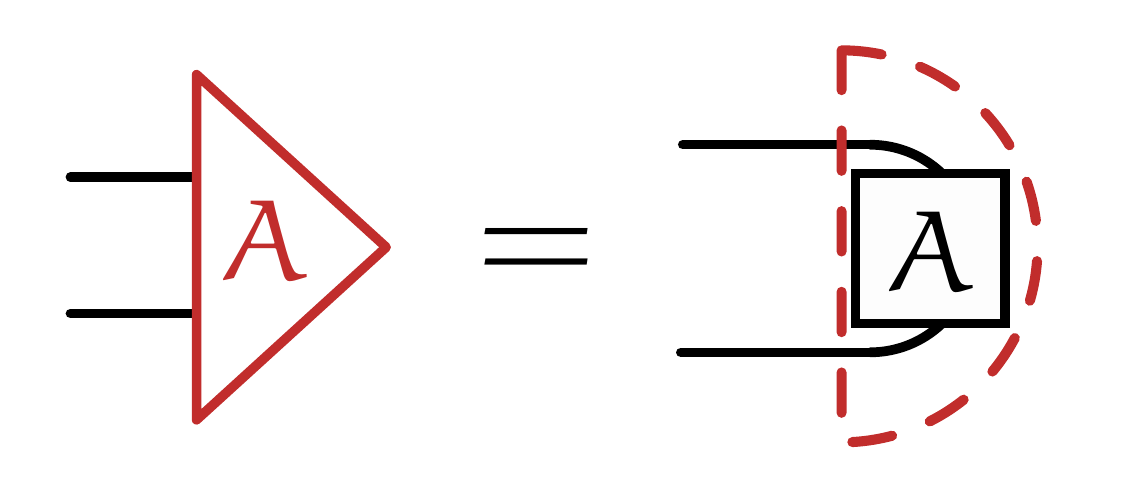}
  \end{center}  \vspace{-5pt}
  \emph{We can think of the vectorization $|A\rangle\!\!\rangle$ as a
    ket formed by bending a wire through $A$; we make this precise below.
  } \vspace{-5pt}
}Since it takes $\Delta$ to $-\Delta$, transposition
appears to have a negative eigenvalue. 
To make this idea precise and compare to the general Kraus form
(\ref{eq:op-born}), we must turn maps between matrices
into maps between vectors.
A simple option is to flip a bra into a ket:
\[
  E_{(ij)} = |i\rangle\langle j| \quad \longmapsto \quad
  |E_{(ij)}\rangle\!\!\rangle = |i\rangle \otimes |j\rangle.
\]
where $|\cdot \rrangle$ denotes the \emph{vectorization} or \emph{Choi-Jamiołkowski dual}\sidenote{``Linear transformations
  which preserve trace and positive semidefiniteness of operators''
  (1972), Andrzej Jamiołkowski; ``Completely
  Positive Linear Maps on Complex Matrices'' (1975), Man-Duen Choi.}
of a matrix.
We can extend this to arbitrary $A =
\alpha_{ij}E_{(ij)}$ by linearity:
\begin{align}
  |A\rangle\hspace{-3pt}\rangle =
  \alpha_{ij} |i\rangle \otimes |j\rangle & = A |j\rangle \otimes
  |j\rangle \notag\\ & = (A \otimes I) |I\rangle\hspace{-3pt}\rangle \notag \\
  & =(I \otimes A^\top) |I\rangle\hspace{-3pt}\rangle,   \label{eq:choi3}
\end{align}
using $A|j\rangle = A_{ij}|i\rangle$ and $A^\top|i\rangle =
A_{ij}|j\rangle$.
The vectors $|A\rangle\hspace{-3pt}\rangle$ live in the Hilbert
space $\tilde{\mathcal{H}}\otimes \tilde{\mathcal{H}}$. 

To ``matricize'' a linear transformation $\mathcal{E}:
\mathcal{B}(\tilde{\mathcal{H}})\to \mathcal{B}(\tilde{\mathcal{H}})$,
we define the \emph{Choi matrix} $\mathcal{J}(\mathcal{E}) \in
\mathcal{B}(\tilde{\mathcal{H}}\otimes \tilde{\mathcal{H}})$ with
matrix elements
\begin{align}
  \label{eq:choi1}
 \llangle E_{(ij)} |\mathcal{J}(\mathcal{E})| E_{(\ell m)}\rrangle & =
 \llangle E_{(ij)}|\mathcal{E} [E_{(\ell m)}]\rrangle \\ & = \llangle \mathcal{E}^*[E_{(ij)}]| E_{(\ell m)}\rrangle.
  \label{eq:choi2}
\end{align}
We can now talk rigorously about the eigenvalues of the matrix
$\mathcal{J}(\mathcal{E})$ acting on vectorized $|A\rrangle$. Letting
$\mathcal{E}_\top$ denote the transpose operation, we can rerun our
argument from earlier and find that
\[
  \mathcal{J}(\mathcal{E}_\top) |\Delta\rrangle = -|\Delta\rrangle.
\]
On the other hand, we now show that quantum operations $\mathcal{C}_{\mathfrak{K}}$ in Kraus form
correspond to Choi matrices $\mathcal{J}(\mathcal{C}_{\mathfrak{K}})$ with
positive eigenvalues. Taking (\ref{eq:choi1}) as the definition:
\begin{align}
  \llangle A| \mathcal{J}(\mathcal{C}_{\mathfrak{K}})| A\rrangle & =
                                                                   \llangle
                                                                   A |
                                                                   B^*_{(k)}AB_{(k)}\rrangle
                                                                   = \llangle AB_{(k)} | AB_{(k)}\rrangle \geq 0,
\end{align}
swapping a Kraus operator over with (\ref{eq:choi2}).\marginnote{To be
clear, the adjoint of left-multipication by $B_{(k)}^*$ is
right-multiplication by $B_{(k)}$.}
Thus, $\mathcal{J}(\mathcal{C}_{\mathfrak{K}})$ is a positive matrix
with eigenvalues $\lambda \geq 0$, so the transpose cannot be a
channel!

We can capture these ideas more perspicuously using \gls{awd}s. Recall
the Bell state from Fig. \ref{fig:tensor3}, defined as the sum
$|\Psi_\text{Bell}\rangle=\tfrac{1}{\sqrt{2}}|bb\rangle$ for
$b\in\{0,1\}$.
We have generalized to the (non-normalized) sum $|ii\rangle$, $i\in\mathfrak{I}$.
Suppose $|i\rangle = U_{(i)}|0\rangle$ for some collection of
unitaries $U_{(i)}$. Then we depict the sum
\[
  |I\rrangle = (U_{(i)} \otimes U_{(i)}) |0 0\rangle.
\]
simply as a bent line, as in Fig. \ref{fig:choi1}:

\begin{figure}[h]
  \centering
  \vspace{-3pt} 
  \includegraphics[width=0.33\textwidth]{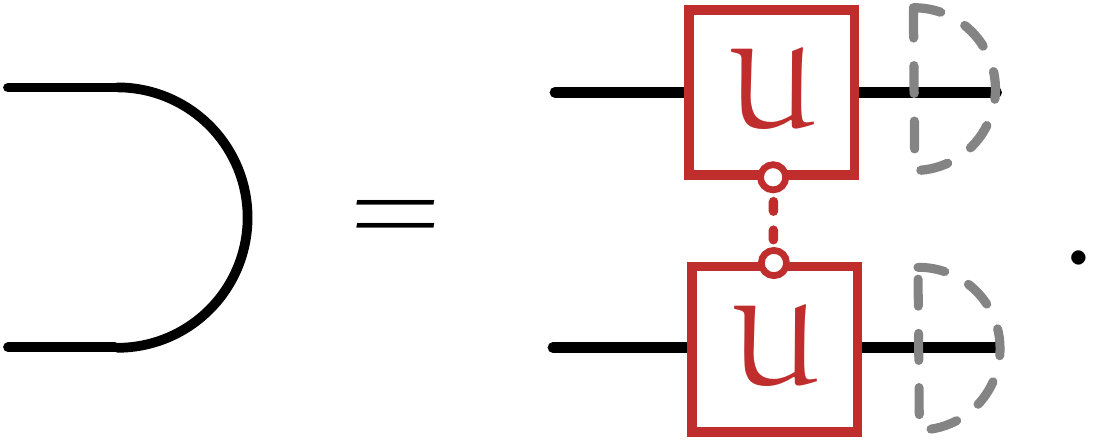}
  \caption{The vectorized identity, aka the Bell state.
  }
  \label{fig:choi1}
  \vspace{-6pt}
\end{figure}

\noindent The vectorization identities of (\ref{eq:choi3}) can then be
expressed as

\begin{figure}[h]
  \centering
  \vspace{-3pt} 
  \includegraphics[width=0.6\textwidth]{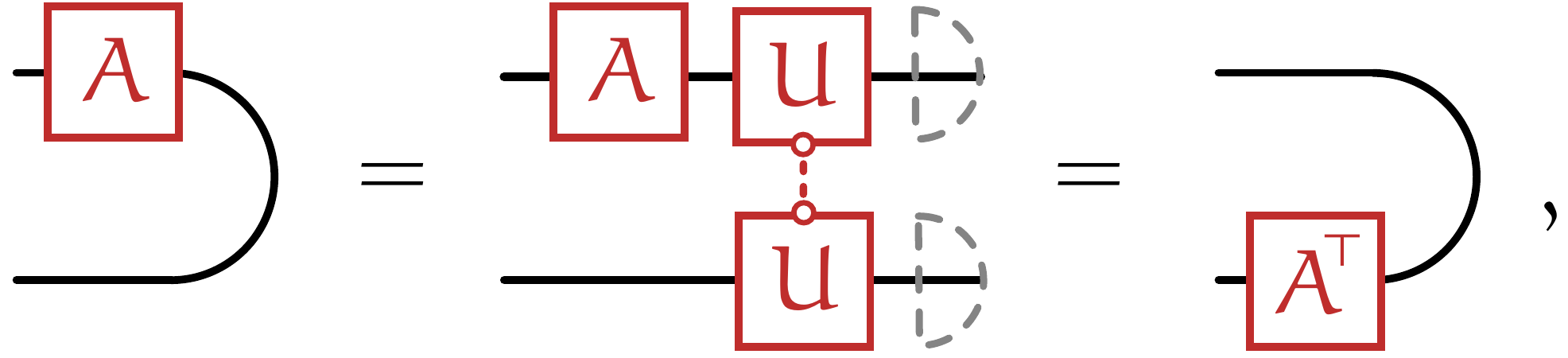}
  \caption{The identities of vectorization $|A\rangle\!\!\rangle$.
  }
  \label{fig:choi2}
  \vspace{-6pt}
\end{figure}

\noindent so sliding an operator around the bend is the same as
transposing.
This is now a vector state, with the Choi matrix capturing the effect
of applying $\mathcal{E}$ on the top lead.

Before moving on, we make two diagrammatic observations.
First, the fact that the transpose corresponds to sliding an operator around the bend suggests the follow notation:

\begin{figure}[h]
  \centering
  \vspace{-3pt} 
  \includegraphics[width=0.38\textwidth]{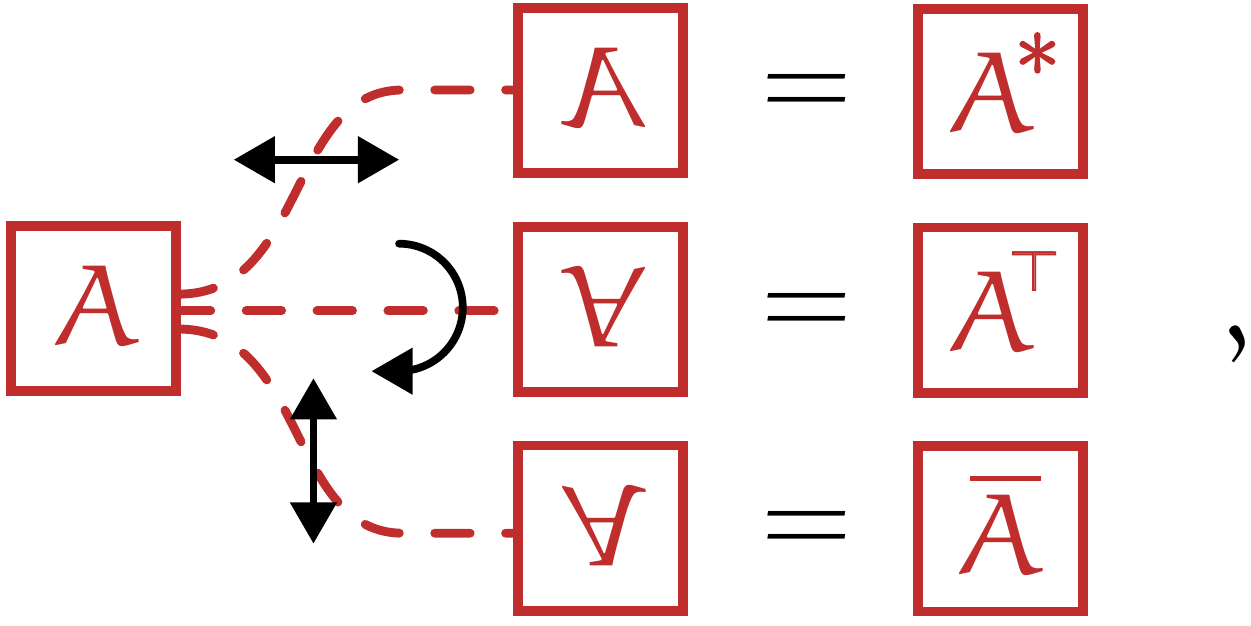}
  \caption{A mapping between involutions of a square and involutions
    of an operator.  See ``Efficient decoding for the Hayden-Preskill
  protocol'' (2017), Beni Yoshida and Alexei Kitaev.
  }
  \label{fig:choi3}
  \vspace{-6pt}
\end{figure}

\noindent This is an isomorphism between symmetries of the square and
symmetries of the operator, in the sense that, e.g. a rotation
following by a horizontal reflection equals a vertical reflection, and $(A^\top)^*=\overline{A}$.

Second, we already saw the bend of Fig. \ref{fig:choi1} appear in the
trace of Fig. \ref{fig:mixed3}. If the notation is consistent, we
should be able to write
\begin{equation}
 \mbox{tr}[\varpi A] = \llangle I | \varpi I\rrangle.\label{eq:44}
\end{equation}
We can easily check this is true:
\[
  \llangle I | \varpi A\rrangle = \langle ii | (\varpi A \otimes I)
  |jj \rangle = \delta_{ij} \langle i|\varpi A|j\rangle =
  \mbox{tr}[\varpi A],
\]
using the expressions in (\ref{eq:cyclic}).
A cute observation: the second line enforces $\delta_{ij}$, so we our
sum $\langle i|\varpi A|i\rangle = \mathcal{C}_{|i\rangle}[\varpi A]$
takes the form of a sum of conjugators, but with bras and kets!
This means that the trace is itself a quantum operation $\mathcal{C}_{\mathfrak{I}}$, and in
particular a channel since the density matrix $\varpi$ is normalized.
The same is true of the partial trace, since $\mathfrak{I}$
can label the basis of a tensor factor rather than the full space.
We will see in a moment how viewing states as channels is the
natural perspective.

A map with a positive Choi matrix 
is called \emph{completely positive (CP)}.
This equivalence between CP maps and Kraus forms, first proved by Choi, leads to the terminology \emph{completely positive
  trace-preserving (CPTP) map} as a synonym for ``channel''.
Though it is sometimes presented as a physical necessity, CP
violations result from certain physical interventions, just like the
loss of purity or trace:\marginnote{
  \vspace{-100pt}
  \begin{center}
    \hspace{15pt}\includegraphics[width=0.95\linewidth]{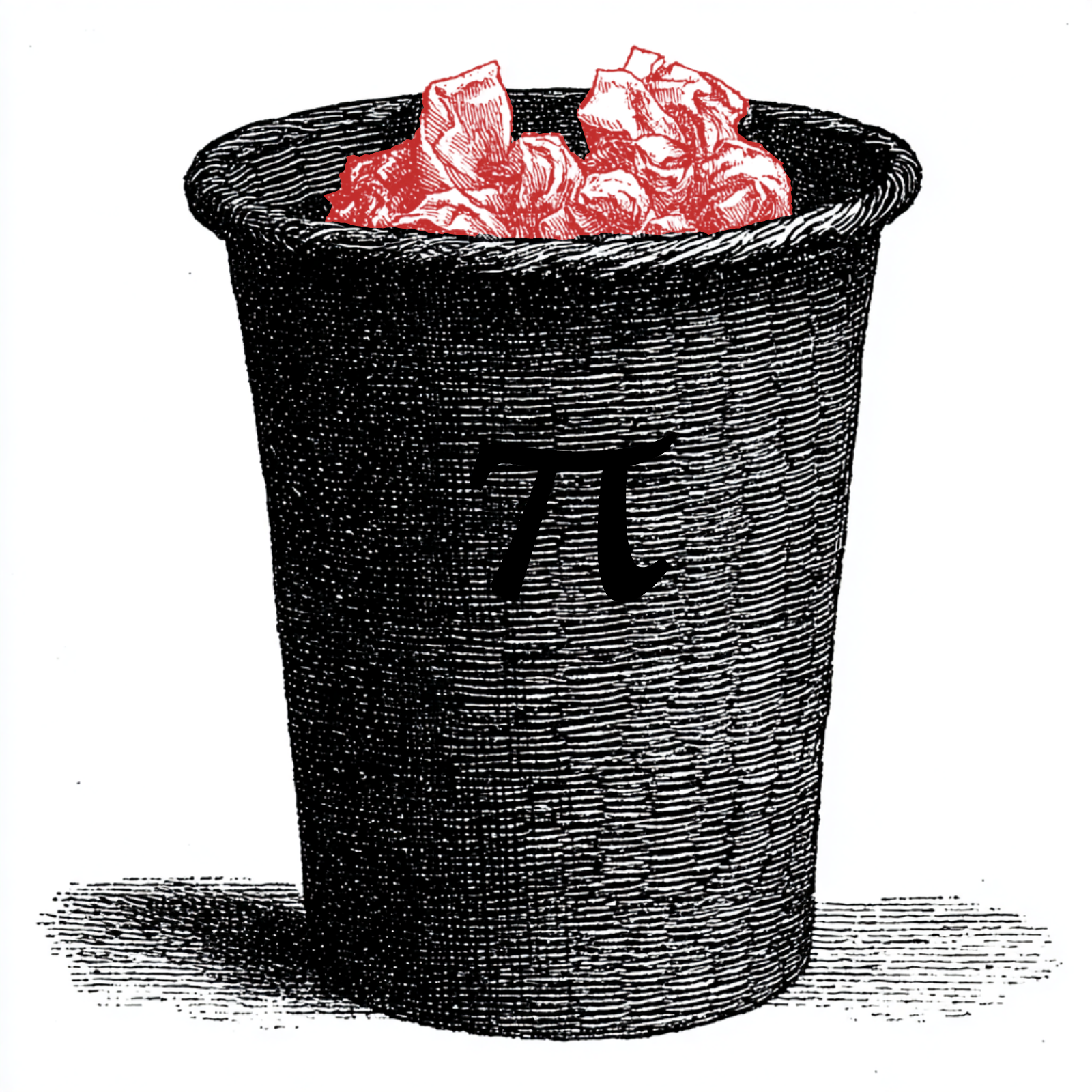}
  \end{center}  \vspace{-5pt}
  \emph{States can be compromised in various ways: classically mixed, partially lost, or correlated with the
    environment.
  } \vspace{-5pt}
}
\vspace{3pt}
\begin{center}
\begin{tabular}{ c|c } 
 \hline
 physical intervention & mathematical effect \\ \hline
 add classical randomness & pure states become mixed \\ 
  destroy subsystems & trace not conserved \\
   reduced dynamics with correlation & CP violations \\ 
 \hline
\end{tabular}
\end{center}
\vspace{4pt}
In particular, when a system is connected to an environment, it is easy for the
\emph{reduced} system dynamics to be non-CP in the presence of
correlations with the environment.\sidenote{On initial
  environment-system correlations and reduced dynamics, see ``Reduced Dynamics Need Not Be Completely Positive'' (1994), Philip
  Pechukas 
Hi!   
  and
  ``Dynamics beyond
  completely positive maps: some properties and applications'' (2006),
  Hilary Carteret, Daniel Terno, and Karol Życzkowski. \vspace{5pt}
}

While complete positivity is a convenience rather
than a necessity, this convenience is not to be underestimated.
A prime example is the \emph{Stinespring dilation
  theorem},\sidenote{``Positive Functions on C${}^*$-Algebras'' (1955),
  W. Forrest Stinespring; \emph{Effects and Operations: Fundamental
    Notions of Quantum Theory} (1983), Karl Kraus.} generalizing both the GNS construction and
Choi's theorem.
To motivate this, recall that a state $\pi: \mathcal{A} \to \mathbb{C}$ is associated with a Hilbert
space $\mathcal{H}_\pi$ where vanishing correlations have been
quotiented out. The algebra $\mathcal{A}$ acts via
matrices $\Phi_\pi(A)$ on $\mathcal{H}_\pi$, and to take
expectations, we can either directly
evaluate $\pi(A)$, or in the Hilbert space
$\mathcal{H}_\pi$, compute $\mathbb{E}[\Phi_\pi(A)] = \langle
0|\Phi_\pi(A)|0\rangle$ where $|0\rangle$ is our fiducial
representative satisfying $\langle 0|0\rangle = 1$. Thus, we find
  \begin{equation}
  \pi(A) = \langle 0|\Phi_\pi(A)|0\rangle 
  \quad \Longrightarrow \quad \pi = \mathcal{C}_{|0\rangle} \circ \Phi_\pi,\label{eq:stinespring-gns}
\end{equation}
since the LHS holds for arbitrary $A$.
We can represent this more explicitly with domains and codomains shown:
\[
  \begin{tikzcd}
    \mathcal{A} \arrow{r}{\Phi_\pi} \arrow[swap]{dr}{\pi} & \mathcal{B}(\mathcal{H}_\pi) \arrow{d}{\mathcal{C}_{|0\rangle}} \\
     & \mathbb{C}
  \end{tikzcd}.
\]
Stinespring's theorem states that, for CP map
$\mathcal{E}:\mathcal{A}\to \mathcal{B}(\mathcal{H})$, there is an
associated Hilbert space $\mathcal{H}_\mathcal{E}$, representation
$\Phi_{\mathcal{E}}: \mathcal{A}\to
\mathcal{B}(\mathcal{H}_\mathcal{E})$, and map $V: \mathcal{H} \to
\mathcal{H}_\mathcal{E}$, exhibiting the same structure
$\mathcal{E}[A] = V^* \Phi_\mathcal{E}(A)V$, or
\begin{equation}
  \begin{tikzcd}
    \mathcal{A} \arrow{r}{\Phi_\mathcal{E}} \arrow[swap]{dr}{\mathcal{E}} & \mathcal{B}(\mathcal{H}_\mathcal{E}) \arrow{d}{\mathcal{C}_{V}} \\
     & \mathcal{B}(\mathcal{H})
  \end{tikzcd}.\label{eq:43}
\end{equation}
If $\mathcal{E}$ is unital, with $\mathcal{E}(I) = I_{\mathcal{H}}$, then $V$ is an \emph{isometry} $V^*V =
I_{\mathcal{H}}$, replacing the condition $\langle 0|0\rangle = 1$. \marginnote{The appendix also shows how Stinespring's
  theorem generalizes Choi's
  equivalence result.}
The
construction is similar in spirit to GNS (\S
\hyperlink{sec:6}{5}).\sidenote{For details, see Nielsen and Chuang
  (2000) or Kadison and Ringrose (1983).}

Thus, a channel
$\mathcal{E}$ can always be decomposed into (a) a representation
$\Phi_{\mathcal{E}}$ of $\mathcal{A}$ acting on an auxiliary Hilbert space
$\mathcal{H}_{\mathcal{E}}$; and (b) conjugation by an isometry $V$ from the target Hilbert
space $\mathcal{H}$ to the auxiliary $\mathcal{H}_{\mathcal{E}}$.
So much for the math. Physically, we act on some bigger\marginnote{If we replace the isometry $V$ with a projector
$\Pi$, we leak probability; if we replace conjugation
$\mathcal{C}_V$ with another map, we lose complete positivity.}
system ($\mathcal{H}_{\mathcal{E}}$) but observe only the smaller one
($\mathcal{H}$). Instead of dwelling further on the subtleties of
information loss, we turn to something a little more exciting: how to win it back again.

\section{15. How to recover}\hypertarget{sec:16}{}

Quantum information is fragile. By performing operations (or more
general non-positive updates), we can easily mix, absorb and entangle observables
with the environment. 
In order to compute usefully and not blow our lab budget, we need
ways to keep information in the system.
We visualize the high-level strategy in Fig. \ref{fig:qec1} (left):

\begin{figure}[h]
  \centering
  \vspace{-5pt} 
  \includegraphics[width=0.65\textwidth]{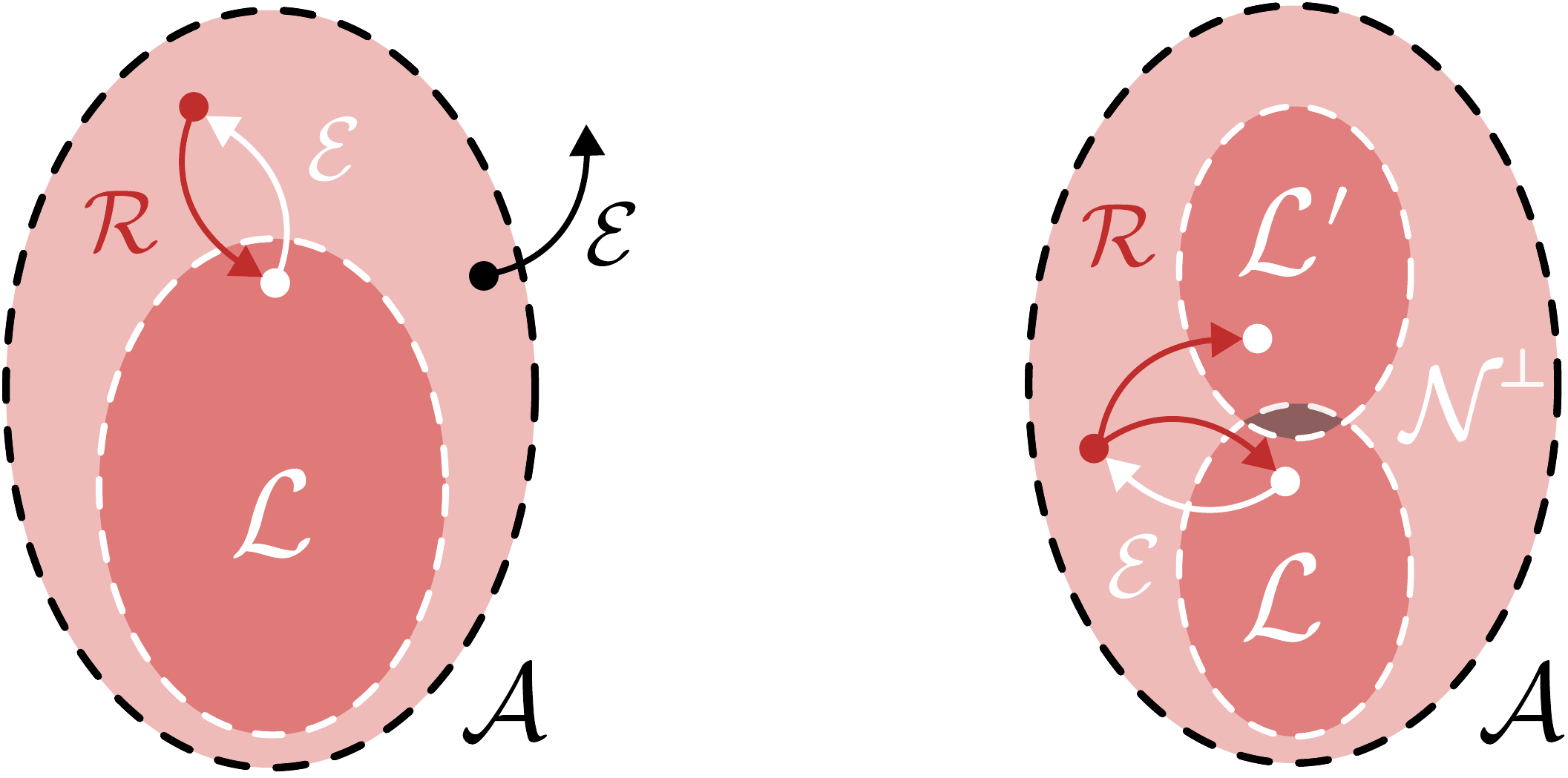}
  \caption{\textsc{Left}. The error process $\mathcal{E}$ kicks the black data point
    out of the system. The white point is perturbed but
    can be recovered with $\mathcal{R}$. \textsc{Right.} A weaker
    condition for recovery is that $\mathcal{R}$ restores a linear
    combination of the original point and an operator in the commutant.
  }
  \label{fig:qec1}
  \vspace{-6pt}
\end{figure}

\noindent Under a physical \emph{error process} $\mathcal{E}$, with
Kraus operators $E_{(k)}$ for $k\in\mathfrak{K}$, some
data 
will be removed from the system altogether and become
unrecoverable. But there may be a smaller, protected subsystem called the \emph{logical
  subalgebra} $\mathcal{L} \subseteq \mathcal{A}$ where information
remains recoverable in
  principle. The map $\mathcal{R}$ that restores the data back to its rightful place
  is called the \emph{recovery map}.\marginnote{
  \vspace{-80pt}
  \begin{center}
    \hspace{-10pt}\includegraphics[width=0.95\linewidth]{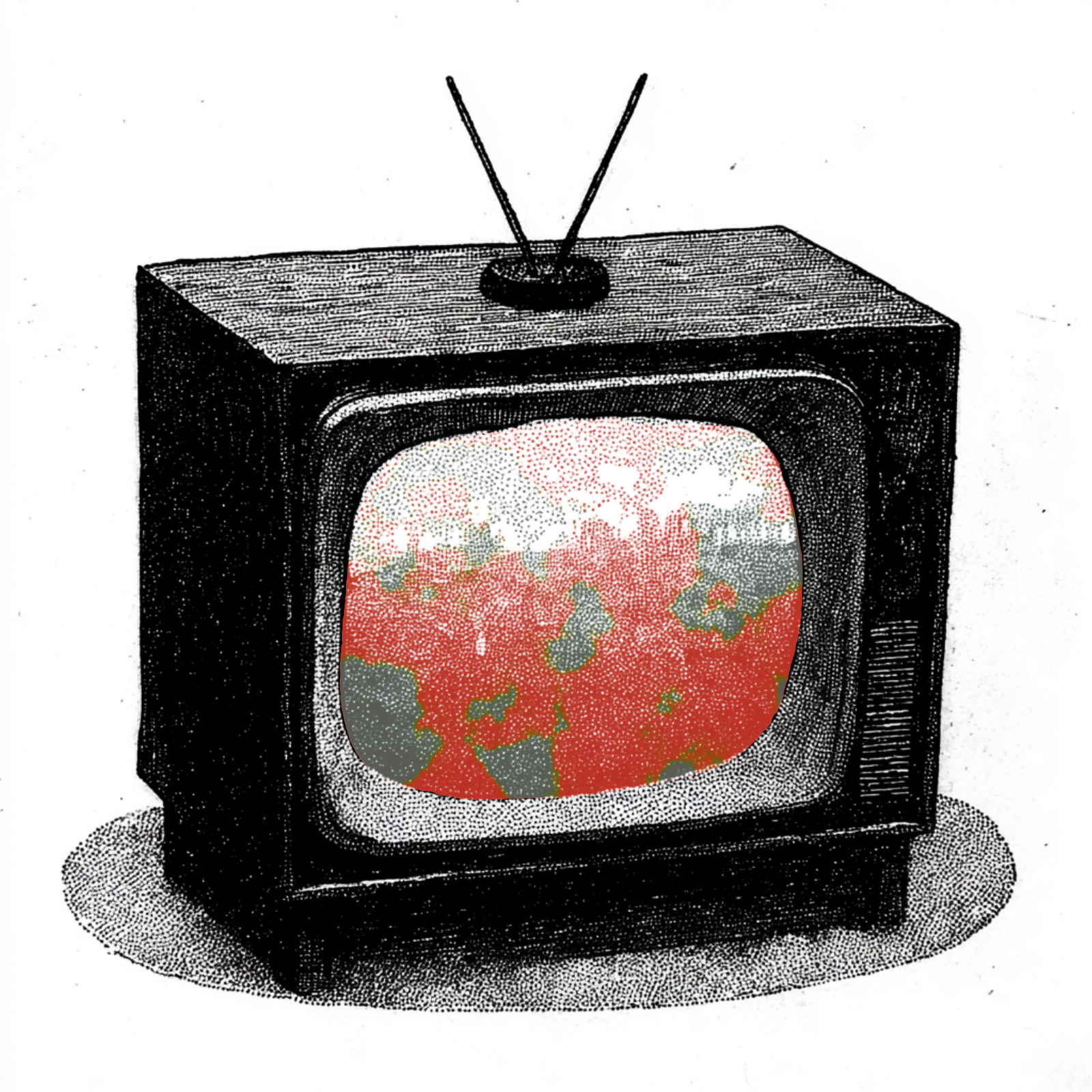}
  \end{center}  \vspace{-10pt}
  \emph{
    Error correction is an attempt to change a noisy channel.
  }
  \vspace{-10pt}
}

\emph{Quantum error correction (QEC)} is the art and science of
building recovery maps. 
Let's try to fill in
the details. For any $A \in \mathcal{L}$, we should be able to undo the
error channel with the recovery map. Equivalently, $\mathcal{R}$ is
the left inverse of $\mathcal{E}$ on $\mathcal{L}$:
\begin{equation}
  (\mathcal{R} \circ \mathcal{E})[A] = A \quad \Longrightarrow \quad
  (\mathcal{R} \circ \mathcal{E})|_{\mathcal{L}} =
  \text{id}_{\mathcal{L}},\label{eq:qec2}
\end{equation}
where $\text{id}_{\mathcal{L}}=\mathcal{C}_{I_\mathcal{L}}$ is the
identity channel.
But there is a subtlety lurking in the restriction $|_{\mathcal{L}}$.
Fig. \ref{fig:qec1} (right) shows a situation where $(\mathcal{R}\circ
\mathcal{E})[A]$ has a component \emph{outside} $\mathcal{L}$. This shouldn't
matter, since all we care about is behaviour \emph{inside} $\mathcal{L}$.
Let's parse this more carefully.

To start with, assume $\mathcal{L}$ and
$\mathcal{L}'$ join to give the full algebra $\mathcal{A}$. By von Neumann's theorem
(\ref{eq:bicommutant}), they are tensor factors $\mathcal{A} \cong \mathcal{L} \otimes
\mathcal{L}'$, so we can ignore $\mathcal{L}'$ using the partial trace.
In general, $\mathcal{L} \vee \mathcal{L}' = \mathcal{N}$ is some
subalgebra of\marginnote{We can define the complement
  $\mathcal{N}^\perp$ as the subspace of lowest dimension such that
  $\mathcal{N}\oplus \mathcal{N}^\perp =
  \mathcal{A}$. Unlike an orthogonal complement, it is not unique since it can ``tilt''.}
$\mathcal{A}$, leaving a non-unique complement $\mathcal{N}^\perp$ so that
\begin{equation}
  \mathcal{A} = \mathcal{L} \otimes \mathcal{L}' \oplus
  \mathcal{N}^\perp.\label{eq:qec0}
\end{equation}
In this case, we need to restrict to the tensor product with a projector $\Pi_{\mathcal{N}}$ before we can use the partial trace.
This modifies (\ref{eq:qec2}) to
\begin{equation}
  \mbox{tr}_{\mathcal{L}'}\big[\Pi_{\mathcal{N}} (\mathcal{R}\circ\mathcal{E})[A]\Pi_{\mathcal{N}}\big] =
  A \text{   for all  } A \in \mathcal{L},\label{eq:qec1.5}
\end{equation}
or slightly more cleanly, $(\mathbb{E}_{\mathcal{L}}\circ
  \mathcal{R}\circ\mathcal{E
  })\big |_{\mathcal{L}} =\text{id}_{\mathcal{L}}$,
where $\mathbb{E}_{\mathcal{L}}=\mbox{tr}_{\mathcal{L}'} \circ
\mathcal{C}_{\Pi_{\mathcal{N}}}$.
Showing the source and target of maps:
\begin{equation}
  \begin{tikzcd}
    \mathcal{L} \arrow{r}{\mathcal{E}} \arrow[swap]{d}{\text{id}} & \mathcal{A} \arrow{d}{\mathcal{R}} \\
    \mathcal{L} \arrow[swap, leftarrow]{r}{\mathbb{E}_{\mathcal{L}}} & \mathcal{A} 
  \end{tikzcd}.\label{eq:qec2}
\end{equation}
When $\mathcal{R}$ exists, we say $\mathcal{E}$ is \emph{correctable}
on $\mathcal{L}$.

\begin{figure}[h]
  \centering
  \vspace{-0pt} 
  \includegraphics[width=0.62\textwidth]{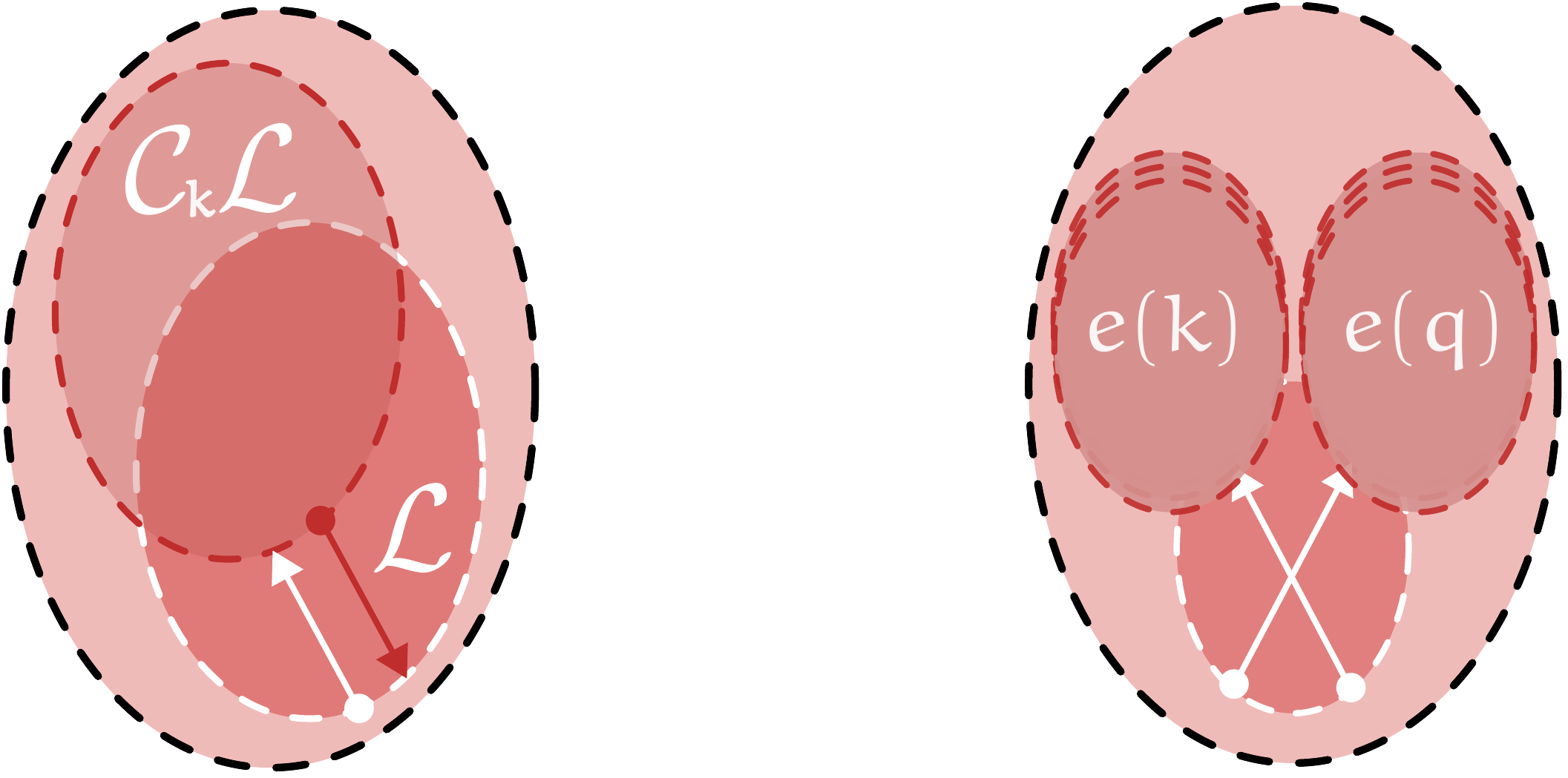}
  \caption{The KL criterion for operator error
    correction. \textsc{Left}. Errors shift $\mathcal{L}$ invertibly
    without collapsing it. \textsc{Right.} Errors must either be
    disjoint or identical, in which case they form a little stack
    called a syndrome.
  }
  \label{fig:qec2}
  \vspace{-7pt}
\end{figure}

What are some reasonable conditions for correctability?
The projective measurements from \S
\hyperlink{sec:14}{13} provide a helpful illustration. 
The spectral theorem gives rise to \emph{orthogonal} projectors; this
is helpful because it uniquely tells you which filter
was applied.
On the other hand, there is no way correct wavefunction collapse since
(nontrivial) projectors cannot be inverted; the cat cannot
  be revived.\sidenote{For a formal proof, see ``On the Hardness of Detecting Macroscopic Superpositions'' (2020), Scott Aaronson, Yosi Atia, and Leonard Susskind.}
We can write orthogonality as
\[
  \Pi_{(\lambda)}^*\Pi_{(\mu)} = \delta_{\lambda\mu}\Pi_{(\hat{\lambda})}.
\]
This suggests correctability requires two things. 
First, the Kraus operators $E_{(k)}$ should be injective when
restricted to $\mathcal{L}$, as in Fig. \ref{fig:qec2}
(left); we take ``injective'' to mean our Kraus operators are \emph{scaled
  isometries}.\marginnote{In other words, $E_{(k)}|_\mathcal{L} =
  \sqrt{\nu_k} V_{(k)}$ for $V_{(k)}: \mathcal{L}\to \mathcal{A}$
  satisfing $V_{(k)}^* V_{(k)} = I_\mathcal{L}$.}
Second,
that they are orthogonal, so we know which error ``fired''. We write both conditions in one fell
swoop:
\[
  \mathbb{E}_{\mathcal{L}}[E_{(k)}^* E_{(q)}] =
  \nu_{k}\delta_{kq}I_{\mathcal{L}},\]
where $\nu_{k} > 0$ allows errors to
rescale parts of the space. 

This is close, but a little too strong: we don't actually need the $E_{(k)}$ to be
orthogonal, provided they \emph{overlap perfectly}, as depicted in
Fig. \ref{fig:qec2} (right). We can apply the same correction to
different errors!

\noindent These error classes are called \emph{syndromes}, and we can define a
function $e: \mathfrak{K} \to \mathfrak{E}$ which assigns a Kraus
index $k \in \mathfrak{K}$ to a syndrome $e(k) \in \mathfrak{E}$.
Then we can write our guess at as
\begin{equation}
  \mathbb{E}_{\mathcal{L}}[E_{(k)}^* E_{(q)}] =
  \nu_{e(k)}\delta_{e(\hat{k})e(q)}I_{\mathcal{L}}.\label{eq:qec3}
\end{equation}
These are the \emph{Knill-Laflamme (KL) conditions},\sidenote{``Theory of Quantum Error
  Correction for General Noise'' (2000), Emanuel Knill, Raymond
  Laflamme and Lorenza Viola. Necessity in the case of OQEC was shown in ``Unified and Generalized Approach to
  Quantum Error Correction'' (2005), David Kribs, Raymond Laflamme,
  David Poulin; sufficiency in ``Algebraic and information-theoretic
  conditions for operator quantum error-correction'' (2005), Michael
  Nielsen and David Poulin.}, and it can be shown they are both
necessary and sufficient for correctability.
It is usually written in the equivalent, but basis-independent, form
  \begin{equation}
  \mathbb{E}_{\mathcal{L}}[E_{(k)}^* E_{(q)}] =
  \nu_{kq}I_{\mathcal{L}}\label{eq:qec4},
\end{equation}
with $\nu_{kq}$ a positive \emph{syndrome matrix}. Since positive
matrices are unitarily diagonalizable, this recovers (\ref{eq:qec3}).

Before we turn to explicit recovery channels, we need to add to our
visual toolbox.
The algebraic structure (\ref{eq:qec0}) relevant to OQEC involves both direct
sums and tensor products, so our diagrams must accomodate both. In
Fig. \ref{fig:qec3}, we show our proposal. We have tensor and direct
sum ``fan-outs'', indicated by $\oplus$ and $\otimes$, and optional
``fan-ins'' indicated by $\bullet$. Below, the initial direct sum
splits the algebra into $\mathcal{A} = \mathcal{N}\oplus
\mathcal{N}^\perp$, and the tensor product splits $\mathcal{N} =
\mathcal{L}\otimes \mathcal{L}'$:

\begin{figure}[h]
  \centering
  \vspace{-0pt} 
  \includegraphics[width=0.25\textwidth]{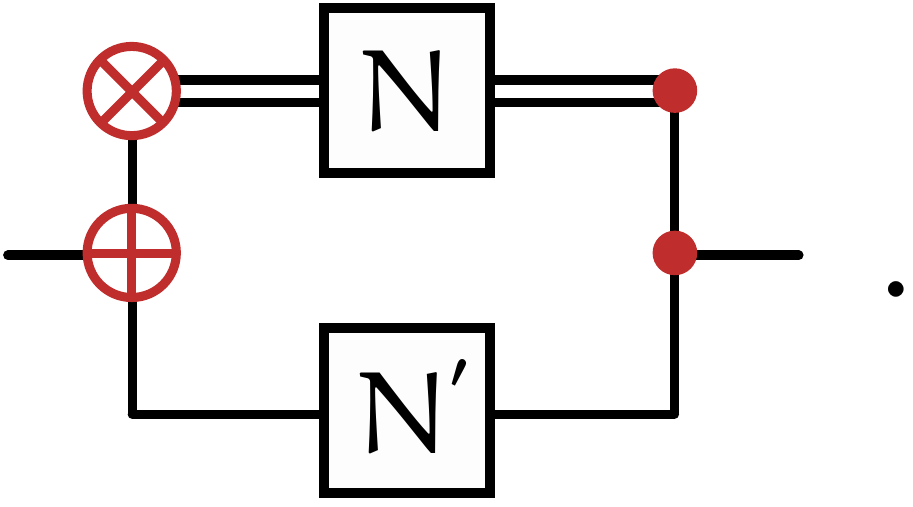}
  \caption{Fan-out notation for direct sums and tensor products.}
  \label{fig:qec3}
  \vspace{-4pt}
\end{figure}

\begin{figure}[h]
  \centering
  \vspace{-0pt} 
  \includegraphics[width=0.56\textwidth]{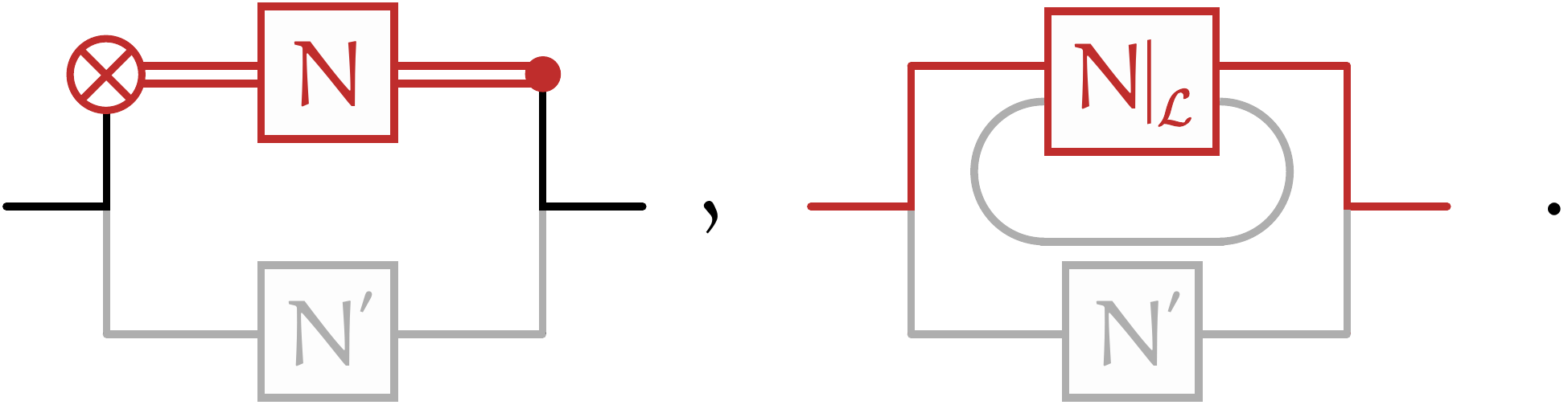}
  \caption{Undoing the operations of Fig. \ref{fig:qec3}. \textsc{Left.} Projection isolates components of a direct
    sum. \textsc{Right.} The partial trace isolates components of a tensor product.}
  \label{fig:qec4}
  \vspace{-10pt}
\end{figure}

\noindent This helps us illustrate the difference between projection
and partial trace (Fig. \ref{fig:qec4}). Loosely speaking, projection is the inverse
of direct sums, while the partial trace is the inverse of a tensor product. 

\begin{figure}[h]
  \centering
  \vspace{-5pt} 
  \includegraphics[width=0.54\textwidth]{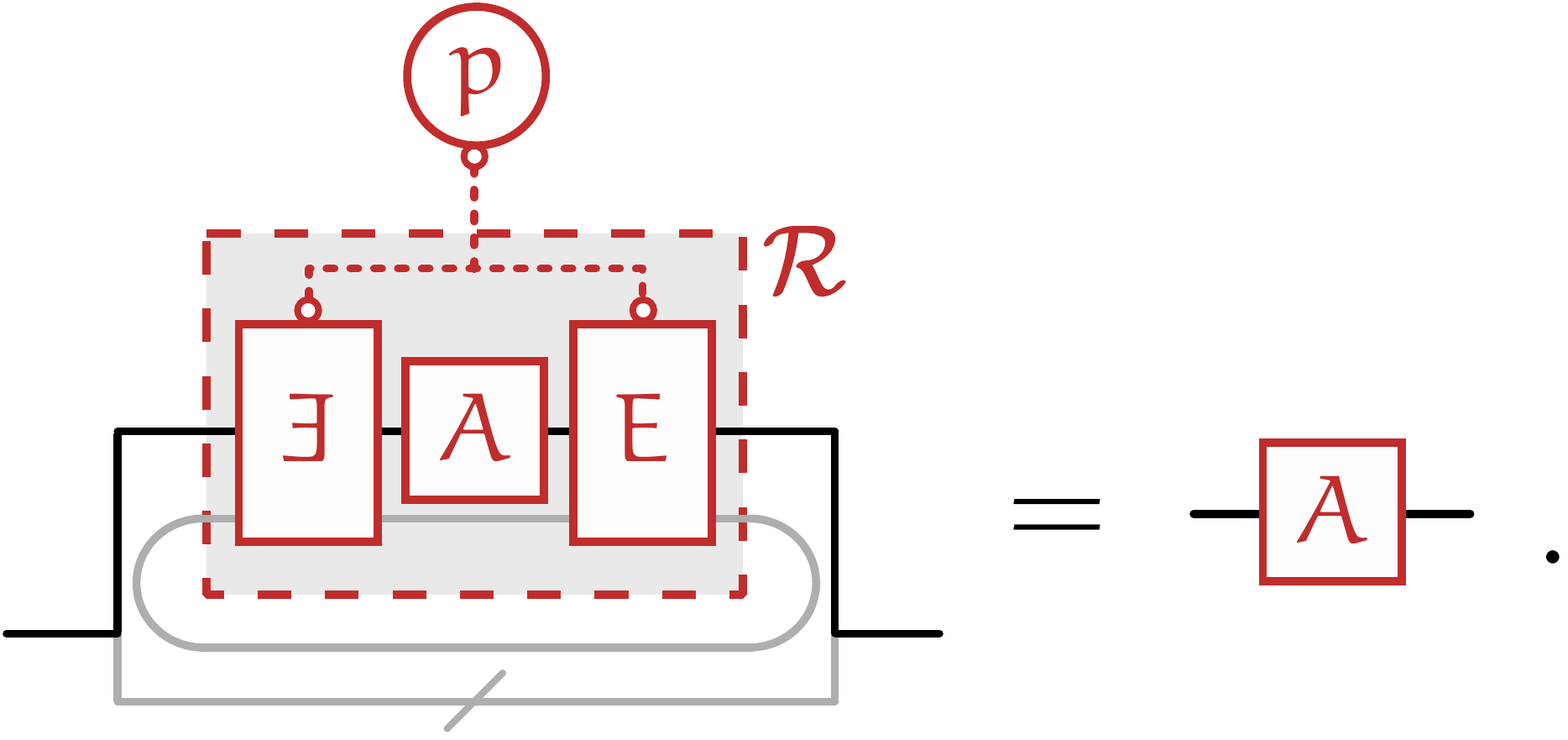}
  \caption{The definition of a correctable channel. We suppress the explicit, state-dependent form of the Kraus coefficients $p_k
    = \mathcal{C}^{(k)}[\pi]$.}
  \label{fig:qec5}
  \vspace{-7pt}
\end{figure}

We encode the correctability
condition (\ref{eq:qec1.5}) using fan-out notation in Fig. \ref{fig:qec5}. The error channel $\mathcal{E}$ is nested inside
the recovery map $\mathcal{R}$; since the channels are linear, they split
over the direct summands, and we can ignore their effect on
$\mathcal{N}^\perp$ when we ignore $\mathcal{N}^\perp$.
We then loop out $\mathcal{L}'$, leaving the operator $A$ on
the $\mathcal{L}$ wire trapped.

\noindent A recovery map $\mathcal{R}$ frees $A$ from this somewhat elaborate cage. The basis-independent KL
conditions (\ref{eq:qec4}) for the existence of $\mathcal{R}$
are expressed more compactly in Fig. \ref{fig:qec6}:

\begin{figure}[h]
  \centering
  \vspace{-5pt} 
  \includegraphics[width=0.51\textwidth]{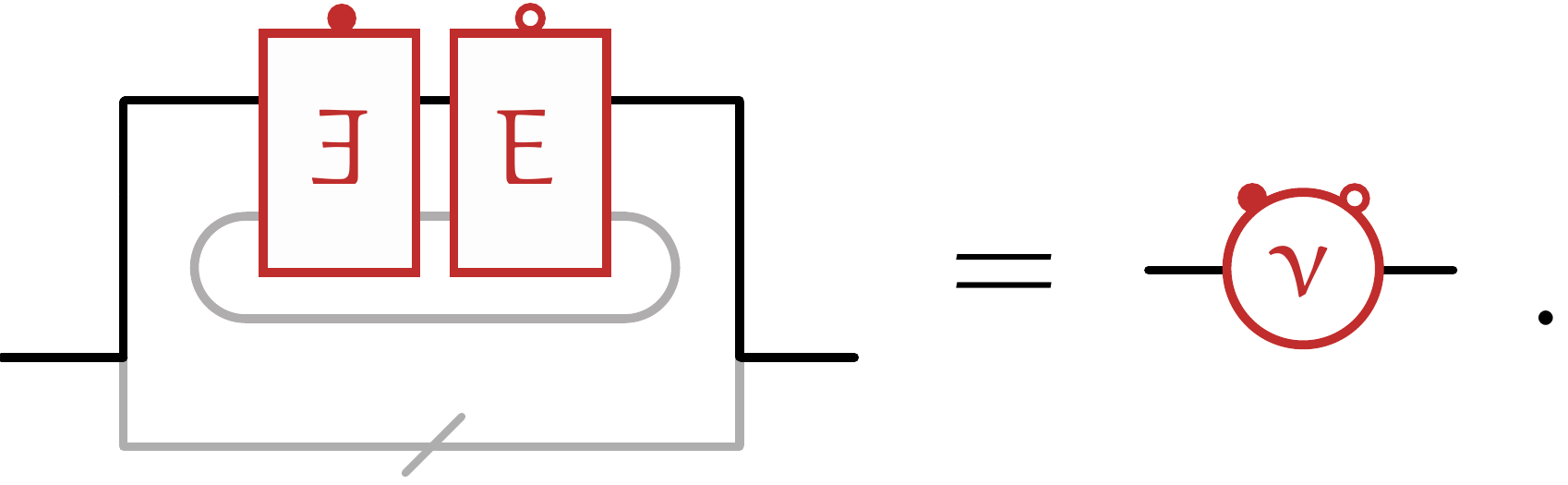}
  \caption{Basis-independent KL conditions for operator error correction.}
  \label{fig:qec6}
  \vspace{-7pt}
\end{figure}

\noindent It's not at all obvious from these diagrams that Fig. \ref{fig:qec6}
should be equivalent to the existence of a box that realizes
Fig. \ref{fig:qec5}, but a purely diagrammatic proof is possible.

Given a set of Kraus operators $E_{(k)}$ satisfying the KL
conditions, it's easy to build the recovery map: if we measure
$\mathcal{E} \measure k$, we apply some local inverse $E^*_{(k')}$ where
$e(k')=e(k)$ is some fiducial representative from the syndrome
$e(k)$. In other words,
\begin{equation}
  \label{eq:45}
  \mathcal{R} = \mathcal{C}_{\mathfrak{E}} =
  \sum_{e\in\mathfrak{E}}\mathcal{C}_e, \quad \mathcal{C}_e[A] =
  E_{e}^* A E_{\hat{e}}.
\end{equation}
What is much less obvious is how to construct channels
satisfying (\ref{eq:qec4}) and match them to the
real-life patterns of computational error. We'll leave the error characterization to
the experimentalists, but construct general
\emph{quantum error-correcting codes (QECCs)}, which explicitly
realize (\ref{eq:qec4}), 
in the next section.

\section{16. How to communicate}\hypertarget{sec:17}{}

We started with Shannon's graduate work on digital circuits.
It is fitting that we finish with his ``mature'' work on the theory
of communication.
Published in 1948 in the Bell Labs technical journal and turned into a
book with Warren Weaver the next year,\sidenote{``A
  Mathematical Theory of Communication'' (1948), Claude E. Shannon; \emph{A
  Mathematical Theory of Communication} (1949), Claude E. Shannon and
Warren Weaver.} it is widely regarded as the
founding document of the digital age. One of its key contributions is
a formal picture of communication called the \emph{Shannon} or
\emph{Shannon-Weaver model}, shown in Fig. \ref{fig:qecc0}:

\begin{figure}[h]
  \centering
  \vspace{-15pt} 
  \includegraphics[width=0.6\textwidth]{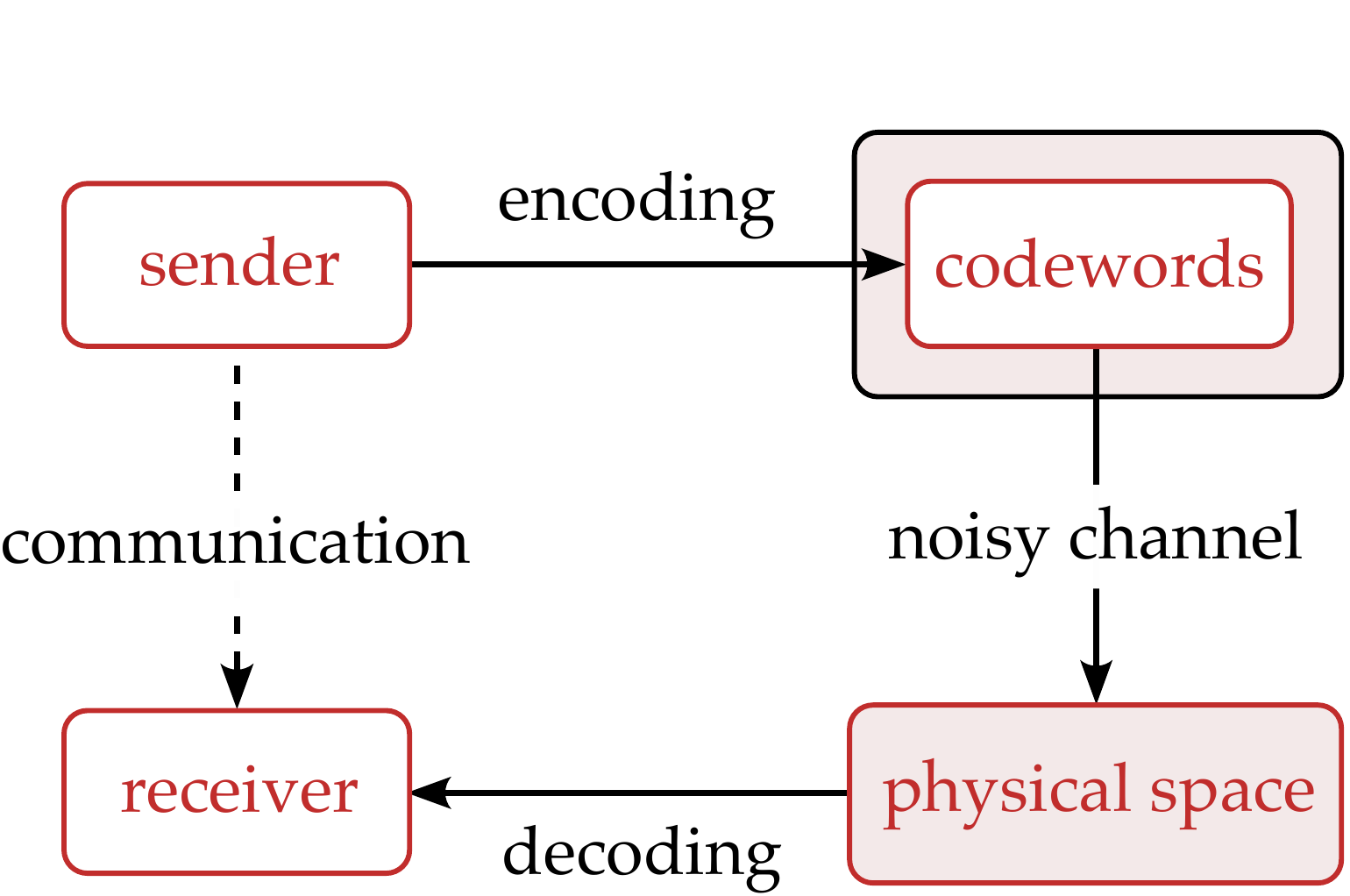}
  \caption{Components of the Shannon-Weaver model.}
  \label{fig:qecc0}
  \vspace{-2pt}
\end{figure}

\noindent Although largely self-explanatory, we give a brief
gloss. The sender wants to convey a message in some source alphabet to
the receiver; they must use a noisy channel, so their strategy is to
first encode letters into longer strings 
called codewords, which live in some protected logical region of
physical message space.
Codewords are transmitted through the
noisy channel and decoded at the receiving end.

The \emph{repetition code} illustrates these ideas in a crucial way. Suppose someone
wants to send a sequence of
binary digits $b \in \mathbb{B}$, and encodes each bit with its
$n$-fold repetition $c_n(b) \in
\mathbb{B}^n$:\marginnote{
  \vspace{-0pt}
  \begin{center}
    \hspace{-20pt}\includegraphics[width=0.65\linewidth]{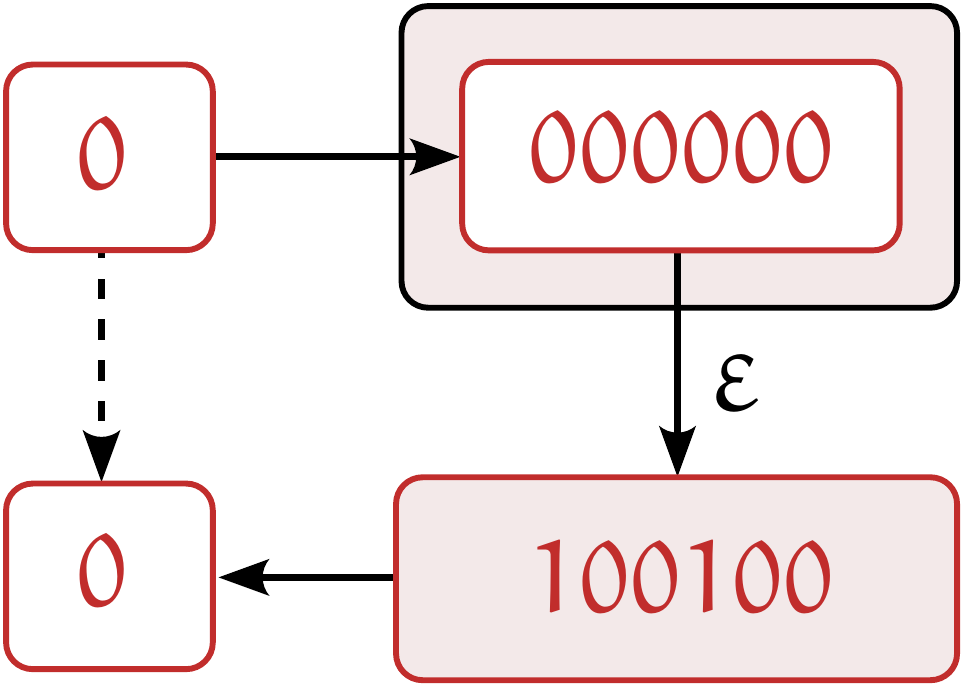}
  \end{center}  \vspace{-0pt}
  \emph{Shannon-Weaver model of repetition.
  } \vspace{10pt}
}
\[
  b \overset{}{\mapsto} c_n(b) = bbb\ldots b = b^n.
\]
The strings $c_n(b)=b^n$ are the codewords. A reasonable choice of decoder is
majority vote for each $n$-bit block, or equvalently, the closet
codeword in Hamming distance\marginnote{The \emph{Hamming norm} is
another name for the $\ell_1$ norm on $\mathbb{B}^n$.} 
\[
 c \overset{}{\mapsto} d_n(c) = \argmin_b \Vert c_n(b) -c \Vert_1.
\]
If bits are independently corrupted with probability $\varepsilon$, then 
applying Hoeffding's inequality\sidenote{``Probability inequalities for sums of
  bounded random variables'' (1963), Wassily Hoeffding. Here, we apply
the inequality to the number of bitflips, which for our channel is
binomially distributed with mean $n\varepsilon$.} to the number of
bitflips gives the following bound on decoding errors:
\[
  \mathbb{P}[\text{decoding error}] \leq e^{-n\left(\tfrac{1}{2}-\varepsilon\right)}.
\]
Provided $\varepsilon < \tfrac{1}{2}$, the probability of such an error
decreases exponentially with $n$, and can be made arbitrarily unlikely.

\begin{figure}[h]
  \centering
  \vspace{-8pt} 
  \includegraphics[width=0.42\textwidth]{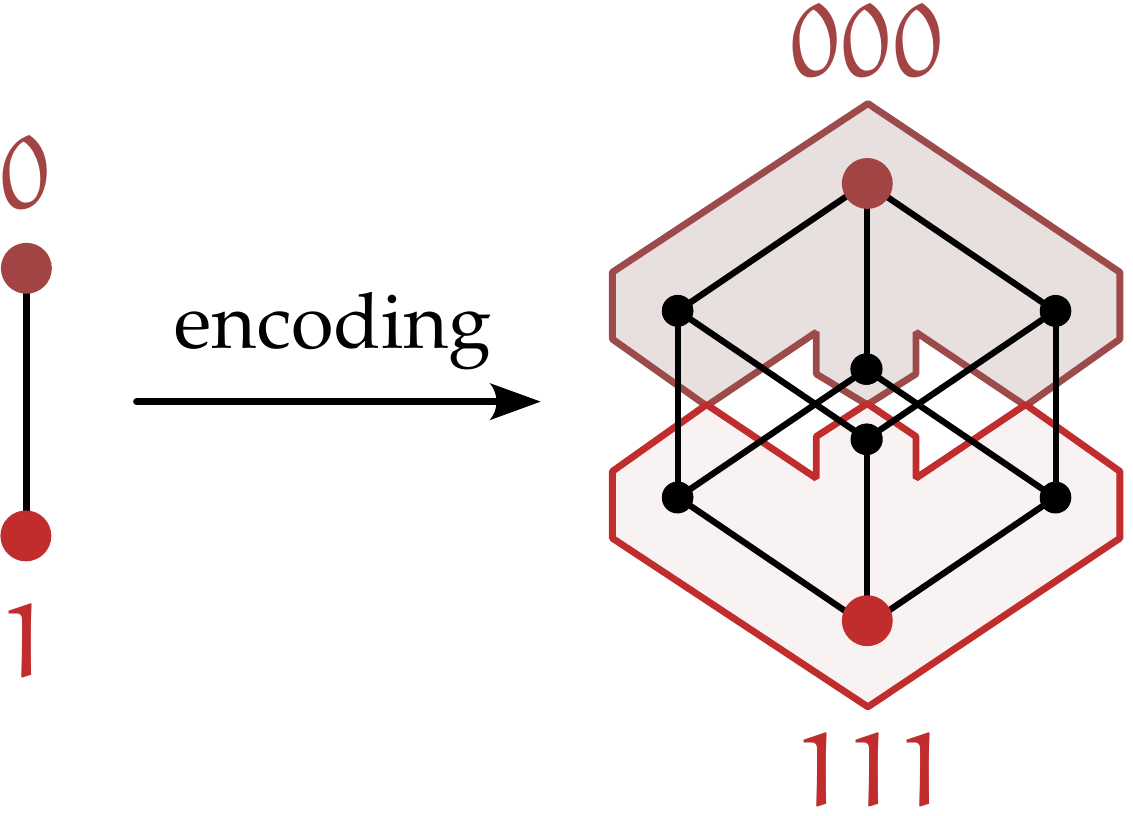}
  \caption{The repetition codes maps bits $b$ to complementary
    code neighbourhoods $\mathbb{H}_b^n$ on a hypercube $\mathbb{B}^n$.}
  \label{fig:qecc1.5}
  \vspace{-5pt}
\end{figure}

The repetition code is simple but deeply instructive. We adapt it to
the quantum setting by first making some general observations.
First, the decoding step suggests that, instead of identifying a bit
with a point $c_n(b) \in \mathbb{B}^n$, we should morally view it as
the \emph{code neighbourhood} $\mathbb{H}_b^n = d_n^{-1}(b)$. Below
(Fig. \ref{fig:qecc1.5}), we illustrate with neighbourhoods for the
$n=3$ repetition code.
The codewords are important, but their role is to define the
boundaries of the
neighbourhood: like a seed in a Voronoi
diagram,\sidenote{``Nouvelles
  applications des paramètres continus à la théorie des formes
  quadratiques'' (1908), Georges Voronoi.  
} we partition the hypercube according to whichever
codeword is closest.

The second observation is that this partition is \emph{symmetric}.
 Each code neighborhood has the same size and shape, so we can factorize 
 the auxiliary physical space $\mathbb{B}^n$ into a product of
the ``Hamming ball'' $\mathbb{H}^n$, and the source alphabet
 $\mathbb{B}$, as in Fig. \ref{fig:qecc2}:

\begin{figure}[h]
  \centering
  \vspace{-5pt} 
  \includegraphics[width=0.67\textwidth]{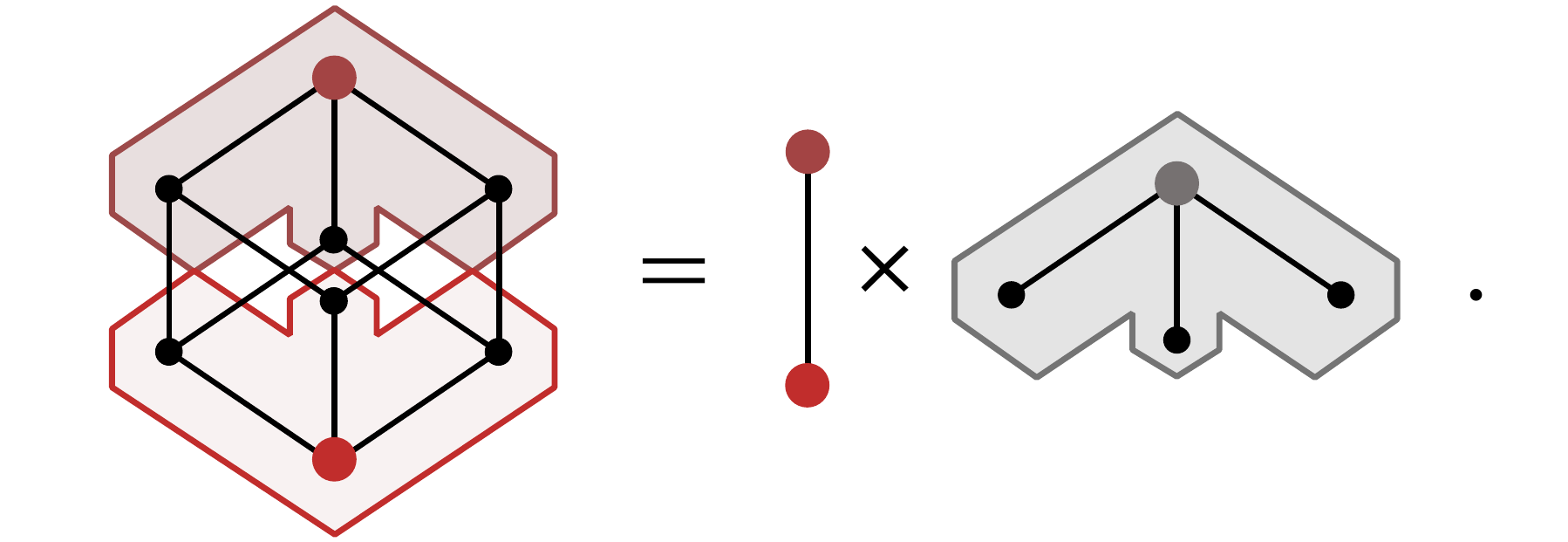}
  \caption{Factorizing a hypercube into the Booleans and a ``Hamming
    ball'' $\mathbb{H}^n$, centered on a codeword.}
  \label{fig:qecc2}
  \vspace{-6pt}
\end{figure}

\noindent As an equation, we can write
\[
  \text{physical space $=$ alphabet
    $\times$ code neighbourhood}.
\]
The alphabet is the set of things we want to be able to send.
The neighbourhood tells us the sorts of errors we can make without
corrupt the information we were trying to encode.
The physical message space is the product of the two.

Let's now extend the reptition code from classical alphabets to
``quantum'' alphabets. Instead of classical letters,
we can send letters from some alphabet of \emph{operators}
$\Sigma \subseteq \mathcal{A}$. For instance, we could try to send Pauli operators $\Sigma =
\{X, Z\} \subseteq \mathcal{A}_\text{Pauli}$, and choose
\[
  B \mapsto C_n[B] = B^{\otimes n} \in \mathcal{A}_\text{phys} =
  \mathcal{A}_\text{Pauli}^{\otimes n},
\]
for $B \in \Sigma$.
In general, $\mathcal{A}_\text{phys}$ will denote the \emph{physical
  algebra} we encode our operators in.
The decoding step is more involved. We choose neighbourhoods shaped like
a \emph{group}.\sidenote{For a short introduction to group theory, see
e.g. \emph{Algebra} (1998), Michael Artin.} 
For now, we'll just define a \emph{stabilizer group} as
a subset $\mathcal{S} \subseteq \mathcal{U}(\mathcal{A}^{\otimes n})$ such that
\begin{itemize}[itemsep=-2pt]
\item \emph{closure}: if $S, S' \in \mathcal{S}$ then $S \cdot S' \in
  \mathcal{S}$, or equivalently $\mathcal{S}\cdot \mathcal{S} = \mathcal{S}$;
\item \emph{inversion}: for $S \in \mathcal{S}$, then $S^*=S^{-1}
  \in\mathcal{S}$, or equivalently $\mathcal{S}^* = \mathcal{S}$;
\item \emph{commutativity}: for $S, S' \in \mathcal{S}$, $SS' = S' S$.
\end{itemize}
The name comes from the fact that it stabilizes (leaves invariant) the
shape of the neighbourhood.\newpage


Suppose that the stabilizer group is generated\marginnote{By ``linear
  generation'', we mean 
  $\langle\mathcal{X}_\mathcal{S}\rangle_{\text{phys}}^\times =
  \mathcal{S}$, where $\langle \cdot \rangle_{\text{phys}}$ is the
  subalgebra of $\mathcal{A}_\text{phys}$ generated by the argument,
  and superscript $\times$ indicates invertible elements.} by a set of operators
$\mathcal{X}_{\mathcal{S}}$.
Since $\mathcal{S}$ is commutative, each generator $S \in
\mathcal{X}_{\mathcal{S}}$ commutes with every other, so we can
simultaneously measure them. 
We want to make sure these measurements also don't disturb the
information we're trying to send, so each
$S$ must commute with every element of $\Sigma$.
The set of operators that multiplicatively
commute\marginnote{The commutant is closely related to the
  centralizer;
in fact, it generates it in the sense that
$\langle\mathcal{A}'\rangle^\times = Z(\mathcal{A})$.
}
with every element of $\mathcal{S}$
is called the \emph{centralizer}, and denoted
\begin{equation}
Z(\mathcal{S}) = \left\{A \in \mathcal{A}^{\times}_\text{phys}: [A,
  S]_\times = I \text{ for all } S\in\mathcal{S}\right\}.\label{eq:51}
\end{equation}
So, for stabilizer measurements to play nicely with encoding, we
need the centralizer to contain our alphabet, $\Sigma \subseteq Z(\mathcal{S})$.

We can expand $\Sigma$ to a group $\Sigma^\times$ by multiplying
and taking inverses. This group still lives inside the centralizer, $\Sigma^\times
\subseteq Z(\mathcal{S})$, since it was made of things that commuted
with $\mathcal{S}$.
But if $\Sigma^\times$ is strictly smaller than $Z(\mathcal{S})$,
there is arbitrage; we can either transmit more
operators (make $\Sigma$ bigger) or make more mutually compatible measurements
(make $Z(\mathcal{S})$ smaller).
So let's assume they are equal!
We let $\Pi_{\mathcal{S}}$ denote the projector onto the subspace
fixed by $\mathcal{S}$, usually called the \emph{code subspace}:
\begin{equation}
  \label{eq:stab-proj}
  \Pi_{\mathcal{S}} = \frac{1}{|\mathcal{S}|}\sum_{S\in \mathcal{S}} S.
\end{equation}
We prove this has the desired effect in Appendix
\hyperlink{app:stormer}{B}.

So, we have operators $\Sigma$ we can send and measurements
$\mathcal{X}_{\mathcal{S}}$ we can make. What errors can we correct?
We will show that the KL criterion
(\ref{eq:qec3}) is satisfied when the component Kraus
operators $E$ of an error channel $\mathcal{E}$ \emph{projectively
  commute} with each
element of $\mathcal{S}$, i.e. they commute up to a phase.
We call the set of all such elements the \emph{projective centralizer}:
\begin{equation}
  \label{eq:proj-cent}
  PZ(\mathcal{S}) = \{E \in \mathcal{A}^\times_\text{phys}: [E,
  S]_\times \in U(1)I \text{ for all } S \in \mathcal{S}\}.
\end{equation}
We summarize the phases with an $E$-dependent function\marginnote{
  \begin{center}
    \vspace{-100pt}
    \includegraphics[width=1.1\linewidth]{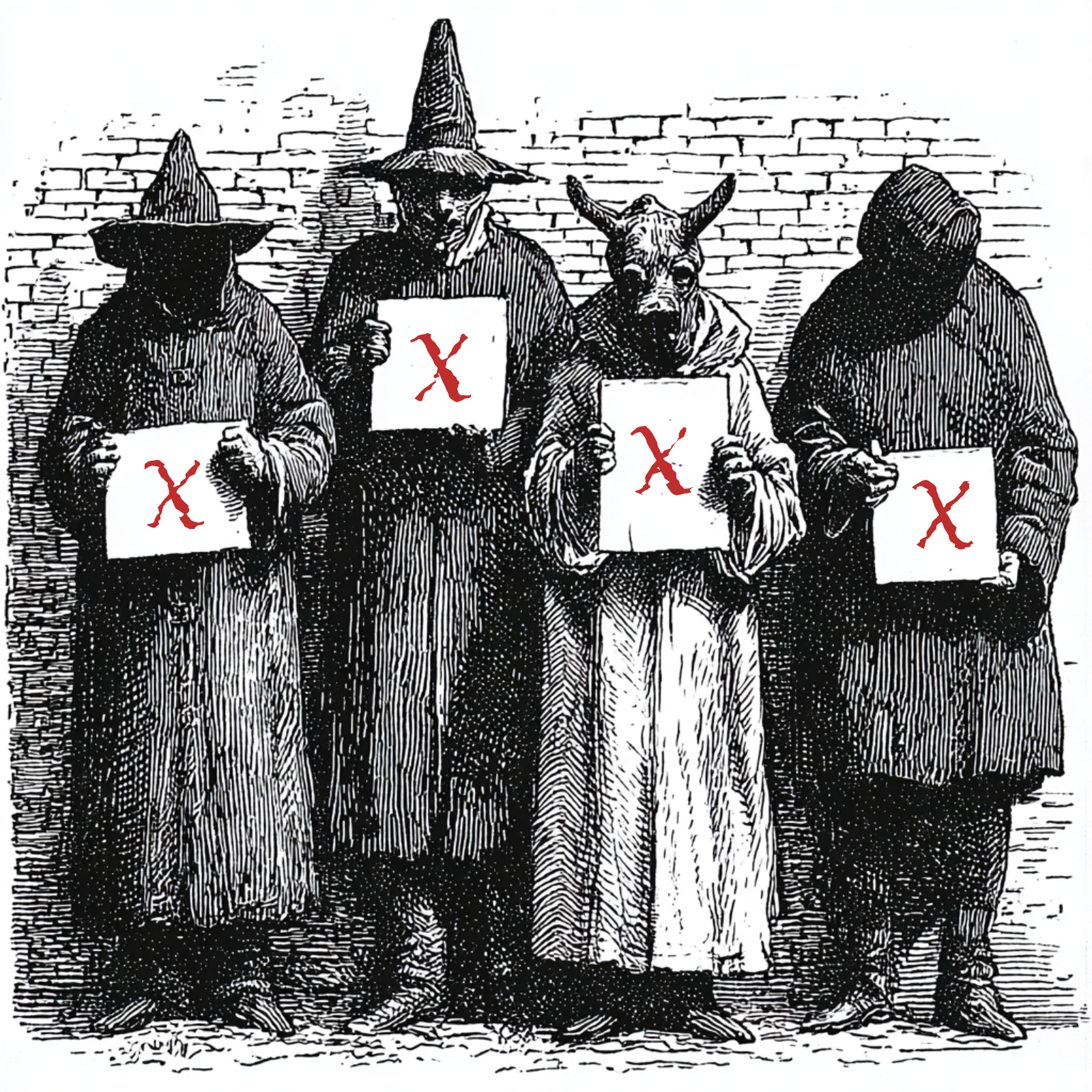}
  \end{center}  \vspace{-5pt}
  \emph{Identifying distinct characters.
  } \vspace{10pt}
}
\begin{equation}
[E,S]_\times = \chi_E(S) I, \quad \chi_E(S)\in U(1),\label{eq:stab-char}
\end{equation}
which we call a \emph{stabilizer character}. 
It follows that $\chi_E$ is multiplicative:
\begin{align}
\chi_E(SS') =[E,SS']_\times =[E,S]_\times[E,S']_\times = \chi_E(S) \chi_E(S').
\end{align}
We let $\widehat{\mathcal{S}}$ denote the set of all multiplicative
functions $\chi: \mathcal{S}\to U(1)$. This can be turned into a group
(isomorphic to $\mathcal{S}$ itself\sidenote{By Pontryagin
  duality. See e.g. \emph{Fourier Analysis on Groups} (1962), Walter
  Rudin.}) with group operation the pointwise product,
$[\chi\chi'](s) = \chi(s)\chi'(s)$.

As with the Jordan characters (\ref{eq:jordan-mult}) we encountered
earlier, multiplicativity corresponds to well-defined
measurements, in this case of the elements of $\mathcal{S}$.
The moral is that correctable errors are \emph{patterns of sharp stabilizer measurement}.
These patterns are slightly redundant, due to $\mathcal{S}$ itself.
Suppose $E' = ES$ for some $S
\in\mathcal{S}$, with associated characters $\chi_{E'}$ and
$\chi_{E}$. We find
\begin{align*}
  \chi_{E'}(S')  = [E', S']_\times 
        & = [ES, S'] \\
        & = ES'S(S'ES)^{-1} \\ &
   = \chi_E(S') S'ES(S'ES)^{-1} = \chi_E(S'),
\end{align*}
using commutativity on the middle line.
Since this holds for arbitrary $S'$, the characters are equal, $\chi = \chi'$.
These operators are in the same coset
$E\mathcal{S} $; it turns out this condition is not
only sufficient, but necessary\marginnote{This follows from the fact
  that only one character is trivial.} for equality of characters, so distinct
characters correspond to the quotient
\begin{equation}
  \label{eq:PC/S}
  \mathfrak{E} = PZ(\mathcal{S})/ \mathcal{S},
\end{equation}
where $\mathfrak{E}$ anticipates that these will serve as
our syndromes. We now have the stabilizer genuinely serving as a
neighbourhood, just like the repetition code, though there is no
reason for the associated alphabet $\mathfrak{E}$ to equal $\Sigma$,
or equivalently, for $PZ(\mathcal{S})/\mathcal{S} \cong Z(\mathcal{S})$.

Let's verify the Knill-Laflamme conditions are met.\marginnote{
  \begin{center}
    \vspace{-0pt}
    \includegraphics[width=0.45\linewidth]{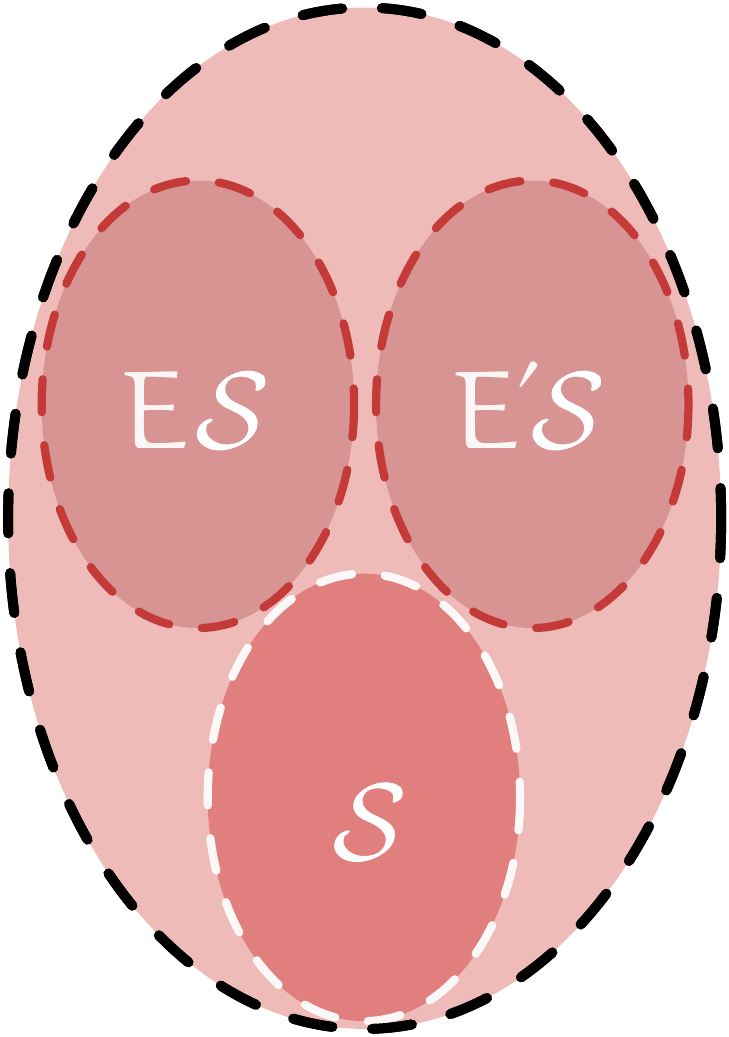}
  \end{center}  \vspace{-0pt}
  \emph{A cartoon version of (\ref{eq:PC/S}): we split the projective
    centralizer into blobs of shape $\mathcal{S}$.
  } \vspace{10pt}
}
Since characters are multiplicative and projectors idempotent, evaluating the latter on the former gives
  \begin{equation}
  \chi(\Pi_\mathcal{S}) = \chi(\Pi_\mathcal{S}^2) =
  \chi(\Pi_\mathcal{S})^2 \quad \Longrightarrow \quad
  \chi(\Pi_\mathcal{S}) = 0, 1\label{eq:charsum}
\end{equation}
for all $\chi \in  \widehat{\mathcal{S}}$.
The only way to obtain unity in (\ref{eq:charsum}) is to have constant
$\chi(S) = 1$, since otherwise we get less than $1$. But since the
only other option is $0$, the sum
over any nonconstant character vanishes!
Earlier, we considered a physical algebra of the form
\[
  \mathcal{A}_\text{phys} = \mathcal{L} \otimes \mathcal{L}' \oplus
  \mathcal{N}^\perp.
\]
In this case, our physical space takes the form
$\mathcal{A}_\text{phys} = \mathcal{L} \oplus \mathcal{L}^\perp$ for
$\mathcal{L} = Z(\mathcal{S})$,
so we need only use the projector $\mathbb{E}_\mathcal{S} =
\mathcal{C}_{\Pi_\mathcal{S}}$.
We can explicitly
  describe the remaining summands $\mathcal{L}^\perp$ using the
  \emph{Peter-Weyl decomposition}.\sidenote{``Die Vollständigkeit der
  primitiven Darstellungen einer geschlossenen kontinuierlichen
  Gruppe'' (1927), Fritz Peter and Hermann Weyl.} Projecting onto
these instead of the centralizer gives rise to \emph{Clifford codes}. 

The expected overlap of two operators $E, E'$ which projectively commute with $\mathcal{S}$ is then\marginnote{\vspace{15pt}\\ \noindent Since $\chi\chi' $ is multiplicative,
(\ref{eq:charsum}) applies and $[\overline{\chi}\chi'](\Pi_\mathcal{S}) = 1$ just
in case $\chi = \chi'$.}
\begin{align}
  \mathbb{E}_\mathcal{S}[E^* E'] & = \Pi_\mathcal{S}
                                              E^*
                                              E'\Pi_\mathcal{S}
                                              \notag \\ & =
                                           \big[\chi'\overline{\chi}\big](\Pi_\mathcal{S}) E^*E' \Pi_\mathcal{S} \notag
  \\ & =\delta_{\chi, \chi'}E^* E'
       \Pi_\mathcal{S}  \label{eq:pre-stab} \\
  & = \delta_{\chi,\chi'}\nu_E I_\mathcal{S},   \label{eq:klv-stab}
\end{align}
where (\ref{eq:klv-stab}) assumes $E, E'$ are equivalent scaled
isometries on $\mathcal{L}$ 
when they have the same character.
In particular, if distinct Kraus operators belong to different classes
in (\ref{eq:PC/S}),
then we only need to make sure they are scaled isometries to satisfy
the KL criterion (\ref{eq:qec3}).


\section{17. A game of codes}\hypertarget{sec:18}{}


%

We'll finish by looking at the simplest nontrivial examples in detail,
and throw in some diagrammatic shorthands at no extra cost.
To warm up, let's encode $m=1$ copy of the Pauli algebra in $n=2$ copies,
called the $[[n, m]] = [[2, 1]]$ code for short.
The codewords are $C_2[L] = L
\otimes L$ for $L \in \{X, Z\}$,
and $C_n$ is the \emph{repetition map}. We denote this
using a box with tensor power superscript, as in Fig. \ref{fig:rep-chan1}:

\begin{figure}[h]
  \centering
  \vspace{-5pt}
  \includegraphics[width=0.33\textwidth]{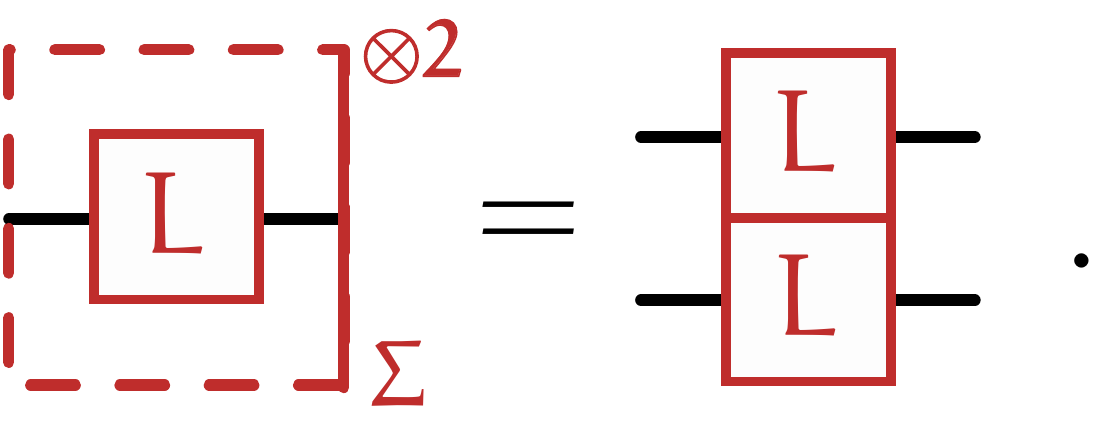}
  \caption{The twofold repetition map.}
  \label{fig:rep-chan1}
  \vspace{-5pt}
\end{figure}

\noindent The solid line on the right of the box is a ``filter'', restricting the inputs to
elements $L \in \Sigma$, with the filtering set $\Sigma$ in the
subscript.
To extend this to a channel, i.e. a \emph{linear} map on
inputs, we require that $\Sigma$ be linearly independent, so that
\emph{coherent repetition}
\begin{equation}
  C_n\big[\lambda_i L_i\big] = \lambda_i L_i^{\otimes  n}\label{eq:62}
\end{equation}
is well-defined. Note that is \emph{not} the same as the $n$-fold
tensor product of an arbitrary input, which leads to a nonlinear map
(no cloning). In the language of awds, our coherent repetition channel is

\begin{figure}[h]
  \centering
  \vspace{-5pt}
  \includegraphics[width=0.35\textwidth]{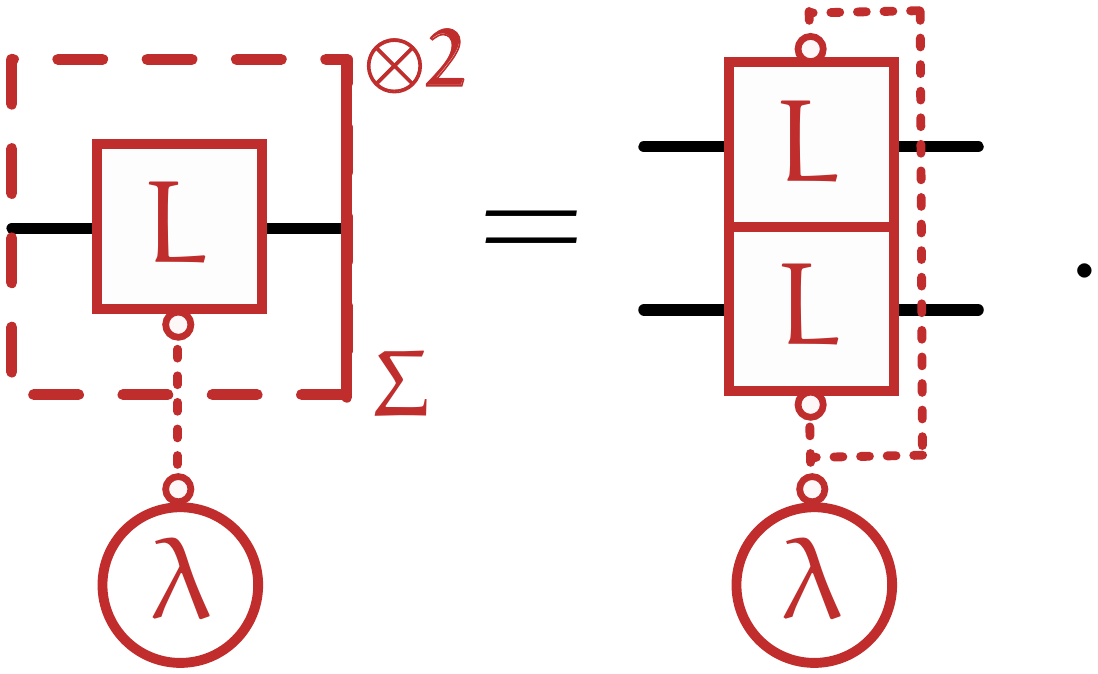}
  \caption{The two-fold coherent repetition channel.}
  \label{fig:rep-chan2}
  \vspace{-6pt}
\end{figure}

We now turn to the stabilizer $\mathcal{S} \subseteq
\mathcal{A}_\text{phys}=\mathcal{A}_\text{Pauli}^{\otimes 2}$. We need it to
(a) commute with $\Sigma$, and (b) commute with itself. If we stick to
tensor products of operators $I, X, Y, Z$, the only nontrivial combination that meet requirements (a) and (b) is $Y
\otimes Y$, since only $Y$ anticommutes with both $X$ and $Z$.
Thus, we have a lone generator $Y \otimes Y$.
We'll use awd notation for multiplicative commutators,

\begin{figure}[h]
  \centering
  \vspace{-5pt}
  \includegraphics[width=0.45\textwidth]{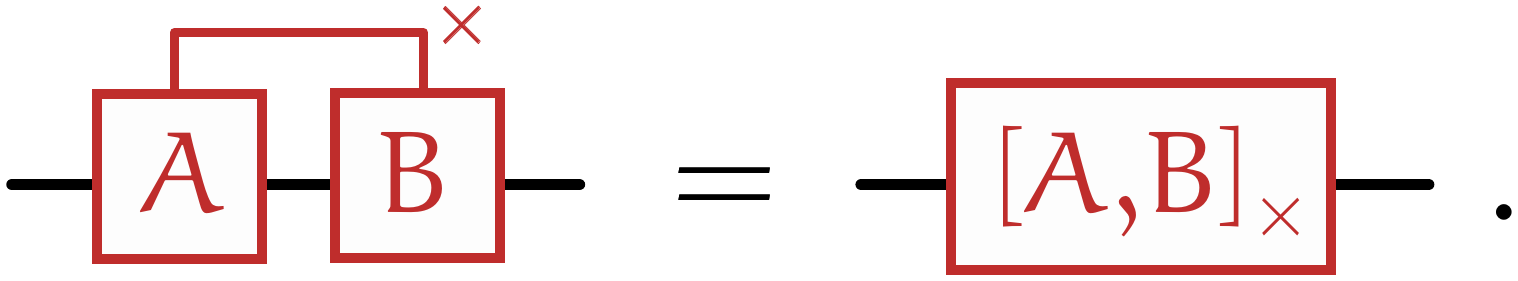}
  \label{fig:rep-chan3}
  \vspace{-8pt}
\end{figure}

\begin{figure}[h]
  \centering
  \vspace{-10pt}
  \includegraphics[width=0.45\textwidth]{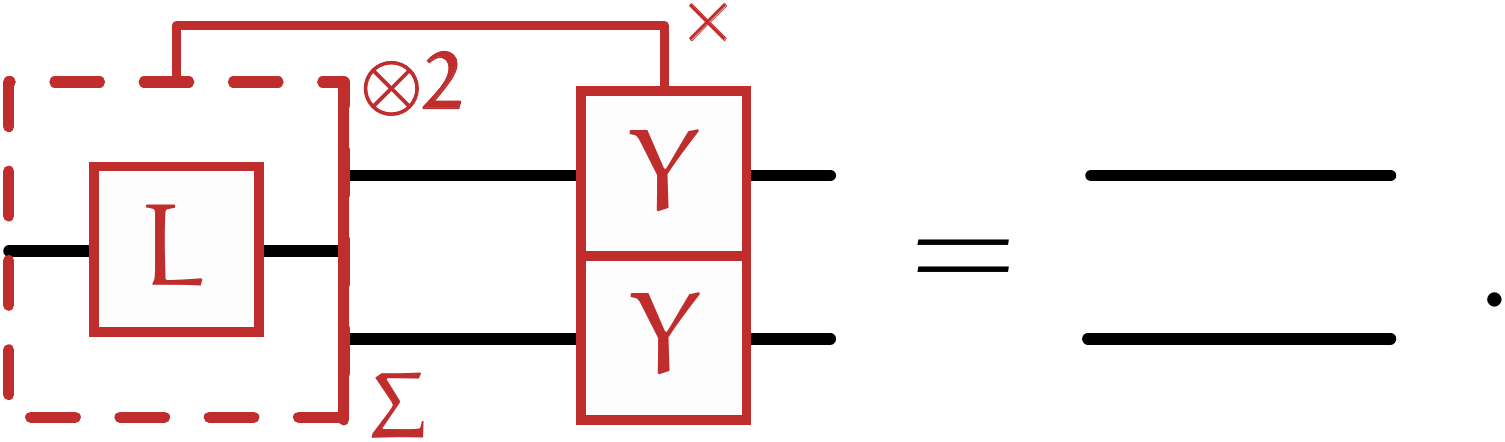}
  \caption{Codewords and stabilizer generators annihilate.}
  \label{fig:rep-chan4}
  \vspace{-10pt}
\end{figure}

\noindent
We omit the $\times$ to indicate the usual additive commutator.
The fact that codewords commute with
stabilizer elements is expressed by
Fig. \ref{fig:rep-chan4}. Borrowing some intuition from particle
physics, we can think of them as ``annihilating'' each
other.

The next step is to find the projective stabilizer, and identify
stabilizer characters according to (\ref{eq:proj-cent}).
For our single stabilizer generator $Y \otimes Y$, this is fairly
easy.
First, the elements which give phase $\chi = +1$ when commuted with
$Y \otimes Y$:
\[
\Sigma \otimes \Sigma, \quad X\Sigma \otimes X \Sigma,
\]
where\marginnote{Technically, we can view $I\otimes Y$ and $Y
  \otimes I$ as two errors, since $Y \propto XZ$. 
} a tensor product of sets indicates the set of tensor
products.
This tells us that an error on both copies ($\Sigma \otimes \Sigma$)
yields the same character as no error ($X\Sigma \otimes
X\Sigma$). Put differently, the $[[2, 1]]$ code cannot \emph{detect}
two errors.
Similarly, the elements with commutator $\chi = -1$ are
\[
  X\Sigma \otimes \Sigma, \quad \Sigma \otimes X\Sigma.
\]
Thus, the $[[2,1]]$ code can detect a single error on either copy of
the Pauli algebra, but because the errors yield the same character, we
cannot tell which copy is corrupted! So it can \emph{detect but not
  correct} a single error. Since two errors suffice to undetectably corrupt logical
data, we say this code has \emph{distance} $d=2$ between codewords, and append this to our shorthand, $[[n, m, d]]=[[2,
1, 2]]$.




This isn't the greatest code;\marginnote{
  \begin{center}
    \vspace{-80pt}
    \includegraphics[width=0.95\linewidth]{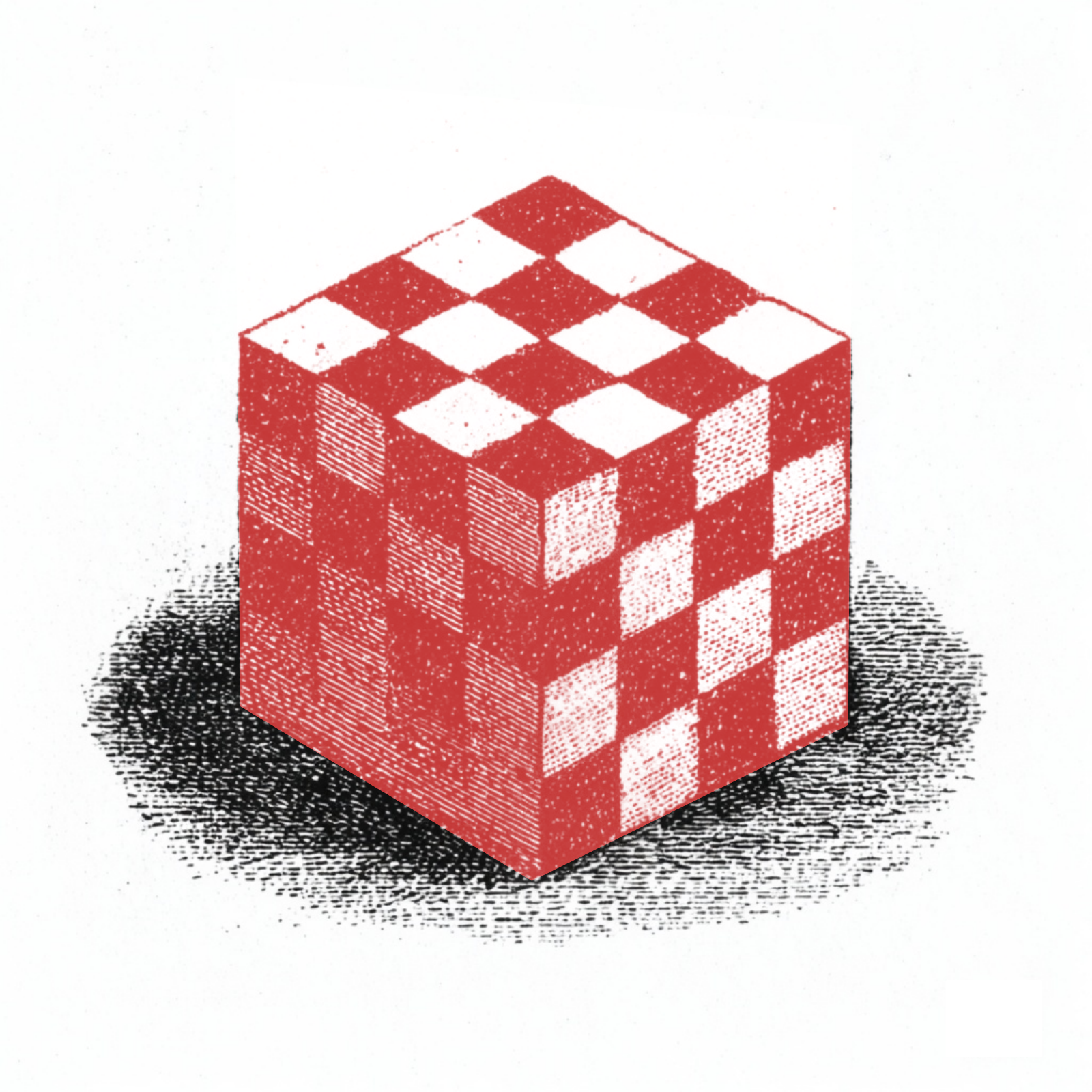}
  \end{center}  \vspace{-20pt}
  \emph{A game of codes, also called higher-dimensional checkers.} 
} error-correction is a high-stakes game of checkers against Nature,
and we need to do better! First,
it doesn't faithfully represent the Pauli algebra, since the codewords
\emph{commute} rather than \emph{anticommute}:
\[
  C_2[X]C_2[Z] = (-1)^2 C_2[Z]C_2[X] = C_2[Z]C_2[X].
\]
For this reason, it makes more sense to consider odd $n$.
The second problem is that we can't actually correct any errors!
For $n=3$, a very similar construction uses stabilizer generators 
$Y \otimes  Y \otimes I$ and $I \otimes Y \otimes Y$, and allows us to
not only detect, but correct a single $X$ (\emph{bit flip}) error or a
single $Z$ (\emph{phase}) error, but not both.

We'll set our sights a little higher, and construct an $n=5$ code which can correct both.
This is essentially possible because, when we encode $X \mapsto C_5[X]$ and $Z \mapsto
C_5[Z]$, a new option 
emerges for stabilizer generators: build them of $X$ and $Z$ rather than $Y$.
For instance, 
\[
  S_{(0)} = I \otimes X \otimes Z \otimes Z \otimes X
\]
always gets a factor of $(-1)^2 = 1$ when multiplicatively
commuted with a codeword.\marginnote{
  \begin{center}
    \vspace{-120pt}
    \includegraphics[width=0.6\linewidth]{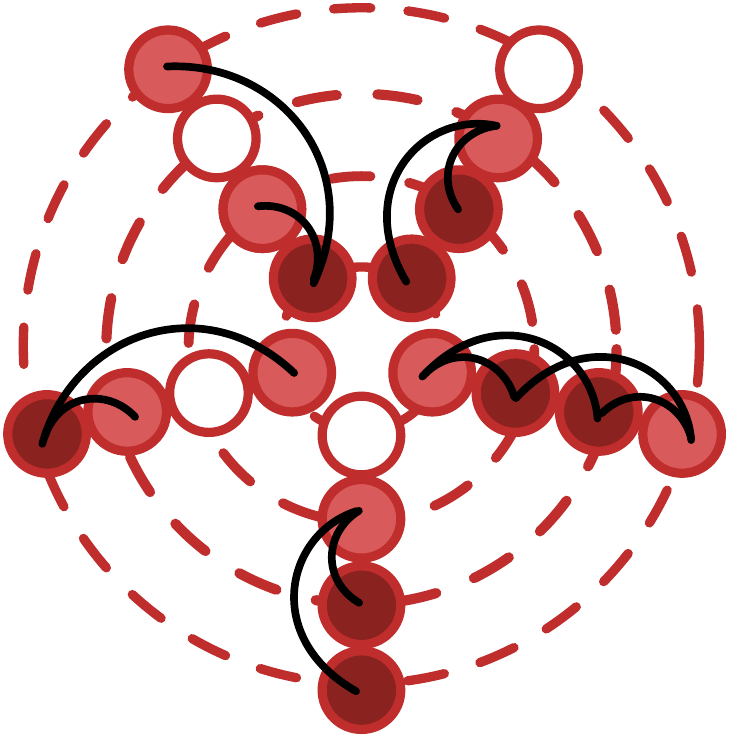}
  \end{center}  \vspace{-0pt}
  \emph{Proof by necklace that $S_{(k)}$, $k=0,1,2,3$,
    form a commutative group: any pair of circles has precisely two light/dark
    red ($XZ$) clashes, indicated by black arcs.
  } \vspace{10pt}
}
To form a commutative group, we need them to have an even number of $XZ$
overlaps; given $S_{(0)}$, only its cyclic shifts $S_k$ have this property.
More formally, for $k = 1,2, 3$, we cyclically permute forward with
the operator $T$:
\[
  S_{(k)} = T^{k} S_{(0)}, \quad T\left[\bigotimes_{\ell \in \mathfrak{L}} A^{(\ell)}\right]
  = \bigotimes_{\ell \in \mathfrak{L}} A^{(\ell + 1)},
\]
so $T$ increments each tensor label (assuming we
index cyclically). Note that $S_{(4)} = S_{(0)}S_{(1)}S_{(2)}S_{(3)}$ is automatically included.

Next, we have rustle up some errors.
We'll check that we can distinguish $X$, $Y$ and $Z$ errors on the first
Pauli algebra; the character pattern and symmetry will show that these
errors can be distinguished, hence corrected, on any copy.
Let's first set
\[
  E_{(1,i)} = \sigma_{(i)} \otimes I \otimes I \otimes I \otimes I = \sigma_{(i)}^{(1)}.
\]
It's easy to read off the associated character\marginnote{
  \begin{center}
    \vspace{-70pt}
    \includegraphics[width=0.7\linewidth]{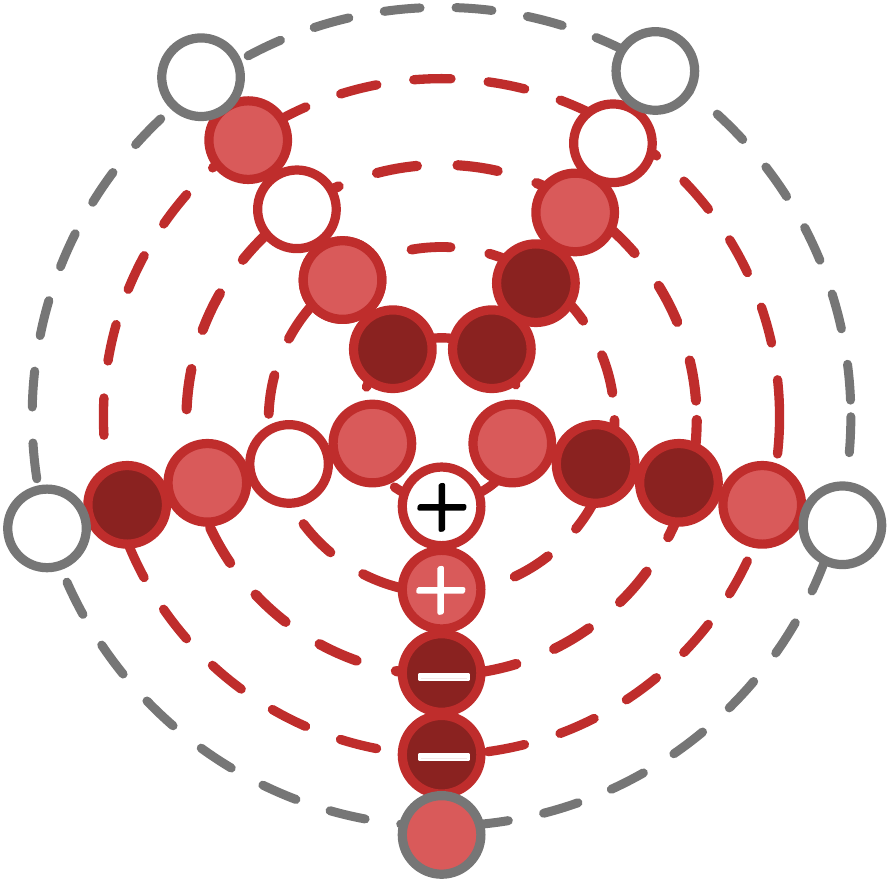}
  \end{center}  \vspace{-0pt}
  \emph{Entries of the character associated to $E_{(1,1)}$, read
    vertically down.} \vspace{10pt}
}
\[
  \chi_{(1,i)}(S_{(k)}) = 
  \begin{cases}
    +1, +1, -1, -1, +1 & i = 1 \\
    +1, -1, -1, -1, -1 & i = 2 \\
    +1, -1, +1, +1, -1 & i = 3,
  \end{cases}
\]
where we range over $k=0,1,2,3,4$.
Since $X$, $Y$ and $Z$ all have different characters, they are all
distingushable; moreover, changing the Pauli algebra on which the error is
applied simply shifts the all-$1$ column, so single Pauli errors on different
algebras can be told apart. Since we can detect $X$, $Y$ and $Z$, we
can detect \emph{any} single-Pauli error. This is called the
\emph{five-qubit stabilizer code},\sidenote{``Perfect Quantum Error Correcting Code" (1996), Ray
  Laflamme, Cesar Miquel, Juan Pablo Paz, and Wojciech Zurek;
  ``Mixed-state entanglement and quantum error correction'' (1996),
  Charles Bennett, David DiVincenzo, John Smolin, and William
  Wootters.} or the $[[5,1,3]]$ since it encodes $m=1$ copy of the
Pauli algebra into $n=5$, with a distance of $d=3$ between them. Our
checkers game just leveled up.

\section{Exit through the gift shop}\hypertarget{sec:19}{}

We hope you enjoyed your visit to the quantum quarter of
Disneyland, jumped on a few rides, and laughed at the marginalia. But
you may be looking for something useful to remember your visit by. In this final
section,
we'll preview applications that take us beyond
neatness of analogy, or stability of foundation, and into the realm of genuine
real-world use. We defer a detailed treatment to future instalments of SIQP.

\subsection*{Harmonic oscillators} Abstract wiring diagrams were initially
introduced as an alternative to
circuit diagrams. But we were secretly building a programming language
all along! \texttt{Yaw} is a high-level, functional framework quantum for
programming whose formal semantics is based on awds. This makes it
considerably more flexible than any existing quantum language, a fact we
illustrate with two examples.

First of all, recall the harmonic oscillator, with ladder operators
$a, a^*$ satisfying the \emph{canonical commutation relations}
\begin{equation}
  [a, a^*] = i.\label{eq:ccr2}
\end{equation}
This is pictured in Fig. \ref{fig:weyl-ccr} (above).
As we discussed in \S \hyperlink{sec:3}{2}, this does not lead to a
well-defined C${}^*$-algebra, so we need to employ Weyl's trick (Appendix
\hyperlink{app:functional}{A}) of replacing (\ref{eq:ccr2}) with the
exponentiated form
\begin{equation}
  \label{eq:ccr3}
  [e^{ita}, e^{is a^*}]_\times = e^{-ist} .
\end{equation}
We can easily capture this with an awd, Fig. \ref{fig:weyl-ccr}
(below):

\begin{figure}[h]
  \centering
  \vspace{-5pt}
  \includegraphics[width=0.5\textwidth]{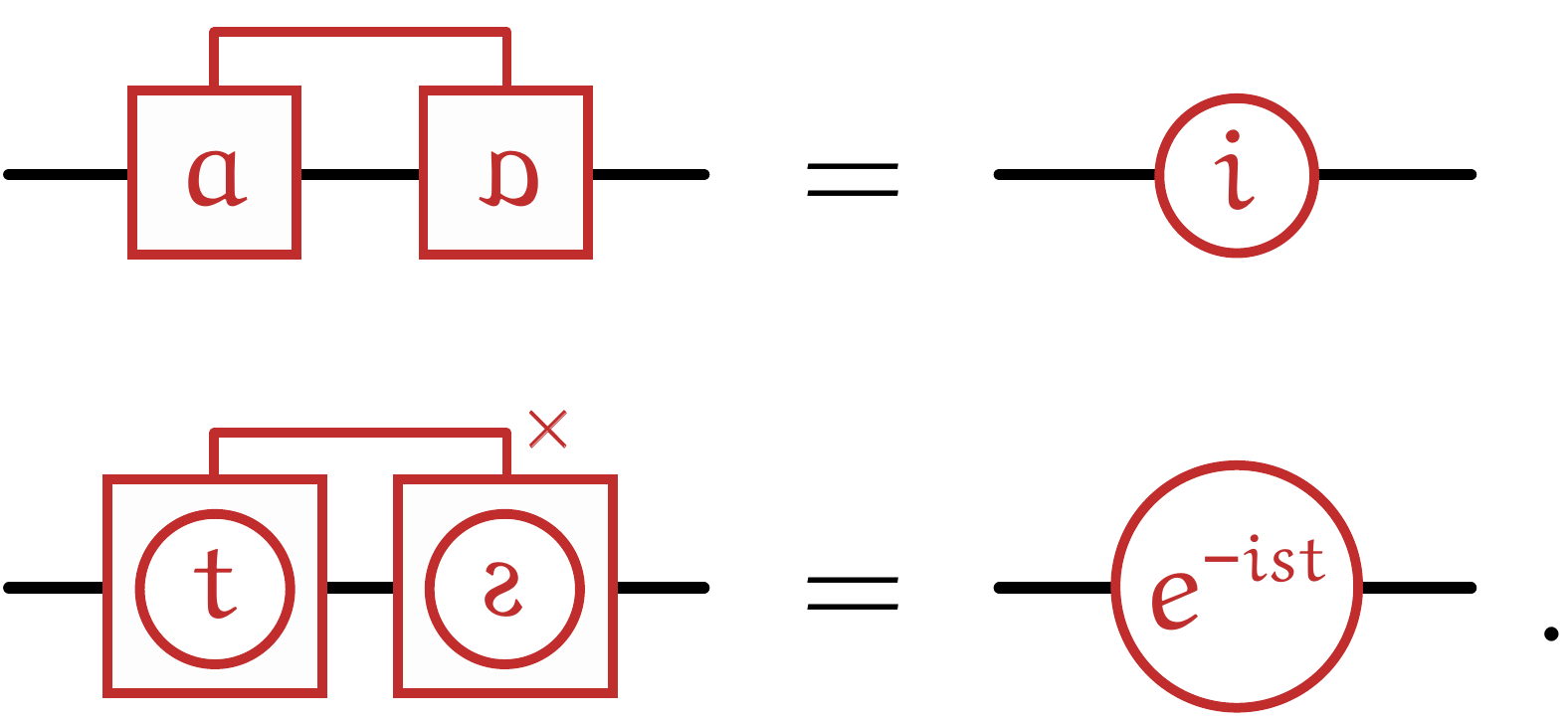}
  \caption{\textsc{Above}. The canonical commutation relation for
    ladder operators. \textsc{Below.} Weyl's exponentiated comutation relation.}
  \label{fig:weyl-ccr}
  \vspace{-5pt}
\end{figure}

\noindent
Here, a square with a circle is just an exponentiated $a$ or $a^*$
depending on the orientation of the label. These objects are
infinite-dimensional, but an advantage of the functional paradigm is \emph{lazy
  evaluation}, which lets us handle infinite data
structures.\sidenote{``A lazy evaluator'' (1976), Peter Henderson and
  James Morris;  ``Cons should not evaluate its arguments" (1976), Dan
  Friedman and David Wise.}

The fact that we can easily specify both continuous- and
discrete-variable algebras in the same computations lets us perform
``mixed-variable'' protocols, such as Brenner et al.'s remarkable algorithm\sidenote{``Factoring an integer with
three oscillators and a qubit'' (2024), Lukas Brenner, Libor Caha,
Xavier Coiteux-Roy, and Robert Koenig.} for factorizing an arbitrary
integer using three oscillators and a qubit. To the best of our knowledge,
hybrid protocols like this one cannot be implemented on any other platform.

The Weyl commutation relations (\ref{eq:ccr3}) are not only useful for
harmonic oscillators. The familiar defining relations of the Pauli
algebra can be written
\[
  X^2 = Z^2 = I, \quad [X, Z]_\times = -1,
\]
with $X$ and $Z$ playing the role of $a$ and $a^*$ respectively, and
we note that $-1$ is a square root of unity.
Replacing $2$ with $d$, and square roots with $d$th root, we get the
\emph{generalized Pauli algebra}
\begin{equation}
  \label{eq:gpa}
  X^d = Z^d = I, \quad [X, Z]_\times = e^{2\pi i/d} = \omega_d.
\end{equation}
Under the GNS construction, this is isomorphic to the algebra of
$d\times d$ complex matrices $\mathsf{M}_d(\mathbb{C})$.
We use white and red squares for $X$ and $Z$, and
a white circle for $\omega_d$, so the governing relations are:

\begin{figure}[h]
  \centering
  \vspace{-5pt}
  \includegraphics[width=0.6\textwidth]{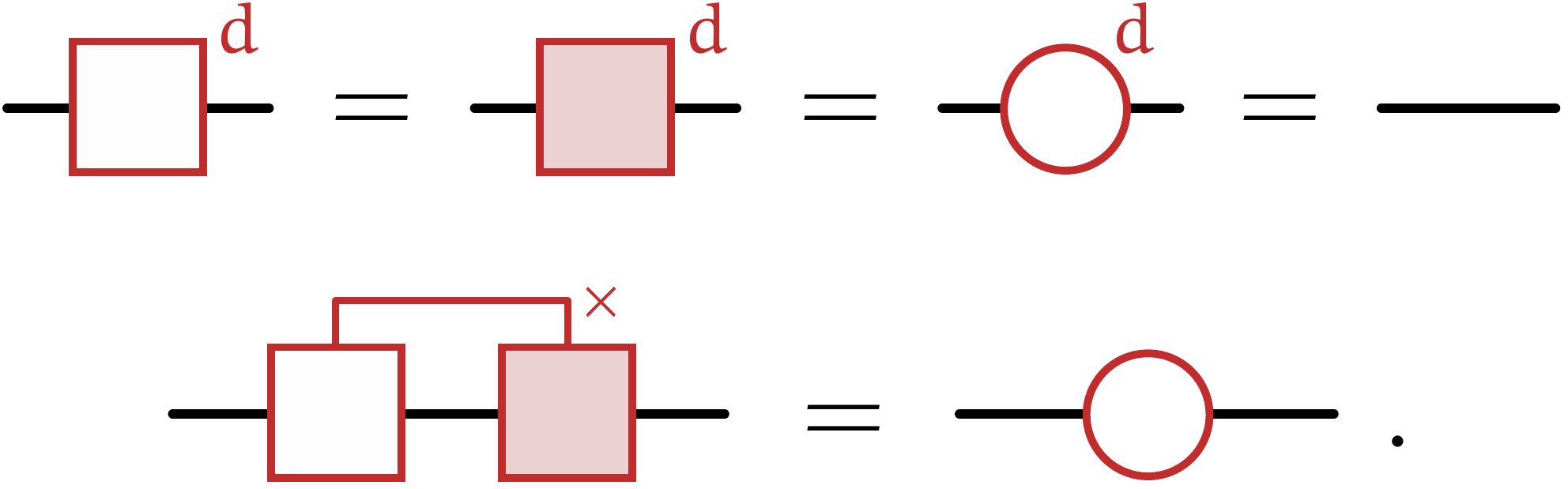}
  \caption{\textsc{Above}. The canonical commutation relation for
    ladder operators. \textsc{Below.} Weyl's exponentiated comutation relation.}
  \label{fig:gpa}
  \vspace{-5pt}
\end{figure}

\noindent This is a compact and appealing presentation of the algebra
of qudits.

\subsection*{Stabilizer circuits}

Compactness helps minimize cognitive load, but has other
advantages. Suppose we wish to multiply two Pauli strings:
\begin{equation}
  A = \bigotimes_{\ell\in\mathfrak{L}} \sigma_{(i_\ell)}, \quad B =
  \bigotimes_{\ell\in\mathfrak{L}} \sigma_{(j_\ell)}.\label{eq:pauli-word}
\end{equation}
Since each $i_\ell, j_\ell \in \{0, 1, 2, 3\}$, it takes two bits to
specify the Pauli at a given location, and for $n = |\mathfrak{L}|$,
this means $4n$ bits total to store both $A$ and $B$. Multiplication
is ``diagonal'' in the Pauli basis, with $n$ products evaluated
according to (\ref{eq:pauli-rel}). Thus, multiplying the operators
takes $\mathcal{O}(n)$ space and $\mathcal{O} (n)$ time.
This is exponentially faster than a naive attempt using matrices,
where storage and steps both scale with the dimension,
$\mathcal{O}(2^n)$.

It's possible to push this much further.
Instead of a single string of Pauli operators, we can utilize
stabilizer groups---the basis of the error-correction games we played
above---to characterize and efficiently simulate the wider class of
\emph{Clifford operations}.\marginnote{Aka the projective centralizer
  of the group of Pauli strings, aka the muggle (non-magical) operations.}
This is the celebrated
\textsc{Gottesman-Knill Theorem}; the \emph{tableau algorithm}
for simulating stabilizer circuits, devised by Aaronson and Gottesman,\sidenote{``Improved simulation of
  stabilizer circuits'' (2004), Scott Aaronson and Daniel
  Gottesman. 
} takes $\mathcal{O}(n^2)$ time and space.
The tableau method also goes beyond stabilizer circuits. For instance,
adding $d$ non-stabilizer operations, which each distribute across
at most $b$ algebras, costs
\[
  \mathcal{O}(4^{2bd}n + n^2)
\]
in both time and space complexity.
This definitely adds some overhead, but is not automatically exponential in $n$.

Some other quantum languages tack stabilizer simulation on, at least for
qubits. But tacked on things can fall off again; put differently, core
language features dictate what a language is good at. 
We have chosen to make stabilizer-based simulation idiomatic on the premise that
\emph{integrated error correction} is the medium-term fate of quantum
programming. Another crucial distinction is that our stabilizer
simulation is broader than the Pauli group.
Above, we gave an algebra-agnostic version of stabilizer codes, and in future
work provide an algorithm for classically simulating these
operations.
This fundamentally different
from existing languages!

\subsection*{Tricks with measurement}


Stabilizer operations are not only reserved for error correction and
simulation. They are also useful for near-term applications, most notably
\textsc{Classical Shadows Method}\sidenote{``Predicting many
  properties of a quantum system from very few measurements'' (2020),
  Hsin-Yuan Huang, Richard Kueng, and John Preskill.} which
which cobbles together a few 
measurements to approximate the expectations of
exponentially many more.
Say we start with an unknown state $\pi$ and we want to find the means
$\pi(A_i)$ for some collection of operators $A_i, i \in \mathfrak{I}$.
The intuition is that traditional tomography, which attempts to
reconstruct $\pi$, is overkill; 
data about
expectations should lie in a smaller space because expectations are related!

The\marginnote{
   \vspace{-53pt}
  \begin{center}
    \hspace{-
      0pt}\includegraphics[width=1.05\linewidth]{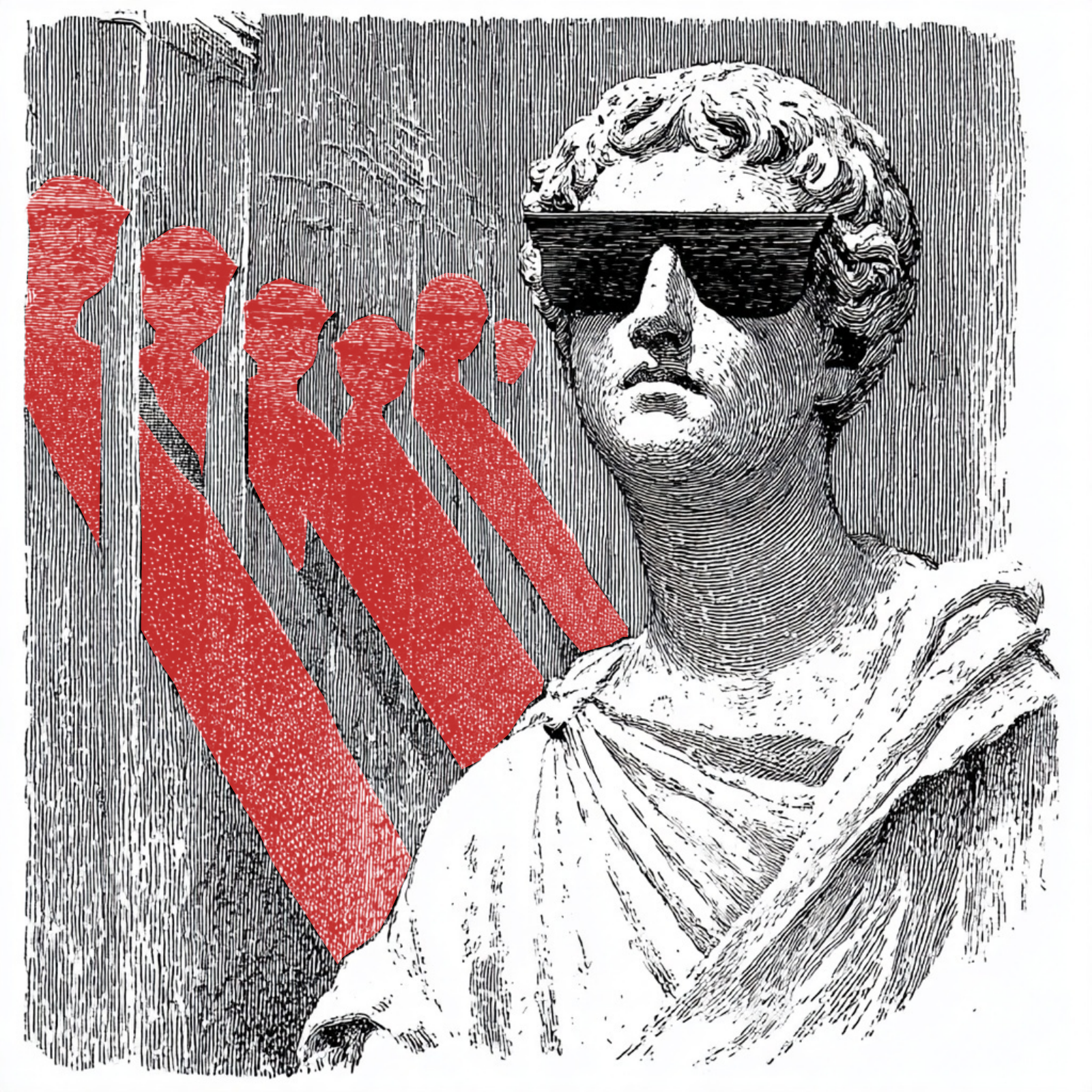}
  \end{center}  \vspace{-0pt}
  \emph{Making tomography cool again.
  } \vspace{10pt}
} technical insight is that we can fix a PVM $\Lambda$, with
a complete, orthogonal set of projectors $\Pi_\lambda$ labelled by $\lambda \in \mathfrak{F}$,
and conjugate it by a random unitary $U \sim \text{Uni}(\mathcal{G})$, where
$\mathcal{G}\subseteq \mathcal{U}(\mathcal{A})$ is a finite group of unitaries. This has the effect
of measuring our input state $\pi$ according to the Born-Lüders rules
\[
  \pi \mapsto p_{U,\lambda}^{-1} \big(C^{\Pi_{\hat{\lambda}}}\circ C^{\hat{U}}\big)[\pi]
  = \pi_{U,\lambda}, \quad p_{U,\lambda} =
  \mathcal{C}^U[\pi] (\Pi_\lambda).
\]
Since both the measurement and the conjugation are random, we can form
the expectation (using the uniform distribution over $\mathcal{G}$):
\[
  \mathcal{M}(\pi) =
  \mathbb{E}_{U,\lambda}\big[\pi_{\hat{U},\hat{\lambda}}\big] =
  \frac{1}{|\mathcal{G}|}\big(C^{\mathfrak{F}}\circ C^{\mathcal{G}}\big)[\pi].
\]
When the group is large enough that this set of random measurements is
tomographically complete, we can reverse $\mathcal{M}$. We then treat
specific random post-measurement states $\pi_{U,\lambda}$ as samples
of this process, and invert to get
\emph{classical shadows}\marginnote{``Classical'' since this is all done
  using classical post-processing.} $\mathcal{M}^{-1}(\pi_{U,\lambda})$ of $\pi$ we can use for esimating
our desired averages $\pi(A_i)$.

By batching (to deal with
outliers) and taking a median of batch-averaged shadows, we can
estimate the averages to additive precision $\epsilon$ using $\Theta(C_{\mathcal{G}}\log
|\mathfrak{I}|/\epsilon^2)$ shadows, where $C_{\mathcal{G}}$ is the
constant
\begin{align*}
  C_{\mathcal{G}} & = \max_{i\in\mathfrak{I}}\max_{\kappa\in
    S(\mathcal{A})} \mathbb{E}_{U,\lambda | \kappa} \left[
    \widetilde{\mathcal{M}}^{-1}(A_i)^2[\mathcal{C}_U
    \Pi_\lambda]\right] \\ & = \max_{i\in\mathfrak{I}} \Vert A_i\Vert^2_\text{shadow}.
\end{align*}
Here, $\kappa \in S(\mathcal{A})$ is an arbitrary state, and
$\widetilde{\mathcal{M}}^{-1}$ the result of pulling the shadow map
onto density matrices, then pushing the results back to the space of
(not necessarily positive) linear functionals on $\mathcal{A}$.
We also define the \emph{shadow metric} $\Vert
\cdot\Vert^2_\text{shadow}$, which measures the complexity of an
expectation with respect to our chosen shadows.

All of this can be made analytically explicit in the case of random
Cliffords and Paulis, highlighted for different reasons above. The algebraic framework naturally suggests several extensions:
\begin{itemize}[itemsep=-2pt]
\item the stabilizer connection should make it possible to explore
  randomization schemes related to early fault-tolerant architectures;
\item the ability to work with non-qubit analogues of the Clifford
  and Pauli groups may lead to hardware-adapted shadow ansatzae;
\item the expressivity of the language should make it easier to vary,
  extend, and modularize the protocol, e.g. swapping out classical
  shadows for estimation via regularized least squares.\sidenote{``On the connection between least squares, regularization, and
classical shadows'' (2024), Zhihui Zhu, Joseph Lukens, and Brian
Kirby.}
\end{itemize}
We leave these developments to future work.
\newpage


\addtocontents{toc}{\protect\vspace{-20pt}\protect\contentsline{part}{\textsc{\Large{appendices}}}{}{}}


\section{A. Commutative C${}^*$-algebras}\hypertarget{app:functional}{}


\subsection*{A.1. Continuous functional calculus}

Recall that $\sigma(A)$ is the spectrum of an element $A \in
\mathcal{A}$. Define a commutative C${}^*$-algebra of continuous
functions on the spectrum,
\[
  \text{C}^*(\sigma(A)) = C^0(\sigma(A)).
\]
When $A$ is normal, $AA^* = A^*A$, there is a unique assignment
\begin{equation}
  \Phi_A: \text{C}^*(\sigma(A)) \to \mathcal{A} \quad \text{with}\quad
  \Phi_A(\mathbf{1}) = I_{\mathcal{A}},
  \Phi_A(\text{id}_{\sigma(A)})=A,\label{eq:cfc}
\end{equation}
where $\mathbf{1}(\lambda) = 1$ is constant and
$\text{id}_{\sigma(A)}(\lambda)=\lambda$ is the identity.
This allows us to define $f(A) = \Phi_A(f)$, which is
called the  \emph{continuous functional calculus} for
$A$.\sidenote{For a proof, see Theorem 4.1.3 of \emph{Fundamentals of the Theory of
    Operator Algebras I} (1983), Richard Kadison and John Ringrose.} 


The continuous functional calculus lets us prove the spectral theorem
without any need for Hilbert space.
For normal $A$, suppose the spectrum consists of $K$ disjoint closed sets,
$\sigma(A) = \bigsqcup_{k\in \mathfrak{K}} \sigma_k$. Then there are associated
projectors $\Pi_{(k)}$ with the following properties:
\begin{itemize}[itemsep=-2pt]
\item each projector has spectrum $\sigma(\Pi_{(k)}) = \sigma_k$;
\item projectors commute with $A$, $\Pi_{(k)} A = A\Pi_{(k)}$;
\item projectors are orthogonal, $\Pi_{(j)}\Pi_{(k)} =
  \delta_{jk}\Pi_{(k)}$; and
\item projectors resolve the identity, $\sum_{k\in\mathfrak{K}}
  \Pi_{(k)} = I$.
\end{itemize}
The proof is straightforward. We define a \emph{characteristic
  function} $\mathbf{1}_{k}(\lambda) =
\mathbb{I}[\lambda\in\sigma_k]$, which is continuous since the inverse
image $\{1\} = \mathbf{1}_{k}^{-1}(\sigma_k)$ of closed set $\sigma_k$
is closed. Similarly, $\{0\} = \mathbf{1}_{k}^{-1}(\sigma_k^c)$ is
closed, with
\[
  \sigma_k^c = \bigsqcup_{j\neq k} \sigma_k
\]
a finite union of closed sets by assumption, hence closed.
The remaining properties follow from applying the
continuous functional calculus $\Phi_a$ to the characteristic
functions $\mathbf{1}_{k}$.
Thus, we have the usual spectral theorem for normal operators, with a
minimal amount of formal baggage.

\subsection*{A.2. Weyl commutation relations}

Another application of the continuous functional calculus is taming
wild commutation relations, such as the \emph{canonical
  commutation relation (CCR)} for self-adjoint $Q$, $P$:
\begin{equation}
  [Q, P] = i I,\label{eq:ccr1}
\end{equation}
in units where $h = 2\pi$.
This CCR cannot be satisfied by trace class operators $Q$ and $P$, since
taking the trace of the LHS gives zero (using cyclicity of trace)
while the right does not. A similar argument shows that $Q$ and $P$
are unbounded, and therefore cannot live in any C${}^*$-algebra, whose
members act as bounded linear operators on some GNS Hilbert space.
How do we accomodate them?

Although unbounded, $P$ and $Q$ are nevertheless normal, and we can
define the exponentials $q(t) = e^{iQt}$ and $p(t)=e^{iPt}$ via the
continuous functional calculus.
The \emph{Weyl CCR} is the exponentiated or braided form of (\ref{eq:ccr1}):
\begin{equation}
  \label{eq:ccr2}
  [q(t), p(s)]_\times = e^{-ist}.
\end{equation}
If the \emph{Baker-Campbell-Hausdorff (BCH)} formula holds, then
(\ref{eq:ccr1}) follows from (\ref{eq:ccr2}) by expanding in the
coincident limit $s \to t$.\sidenote{See for instance \emph{Quantum Computation and
    Quantum Information} (2000), Michael Nielsen and Isaac Chuang.}
If BCH doesn't hold, all bets are off; if it does, the question of
\emph{uniqueness} remains, i.e. whether the operators satisfying
(\ref{eq:ccr2}) are uniquely determined to be $q(t)$ and $p(s)$.
It's clear that any unitarily equivalent operators also satisfy
(\ref{eq:ccr2}),
\[
  [U^* q(t) U, U^* p(s) U]_\times = e^{-ist} U^*U = e^{-ist}.
\]The \emph{Stone-von Neumann
  theorem}\sidenote{For a (perhaps excessively) pedagogical treatment,
  we refer the reader to Theorem 5.6.36 of \emph{Fundamentals of the Theory of
    Operator Algebras II} (1997), Richard Kadison and John Ringrose.} asserts the
converse, i.e. any pair of one-parameter families $q'(t)$ and $p'(s)$
must be unitarily equivalent to $q(t)$ and $p(s)$.

\section{B. Proof details}\hypertarget{app:stormer}{}

\subsection*{B.1. Jordan characters}

Building on a result of Kadison and Singer, Størmer\sidenote{``Extensions of pure states'' (1959), Kadison and
  Singer; ``A characterization of pure states of
  \gls{cstar}s'' (1967), Erling Størmer.} proved the following lemma: if $\mathcal{D}_{\pi}'
\supseteq \mathcal{D}_\pi$ and $\pi$ is pure, then either
$\mathcal{D}_{\pi}' = \mathcal{A}_\text{sa}$ (the self-adjoint
elements of $\mathcal{A}$) or $\mathcal{D}_{\pi}'
=\mathcal{D}_{\pi}$. We won't reproduce the proof, but we do note two
corollaries: (a) pure states have maximal definite sets,
assuming that no state is definite on all $\mathcal{A}_\text{sa}$,
i.e. no one-dimensional representations; and
(b) pure states are rigid on these definite sets.
\begin{itemize}[itemsep=-2pt]
\item[(a)] If $\mathcal{D}_{\pi'} \subseteq \mathcal{D}_{\pi}$ and
  $\pi$ is pure, then $\mathcal{D}_{\pi'} = \mathcal{D}_{\pi}$,
  assuming the algebras does not possess characters, i.e. one-dimensional representations.
\item[(b)] Suppose $\pi'$ is pure and agrees with pure $\pi$ on
  $\mathcal{D}_{\pi'}$. Then for any $A \in \mathcal{D}_{\pi'}$, we have
  \[
    \pi(A^2) = \pi'(A^2) = \pi'(A)^2 = \pi(A)^2,
  \]
  and hence $A \in \mathcal{D}_{\pi'}$. Since $\mathcal{D}_{\pi}
  \supseteq \mathcal{D}_{\pi'}$, Størmer's lemma implies
  $\mathcal{D}_{\pi} = \mathcal{D}_{\pi'}$ with identical values, and
  hence the kernels agree, $\mathcal{K}_{\pi} =
  \mathcal{K}_{\pi'}$. This forces the same GNS Hilbert space up to a unitary,
  $\mathcal{K}_\pi= U^*\mathcal{K}_{\pi'} U$ and hence $[A]_\pi =
  [U^*AU]_{\pi'}$ for all $A \in \mathcal{A}$. 

  \hspace{10pt} 
  Sharpness implies that, for all $\Gamma \in
  \mathcal{D}_{\pi'}$,
  \[
    [\Gamma \,U ]_{\pi'} = [U\Gamma]_{\pi'}
  \]
  so the commutator of $U$ and $\Gamma$ is null. It follows that $U$ can be assigned a sharp
  value without disturbing the remaining measurements in
  $\mathcal{D}_{\pi'}$. This contradicts the maximality of $\mathcal{D}_{\pi'}$ unless we
  already have $U \in \mathcal{D}_{\pi'}$. Thus, we have
  \[
    [U^*AU]_{\pi'} = [U^*]_{\pi'}[A]_{\pi'}[U]_{\pi'} =
    e^{-i\theta}[A]_{\pi'}e^{i\theta} = [A]_{\pi'},
  \]
  since $[U]_{\pi'} = e^{i\theta} I$ for unitary sharp $U$.
  It follows that $[A]_{\pi'} = [A]_\pi$ for all $A\in\mathcal{A}$,
  hence $\pi' = \pi$.
\end{itemize}

\subsection*{B.2. Robertson-Schrödinger uncertainty relation}

Definite operators, the
mechanism by which we specify and control state, are the quantum
equivalent of a Boolean switch.
But unlike Boolean circuits, where we are free to simultaneously
configure every switch, making one operator definite will make others
blurry by the uncertainty principle. We can use our correlator to
place some concrete, state-dependent bounds on this blurriness.
First, note that
\[
  G_\pi(\Delta A, \Delta B) = \pi[(\Delta B)^\ast\Delta A ] =
  \pi(B^\ast A) - \pi(B^\ast)\pi(A).
\]
A little algebra gives
\begin{align}
G_\pi(\Delta A, \Delta B) - G_\pi(\Delta B^\ast, \Delta A^\ast) & =
                                                                  \pi
                                                                  (B^\star
                                                                  A) -
                                                                  \pi(AB^\star) \notag
  \\
  & =
                                                                  \pi([B^\star,
                                                                  A]) \label{eq:rs1}
  \\
  G_\pi(\Delta A, \Delta B) + G_\pi(\Delta B^\ast, \Delta A^\ast) & =
                                                                    \pi(B^\ast A) + \pi(A B^\ast) - 2 \pi(B^*)\pi(A) \notag \\
  & = \pi(\{B^\star, A\}) - 2 \pi(B^*)\pi(A). \label{eq:rs2}
\end{align}
where $[A, B] = AB - BA$ is the \emph{commutator} and $\{A, B\} = AB +
BA$ the \emph{anticommutator} of operators.

When $A$ and $B$ are self-adjoint, (\ref{eq:rs1}) and (\ref{eq:rs2})
are respectively proportional to
the imaginary and real part of $G_\pi(\Delta A, \Delta B)$, so
\begin{equation}
|G_\pi(\Delta A, \Delta B)|^2 = \frac{1}{4}|\pi([A, B])|^2 +
\left|\pi(A \circ B ) - \pi(A)\pi(B)\right|^2,\label{eq:rs3}
\end{equation}
where $2 A \circ B = \{A,B\}$ is the Jordan product.
Finally, using the Cauchy-Schwarz inequality (\ref{eq:sesq3}), we can
lower bound the product of variances:
\begin{equation}
  \label{eq:rs4}
\frac{1}{4}|\pi([A, B])|^2 +
\left|\pi(A \circ B) - \pi(A)\pi(B)\right|^2 \leq \Vert
\Delta A\Vert_\pi^2 \Vert\Delta B\Vert_\pi^2.
\end{equation}
In words, the product of variances is bounded below by the squared
commutator plus squared gap between regular and Jordan product.

\subsection*{B.3. Code subspace projection}

We claimed above that the operator (\ref{eq:stab-proj}), given by
\[
  \Pi_\mathcal{S} = \frac{1}{|\mathcal{S}|} \sum_{S\in \mathcal{S}}S,
\]
projects onto the code subspace
\[
  \mathcal{H}_\mathcal{S} = \{\psi \in \mathcal{H}: S \psi = \psi
  \text{ for all } S \in \mathcal{S}\},
\]
where we work in a fixed representation $\mathcal{H}$.
First, we show this is
a projector. It is Hermitian since $\mathcal{S}$ is closed under
adjoints, $S \in \mathcal{S}$ implies $S^* \in \mathcal{S}$:
\[
  \Pi_\mathcal{S}^* = \frac{1}{|\mathcal{S}|} \sum_{S \in \mathcal{S}}
  S^*= \frac{1}{|\mathcal{S}|} \sum_{S \in \mathcal{S}}
  S = \Pi_\mathcal{S}.
\]
Similarly, it is a projector since $S\cdot\mathcal{S} = \mathcal{S}$,
i.e. multiplication by a fixed element is a bijection on $\mathcal{S}$:
\[
  \Pi_\mathcal{S}^2 = \frac{1}{|\mathcal{S}|^2} \sum_{S\in
    \mathcal{S}}\sum_{S'\in S\cdot\mathcal{S}} S' =\frac{1}{|\mathcal{S}|^2}\sum_{S\in
    \mathcal{S}}\sum_{S'\in \mathcal{S}} S' = \frac{1}{|\mathcal{S}|}\sum_{S'\in \mathcal{S}} S' =\Pi_\mathcal{S}.
\]
Thus, $\Pi_\mathcal{S}$ is indeed a projector.

To verify the domain of projection, first note that by the same logic
as above,
\[
  S \Pi_\mathcal{S} = \Pi_\mathcal{S} = \Pi_\mathcal{S} S,
\]
and hence $\Pi_\mathcal{S}$ projects into the set of fixed points of
$\mathcal{S}$.
Finally, we confirm that every element of the code subspace
$\psi\in\mathcal{H}_\mathcal{S}$ is fixed:
\[
  \Pi_\mathcal{S} \psi = \frac{1}{|\mathcal{S}|}\sum_{S\in\mathcal{S}}
  S \psi = \frac{1}{|\mathcal{S}|}\sum_{S\in\mathcal{S}}\psi = \psi.
\]
Thus, $\Pi_\mathcal{S}$ is a projector with range
$\mathcal{H}_\mathcal{S}$.

  \vspace{5pt}
 
\par\noindent\rule{0.7\textwidth}{0.4pt}

\vspace{6pt}
  
\begin{flushleft}
  \textsc{acknowledgments}
\end{flushleft}

  \noindent Thanks to those who have helped me refine my
  ideas over the course of the project, including Shovon Biswas, Leon
  Di Stefano,
  Stepan Fomichev, Achim Kempf,
  Filippo Miatto, and Petar
  Simidzija.
  I'm especially
grateful to Jon Male, Martine Wakeham, and Clara Weill, whose
emotional and financial support made this project
possible.

\newpage



\section[References]{}\hypertarget{References}{}

\nocite{*}

  \bibliography{siqp}
  \bibliographystyle{acm}

  \vspace{5pt}
 
\par\noindent\rule{0.7\textwidth}{0.4pt}

\vspace{6pt}
  
\begin{flushleft}
  \textsc{colophon}
\end{flushleft}

\noindent  This document is typeset using the
  \href{https://tufte-latex.github.io/tufte-latex/}{Tufte-\LaTeX}
  document class, with \emph{Palatino} as the body font,
  \href{https://fonts.google.com/specimen/IBM+Plex+Mono}{\texttt{IBM Plex
      Mono}} for teletype, and $AMS\, Euler$ for math.
  The primary visual inspirations were the
  \href{https://en.wikipedia.org/wiki/Life_Nature_Library}{Life Nature
    Library}, the books of
  \href{https://www.edwardtufte.com/online-course/?gad_source=1&gclid=EAIaIQobChMIvp714YvOjAMVMQ6tBh3SuQNSEAAYASAAEgJhM_D_BwE}{Edward
    Tufte}, and pre-\LaTeX \hspace{1pt}
  mathematics textbooks.
  Illustrations were created with a combination of Midjourney and Inkscape.
  Finally, this is distributed under a
  \href{https://creativecommons.org/licenses/by-nc-nd/4.0/deed.en}{CC
    BY-NC-ND} license; feel free to redistribute in its current form,
  but please ask before excerpting or modification.

\end{document}